\documentclass[11pt,a4paper]{scrartcl}

\usepackage[bf]{caption}
\usepackage[utf8]{inputenc} 
\usepackage[T1]{fontenc} 
\usepackage[UKenglish]{babel} 
\usepackage[section]{placeins} 
\usepackage{a4, 
            amsfonts, 
            amsmath, 
            amssymb, 
            amsthm, 
            appendix, 
            arydshln, 
            booktabs, 
            calligra, 
            etex, 
            etoolbox, 
            fancyvrb, 
            geometry, 
            hyperref, 
            lipsum, 
            listings, 
            longtable, 
            lscape, 
            mathrsfs, 
            pbox, 
            scrlayer-scrpage, 
            subfig, 
            tikz, 
            cite,
            todonotes, 
            xy 
}

\usetikzlibrary{matrix,arrows,automata,calc,chains,intersections,decorations.markings,shapes.geometric,positioning,plotmarks,scopes,patterns,fit}
\xyoption{all}

\definecolor{FireBrick}{rgb}{0.5812,0.0074,0.0083}
\definecolor{RoyalBlue}{rgb}{0.0236,0.0894,0.6179}
\definecolor{RoyalGreen}{rgb}{0.0236,0.6179,0.0894}
\definecolor{RoyalRed}{rgb}{0.6179,0.0236,0.0894}
\definecolor{LightBlue}{rgb}{0.8544,0.9511,1.0000}
\definecolor{Black}{rgb}{0.0,0.0,0.0}
\definecolor{linkColor}{rgb}{0.0,0.0,0.554}
\definecolor{citeColor}{rgb}{0.0,0.0,0.554}
\definecolor{fileColor}{rgb}{0.0,0.0,0.554}
\definecolor{urlColor}{rgb}{0.0,0.0,0.554}
\definecolor{promptColor}{rgb}{0.0,0.0,0.589}
\definecolor{brkpromptColor}{rgb}{0.589,0.0,0.0}
\definecolor{gapinputColor}{rgb}{0.589,0.0,0.0}
\definecolor{gapoutputColor}{rgb}{0.0,0.0,0.0}
\definecolor{FuncColor}{rgb}{0.0,0.0,0.0}
\definecolor{Chapter}{rgb}{0.0,0.0,0.0}
\definecolor{DarkOlive}{rgb}{0.1047,0.2412,0.0064}
\definecolor{darkgreen}{rgb}{0.05,0.6,0.1}
\definecolor{LightCyan}{rgb}{0.88,1,1}

\newcommand{\bfR}{\mathbf{R}}

\renewcommand{\tilde}{\widetilde}
\newcommand{\vect}[1]{\boldsymbol{#1}}
\newcommand{\Mint}{\int \limits}

\renewcommand{\tilde}{\widetilde}
\renewcommand{\hat}{\widehat}
\renewcommand{\bar}{\overline}

\newcommand{\ie}{i.e.\ }
\newcommand{\eg}{e.g.\ }
\newcommand{\fp}{f.p.\ }
\newcommand{\cf}{c.f.\ }

\renewcommand{\bf}[1]{\textbf{#1}}

\renewcommand{\rm}[1]{\textrm{#1}}

\renewcommand{\[}{\begin{equation}}
\renewcommand{\]}{\end{equation}}

\newcommand{\be}{\begin{equation}}
\newcommand{\ee}{\end{equation}}

\newcommand{\beq}{\begin{equation}}
\newcommand{\eeq}{\end{equation}}

\newcommand{\bal}{\begin{aligned}}
\newcommand{\eal}{\end{aligned}}

\newcommand{\bea}{\begin{eqnarray}}  
\newcommand{\eea}{\end{eqnarray}}

\newcommand{\ebox}[1]{
\begin{center}
\fbox{\begin{minipage}[c]{0.9 \textwidth} #1 \end{minipage}}
\end{center}
}

\DeclareMathOperator{\Hom}{\mathscr{H}\text{\kern -3pt {\calligra\large om}}\,}

\makeatletter
\def\enumfix{%
\if@inlabel
 \noindent \par\nobreak\vskip-\topsep\hrule\@height\z@
\fi}
\let\olditemize\itemize
\def\itemize{\enumfix\olditemize}
\makeatother

\makeatletter
\def\enumfix{%
\if@inlabel
 \noindent \par\nobreak\vskip-\topsep\hrule\@height\z@
\fi}
\let\oldenumerate\enumerate
\def\enumerate{\enumfix\oldenumerate}
\makeatother

\newtheoremstyle{break}  
  {\topsep}   
  {\topsep}   
  {}  
  {0pt}       
  {\bfseries} 
  {:}         
  {\newline}  
  {}          

\theoremstyle{break}

\makeatletter
\let\@addpunct\@gobble
\makeatother

\lstnewenvironment{mat}
{\lstset{language=mathematica,mathescape,columns=flexible}}
{}

\sbox0{\small\ttfamily A}
\edef\mybasewidth{\the\wd0 }
\lstnewenvironment{gap}
{\lstset{language=GAP,
	frame=single,
	basicstyle=\ttfamily\small,
	numbers=left,
	numberstyle=\small,
	keywordstyle=\color{red},
	stringstyle=\color{blue},
	commentstyle=\color{green!70!black},
	columns=fullflexible,
	basewidth=\mybasewidth,
	keepspaces=true,
	stepnumber=1,
	showstringspaces=false,
	breaklines=true}}
{}

\def\hybrid{\topmargin -20pt    \oddsidemargin 0pt
        \headheight 15.2pt \headsep 0pt
        \textwidth 6.25in       
        \textheight 9 in       
        \marginparwidth .875in
        \parskip 5pt plus 1pt 
        \jot = 1.5ex
   }
\hybrid

\numberwithin{equation}{section}
\numberwithin{table}{section}

\def\sectionautorefname{Section}
\def\subsectionautorefname{Section}

\addto\extrasUKenglish{%
  \renewcommand{\sectionautorefname}{Section}%
}

\makeatletter
\patchcmd{\hyper@makecurrent}{%
    \ifx\Hy@param\Hy@chapterstring
        \let\Hy@param\Hy@chapapp
    \fi
}{%
    \iftoggle{inappendix}{
        \@checkappendixparam{chapter}%
        \@checkappendixparam{section}%
        \@checkappendixparam{subsection}%
        \@checkappendixparam{subsubsection}%
        \@checkappendixparam{paragraph}%
        \@checkappendixparam{subparagraph}%
    }{}%
}{}{\errmessage{failed to patch}}

\newcommand*{\@checkappendixparam}[1]{%
    \def\@checkappendixparamtmp{#1}%
    \ifx\Hy@param\@checkappendixparamtmp
        \let\Hy@param\Hy@appendixstring
    \fi
}
\makeatletter

\newtoggle{inappendix}
\togglefalse{inappendix}

\apptocmd{\appendix}{\toggletrue{inappendix}}{}{\errmessage{failed to patch}}
\apptocmd{\subappendices}{\toggletrue{inappendix}}{}{\errmessage{failed to patch}}

\begin{document}

\let\subsectionautorefname\sectionautorefname
\let\subsubsectionautorefname\sectionautorefname

\baselineskip=14pt
\parskip 5pt plus 1pt

\vspace*{-1.5cm}
\begin{flushright}    
  {\small

  }
\end{flushright}

\vspace{2cm}
\begin{center}        
  {\LARGE Gauge Backgrounds and Zero-Mode Counting in F-Theory}
\end{center}

\vspace{0.75cm}
\begin{center}        
Martin Bies\textsuperscript{1}, Christoph Mayrhofer\textsuperscript{2}, and Timo Weigand\textsuperscript{1,3}
\end{center}

\vspace{0.15cm}
\begin{center}        
  \emph{\textsuperscript{1} Institut f\"ur Theoretische Physik, Ruprecht-Karls-Universit\"at, \\
             Philosophenweg 19, 69120 
             Heidelberg, Germany}
             \\[0.15cm]
  \emph{\textsuperscript{2} Arnold Sommerfeld Center for Theoretical Physics, \\
             Theresienstra{\ss}e 37, 80333 M\"unchen, Germany}
             \\[0.15cm]
  \emph{\textsuperscript{3} CERN, Theory Division, \\
             CH-1211 Geneva 23, Switzerland}
             \\[0.15cm]
   
 \end{center}

\vspace{2cm}

\begin{abstract}
Computing the exact spectrum of charged massless matter is a crucial step towards understanding the effective field theory describing F-theory vacua in four dimensions. In this work we further develop a coherent framework to determine the charged massless matter in F-theory compactified on elliptic fourfolds, and demonstrate its application in a concrete example. The gauge background is represented, via duality with M-theory, by algebraic cycles modulo rational equivalence. Intersection theory within the Chow ring allows us to extract coherent sheaves on the base of the elliptic fibration whose cohomology groups encode the charged zero-mode spectrum. 
The dimensions of these cohomology groups are computed with the help of modern techniques from algebraic geometry, which we implement in the software \texttt{gap}. We exemplify this approach in models with an Abelian and non-Abelian gauge group and observe jumps in the exact massless spectrum as the complex structure moduli are varied. An extended mathematical appendix gives a self-contained introduction to the algebro-geometric concepts underlying our framework.


\end{abstract}

\thispagestyle{empty}
\clearpage


\newpage
\setcounter{tocdepth}{2}
\tableofcontents


\section{Introduction} \label{sec:Intro}

String theory encodes the consistent coupling of gauge dynamics to gravity like no other framework for quantum gravity available to date.
In compactifications of string theory to lower dimensions, a significant portion of this information is encapsulated in the geometry of the compactification space. 
A particularly coherent approach to studying the ensuing relations between geometry, gauge theory and gravity has emerged in the context of F-theory \cite{Vafa:1996xn,oai:arXiv.org:hep-th/9602114,oai:arXiv.org:hep-th/9603161}. In this spirit F-theory compactifications to six dimensions have been under close scrutiny over the past years with the aim of establishing an ever more accurate dictionary between the properties of the effective field theory and the geometry and topology of elliptically fibred Calabi-Yau three-folds. This has culminated so far in the classification of end points of Higgs branches in six-dimensional $(1,0)$ theories \cite{Morrison:2012np} as well as of the possible $(1,0)$ superconformal field theories with a tensor branch \cite{Heckman:2013pva,Heckman:2015bfa} arising from F-theory.

When we try to extend the programme of understanding the effective field theories of F-theory compactifications to lower dimensions, more intricate structures are encountered  which play no role in six-dimensional F-theory vacua. F-theory compactifcations to four dimensions are clearly motivated not only by their possible connections to particle physics \cite{Donagi:2008ca,Beasley:2008dc,Weigand:2010wm,Maharana:2012tu}, but also because they may open yet another door towards understanding the structure of four-dimensional ${\cal N}=1$ supersymmetric quantum field theories. F-theory compactifications to two dimensions \cite{Schafer-Nameki:2016cfr,Apruzzi:2016iac} give a framework for studying new examples of chiral $(0,2)$ quantum field theories and SCFTs.

The perhaps most important difference in compactifications to four and two dimensions compared to their six-dimensional cousins is the appearance of non-trivial gauge backgrounds, which are of the utmost relevance for the very definition of the string vacuum. First, such backgrounds can generate potential terms in the effective action. The induced   D-terms and F-terms are moduli dependent and play an essential role in stabilising the latter. Second, and not unrelatedly, the spectrum of charged massless matter fields depends on the gauge background. The programme of understanding F-theory compactifications to four and two dimensions therefore hinges on our ability to extract this information from a given flux background. The present article reports on what we believe is important progress in this direction.

The first step is to represent the gauge background in a globally defined F-theory vacuum in a computationally accessible way. For definiteness, let us from now on focus on F-theory compactified to four dimensions on an elliptically fibred Calabi-Yau 4-fold $\hat{Y_4}$. By duality with M-theory massless matter states arise from the excitations of M2-branes wrapping vanishing cycles in the elliptic fibre. Since the M2-branes couple to the M-theory 3-form potential $C_3$, our task is to represent the gauge data encoded in this anti-symmetric gauge field. Mathematically, such gauge backgrounds are encapsulated in the Deligne cohomology group $H^4_D( \hat{Y_4}, \mathbb{Z}(2))$, whose construction is reviewed \eg in \cite{EsnaultDeligne}.\footnote{Deligne cohomology has been explored in M-theory and F-theory in \cite{Diaconescu:2003bm,Freed:2004yc,oai:arXiv.org:hep-th/0409158} and, respectively, \cite{Curio:1998bva,Donagi:1998vw,Donagi:2011jy, Clingher:2012rg,Anderson:2013rka,Bies:2014sra}. An equivalent formulation can be given in terms of the theory of Cheeger-Simons differential characters \cite{Cheeger-Simons} as investigated in the context of F/M-theory \eg in \cite{oai:arXiv.org:1203.6662}.} The Deligne cohomology contains both information about the gauge flux, \ie the background value of the M-theory 4-form field strength $G_4$, and about the flat, but topologically non-trivial configurations of the gauge potential. In \cite{Bies:2014sra} it has been described how this data is most conveniently represented by equivalence classes of algebraic 2-cycles modulo rational equivalence, \ie by elements of the Chow group $\mathrm{CH}^2(\hat{Y_4} )$. This construction rests on the existence (but fortunately not the details) of a refined cycle map, known in the mathematics literature (\eg \cite{EsnaultDeligne}) to be a ring homomorphism from the Chow group to the Deligne cohomology group which is surjective over $\mathbb Q$ if the Hodge conjecture holds. The advantage of this approach is that, unlike   $H^4_D(\hat{Y_4},\mathbb{Z}(2))$, the group of Chow classes is accessible very directly and in a constructive, geometric way suitable for explicit computations.\footnote{This comes at the precise of introducing some redundancy in the description of gauge backgrounds via Chow groups as the refined cycle map is not injective. This will not pose any problems in the applications of this article.} 

The second step is then to extract from this geometric data the gauge bundle to which the zero-modes of wrapped M2-branes couple \cite{Bies:2014sra}.
The zero-modes localised on intersection curves of two 7-branes  on the base of $\hat{Y_4}$ are known to transform as cohomology classes of certain gauge bundles twisted with the spin bundle of the matter curve. This follows already from the local approach to F-theory by studying the topologically half-twisted field theory on the worldvolume of the 7-branes \cite{Donagi:2008ca,Beasley:2008dc}. The zero-modes in the topologically twisted field theory are to be identified with the zero-modes arising by quantisation of the moduli space of wrapped M2-branes in the spirit of \cite{Witten:1996qb}. The 1-form gauge potential to which these excitations couple is obtained by integrating $C_3$ over the cycle wrapped by the M2-brane in the fibre. Mathematically, this operation has a clear and well-defined meaning in the language of the intersection product within the Chow ring \cite{Bies:2014sra}, as we review in \autoref{subsec:ReviewChowGroupsAndG4Fluxes}. Having established the gauge bundle whose cohomology groups count the massless matter states using this machinery, the third and final step consists in evaluating these cohomology groups. 

In this work we make substantial progress along all three of these steps. In \autoref{sec:SystGaugeBack} we systematically explore the gauge backgrounds underlying so-called vertical gauge fluxes in F-theory. The idea is that each matter surface defines by itself a Chow class and hence a gauge background. If this gauge background is to preserve the non-Abelian gauge group in the F-theory limit extra modification terms have to added. The resulting Chow class defines a \emph{matter surface flux}.\footnote{More appropriately, the term matter surface \emph{flux} should be reserved for the cohomology class in $H^{2,2}(\hat{Y_4})$ associated with this Chow class in $\mathrm{CH}^2(\hat{Y_4})$ as this is what represents the flux $G_4$, while the Chow class in general encodes much more information about the gauge background than merely the curvature.} The relation between matter surfaces and gauge backgrounds has first been observed  in \cite{Marsano:2011hv} and further described in \cite{Borchmann:2013hta}. If the homology class of the matter surface is vertical, then the associated flux can alternatively be described as a linear combination of a basis of $H^{2,2}_\mathrm{vert}(\hat{Y_4})$, as investigated systematically in \cite{Braun:2011zm,Marsano:2011hv,Krause:2011xj,oai:arXiv.org:1111.1232,oai:arXiv.org:1202.3138,Cvetic:2013uta,Borchmann:2013hta,Bizet:2014uua,Braun:2014xka,Cvetic:2015txa,Lin:2015qsa,Jockers:2016bwi,Lin:2016vus}. 

Equipped with this representation of gauge background data, we systematically develop the intersection theoretic operations \cite{Bies:2014sra} which allow us to extract the relevant gauge bundles on the matter curves. We observe that, as a consequence of non-trivial relations among Chow classes, the intersections of interest can be chosen to be transverse. The Chow relations in question are in fact deeply related to the absence of gauge anomalies in F-theory and are hence of interest by themselves. We will develop this interesting point further in \cite{Bies:2017-2}. As it turns out, the relevant gauge bundle on the matter curve can in general not be obtained as the pullback of a line defined on an ambient space of the curve such as the 7-brane divisor or the base of the fibration. This means that its pushforward to the base merely defines a (proper) coherent sheaf. 

The remaining task is hence to develop the suitable machinery which allows us to compute the cohomology groups of such coherent sheaves and thereby the massless matter content of an F-theory compactification. A general framework addressing precisely this point has evolved in computational algebraic geometry \cite{2010arXiv1003.1943B, 2012arXiv1202.3337B, 2012arXiv1210.1425B, 2012arXiv1212.4068B, 2014arXiv1409.6100B, BL_GabrielMorphisms}. The idea is very simple: The coherent sheaf in question is defined in terms of the (Chow) class of certain points on $B_3$. It is precisely these point classes which our intersection theoretic operations provide. The point classes are given very explicitly in terms of the vanishing locus of a set of functions. This defines an ideal within the coordinate ring of the space. In algebraic geometry, such ideals can be translated into sheaves, more precisely into their associated ideal sheaves. This data is represented with the help of an object known as an \fp module, which, when the dust has settled, is nothing but a matrix encoding the relations between the functions generating the vanishing ideal. Finally, the cohomology groups we are after translate into suitable extension modules associated with this module. We describe these steps in \autoref{sec:ComputingTheSpectra}. In order to make this article self-contained and more accessible to non-experts we are including an extended \autoref{sec:ATasteOfAlgebraAndAlgebraicGeometry} which provides the required mathematical background and fills in the more technical details.

The computation of the extension groups can be performed algorithmically with the help of the computer programme \texttt{gap} \cite{GAP4} and is phrased in the language of \emph{categorical programming} of \texttt{CAP} \cite{CAP, PosurDoktor, GutscheDoktor}. We exemplify the use of this technology by computing the massless spectrum in a family of four-dimensional F-theory compactifications with gauge group $SU(5) \times U(1)$ over base $\mathbb P^3$. In particular we observe an explicit dependence of the massless matter spectrum on the complex structure moduli defining the elliptic fibration. 

This article is organized as follows: We begin by reviewing, in \autoref{sec:FluxesSpectraChow}, how Chow groups encode the gauge background in F-theory \cite{Bies:2014sra}, mathematical details being relegated to \autoref{subsec:ReviewChowGroupsAndG4Fluxes}. In \autoref{sec:SystGaugeBack} we systematically describe the vertical gauge backgrounds including in particular the matter surface fluxes. We also flesh out how to concretely apply intersection theory within the Chow ring to deduce the gauge bundles to which the massless zero-modes couple. To exemplify this general technology we classify, in \autoref{sec:ToricFTheoryGUTModels}, the matter surface fluxes in an F-theory fibration with gauge group $SU(5) \times U(1)$. Even though the geometry of this fibration has already been worked out in \cite{Krause:2011xj}, we collect all the required data, in particular the explicit structure of the fibre in various codimensions, in \autoref{sec:FibreStructure} in order to make this article self-contained. Section \ref{sec:MasslessSpectraOfCurveSupportFluxesII} contains the intersection theoretic manipulations for the vertical gauge backgrounds of this model, performed in two different ways. Further computational technicalities are collected in  \autoref{sec:MasslessSpectraOfCurveSupportFluxesTedious}. The result is an explicit parametrization of the gauge bundles on the matter curves in terms of vanishing ideals. In \autoref{sec:ComputingTheSpectra} and \autoref{sec:ATasteOfAlgebraAndAlgebraicGeometry} we explain how this data can be translated into a so-called \fp graded module such that the computation of the cohomology groups can be performed with the help of computer algebra. Applying this technology to our example geometry we compute the exact charged massless matter for different choices of complex structure and observe jumps in the number of vector-like pairs as we wander in the moduli space. We conclude with a list of further directions of research in \autoref{sec:Conclusion}.

\section{\texorpdfstring{$\mathbf{G_4}$}{G4}-Fluxes, Massless Spectra and Chow Groups}  \label{sec:FluxesSpectraChow}

To set the stage we begin with a brief review of the essentials of F-theory compactifications to four dimensions, with special emphasis on the formulation of the gauge background following \cite{Bies:2014sra}. In particular we make the connection between the well established counting of charged bulk and localised zero-modes in the local topologically twisted field theory approach to F-theory and the global definition of the gauge background via Chow groups. More details can be found in \autoref{subsec:ReviewChowGroupsAndG4Fluxes}.

\subsection{F-Theory Compactifications on Smooth 4-Folds} \label{F-CompGen1}

We are compactifying F-theory to four dimensions on an elliptically fibred Calabi-Yau 4-fold $Y_4$ with projection map $\pi \colon Y_4 \rightarrow B_3$. 
By definition, the generic fibre is a smooth elliptic curve. It degenerates and becomes singular over the discriminant locus $\Delta \subseteq B_3$. A smooth resolution 
\[ \hat{\pi} \colon \hat{Y}_4 \twoheadrightarrow B_3 \label{hatY4hatpi} \]
of the singularities of $Y_4$
can always be obtained by replacing singular fibres by a finite number of $\mathbb{P}^1$s. 

Throughout this work we assume the existence of a smooth resolution $\hat{Y}_4$ which satisfies the Calabi-Yau condition.\footnote{This condition is violated in the presence of $\mathbb Q$-factorial terminal singularities as studied systematically in \cite{Arras:2016evy} and references therein. In such instances, we are forced to work on a singular space if we want to keep the Calabi-Yau property.} We furthermore assume that  the resolved fibration (\ref{hatY4hatpi})  admits a section, typically referred to as the \emph{zero section} $S_0$ with associated (co)homology class $[ S_0 ] \in H^{1,1} ( \hat{Y}_4 )$.\footnote{The existence of a section as such is not a restriction because one can always pass to the associated Jacobian fibration even if a given fibration has no section. However, typically the Weierstrass model describing the Jacobian  has $\mathbb Q$-factorial terminal singularities in codimension-two which cannot be resolvant crepantly \cite{Braun:2014oya}. In this sense, the prime assumption is really the absence of terminal singularities such that a smooth Calabi-Yau fibration exists.} Resolutions along these lines have been constructed for four-dimensional F-theory compactifications for instance in \cite{Blumenhagen:2009yv,Grimm:2009yu,Chen:2010ts,oai:arXiv.org:1011.6388,Knapp:2011wk,Esole:2011sm,Marsano:2011hv,Krause:2011xj,oai:arXiv.org:1111.1232,oai:arXiv.org:1202.3138,Lawrie:2012gg,Hayashi:2013lra,Hayashi:2014kca,Esole:2014bka} and references therein. F-theory compactifications that do not require the existence of a section are discussed \eg in \cite{Braun:2014oya,Anderson:2014yva,Klevers:2014bqa,Garcia-Etxebarria:2014qua,Mayrhofer:2014haa,Mayrhofer:2014laa,Cvetic:2015moa,Lin:2015qsa,Kimura:2016crs}.

The discriminant $\Delta$ over which the fibre degenerates is decomposed of irreducible varieties $\Delta_I$,
\[ \Delta = \bigcup_{I \in {\cal I}}{\Delta_I} \,. \]
Over the irreducible components $\Delta_I$, the extended Dynkin diagram of a Lie algebra $\mathfrak{g}_I$ is represented by the topology of the $\mathbb{P}^1$s in the resolved fibre.\footnote{If the fibre is of Kodaira type I$_1$ or II, no resolution is required and the gauge algebra $\mathfrak{g}_I$ is trivial.} As a consequence, vector multiplets in the adjoint representation of $\mathfrak{g}_I$ are found to propagate on $\Delta_I$. These gauge degrees of freedom can be described by a topologically twisted 8-dimensional $\mathcal{N} = 1$ supersymmetric gauge theory \cite{Donagi:2008ca,Beasley:2008dc}, where the gauge background data is provided by vector bundles on $\Delta_I$. 

Additional localised matter arises on the so-called \emph{matter curves}. Such a matter curve $C_{IJ} \subseteq B_3$ is a codimension-2 subvariety along which an intersection $\Delta_I \cdot \Delta_J$ of two of the irreducible components $\Delta_I$ of the discriminant $\Delta$ occurs. This includes possible self-intersections, \ie $I = J$. Typically, the fibre structure over the matter curve $C_{IJ}$ represents the extended Dynkin diagram of a Lie algebra $\mathfrak{h}_{IJ}$ into which the Lie algebras $\mathfrak{g}_I$ and $\mathfrak{g}_J$ are embedded.\footnote{Due to monodromy effects, the  $\mathbb P^1$s associated with one or more nodes of the affine Dynkin diagram may be absent in the resolved geometry.}

In the limit of vanishing fibre volume, M2-branes that wrap suitable linear combinations of $\mathbb P^1$s in the fibre over $C_{IJ}$ give rise to massless matter in some representation $\mathbf R$ of $\mathfrak{g}_{I}\oplus \mathfrak{g}_J$. Therefore, to each state of the representation $\mathbf R$  one can associate a complex 2-cycle. More precisely, let us label by $a = 1, \ldots \mathrm{dim}(\mathbf{R})$ the different states in representation $\mathbf{R}$ and by $\beta^a(\bfR)$ the associated weight vector. Then to each $\beta^a(\bfR)$ there exists a complex 2-cycle $S^a_\bfR$, termed \emph{matter surface}, which is given as the linear combination of the fibre $\mathbb{P}^1$s over the \emph{matter curve} $C_{IJ}$ associated with the corresponding state. Therefore one usually denotes this matter curve as $C_\bfR$, and we will indeed adapt this notation in the remainder of this article. We will give more details on the localised massless states in \autoref{subsubsec:LocalisedMatterAndTheSpinBundle}.

In addition to the geometric information encoded in the fibration $\hat{\pi} \colon \hat{Y}_4 \twoheadrightarrow B_3$, defining  an F-theory compactification to four dimensions requires specifying the gauge background. By duality with M-theory, the gauge background is encoded in the background data of the M-theory 3-form $C_3$ and its field strength $G_4$. There exists a precise mathematical characterisation of this data in terms of a \emph{Deligne cohomology} class of $\hat{Y_4}$, \cf \autoref{subsec:ReviewChowGroupsAndG4Fluxes}.\footnote{Note that there is also an equivalent formulation in terms of the theory of Cheeger-Simons differential characters \cite{Cheeger-Simons}, as employed \eg in \cite{oai:arXiv.org:1203.6662} in F/M-theory.} The Deligne cohomology class $H^4_D(\hat{Y_4}, \mathbb Z(2))$ \cite{BeilinsonConjectures}  has been explored to describe the necessary gauge data both in M-theory and F-theory in \cite{Diaconescu:2003bm,Freed:2004yc,oai:arXiv.org:hep-th/0409158,oai:arXiv.org:1203.6662} and  \cite{Curio:1998bva,Donagi:1998vw,Donagi:2011jy, Clingher:2012rg,Anderson:2013rka,Bies:2014sra}, respectively. 

The key observation is that to each 3-form gauge background we can on the one hand associate a quantised field strength $G_4$, or flux, but in addition it is necessary to specify the holonomies of $C_3$ around non-trivial 3-cycles which correspond to flat gauge backgrounds, or Wilson `lines'.
This implies that the full gauge background data, as given by elements of $H^4_D(\hat{Y_4}, \mathbb Z(2))$,  fits into the short exact sequence\footnote{The existence of such an object follows from the Deligne-Beilinson complex \cite{BeilinsonConjectures}. See also \cite{Bies:2014sra} for a review and further references.}
\[ 0 \longrightarrow J^2 \left( \hat{Y}_4 \right) \longrightarrow H_D^4 \left( \hat{Y}_4, \mathbb{Z} \left( 2 \right) \right) \stackrel{\hat{c}_2}{\longrightarrow} H_{\mathbb{Z}}^{2,2} \left( \hat{Y}_4 \right) \longrightarrow 0 \,. \label{SESDeligne} \]
An element in  $H_D^4 ( \hat{Y}_4, \mathbb{Z} ( 2 ) )$  is mapped to an integral $( 2,2 )$-form via a surjective map $\hat{c}_2$. The so-obtained $( 2,2 )$-form is interpreted as the $G_4$ flux associated with the gauge background. Since the map $\hat{c}_2$ is only surjective, but not injective, it has a kernel, given by the intermediate Jacobian $J^2 ( \hat{Y}_4 )$. This is the space of flat 3-form connections which, by definition,  have vanishing gauge flux even though such configurations are non-trivial topologically.

Elements of $H^4_D(\hat{Y_4}, \mathbb Z(2))$ can in turn be represented by elements $A \in \text{CH}^2 ( \hat{Y}_4 )$, \ie by complex 2-cycles modulo rational equivalence \cite{Bies:2014sra}. As we review further in \autoref{subsec:ReviewChowGroupsAndG4Fluxes}, rational equivalence for complex $p$-cycles is the direct generalisation of linear equivalence for divisors. This means that two $p$-cycles on $\hat{Y_4}$ are rationally equivalent if their difference can be written as the zeroes or poles of a meromorphic function on a $(p+1)$-dimensional irreducible subvariety of $\hat{Y_4}$. The group of $p$-cycles up to such equivalence is the Chow group $\mathrm{CH}_p(\hat{Y_4})$, while we denote the group of codimension $p$ cycles up to rational equivalence as $\mathrm{CH}^p(\hat{Y_4})$. 
In particular, the divisor class group $\mathrm{Cl}(\hat Y_4)$ hence coincides with $\mathrm{CH}^1(\hat{Y_4})$ on the smooth variety $\hat{Y_4}$.

What makes Chow groups relevant to characterise the gauge background on $\hat{Y_4}$ is the existence of a group homomorphism called \emph{refined cycle map} (see \eg p.~123 in \cite{GreenMurreVoisin})
\[ \hat{\gamma}_p \colon \mathrm{CH}^p \left( \hat{Y}_4 \right) \to H^{2p}_D \left( \hat{Y}_4, \mathbb{Z} \left( 2 \right) \right) \]
which gives rise to a ring homomorphism
\[ \hat{\gamma} \colon \mathrm{CH}^{\bullet} \left( \hat{Y}_4 \right) \to H^{\bullet}_D \left( \hat{Y}_4, \mathbb{Z} \left( 2 \right) \right) \; , \; x \mapsto \bigoplus_{p \in \mathbb{N}_{\geq 0}}{\hat{\gamma_p} \left( x \right)} \, . \]
The case $p = 2$ is of particular importance to us because the map
\[ \hat{\gamma}_2 \colon \mathrm{CH}^2 \left( \hat{Y}_4 \right) \to H^{4}_D \left( \hat{Y}_4, \mathbb{Z} \left( 2 \right) \right) \label{refcyclemap} \]
allows us to associate to each class of complex 2-cycles an element in $H^4_D(\hat{Y_4}, \mathbb Z(2))$ defining the gauge background. 

Fortunately, the concrete form of $\hat{\gamma}$ is not required for our computations, as we will see. This map is surjective over $\mathbb Q$ if and only if the Hodge conjecture holds. It is in general not injective. This means that two different Chow classes may in principle map to the same element in $H^4_D ( \hat{Y}_4, \mathbb{Z} ( 2 ) )$. 
This is not a big drawback for our purposes. What will be crucial is rather the fact that two algebraic cycles which are rationally equivalent are guaranteed to map to the same element in $H^4_D ( \hat{Y}_4, \mathbb{Z} ( 2 ) )$. This means that manipulations modulo rational equivalence do not change the 3-form background parametrised by an algebraic 2-cycle. This fact comes in particularly handy whenever the 2-cycles representing the gauge background are in turn obtained by pullback from an ambient space on which the Chow group is explicitly known. An example would be a situation where $\hat{Y_4}$ is embedded into a smooth toric ambient space $X_\Sigma$. On such smooth $X_\Sigma$, rational equivalence and homological equivalence coincide. Therefore we can manipulate the underlying 2-cycle as long as we keep its homology class on $X_\Sigma$ fixed, and are guaranteed that the gauge data remains unchanged. This important property and a more precise summary of the formalism of \cite{Bies:2014sra} are reviewed further in \autoref{subsec:ReviewChowGroupsAndG4Fluxes}. Fans of commutative diagrams can find all information encrypted in \autoref{fig:C3BackgroundFromChow} therein.

The cohomology class associated with such a 2-cycle class in $\mathrm{CH}^2(\hat{Y_4})$ is identified with the 4-form flux $G_4 \in H^{2,2} ( \hat{Y}_4 )$ \cite{oai:arXiv.org:hep-th/9605053,oai:arXiv.org:hep-th/9606122,oai:arXiv.org:hep-th/9908088}. However, unlike the parametrisation via  $\mathrm{CH}^2(\hat{Y_4})$,  keeping track only of $G_4$ merely accounts for \emph{part} of the information of the gauge background and is hence in general incomplete.
Such a (class of) differential forms is then supposed to have ``one leg along the fibre'' \cite{oai:arXiv.org:hep-th/9908088}. Phrased explicitly, this means that the two transversality constraints 
\[ G_4 \cdot \pi^*\omega_4 = 0 \,, \qquad  G_4 \cdot [S_0] \cdot \pi^*\omega_2 =0 \label{transversality-gen1} \]
must be enforced for every element $\omega_4 \in H^4(B_3)$, $\omega_2 \in H^2(B_3)$. Recall that $[S_0] \in H^{1,1}(\hat{Y_4})$ is the class of the zero section of the fibration $\pi \colon \hat{Y}_4 \twoheadrightarrow B_3$, whose existence we assumed as stated above.\footnote{The generalisation of this condition in absence of a section has been described in \cite{Lin:2015qsa}.} Furthermore, in combination with cohomlogy classes, the operation $\cdot$ denotes the intersection product in the cohomology ring, here on $\hat Y_4$. We will come back to these conditions in \autoref{GaugeInvTrans}.

The  class $G_4$ of differential forms is actually required to be an element of $H^{2,2}_{{\mathbb Z}/2}(\hat{Y_4})$ as a consequence of the quantisation condition \cite{oai:arXiv.org:hep-th/9609122}
\[ G_4 + \frac{1}{2} c_2(\hat{Y_4}) \in H^{2,2}_{{\mathbb Z}}(\hat{Y_4}) \, . \label{FW2} \]

Once such a $G_4 \in H^{2,2}_{ \mathbb{Z}/2} ( \hat{Y}_4 )$ is specified, the chiral index of massless matter in representation $\bfR$ localised on the matter curve $C_\bfR$ can be obtained by evaluating the integral \cite{Donagi:2008ca, oai:arXiv.org:0904.1218, Braun:2011zm,Marsano:2011hv,Krause:2011xj, oai:arXiv.org:1111.1232,oai:arXiv.org:1202.3138,oai:arXiv.org:1203.6662},
\[ \chi_{\bfR} = \Mint_{S^a_\bfR}{G_4} \, . \label{chiRsecgen}\]
More precisely, (\ref{chiRsecgen}) counts the chiral index of the state with weight $\beta^a(\bfR)$. If the gauge flux does not break the non-Abelian gauge algebra this result is independent of the particular state $\beta^a(\bfR)$, or equivalently of the matter surface $S^a_\bfR$, $a=1, \ldots , \bfR$. Note that $\chi_\bfR \in \mathbb{Z}$ as a consequence of \cite{oai:arXiv.org:1011.6388,oai:arXiv.org:1202.3138,oai:arXiv.org:1203.4542}
\[ \frac{1}{2} \Mint_{S^a_\bfR}{c_2(\hat{Y_4})} \in {\mathbb Z} \, . \]

For conceptual clarity, we would like to end this review with a word of caution. As we have stressed, we are working on a smooth resolution $\hat{Y_4}$ of the elliptic fibration, which allows us to avoid dealing with singularities. Resolving $\hat{Y_4} \rightarrow Y_4$ by blowing-up the singularities in the fibre is well-known to correspond to moving to the Coulomb branch of the dual 3d M-theory vacuum. The non-Abelian gauge symmetry is restored in the dual 4d F-theory vacuum after taking the fibre volume to zero. Nonetheless, this means that by working with $\hat{Y_4}$ we are only able to detect Abelian gauge backgrounds. Non-Abelian gauge bundles on the 7-branes, by contrast, cannot be encoded in the Deligne cohomology or the Chow groups of the smooth space. This is only a minor drawback for the applications envisaged in this paper because non-Abelian gauge backgrounds would break the gauge algebra on the 7-branes, while Abelian gauge backgrounds are already sufficient to generate a chiral spectrum. Nonetheless it would be interesting to explore such more general backgrounds directly in M-theory, \eg by studying the Deligne cohomology of the singular 4-fold $Y_4$ (see \eg \cite{Anderson:2013rka}) or by further developing alternative techniques such as \cite{Collinucci:2014taa,Collinucci:2016hpz}.

\subsection{Cohomologies in Local Setups via Topological Twist} \label{subsec:SheafCohomologyAndMasslessMatter}

So far we have reviewed the definition of an F-theory vacuum in terms of a globally consistent elliptic fibration with 3-form background. Locally, the gauge theory on the 7-branes enjoys a description as a partially topologically twisted 8d Super-Yang-Mills theory \cite{Donagi:2008ca,Beasley:2008dc}. In this language, the definition both of the gauge background and the computation of the exact massless matter content is well established. As in \cite{Bies:2014sra}, our strategy is to extract the local gauge data from the globally defined geometry and 3-form background, which then allows us to determine the massless matter content. To prepare ourselves for this task,  we now briefly collect the well-known description of massless matter in the local approach via topologically twisted gauge theory.

\subsubsection{Bulk Matter} \label{subsubsec:BulkMatter}

Let us begin by recalling the nature of the so-called bulk 7-brane matter on a non-Abelian brane stack. A stack of 7-branes wrapping the component $\Delta_I$ of the discriminant locus carries, in addition to a gauge multiplet, massless 4d ${\cal N}=1$ chiral multiplets in the adjoint representation of the non-Abelian gauge algebra $\mathfrak{g}_I$ underlying the gauge group $G_I$. As stressed at the end of the previous section, we restrict ourselves to Abelian gauge backgrounds. Locally, these are described by a line bundle $L_I$ in the Picard group $\mathrm{Pic}(\Delta_I)$ on the 7-brane stack on $\Delta_I$.
Embedding the structure group $U(1)_I$ of $L_I$ into $G_I$ breaks $G_I \rightarrow H_I \times U(1)_I$, which induces the decomposition 
\[ {\textbf{adj}}(G_I) \rightarrow \bigoplus_{m_I} {\mathbf r}_{  m_I} \label{adjdecom-gen} \]
into irreducible representations of $H_I$. Then the massless bulk matter in representation ${\mathbf r}_{m_I}$ on $\Delta_I$ is counted by the cohomology groups \cite{Katz:2002gh,Donagi:2008ca,Beasley:2008dc,Blumenhagen:2008zz}
\begin{align}
\begin{split} \label{cohoms-local-bulk}
H^0(\Delta_I, L_{m_I}) \quad \phantom{\oplus} \qquad &  \\
H^1(\Delta_I, L_{m_I}) \quad \oplus \qquad &H^0(\Delta_I, L_{m_I} \otimes K_{\Delta_I}) \\
H^2(\Delta_I, L_{m_I}) \quad \oplus \qquad &H^1(\Delta_I, L_{m_I} \otimes K_{\Delta_I})  \\
                              &         H^2(\Delta_I, L_{m_I} \otimes K_{\Delta_I}) \, .
\end{split}
\end{align}
Here the relevant line bundle $L_{m_I}$ is given by 
\[ L_{m_I} = L_I^{q_I({\mathbf r}_{m_I})} \label{Lmi} \]
with $q_I({\mathbf r}_{m_I})$ the $U(1)_I$ charge of representation ${\mathbf r}_{m_I}$. The second and third line count chiral and, respectively, anti-chiral ${\cal N}=1$ multiplets in representation $\mathbf{r}_{m_I}$. The first and fourth line vanish for supersymmetric fluxes \cite{Blumenhagen:2008zz} and can hence be discarded.
The chiral index resulting from this spectrum is
\[ \chi_{{\mathbf r}_{m_I}} = - {c_1(\Delta_I) \cdot_{\Delta_I} c_1(L_{m_I})} \, . \]

 \subsubsection{Localised Matter and The Definition of The Spin Bundle} \label{subsubsec:LocalisedMatterAndTheSpinBundle}

Consider next the massless matter localised on a curve $C_{\bfR}$, and denote by $L_{\mathbf R} \in \text{Pic} ( C_{\mathbf{R}} )$ the local gauge background to which such matter in the twisted theory on $\mathbb R^{1,3} \times C_{\bf R}$ is coupled. The number of massless matter multiplets in representation ${\bf R}$ is given by the dimensions of the cohomology groups \cite{Donagi:2008ca,Beasley:2008dc}
\bea \label{mattercohoma}
H^{i}(C_{\mathbf{R}}, L_{\mathbf{R}} \otimes \sqrt{K_{C_{\mathbf{R}}}}) 
\eea
with $i=0$ and $i=1$ counting chiral and anti-chiral ${\cal N}=1$ multiplets, respectively. It has already been asserted in \cite{Bies:2014sra} that the correct choice of spin bundle $\sqrt{K_{C_{\mathbf{R}}}}$ appearing in (\ref{mattercohoma}) is the one induced by the holomorphic embedding of $C_{\bf R}$ into the base $B_3$, in the following sense: Let $D_a, D_b \in \text{Cl} ( B_3 )$ and $C = D_a \cap D_b \subseteq B_3$ a curve, then the adjunction formula for $C$ and $D_a$ gives
\[ K_C = \left. K_{D_a} \right|_C \otimes N_{C/D_a} = \left. \left( K_{B_3} \otimes {\cal O}_{B_3}(D_a) \otimes {\cal O}_{B_3}(D_b) \right) \right|_C \equiv \left. M \right|_C \,. \]
The line bundle $M \in \text{Pic} ( B_3 )$ is uniquely determined by its first Chern class since $b^1 ( B_3 ) = 0$. The latter is necessary for an elliptically fibred Calabi-Yau 4-fold $Y_4 \twoheadrightarrow B_3$ to exist, for which $b^1(Y_4) =0$. If both $D_a$ and $D_b$ are spin\footnote{This means there exists  a $\mathbb{Z}$-Cartier divisor $k_a$ such that $2 k_a$ is the canonical divisor on $D_a$.}, $c_1(M)$ is an even class in $H^2(B_3,\mathbb Z)$ and $\sqrt{M}$ is the unique line bundle on $B_3$ with first Chern class $\frac{1}{2} c_1(M) \in H^2(B_3,\mathbb Z)$. The spin structure to take in (\ref{mattercohoma}) is then 
\bea \label{spinstructureM}
\sqrt{ K_{C_\mathbf{R}}} = \left. \sqrt{M} \right|_{C_{\mathbf{R}}}.
\eea
If $D_a$ or $D_b$ are not spin, then the Freed-Witten quantisation condition ensures that $L_{\mathbf R} \otimes \sqrt{K_{C_{\mathbf R}}}$ can be split up as a product of two integral bundles $L_1 \otimes L_2$ such that $L_2$ is a product of line bundles obtained as the pullback of well-defined bundles on $B_3$.  The pullback bundles underlying $L_2$ include the contribution from the spin structure.

\subsection{From \texorpdfstring{$\mathbf{C_3}$}{C3}-Backgrounds to Line Bundles }  \label{sec:FromC3toL}

Our next task is now to extract the line bundles appearing in the cohomologies (\ref{cohoms-local-bulk}) and (\ref{mattercohoma}) from the gauge background on the globally defined 4-fold $\hat{Y_4}$. The general idea is that the matter states are associated with the quantised moduli space of M2-branes \cite{Witten:1996qb} wrapping suitable components of the fibre, as reviewed in \autoref{F-CompGen1}. An M2-brane couples in a standard way to the 3-form background via the Chern-Simons action $S_{\rm CS} = 2 \pi \int_{\mathrm{M2}}{C_3}$. We therefore need to integrate the 3-form gauge background over the fibral curve wrapped by the M2-brane associated with a given state. The resulting object then describes the Abelian gauge background to which the M2-brane excitations along the base couple. 

The formalism of Chow groups allows us to perform this operation of integration over the fibre in a manner which is guaranteed to keep the full amount of information about the gauge background \cite{Bies:2014sra}. Indeed, by using intersection theory within the Chow ring we are able to extract a line bundle either on the 7-brane $\Delta_I$ or on the matter curve $C_{\mathbf{R}}$ on the base. The compatibility of the refined cycle map (\ref{refcyclemap}) with intersection within the Chow ring and with pushforward under the fibration $\pi$ ensures that this procedure correctly recovers the information both about the first Chern class and about the flat holonomies of the line bundle on the base.

\subsubsection{Localised Matter}

For localised matter this formalism has already been carried out in \cite{Bies:2014sra}. Consider a matter surface $S^a_\bfR$, given by a fibration of rational curves over the matter curve $C_\bfR$ on the base $B_3$. Furthermore fix a 2-cycle class $A \in \mathrm{CH}^2(\hat{Y_4})$ to represent the 3-form background as reviewed in \autoref{F-CompGen1}. We can then form the intersection product $S^a_\bfR \cdot_{\iota_{\bfR,a}} A$, loosely speaking by pulling back $A$ to $S^a_\bfR$. The notation and details of this construction are explained in \autoref{app:intersection}. Since the pullback is compatible with rational equivalence and preserves codimension, we can view this as an object in $\mathrm{CH}^2(S^a_\bfR)$, \ie an element of the class of points on $S^a_\bfR$. The operation of integration along the fibre motivated above corresponds to projection onto the base by considering the object
\[ D\left(S^a_\bfR, A\right) := \pi_{\bfR \ast} (S^a_\bfR \cdot_{\iota_{\bfR,a}} A) \in \mathrm{CH}_0(C_\bfR) \,. \label{projectionformula1} \]
Here $\pi_{\bfR,a}$ denotes the projection from the surface $S^a_\bfR$ to $C_\bfR$. We have therefore obtained a Chow class of points on $C_{\mathbf{R}}$. Since 
\[ \mathrm{CH}_0(C_\bfR) \simeq \mathrm{CH}^1(C_\bfR) \simeq \mathrm{Pic}(C_\bfR) \]
this defines a line bundle 
\[ L\left(S^a_\bfR, A\right) = {\cal O}_{C_\bfR} \left( D\left(S^a_\bfR, A\right) \right) \]
on $C_\bfR$. By construction, this is the line bundle induced by the gauge background to which the quantum excitations of the M2-brane wrapping the fibre of $S^a_\bfR$ couple. If the gauge background preserves the non-Abelian gauge symmetry in the F-theory limit, then for fixed $\mathbf{R}$ this construction gives the same line bundle for each choice of $S^a_\bfR$, $a= 1, \ldots, \mathrm{dim}(\bfR)$, and we can omit the index $a$. By comparison with (\ref{mattercohoma}), the massless chiral matter in representation $\bfR$ is counted by
\[ H^i \left(C_\bfR, {\cal L}\left(S_\bfR, A\right) \right)\,, \qquad \quad {\cal L}\left(S_\bfR, A\right) = {L}\left(S_\bfR, A\right) \otimes \sqrt{K_{C_\bfR}} \, .  \label{cohomoswithflux}\]
A formalisation of these steps based on an accurate application of the intersection product within the Chow ring has been given in \cite{Bies:2014sra} and is reviewed in \autoref{app:intersection}.

Note that (\ref{cohomoswithflux}) counts the exact massless matter spectrum modulo potential spacetime instanton corrections in F-theory: For instance, 
if the matter is charged only under an Abelian gauge group which acquires a St\"uckelberg mass, 
D3/M5-instantons can still generate non-perturbative mass terms which are exponentially suppressed by the K\"ahler moduli in the sense of \cite{Blumenhagen:2006xt,Ibanez:2006da,Florea:2006si,Haack:2006cy,Blumenhagen:2009qh}. These corrections are not accounted for by the topological field theory derivation of (\ref{cohomoswithflux}).

\subsubsection{Bulk Matter}\label{bulk-global}

In a similar manner, we can now define the line bundles on a stack of 7-branes wrapping $\Delta_I$ to which the bulk matter couples. The fibre of $\hat{Y_4}$ over a generic point on $\Delta_I$ takes the form of a connected sum of rational curves $\mathbb P^1_{i_I}$ intersecting like the nodes of the affine Dynkin diagram of $\mathfrak{g}_I$. Fibring $\mathbb P^1_{i_I}$ over $\Delta_I$ defines the so-called resolution divisor $E_{i_I}$. 

Let us now consider the Chow class $E_{i_I} \in \mathrm{CH}^1(\hat{Y_4})$ associated with $E_{i_I}$, and fix a gauge background $A \in \mathrm{CH}^2(\hat{Y_4})$. Recall that pullback is well-defined for $\mathrm{CH}^{\bullet}(\hat{Y_4})$. Therefore we can pull $A$ back to $E_{i_I}$ by considering the intersection product. This produces an element $E_{i_I} \cdot_{\iota_{i_I}} A \in \mathrm{CH}^2(E_{i_I}) \simeq \mathrm{CH}_1(E_{i_I})$. Integration over the fibre amounts to projection onto the base of $E_{i_I}$ via the pushforward of $\pi_{i_I} \colon E_{i_I} \, {\twoheadrightarrow} \, \Delta_I$. This results in
\[ \pi_{i_I \ast}(E_{i_I} \cdot_{\iota_{i_I}} A) \in \mathrm{CH}_1(\Delta_I) \,. \]
On the complex 2-cycle $\Delta_I$, we have that
\[ \mathrm{CH}_1(\Delta_I) \simeq \mathrm{CH}^1(\Delta_I)  \simeq \mathrm{Pic}(\Delta_I) \,. \]
We therefore identify
\[ L_{i_I} ={\cal O}\left( \pi_{i_I \ast}(E_{i_I} \cdot_{\iota_{i_I}} A)\right) \]
as the line bundle on $\Delta_I$ induced by the gauge background with structure group $U(1)_{i_I}$. This process is to be repeated for all $E_{i_I}$, $i_I=1, \ldots, \mathrm{rk}(\mathfrak{g}_I)$.
If the structure group $U(1)_I$ of the line bundle $L_I$ appearing  in \autoref{subsubsec:BulkMatter} is given by the linear combination $L_I = \sum_{i_I} a_{i_I} \,  U(1)_{i_I}$, then 
\[ L_I = \bigotimes_{i_I} L_{i_I}^{a_{i_I}} \in \mathrm{Pic}(\Delta_I) \,. \]

\section{Systematics of Gauge Backgrounds } \label{sec:SystGaugeBack}

We now  systematically describe the representation of a gauge background by an element $A \in \mathrm{CH}^2(\hat{Y_4})$. As reviewed, the cohomology class $G_4:=[A] \in H^{2,2}(\hat{Y_4})$ defines a 4-form flux on $\hat{Y_4}$. The cohomology group  $H^{2,2}(\hat{Y_4})$ enjoys a decomposition into three orthogonal subspaces
\[ H^{2,2}(\hat{Y_4},\mathbb R) =  H^{2,2}_\mathrm{vert}(\hat{Y_4},\mathbb R)  \oplus H^{2,2}_\mathrm{hor}(\hat{Y_4},\mathbb R)  \oplus H^{2,2}_\mathrm{rem}(\hat{Y_4},\mathbb R) \,. \]
The primary vertical subspace $H^{2,2}_\mathrm{vert}(\hat{Y_4},\mathbb R)$ is generated by elements of $H^{1,1}(\hat{Y_4}) \wedge H^{1,1}(\hat{Y_4})$, while the primary horizontal subspace $H^{2,2}_\mathrm{hor}(\hat{Y_4},\mathbb R)$ is the subspace of $H^{2,2}(\hat{Y_4})$ obtained by variation of Hodge structure starting from the unique $(4,0)$ form \cite{Greene:1993vm}. Finally, $H^{2,2}_\mathrm{rem}(\hat{Y_4},\mathbb R)$ denotes the remainder piece which is neither vertical nor horizontal \cite{Braun:2014xka}.

We first revisit, in \autoref{GaugeInvTrans}, the constraints on a consistent gauge background posed by gauge invariance and transversality, directly at the level of algebraic cycles rather than of cohomology.
To systematise the geometric description of vertical gauge backgrounds we then introduce the notion of a matter surface flux in \autoref{sec-systematicsvertical}.
In \autoref{subsec:NewStrategyForMasslessSpectra} we detail the intersection theoretic operations which extract from such backgrounds the sheaves whose cohomology groups count the massless matter.

\subsection{Gauge Invariance and Transversality} \label{GaugeInvTrans}

The condition for a flux $G_4 \in H^{2,2}_{\mathrm{vert}}(\hat{Y_4})$ not to break non-Abelian gauge symmetries in F-theory is typically formulated in the literature as
\[ G_4 \cdot [E_{i_I}] \cdot [D^\mathbf{b}_\alpha] = 0 \qquad \forall \quad [D^\mathbf{b}_\alpha] \in H^{1,1}(B_3) \, . \label{gauge-inv-verta} \]
However, this condition is not sufficient to ensure gauge invariance of a flux in $H^{2,2}_{\mathrm{rem}}(\hat{Y_4})$: Due to the orthogonality of $H^{2,2}_{\mathrm{rem}}(\hat{Y_4})$ and $H^{2,2}_{\mathrm{vert}}(\hat{Y_4})$ any flux $H^{2,2}_{\mathrm{rem}}(\hat{Y_4})$ satisfies (\ref{gauge-inv-verta}) even though it might break the non-Abelian gauge group on a 7-brane, a prime example being the so-called hypercharge flux in F-theory models based on gauge group $SU(5)$. Furthermore, we seek to find a condition not only for the gauge flux, but more generally for the Chow class $A$ with $[A] = G_4$ representing the full gauge background.

From the perspective of the topologically twisted theory on the 7-brane $\Delta_I$, the condition for gauge invariance is that the gauge bundle embedded into the structure group of the non-Abelian group should be the trivial bundle on $\Delta_I$. Hence given an element $A \in \mathrm{CH}^2(\hat{Y_4})$ with $[A] = G_4$ the correct condition to impose is
\[ \pi_{i_I\ast} \left(  E_{i_I} \cdot_{i_I} A  \right) = 0 \in \mathrm{CH}_1(\Delta_I)    \simeq \mathrm{CH}^1(\Delta_I) \qquad \forall \, E_{i_I} \, . \label{condition-gauge-weaker} \]

The condition for a 4-form flux $G_4 \in H^{2,2}(\hat{Y_4})$ to descend to a well-defined gauge flux in F-theory is typically formulated as the transversality conditions
\[ G_4 \cdot [D^\mathbf{b}_\alpha] \cdot [D^\mathbf{b}_\beta] = 0 \qquad \forall \quad [D^\mathbf{b}_\alpha], [D^\mathbf{b}_\beta] \in H^{1,1}(B_3) \label{verticality-hom-1} \]
and
\[ G_4 \cdot [D^\mathbf{b}_\alpha] \cdot [S_0] = 0 \qquad  \forall  \quad [D^\mathbf{b}_\alpha]   \in H^{1,1}(B_3) \, . \label{verticality-hom-2} \]
One possible derivation of these constraints is via the observation that they are equivalent to the absence of certain Chern-Simons-terms in the dual 3d M-theory vacuum of the form $\int_{\mathbb R^{1,2}}{A_\alpha \wedge F_\beta}$ and, respectively, $\int_{\mathbb R^{1,2}}{A_\alpha \wedge F_0}$. Here $A_\alpha$ and its field strength refer to the $h^{1,1}(B_3)$ vector multiplets associated with the K\"ahler moduli of the base $B_3$ in the 3d ${\cal N}=2$ theory, and $A_0$, $F_0$ refer to the vector multiplets associated with the Kaluza-Klein $U(1)$ associated with circle reduction of the 4d F-theory to the 3d M-theory \cite{oai:arXiv.org:1111.1232}. Now, since these Chern-Simons terms are not generated at one loop in the transition from the 4d F-theory to the 3d M-theory vacuum, and they do not either descend from classical terms in 4d upon circle reduction, they must be absent in the M-theory effective action in order for the 3d vacuum to lift to a Poincar\'e invariant F-theory vacuum \cite{oai:arXiv.org:1111.1232}.

While conceptually very clear, this derivation is only sensitive to the intersection product in cohomology. In particular, these conditions are again trivially satisfied by a gauge flux $G_4$ not in $H^{2,2}_{\mathrm{vert}}(\hat{Y_4})$. An alternative derivation of condition (\ref{verticality-hom-1}) is to require that $G_4$ do not affect the chirality of states wrapping the full fibre as these correspond to the higher KK modes in M-theory. Their chirality must therefore equal that of the zero-mode in order for the 4d F-theory vacuum to be Lorentz invariant. In particular, consider a matter surface $S^a(\mathbf{R})$ over a curve $C_\mathbf{R}$ in the base. The surface $S^a_n(\mathbf{R})$ is defined by adding to the fibral curves over $C_\mathbf{R}$ $n$ multiples of the full fibre $F$. M2-branes wrapping the fibre of $S^a_n(\mathbf{R})$ correspond to the $n$-th KK state associated with the 4d F-theory multiplet with weight $\beta^a(\mathbf{R})$. The condition (\ref{verticality-hom-1}) guarantees that
\[ G_4 \cdot [S^a_n(\mathbf{R})] = G_4 \cdot [S^a(\mathbf{R})]   \qquad \forall \, n \, . \]
Hence the 3d chirality of the spectrum of KK modes with weight $\beta^a(\mathbf{R})$ equals that of the KK zero-mode in the 3d ${\cal N}=2$ M-theory vacuum.\footnote{Clearly there is no notion of chirality in 3d in the sense of Weyl spinors. Rather, a 3d ${\cal N}=2$ theory is defined to be vectorlike if the number of 3d ${\cal N}=2$  chiral multplets in representation $\mathbf{R}$ and $\mathbf{\bar R}$ coincides, and chiral otherwise.}

However, once we specify the gauge background beyond its flux, we require not only that the 3d net chirality of KK modes should agree, but rather that the exact number of 3d KK states in representation $\mathbf{R}$ and $ \mathbf{\bar R}$ must match that of the zero modes. This is guaranteed if\footnote{Here and in the sequel, whenever the precise meaning is clear by the context,  we abbreviate the intersection product within the Chow ring simply by $\cdot$ and the projection just by $\pi$ to avoid too heavy notation.}
\[ \pi_*( S^a_n(\mathbf{R}) \cdot A) = \pi_*( S^a(\mathbf{R}) \cdot A ) \qquad \forall n \, . \]
Given the construction of $S^a_n(\mathbf{R})$ described above, a sufficient condition generalizing (\ref{verticality-hom-1}) is to require that
\[ \pi_*(  \pi^{-1} C \cdot A  ) = 0 \qquad \forall \, C \in \mathrm{CH}_1(B_3) \, . \label{TransverseChow1} \]
From a conceptual point of view this is therefore proposed as the condition replacing (\ref{verticality-hom-1}) at the level of Chow groups. It would be interesting to investigate if there exist gauge backgrounds whose associated flux satisfies (\ref{verticality-hom-1}) even though its underlying Chow class violates (\ref{TransverseChow1}).

Less clear is the correct interpretation of (\ref{verticality-hom-2}) at the level of Chow groups. In view of (\ref{condition-gauge-weaker}) and (\ref{TransverseChow1}), a natural generalisation would be to require that
\[ \pi_*( S_0 \cdot A  ) = 0 \, . \label{TransverseChow2} \]
Again, we leave it for future investigations to determine if there are any non-trivial examples of gauge backgrounds which distinguish between (\ref{TransverseChow2}) and (\ref{verticality-hom-2}). For the cycles considered in this article, both conditions lead to equivalent constraints.

\subsection{Systematics of Vertical Gauge Backgrounds} \label{sec-systematicsvertical}

We will focus in this article on gauge backgrounds whose associated $G_4$ flux lies in $H^{2,2}_{\mathrm{vert}}(\hat{Y_4},\mathbb R)$, subject to the transversality conditions (\ref{transversality-gen1}).\footnote{For an incomplete list of more recent works on aspects of fluxes in $H^{2,2}_{\mathrm{hor}}(\hat{Y_4},\mathbb R)$ and the induced superpotential see \eg \cite{Grimm:2009sy,Grimm:2009ef,Alim:2010za,Braun:2011zm,oai:arXiv.org:1203.6662,Bizet:2014uua} and references therein.} These can be classified, on a fibration over a generic base space $B_3$, as follows:

\paragraph{Matter Surface Flux}

Consider a matter surface and its associated element $S^a_\bfR \in \mathrm{CH}^2(\hat{Y_4})$. By construction $G_4=[S^a_\bfR]$ satisfies (\ref{TransverseChow1}) and 
(\ref{TransverseChow2}). If we are interested in describing a gauge background which does not break the non-Abelian gauge algebra in the F-theory limit, we must modify the Chow class $S^a_\bfR$ by adding suitable terms of the form
\begin{align}
\begin{split}\label{def-mattersurfaceflux}
A({\bfR}) &= S^a_\bfR + \Delta^a(\bfR) \in \mathrm{CH}^2(\hat{Y_4}) \\
\Delta^a(\bfR) &= \left(\beta^a({\mathbf{R}})^T_{i_I} C^{-1}_{i_I j_J} \right) \, \left. E_{j_J} \right|_{  C_{\mathbf{R}} } \,.
\end{split}
\end{align}
Here $E_{i_I}$ denotes the resolution divisors for the gauge algebra $\mathfrak{g}_I$, and $E_{j_J} |_{  C_{\mathbf{R}} }$ their restriction to the curve $C_{\mathbf{R}}$. Furthermore, $C^{-1}_{i_I j_J} = \delta_{IJ} \, C^{-1}_{i_I j_I}$ with $C^{-1}_{i_I j_I}$ the inverse of the Cartan matrix of $\mathfrak{g}_I$. 
The expression (\ref{def-mattersurfaceflux}) is indeed gauge invariant because the intersection of any $E_{i_I}$ with $S^a(\mathbf{R})$ in the fibre reproduces the entry $\beta^a({\mathbf{R}})_{i_I}$ in the weight vector, and likewise intersection of $E_{i_I}$ with the component $E_{j_J}$ in the fibre reproduces the negative of the corresponding Cartan matrix entry. 

One can convince oneself that the final result after adding the correction terms $\Delta^a(\bfR)$ to  $S^a_\bfR$ is independent of the index $a = 1, \ldots, \mathrm{dim}(\bf R)$, which is why  $A_{\bfR} $ carries no such index. We term gauge backgrounds of this form \emph{matter surface fluxes}. As long as $[S^a_\bfR] \in H^{2,2}_{\mathrm{vert}}(\hat{Y_4})$\footnote{This property is satisfied for most models discussed so far in the literature, an exception being \cite{Braun:2014pva}.}, also $A(\bfR) \in H^{2,2}_{\mathrm{vert}}(\hat{Y_4})$. The first example of such a flux has been given, at the level of cohomology, in \cite{Marsano:2011hv}, and this approach to fluxes has been developed systematically in \cite{Borchmann:2013hta}. 

\paragraph{$\mathbf{U(1)}$ Flux}
A second type of vertical gauge background arises in the presence of extra independent rational sections $S_X$. Via the Shioda map, each independent such section is associated with a divisor class $U_X \in \mathrm{CH}^1(\hat{Y_4})$ such that $C_3 = {\mathbb A}_A \wedge [U_X] + \ldots$ defines a $U(1)_A$ gauge potential ${\mathbb A}_X$. Given any divisor class ${ F} \in \mathrm{ CH}^1(B_3)$ the object
\[ A_X ( F) = { F} \cdot U_X \in \mathrm{CH}^2(\hat{Y_4}) \label{AX-general} \]
defines a gauge background which is automatically vertical and respects the non-Abelian gauge algebras. At the level of fluxes these backgrounds have been introduced in \cite{Grimm:2010ez,Krause:2011xj} (see also \cite{Braun:2011zm}).

\paragraph{Cartan Flux}
Third, we can consider the gauge background for the Cartan $U(1)_{i_I}$ by restricting the resolution divisor $E_{i_I}$ to any curve $C \subseteq \Delta_I$. This defines the element
\[ A_{i_I}(C) = \left. E_{i_I}\right|_{C} \in \mathrm{CH}^2(\hat{Y_4}) \, . \label{AiC-def} \]
While automatically vertical, this gauge background clearly breaks the gauge algebra $\mathfrak{g}_I$. If the curve class $[C]$ is in the image of $H^{1,1}(B_3)$ restricted to $\Delta_I$, the class $[A_{i_I}(C)]$ lies in $H^{2,2}_\mathrm{vert} (\hat{Y_4})$. Indeed, if we denote by $[D_C] \in H^{1,1}(B_3)$ the divisor class such that $[C] = [\Delta_I] \cdot [D_C]$, then  $[A_{i_I}(C)] = [D_C] \cdot [E_{i_I}]$. More generally, the curve class $[C] \in H_{2}(\Delta_I)$ might contain components trivial in $H_2(B_3)$, but this still defines a valid flux \cite{Mayrhofer:2013ara,Braun:2014pva}. In this case, $[A_{i_I}(C)]$ contains a component in $H^{2,2}_{\mathrm{rem}}(\hat{Y_4})$ \cite{Braun:2014xka}.

Let us stress again that these three types of vertical fluxes and their underlying elements in $\mathrm{CH}^2(\hat{Y_4})$ are the ones which exist for a generic choice of base $B_3$. However, they are not all independent due to a number of non-trivial relations within the Chow group, which descend to corresponding relations in $H^{2,2}(\hat{Y_4})$ via the cycle map. In particular, in \cite{Bies:2017-2} we prove that anomaly cancellation implies that the matter surface fluxes satisfy a set of linear relations in  $H^{2,2}(\hat{Y_4})$, thereby generalising previous observations in \cite{Lin:2016vus}. We furthermore exemplify that this relation holds at the level of the Chow group, and conjecture this to hold true more generally. 

An equivalent approach to classifying gauge backgrounds with $G_4 \in H^{2,2}_{\mathrm{vert}}(\hat{Y_4})$ is by systematically forming all intersections of two elements in $\mathrm{CH}^1(\hat{Y_4})$ and determining the linearly independent combinations whose cohomology classes satisfy verticality and gauge invariance. This approach requires finding a basis of $H^{2,2}_{\mathrm{vert}}(\hat{Y_4})$ by analysing the relations in the intersection ring. The first systematic such classification for F-theory fibrations over general base has been carried out in \cite{oai:arXiv.org:1202.3138}. This approach is widely used in the literature, including \cite{Krause:2011xj,oai:arXiv.org:1111.1232,Cvetic:2013uta,Bizet:2014uua,Cvetic:2015txa,Lin:2015qsa,Lin:2016vus}.  

\subsection{Intersection Theory for Vertical Gauge Backgrounds}  \label{subsec:NewStrategyForMasslessSpectra}

We now describe how to perform the programme outlined in section \autoref{sec:FromC3toL}  of extracting the line bundle from the various types of gauge backgrounds. The advantage of representing the gauge background by the Chow group class of an algebraic cycle is that transverse intersection products can be evaluated very intuitively. Such transverse intersections factorise into a piece in the fibre and a piece in the base, and the projection onto the base is easily evaluated. Non-transverse intersections, on the other hand, can be expressed as sums or differences of transverse intersections by making use of the relations \emph{within the Chow group} between the Chow classes representing the gauge background. It is here where the formalism of \autoref{F-CompGen1} becomes most crucial: The fact that the refined cycle map (\ref{refcyclemap}) is defined at the level of Chow groups guarantees that using such relations within $\mathrm{CH}^2(\hat{Y_4})$ does not alter the gauge background and is hence permissible in evaluating the intersections.

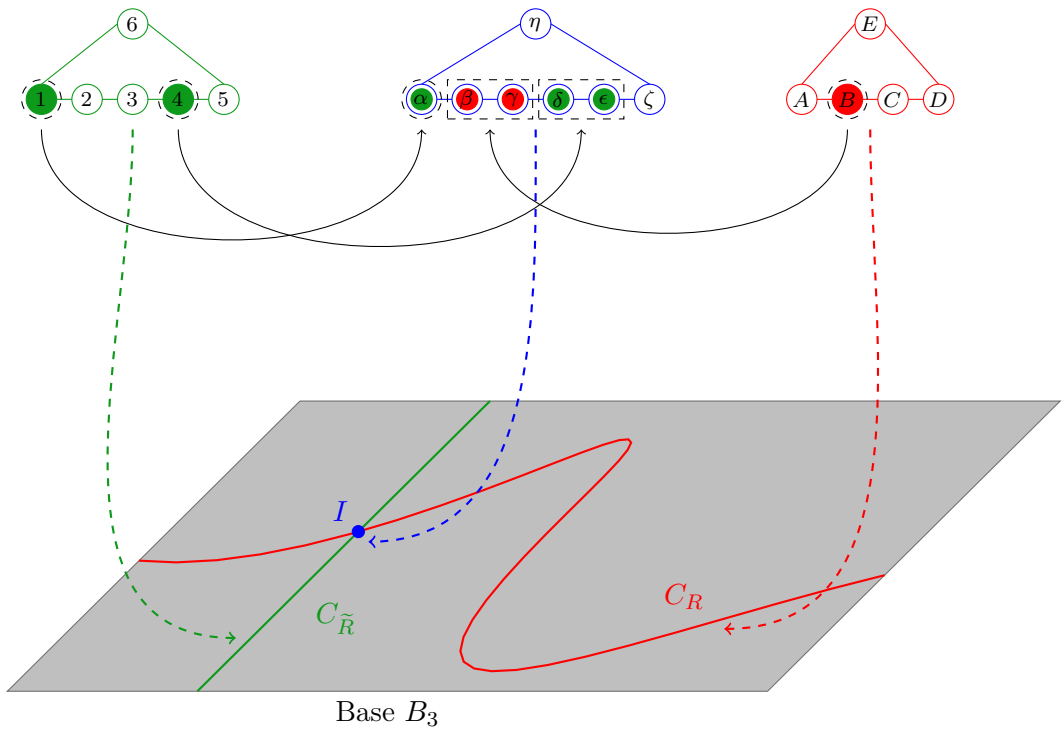
\begin{figure}[tb]
\begin{center}
\begin{tikzpicture}[scale=2]

    \def\l{5}
    \def\h{2}
    
    \draw (0,0,0)--(0,0,\l)--(\l,0,\l)--(\l,0,0)--(0,0,0) [opacity = 0.5, fill=gray];
    \draw[black] (\l/2,0,\l) node[below] {Base $B_3$};    
        
    \draw [darkgreen, thick,  domain=-\l/2:\l/2, samples=40] plot ({\l/4}, {0}, {\l/2-\x} );
    \draw[darkgreen] (9*\l/24,0,3*\l/4) node[left] {$C_{\tilde{R}}$};    

    \draw [red, thick,  domain=0:\l, samples=40] plot ({\x + 0.8 * sin(360*(\x/\l))}, {0}, {0.5*\x + 0.25*\l + 1.5 * cos( 480*(\x/\l)} );
    \draw[red] (10*\l/12,0,3*\l/4) node[above left] {$C_{R}$};

    \draw[darkgreen, fill] (-\l/4 - 0.45,\h,0) circle(0.1);
    \node[anchor=center] at (-\l/4 - 0.45,\h,0) (P1Cm) {$_1$};
    \draw[darkgreen] (-\l/4 - 0.35,\h,0)--(-\l/4-0.25,\h,0);
    \draw[darkgreen] (-\l/4 - 0.15,\h,0) circle(0.1);
    \node[anchor=center] at (-\l/4 - 0.15,\h,0) (P2Cm) {$_2$};
    \draw[darkgreen] (-\l/4 - 0.05,\h,0)--(-\l/4+0.05,\h,0);
    \draw[darkgreen] (-\l/4 + 0.15,\h,0) circle(0.1);
    \node[anchor=center] at (-\l/4 + 0.15,\h,0) (P3Cm) {$_3$};
    \draw[darkgreen] (-\l/4 + 0.25,\h,0)--(-\l/4+0.35,\h,0);
    \draw[darkgreen,fill] (-\l/4 + 0.45,\h,0) circle(0.1);
    \node[anchor=center] at (-\l/4 + 0.45,\h,0) (P4Cm) {$_4$};
    \draw[darkgreen] (-\l/4 + 0.55,\h,0)--(-\l/4+0.65,\h,0);
    \draw[darkgreen] (-\l/4 + 0.75,\h,0) circle(0.1);
    \node[anchor=center] at (-\l/4 + 0.75,\h,0) (P5Cm) {$_5$};
    \draw[darkgreen] (-\l/4 + 0.25, \h + 0.5,0)--(-\l/4+0.75,\h+0.1,0);
    \draw[darkgreen] (-\l/4 + 0.05, \h + 0.5,0)--(-\l/4-0.45,\h+0.1,0);
    \draw[darkgreen] (-\l/4+0.15,\h + 0.5,0) circle(0.1);
    \node[anchor=center] at (-\l/4 + 0.15,\h + 0.5,0) (P6Cm) {$_6$};
    
    \draw[red] (3*\l/4 - 0.45,\h,0) circle(0.1);
    \node[anchor=center] at (3 * \l/4 - 0.45,\h,0) (P1C) {$_A$};
    \draw[red] (3*\l/4 - 0.35,\h,0)--(3*\l/4-0.25,\h,0);    
    \draw[red,fill] (3*\l/4 - 0.15,\h,0) circle(0.1);
    \node[anchor=center] at (3 * \l/4 - 0.15,\h,0) (P2C) {$_B$};
    \draw[red] (3*\l/4 - 0.05,\h,0)--(3*\l/4+0.05,\h,0);
    \draw[red] (3*\l/4 + 0.15,\h,0) circle(0.1);
    \node[anchor=center] at (3 * \l/4 + 0.15,\h,0) (P3C) {$_C$};
    \draw[red] (3*\l/4 + 0.25,\h,0)--(3*\l/4+0.35,\h,0);
    \draw[red] (3*\l/4 + 0.45,\h,0) circle(0.1);
    \node[anchor=center] at (3 * \l/4 + 0.45,\h,0) (P4C) {$_D$};
    \draw[red] (3*\l/4 + 0.1, \h + 0.5,0)--(3*\l/4+0.45,\h+0.1,0);
    \draw[red] (3*\l/4 - 0.1, \h + 0.5,0)--(3*\l/4-0.45,\h+0.1,0);    
    \draw[red] (3*\l/4,\h + 0.5,0) circle(0.1);
    \node[anchor=center] at (3 * \l/4,\h+0.5,0) (P5C) {$_E$};

    \draw[blue] (\l/4 - 0.45,\h,0) circle(0.1);
    \draw[fill, darkgreen] (\l/4 - 0.45,\h,0) circle(0.07);
    \node[anchor=center] at (\l/4 - 0.45,\h,0) (P1I) {$_\alpha$};
    \draw[blue] (\l/4 - 0.35,\h,0)--(\l/4-0.25,\h,0);
    \draw[blue] (\l/4 - 0.15,\h,0) circle(0.1);
    \draw[fill, red] (\l/4 - 0.15,\h,0) circle(0.07);
    \node[anchor=center] at (\l/4 - 0.15,\h,0) (P2I) {$_\beta$};
    \draw[blue] (\l/4 - 0.05,\h,0)--(\l/4+0.05,\h,0);
    \draw[blue] (\l/4 + 0.15,\h,0) circle(0.1);
    \draw[fill, red] (\l/4 + 0.15,\h,0) circle(0.07);
    \node[anchor=center] at (\l/4 + 0.15,\h,0) (P3I) {$_\gamma$};
    \draw[blue] (\l/4 + 0.25,\h,0)--(\l/4+0.35,\h,0);
    \draw[blue] (\l/4 + 0.45,\h,0) circle(0.1);
    \draw[fill, darkgreen] (\l/4 + 0.45,\h,0) circle(0.07);
    \node[anchor=center] at (\l/4 + 0.45,\h,0) (P4I) {$_\delta$};
    \draw[blue] (\l/4 + 0.55,\h,0)--(\l/4+0.65,\h,0);
    \draw[blue] (\l/4 + 0.75,\h,0) circle(0.1);
    \draw[fill, darkgreen] (\l/4 + 0.75,\h,0) circle(0.07);
    \node[anchor=center] at (\l/4 + 0.75,\h,0) (P5I) {$_\epsilon$};
    \draw[blue] (\l/4 + 0.85,\h,0)--(\l/4+0.95,\h,0);
    \draw[blue] (\l/4 + 1.05,\h,0) circle(0.1);
    \node[anchor=center] at (\l/4 + 1.05,\h,0) (P6I) {$_\zeta$};
    \draw[blue] (\l/4 + 0.4, \h + 0.5,0)--(\l/4+1.05,\h+0.1,0);
    \draw[blue] (\l/4 + 0.2, \h + 0.5,0)--(\l/4-0.45,\h+0.1,0);
    \draw[blue] (\l/4+0.3,\h + 0.5,0) circle(0.1);
    \node[anchor=center] at (\l/4 + 0.3,\h+0.5,0) (P7I) {$_\eta$};

    \draw[fill, blue] (\l/4,0,0.45*\l) circle (0.04) node[above left] {$I$};
    
    \node[anchor=south] at (-\l/4 + 0.15,\h - 0.2,0) (P1sCm) {};
    \node[anchor=north] at (\l/4, 0.1, 5*\l/6) (Cm) {};
    \draw[thick, darkgreen, dashed] (P1sCm) edge[out=-90,in=180,->] (Cm);

    \node[anchor=south] at (3*\l/4,\h-0.2,0) (P1sC) {};
    \node[anchor=north] at (10*\l/12,0,3*\l/4) (C) {};
    \draw[thick, red, dashed] (P1sC) edge[out=-90,in=0,->] (C);

    \node[anchor=south] at (\l/4+0.3,\h - 0.2,0) (P1sI) {};
    \node[anchor=north] at (\l/4,0,0.45*\l) (I) {};
    \draw[thick, blue, dashed] (P1sI) edge[out=-90,in=0,->] (I);

    \draw[dashed, black] (\l/4 - 0.45,\h,0) circle(0.13);
    \draw[dashed, black] (-\l/4 - 0.45,\h,0) circle(0.13);
    \draw[dashed, black] (-\l/4 + 0.45,\h,0) circle(0.13);
    \draw[dashed, black] (3*\l/4 - 0.15,\h,0) circle(0.13);
    \draw[rectangle, dashed] (\l/4 - 0.28,\h-0.13,0) rectangle (\l/4 + 0.28,\h+0.13,0);
    \draw[rectangle, dashed] (\l/4 + 0.32,\h-0.13,0) rectangle (\l/4 + 0.88,\h+0.13,0);
    
    \node[anchor=south] at (-\l/4 - 0.45,\h - 0.2,0) (greenP11) {};
    \node[anchor=south] at (-\l/4 + 0.45,\h - 0.2,0) (greenP12) {};
    \node[anchor=south] at (\l/4 - 0.45,\h - 0.2,0) (bgP11) {};
    \node[anchor=south] at (\l/4,\h - 0.2,0) (brP1s) {};
    \node[anchor=south] at (\l/4 + 0.6,\h - 0.2,0) (bgP1s) {};
    \node[anchor=south] at (3*\l/4 - 0.15,\h - 0.2,0) (redP11) {};
    
    \draw[black] (greenP11) edge[out=-90,in=-90,->] (bgP11);
    \draw[black] (greenP12) edge[out=-90,in=-90,->] (bgP1s);
    \draw[black] (redP11) edge[out=-90,in=-90,->] (brP1s);
    
\end{tikzpicture}
\end{center}
\caption{Schematic idea for the identification of the massless spectra of matter surface fluxes.}
\label{figure-0}
\end{figure}

For the matter surface fluxes of the form (\ref{def-mattersurfaceflux}), we therefore proceed as follows:
\begin{enumerate}
 \item Matter surface flux: \\
      Consider a matter curve $C_{\tilde\bfR} \subseteq B_3$ for some representation $\tilde\bfR$, a matter surface $S^a(\tilde\bfR)$ and the associated matter surface flux $A(\tilde \bfR) \in \mathrm{CH}^2(\hat{Y_4})$. It is given by a formal linear sum of $\mathbb{P}^1$- fibrations over $C_{\tilde{\bfR}}$,  depicted in green colour in \autoref{figure-0}.
 \item State: \\ 
      Next we consider a state over the matter curve $C_{{\bfR}}$. Such a state is encoded by a matter surface $S^{a}_\bfR$, whose fibre structure 
      in \autoref{figure-0} is indicated in red.
 \item 'Intersection' of $S^{a}_\bfR$ and $A(\tilde\bfR)$: \\
      We assume that the curves $C_{\tilde{\bfR}}$ and $C_\bfR$ intersect \emph{transversely} in $B_3$. We discuss the case of non-transverse intersections below. \\ 
      For simplicity we assume in \autoref{figure-0} that the two curves intersect in one point $I$ only, with multiplicity $m ( C_{\tilde{\bfR}} \cap C_\bfR, I )$.\footnote{Note that in general multiple such intersection points will exist. Then our strategy has to be repeated for every intersection point.} From knowledge of the splittings in the fibre we have
      \[ \left. A (\tilde \bfR) \right|_{I} = \alpha + \delta + \epsilon, \qquad \left. S^a_{{\bfR}} \right|_{I} = \beta + \gamma \, . \]
      Consequently we can compute the intersection number $n_f$ in the fibre over $I$ as
      \[ n_f \left( I \right) = \left( \left. A (\tilde \bfR) \right|_{I} \right) \cdot \left( \left. S^{a}_\bfR \right|_{I} \right) = \left( \alpha + \delta + \epsilon \right) \left( \beta + \gamma \right) \, . \label{MSFprojform-transverse} \]
      We can therefore consider the divisor
      \[ D\left(S^a_\bfR, A(\tilde\bfR)\right) = \pi_{\bfR \ast} \left(S^a_\bfR \cdot_{\iota_{\bfR,a}} A(\tilde\bfR) \right) = \underbrace{\left( n_f \left( I \right) \cdot m \left( C_{\tilde{\bfR}} \cap C_\bfR, I \right) \right)}_{:= m_I } \cdot I \in \mathrm{CH}^1 \left( C_\bfR \right) \label{intersectionMSF} \]
      and its associated line bundle 
      \[ L\left(S^a_\bfR, A(\tilde\bfR)\right) = \mathcal{O}_{C_\bfR} \left( m_I \cdot I \right) \, . \]
      Then the sheaf cohomologies 
      \[ H^i\left(C_\bfR,  L\left(S^a_\bfR, A(\tilde\bfR)\right)  \otimes \sqrt{K_{C_\bfR}}\right) \]
      count the massless excitations of the state $S^a_\bfR$ in the presence of $A(\tilde\bfR)$.
 \item 'Non-Transverse-Intersections': \\
      If the curves $C_{\tilde{\bfR}}$ and $C_\bfR$ do not intersect transversely, then we can use linear relations among the Chow classes representing the various gauge backgrounds to exchange the specific $A(\tilde\bfR)$ under consideration by a linear combination of other $A(\tilde\bfR)$, for which the relevant intersections are indeed transverse. We will show this in more detail in an example in \autoref{subsec:ExampleMasslessSpectrumOfUniversalFlux}.
\end{enumerate}

For a $U(1)_X$ gauge background described by $A_X(F) = F \cdot U_X$, a very similar logic has already been applied in \cite{Bies:2014sra} to deduce the relevant line bundles on the base. Consider a state in representation $\bfR$ with $U(1)_X$ charge $q_X(\bfR)$. This state is associated with a matter surface $S_\bfR $ over a curve $C_\bfR$ on the base $B_3$.\footnote{Note that we are suppressing the superscript $a$ in $S_\bfR $ since the $U(1)_X$ background is gauge invariant.} We are also introducing the notation
\[ \iota_{C_\bfR}: C_\bfR \hookrightarrow B_3 \]
for the embedding of the curve $C_\bfR$ into $B_3$. Then the line bundle on $C_\bfR$ to which the zero modes couple is given by $L(S_\bfR,A_X(F))= {\cal O}_{C_\bfR}(D(S_\bfR,A_X(F)))$ with
\[ D(S_\bfR,A_X(F)) = \pi_{\bfR \ast} \left(S_\bfR \cdot_{\iota_{\bf R}} A_X(F) \right) = q_X(\bfR) \left( C_\bfR \cdot_{\iota_{C_\bfR}} F  \right)     \in \mathrm{CH}^1(C_\bfR) \, , \label{U1Xint-gen} \]
and the number of zero modes is counted by the dimensions of $H^i(C_\bfR, L(S_\bfR,A_X(F)) \otimes \sqrt{K_{C_\bfR}})$. This result follows from the fact that, as in the construction above, the intersection between the gauge background $A_X(F)$ and the matter surface $S_\bfR$ factors into an intersection in the fibre and in the base.
Projecting onto $C_\bfR$ gives a multiplicity, which by construction of $U_X$ is exactly the $U(1)_X$ charge $q_X(\bfR)$. The term in brackets in (\ref{U1Xint-gen})  describes the intersection on the base.

As a somewhat special case this reasoning can be applied to the Cartan gauge backgrounds (\ref{AiC-def}), where we now invoke the embedding of the matter curve $\iota_{C_\bfR}: C_\bfR \hookrightarrow \Delta_I$ directly into the divisor $\Delta_I$ wrapped by the 7-brane stack to obtain
\[ \pi_{\bfR \ast} \left(S^a_\bfR \cdot_{\iota_{\bf R}} E_{i_I}|_C \right) = \beta^a(\bfR)_{i_I} \left( C_\bfR \cdot_{\iota_{C_\bfR}} C  \right)     \in \mathrm{CH}^1(C_\bfR) \, . \label{Cartanfluxintersection} \]
Here the weight $\beta^a(\bfR)_{i_I}$  is of course by definition  the charge of the state under the Cartan $U(1)_{i_I}$.

We have focused here on the geometric interpretation of the Chow group elements representing the gauge background, for instance as a matter surface flux. Equivalently, one can evaluate the intersection theoretic pairing  if one explicitly presents the vertical gauge data via the intersection of two elements in $\mathrm{CH}^1(\hat{Y_4})$, as remarked at the end of the previous section. The computations within the Chow ring on $\hat{Y_4}$ are significantly simplified if it is possible to express these intersections as the pullback of intersections of elements of $\mathrm{CH}^1(\hat{X_\Sigma})$ of a toric ambient space $X_\Sigma$ of $\hat{Y_4}$. As stressed after (\ref{refcyclemap}) and further in  \autoref{subsec:ReviewChowGroupsAndG4Fluxes}, in this case we can use homological relations on $X_\Sigma$ to evaluate the intersection product. We will encounter this explicitly in the concrete examples studied in the remainder of this article.

\section{F-Theory GUT-Models with \texorpdfstring{$\mathbf{SU ( 5 ) \times U ( 1 )_X}$}{SU(5) x U(1)}-Symmetry} \label{sec:ToricFTheoryGUTModels}

Our next goal is to demonstrate our formalism of computing the line bundles and their sheaf cohomology groups in an explicit example. We shall design the example to be as simple as possible while at the same time exhibiting all ingredients of our general discussion so far. In order to exemplify the role of Abelian fluxes of type (\ref{AX-general}), we need at least one extra rational section. The arguably simplest type of such fibrations is obtained by a $U(1)$ restricted Tate model \cite{Grimm:2010ez}. To study the behaviour of massless matter charged also under a non-Abelian gauge algebra, we model an extra non-Abelian gauge factor, the simplest class being represented by Tate models with an $I_n$ singularity. The existence of additional non-trivial vertical gauge fluxes in such models requires the non-Abelian gauge group to be at least $SU(5)$ \cite{oai:arXiv.org:1202.3138}. In this sense a $U(1)$ restricted Tate model with gauge group $G = SU ( 5 ) \times U ( 1 )_X$ is indeed minimal. The top describing the resolved fibre of this model was originally introduced in \cite{Krause:2011xj}. We will now provide a brief review on the topic. The interested reader is referred to the above references for further details.

\subsection{\texorpdfstring{$\mathbf{SU ( 5 ) \times U ( 1 )_X}$}{SU(5) x U(1)} Fibration} \label{subsec:SpecialFTheoryGUTModel}

Our starting point is a Tate model over a smooth base $B_3$, described as the vanishing locus $V(P_T)$ of a hypersurface equation
\[ P_T   = y^2 + a_1  x y z + a_3  y z^3 - x^3 - a_2  x^2 z^2 - a_4  x z^4 - a_6  z^6 \]
in an ambient space $X_5$. This ambient space $X_5$ is given by a fibration of $\mathbb P_{2,3,1}$ over $B_3$. The homogeneous coordinates on each $\mathbb{P}_{2,3,1}$  fibre are $[ x \colon y \colon z ]$ and the $a_i$ are sections of the $i$-th power of the anti-canonical bundle $\overline{K}_{B_3}$ of the base. 
We are considering fibrations over base spaces $B_3$ compatible with $\hat Y_4$ being Calabi-Yau. In particular this implies $H^1(B_3,\mathbb Q) = 0$. 
In addition, we are restricting ourselves in the sequel to $B_3$ without torsional 1-cycles, \ie $H^1(B_3, \mathbb Z) = 0$. 
By this assumption, the divisor class group on $B_3$ coincides with $H^{1,1}(B_3)$. 

An $SU ( 5 ) \times U ( 1 )_X$ gauge symmetry is engineered by specialising the sections $a_i$ further to
\begin{equation}
  \begin{aligned}
      a_1 &= a_1, \qquad & a_2 &= a_{2,1} \, w, \qquad & a_3 &= a_{3,2}  \, w^2, \\ 
      a_4 &= a_{4,3} \, w^3, \qquad & a_6 &= 0 \, .
  \end{aligned}
\end{equation}
The discriminant component associated with the $SU(5)$ gauge group is
\[ W = V(w) := \{ w=0 \} \subseteq B_3 \,, \]
and setting $a_6 \equiv 0$ implements the extra section associated with $U(1)_X$. To resolve the singularities one introduces blowup coordinates $e_i$, $i=1, \ldots, 4$ and $s$. The coordinate $s$ can be associated with the second section  of this fibration \cite{Krause:2011xj}. The resolved fibration $\hat{\pi} \colon \hat{Y}_4 \twoheadrightarrow B_3$ can be described as hypersurface $V(P_T') \subseteq \hat{X_5}$, where $\hat{X_5}$ is a new ambient space and $P_T^\prime$ the so-called \emph{proper transform} of the Tate polynomial.
Explicitly it is given by
\begin{align}
\begin{split}
P_T^\prime &= y^2 s e_3 e_4 + a_1 x y z s + a_{3,2} y z^3 e_0^2 e_1 e_4 - x^3 s^2 e_1 e_2^2 e_3 \\
           &\quad - a_{2,1} x^2 z^2 s e_0 e_1 e_2 - a_{4,3} x z^4 e_0^3 e_1^2 e_2 e_4 \, ,
\end{split}
\end{align}
where due to the blow-ups $\hat\pi^* w = e_0 e_1 e_2 e_3 e_4$. The resolved fibre ambient space is itself toric with toric coordinates and weights as given in \autoref{TableWeights}.
\begin{table}
\begin{center}
\begin{tabular}{|c|c|cccc|cccc|}
\toprule
& $e_0$ & $e_1$ & $e_2$ & $e_3$ & $e_4$ & x & y & z & s \\
\hline \hline
$\overline{K}_{B_3}$ & $\cdot$ & $\cdot$ & $\cdot$ & $\cdot$ & $\cdot$ & 2 & 3 & $\cdot$ & $\cdot$ \\
W                    & 1 & $\cdot$ & $\cdot$ & $\cdot$ & $\cdot$ & $\cdot$ & $\cdot$ & $\cdot$ & $\cdot$ \\
\hline
$E_1$ & -1        & 1 & $\cdot$ & $\cdot$ & $\cdot$ & -1 & -1 & $\cdot$ & $\cdot$ \\
$E_1$ & -1        & $\cdot$ & 1 & $\cdot$ & $\cdot$ & -2 & -2 & $\cdot$ & $\cdot$ \\
$E_1$ & -1        & $\cdot$ & $\cdot$ & 1 & $\cdot$ & -2 & -3 & $\cdot$ & $\cdot$ \\
$E_1$ & -1        & $\cdot$ & $\cdot$ & $\cdot$ & 1 & -1 & -2 & $\cdot$ & $\cdot$ \\
\hline
Z & $\cdot$   & $\cdot$ & $\cdot$ & $\cdot$ & $\cdot$ & 2 &  3 & 1 & $\cdot$ \\
S & $\cdot$   & $\cdot$ & $\cdot$ & $\cdot$ & $\cdot$ & -1 & -1 & $\cdot$ & 1 \\
\bottomrule
\end{tabular}
\end{center} 
\caption{Toric coordinates and weights for the top describing the resolved fibre ambient space.} 
\label{TableWeights}
\end{table}
Its Stanley-Reisner ideal takes the form
\begin{align}
\begin{split} \label{SRideal}
I_{\text{SR}} \left( \text{top} \right) = & \left\{ xy, x e_0 e_3, x e_1 e_3, x e_4, y e_0 e_3, y e_1, y e_2, z s, z e_1 e_4, z e_2 e_4, \right. \\
                & \qquad \left. z e_3, s e_0, s e_1, s e_4, e_0 e_2, z e_4, z e_1, z e_2, s e_2, e_0 e_3, e_1 e_3 \right\} \, ,
\end{split}
\end{align}
which corresponds to resolution $T_{11}$ in \cite{Krause:2011xj}. 

We will denote by $E_i \in \mathrm{CH}^1(\hat{Y_4})$ the class of algebraic cycles rationally equivalent to the vanishing locus $V ( e_i ) \subseteq \hat{X_5}$.\footnote{We apply similar notations for the vanishing loci associated to the other homogeneous coordinates of the \emph{top}.} These classes correspond to the generators of the Cartan $U (1 )$ symmetries of $SU(5)$. We use $S$ and $Z$ to denote the classes of the extra rational section $V ( s )$ and zero section $V(z)$ in $\text{CH}^1 ( \hat{Y}_4 )$, respectively. These allow us to express the generator of the $U ( 1 )_X$ gauge symmetry as
\[ U_X := - \left( 5 (S- Z- \overline{K})  + 2 E_1 + 4 E_2 + 6 E_3 + 3 E_4 \right) \in \mathrm{CH}^1(\hat{Y_4}) \, . \]

Matter in representations $\mathbf{10}_{1}$, ${\mathbf{5}_{3}}$, ${\mathbf{5}_{-2}}$ and ${\mathbf{1}_{5}}$ localises on the following curves on $B_3$,
\begin{equation} \label{mattercurvesSU5equ}
  \begin{aligned}
     C_{\mathbf{10}_{1}} &= V \left(w, a_{1,0} \right), \qquad & C_{\mathbf{5}_{3}} &= V \left(w, a_{3,2} \right), \\
     C_{\mathbf{5}_{-2}} &= V \left(w, a_1 a_{4,3} - a_{2,1} a_{3,2} \right), \qquad & C_{\mathbf{1}_{5}} &= V \left(a_{4,3}, a_{3,2} \right) \, .
  \end{aligned}
\end{equation}
The subscripts denote the charges of the respective $SU(5)$ representations under the Abelian gauge group factor $U(1)_X$.

The matter surfaces $S^a(\mathbf{R}) \in \text{CH}^2 ( \hat{Y}_4 )$ over these matter curves can be obtained by analysing the fibre structure of $\hat{\pi} \colon \hat{Y}_4 \twoheadrightarrow B_3$ \cite{Krause:2011xj, oai:arXiv.org:1202.3138}. We briefly review this subject in \autoref{sec:FibreStructure}. In particular note that the $U ( 1 )_X$-charge of a state encoded by matter surface $S^a ( \mathbf{R} )$ is encoded in the intersection with $U_X$ in the fibre.

For explicit computation, it will be crucial to make use of the so-called \emph{linear relations} of the $SU ( 5 ) \times U ( 1 )_X$-top. There are three generators for these relations, namely
\begin{align}
\begin{split} \label{linearrelationsambient}
\mathcal{X} - \mathcal{Y} + \mathcal{Z} + \mathcal{E}_0 + \mathcal{E}_1 + \mathcal{E}_2 - \mathcal{W} + \overline{\mathcal{K}}_{B_3} &= 0 \in \text{CH}^1 ( 
      \hat{X_5} ) \\
-3 \mathcal{X} + 2 \mathcal{Y} - \mathcal{S} - \mathcal{E}_1 -2 \mathcal{E}_2 + \mathcal{E}_4 &= 0 \in \text{CH}^1 ( \hat{X_5} ) \\ 
2 \mathcal{X} - \mathcal{Y} - \mathcal{Z} + \mathcal{S} + \mathcal{E}_1 + 2 \mathcal{E}_2 + \mathcal{E}_3 - \overline{\mathcal{K}}_{B_3} &= 0 \in \text{CH}^1 
      ( \hat{X_5} ) \, .
\end{split}
\end{align}
Here $\mathcal{X} \in \text{CH}^1 ( \hat{X_5})$ is the class of algebraic cycles rationally equivalent to $V ( x ) \subseteq \hat{X_5}$, and similarly for the other divisors. In particular we have $ \mathcal{X} |_{\hat{Y_4}} \equiv X \in \text{CH}^1 ( \hat{Y}_4 )$,  this being the class of algebraic cycles in $\hat{Y}_4$ which are rationally equivalent to $ V ( x ) |_{\hat{Y}_4} = V ( x, P_T^\prime ) \subseteq \hat{Y}_4$.

\subsection{(Self-)Intersections of Matter Curves}

To evaluate expressions such as (\ref{intersectionMSF}) we need the (self-)intersections of the matter curves $C_{\mathbf{10}_{1}}$, $C_{\mathbf{5}_{3}}$ and $C_{\mathbf{5}_{-2}}$ in the $SU(5)$ stack divisor $W$. In anticipation of this application, let us derive the relevant results in this section.

Two curves $C_i, C_j \subseteq W$ can be regarded as divisors on $W$, \ie elements of $\mathrm{CH}^1(W)$, and hence their intersection product is a special case of the general intersection product within the Chow ring reviewed in \autoref{app:intersection}. Since it is clear that we will be using the canonical embedding $\iota_{C_i}: C_i \hookrightarrow W$, we will abbreviate $C_i \cdot_{\iota_{C_i}} C_j $ as simply $C_i \cdot  C_j$ and interpret this as the pullback of $C_j$ to $C_i$, \ie as an element in $\mathrm{CH}_0(C_i) \simeq \mathrm{CH}^1(C_i)$ given by
\[ C_i \cdot C_j = \sum_{k = 1}^{N}{m_k \cdot p_k} \in \mathrm{CH}_0(C_i) \, . \]
Here $N \in \mathbb{N}$ is a non-negative integer, $m_k \in \mathbb{Z}$ and $p_k \in C_j$ are points. To this element we associate the line bundle $ {\cal O}_W(C_j) |_{C_j}$, and the intersection number $|C_i \cdot C_j|$ is simply \cite{ATIT, cox2011toric}
\[ |C_i \cdot C_j | = \text{deg} \left( \left. \mathcal{O}_{W} \left( C_i \right) \right|_{C_j} \right) = \Mint_{C_j}{c_1 \left( \left. \mathcal{O}_{W} \left( C_i \right) \right|_{C_j} \right)} = \sum_{k = 1}^{N}{m_k} \, . \]
In this sense, the intersection $D = C_i \cdot C_j$ occurs over the points $p_k$ with multiplicities $m_k$. Note however that this notion of `intersection points' is only well-defined up to linear equivalence of divisors - pick a divisor $D^\prime \neq D$ but with $D^\prime \sim D$, then $D^\prime$ will denote different intersection points with different multiplicities.

\begin{figure}[tb]
\begin{center}
\begin{tikzpicture}[scale=1.7]

    \def\l{5}
    \def\h{2.5}
    
    \draw (0,0,0)--(0,\h,0);
    \draw (\l,0,0)--(\l,\h,0);
    \draw (0,0,\l)--(0,\h,\l);
    \draw (\l,0,\l)--(\l,\h,\l);
    \draw (0,0,0)--(0,0,\l)--(\l,0,\l)--(\l,0,0)--(0,0,0);
    \draw (0,\h,0)--(0,\h,\l)--(\l,\h,\l)--(\l,\h,0)--(0,\h,0);
    
    \draw (0,0,0)--(0,0,\l)--(\l,0,\l)--(\l,0,0)--(0,0,0) [opacity = 0.5, fill=gray];
    \draw[black] (\l/2,0,\l) node[below] {$W$};    
    
    \draw [red, thick,  domain=-\l/2:\l/2, samples=40] plot ({\l/2+\x}, {3*\h/4*sin(50*\x)*sin(50*\x)}, {\l/2-\x} );
    \draw[red] (0.8,\h/2,\l-0.8) node[above left] {$C_{\mathbf{1}_{5}}$};    
    
    \draw [darkgreen, thick,  domain=-\l/2:\l/2, samples=40] plot ({\l/4}, {0}, {\l/2-\x} );
    \draw[darkgreen] (5*\l/12,0,\l/4) node[above left] {$C_{\mathbf{10}_{1}}$};    
    
    \draw [blue, thick,  domain=-\l/2:\l/2, samples=40] plot ({\l/2-\x}, {0}, {\l/2} );
    \draw[blue] (3*\l/4,0,\l/2) node[below] {$C_{\mathbf{5}_{3}}$};    

    \draw [orange, thick,  domain=-1.4:1.4, samples=40] plot ({\l/4*1+\l/4*(1-\x)*(1+\x)}, {0}, {\l/2+\l/8*\x*(1+\x)});
    \draw[orange] (\l/2,0,7*\l/12) node[below] {$C_{\mathbf{5}_{-2}}$};    

    \draw[fill=black] (\l/4,0,3*\l/4) circle (0.04) node[above left] {$Y_1$ \tiny{(\textcolor{darkgreen}{$2$}, \textcolor{orange}{$\frac{1}{2}$})} };
    \draw[fill=black] (\l/4,0,\l/2) circle (0.04) node[above left] {$Y_2$ \tiny{(\textcolor{darkgreen}{$-1$}, \textcolor{blue}{$\frac{1}{3}$}, \textcolor{orange}{$2$})} };
    \draw[fill=black] (\l/2,0,\l/2) circle (0.04) node[below right] {$Y_3$ \tiny{(\textcolor{blue}{$\frac{2}{3}$}, \textcolor{orange}{$\frac{3}{2}$})}};
    
\end{tikzpicture}
\end{center}
\caption{Schematic picture of the intersection properties of the matter curves $C_{\mathbf{10}_{1}}$, $C_{\mathbf{5}_{3}}$, $C_{\mathbf{5}_{-2}}$ and $C_{\mathbf{1}_{5}}$ in the base space $B_3$. We indicate the self-intersection numbers of the matter curves in $W$ by the tiny coloured numbers attached to $Y_i$. The outer box is to represent the base space $B_3$.}
\label{figure-1}
\end{figure}
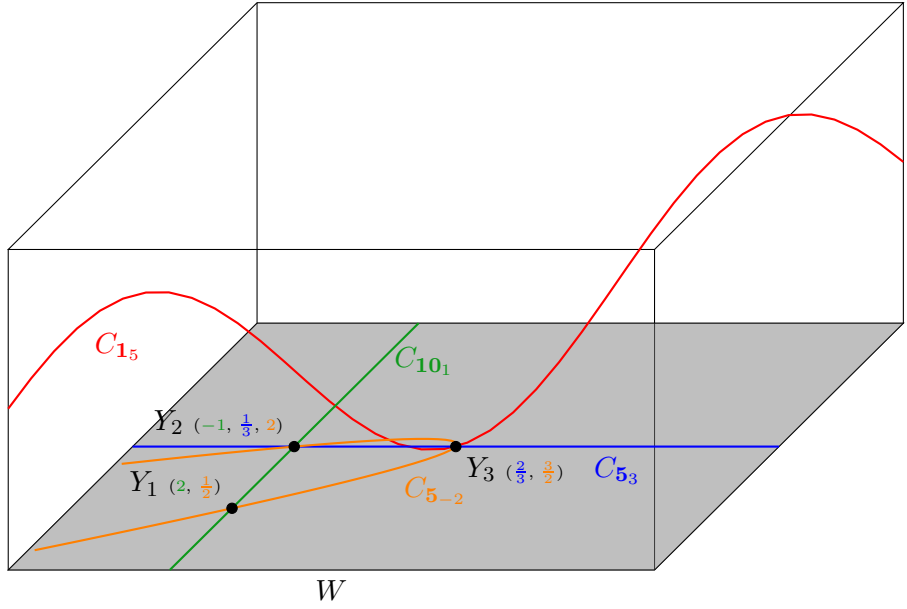

Bearing this in mind, let us work out intersection points and their associated multiplicities for the intersections of the matter curves $C_{\mathbf{10}_{1}}$, $C_{\mathbf{5}_{3}}$ and $C_{\mathbf{5}_{-2}}$. We first define
\[ Y_1 = V \left( w, a_{1,0}, a_{2,1} \right), \quad Y_2 = V \left( w, a_{1,0}, a_{3,2} \right), \quad Y_3 = V \left( w, a_{4,3}, a_{3,2} \right) \, . \label{Yidefinition} \]
Then in view of the explicit realisation (\ref{mattercurvesSU5equ}) of the curves we find for the transverse intersections
\begin{align}
\begin{split} \label{transverseY}
  C_{\mathbf{5}_{3}} \cdot C_{\mathbf{10}_{1}}&= Y_2 \, , \\
  C_{\mathbf{5}_{-2}} \cdot C_{\mathbf{10}_{1}} &= Y_1 + Y_2 \, , \\
  C_{\mathbf{5}_{-2}} \cdot C_{\mathbf{5}_{3}} &= Y_2 + Y_3  \, .
\end{split}
\end{align}
This structure is depicted schematically in  \autoref{figure-1}.
Note that the loci $Y_i$ are in general reducible and consist of the following number of points,
\begin{align}
\begin{split}
n_{Y_1} &=\left[ \overline{K}_{B_3} \right] \cdot_W \left[ 2 \overline{K}_{B_3} - {W} \right] \, , \\
n_{Y_2} &=\left[ \overline{K}_{B_3} \right] \cdot_W \left[ 3 \overline{K}_{B_3} - 2 {W} \right] \, , \\
n_{Y_3} &= \left[ 4 \overline{K}_{B_3} - 3 {W} \right] \cdot_W \left[ 3 \overline{K}_{B_3} - 2{W} \right] \, .
\end{split}
\end{align}

For the self-intersections we proceed very similarly. Let us start with $C_{\mathbf{10}_{1}}$ and note that
\begin{align}
\begin{split}
N_{C_{\mathbf{10}_{1}} \subseteq W} &= \left. \mathcal{O}_{W} \left( \overline{K}_{B_3} \right) \right|_{C_{\mathbf{10}_{1}}} \\
&\cong \left. \mathcal{O}_{W} \left( 2 \cdot \left( 2 \overline{K}_{B_3} - {W} \right) \right) \right|_{C_{\mathbf{10}_{1}}} \otimes \left. \mathcal{O}_{W} \left( -1 \cdot \left( 3 \overline{K}_{B_3} - 2 {W} \right) \right) \right|_{C_{\mathbf{10}_{1}}} \\
&\cong \mathcal{O}_{C_{\mathbf{10}_{1}}} \left( 2 Y_1 - Y_2 \right) \, .
\end{split}
\end{align}
Consequently, up to linear equlvalence, $C_{\mathbf{10}_{1}}$ self-intersects at $Y_1$, $Y_2$ with multiplicites $+2$ and $-1$ respectively. Similarly we have for $C_{\mathbf{5}_{3}}$
\begin{align}
\begin{split}
 N_{C_{\mathbf{5}_{3}} \subseteq W} &= \left. \mathcal{O}_{W} \left( 3 \overline{K}_{B_3} - 2 {W} \right) \right|_{C_{\mathbf{5}_{3}}} \\
 &\cong \left. \mathcal{O}_{W} \left( \frac{1}{3} \overline{K}_{B_3} \right) \right|_{C_{\mathbf{5}_{3}}} \otimes \left. \mathcal{O}_{W} \left( \frac{2}{3} \cdot \left( 4 \overline{K}_{B_3} - 3 {W}  \right) \right) \right|_{C_{\mathbf{5}_{3}}} \\
 &\cong \mathcal{O}_{C_{\mathbf{5}_{3}}} \left( \frac{1}{3} Y_2 + \frac{2}{3} Y_2 \right)
\end{split}
\end{align}
and for $C_{\mathbf{5}_{-2}}$
\begin{align}
\begin{split}
N_{C_{\mathbf{5}_{-2}} \subseteq W} &= \left. \mathcal{O}_{W} \left( 5 \overline{K}_{B_3} - 3 {W} \right) \right|_{C_{\mathbf{5}_{-2}}} \\
&\cong \left. \mathcal{O}_{W} \left( \frac{3}{2} \left( 3 \overline{K}_{B_3} - 2 {W} \right) \right) \right|_{C_{\mathbf{5}_{-2}}} \otimes \left. \mathcal{O}_{W} \left( \frac{1}{2} \cdot \overline{K}_{B_3} \right) \right|_{C_{\mathbf{5}_{-2}}} \\
&\cong \mathcal{O}_{C_{\mathbf{5}_{-2}}} \left( \frac{1}{2} Y_1 + 2 Y_2 + \frac{3}{2} Y_3 \right) \, .
\end{split}
\end{align}
The above manipulations manifestly involve rational coefficients. This is correct by our assumption that $B_3$ does not contain torsional divisors as explained at the beginning of \autoref{subsec:SpecialFTheoryGUTModel}.
Irrespective of the appearance of rational coefficients, note that by construction the line bundles here are integer quantised. 
We summarise our findings in \autoref{table-8}.

\begin{table}
\begin{center}
\begin{tabular}{|c|c|c|c||c|}
\toprule
& $Y_1$ & $Y_2$ & $Y_3$ & overall intersection number \\
\hline \hline
$\mathbf{10}_{1} \cdot \mathbf{10}_{1}$ & $2 \cdot n_{Y_1}$ & $\left(-1 \right) \cdot n_{Y_2}$ & $\cdot$ & $\Mint_{W}{\left[ \overline{K}_{B_3} \right] \wedge \left[ \overline{K}_{B_3} \right]}$ \\
$\mathbf{10}_{1} \cdot \mathbf{5}_{3}$ & $\cdot$ & $1 \cdot n_{Y_2}$ & $\cdot$ & $\Mint_{W}{\left[ \overline{K}_{B_3} \right] \wedge \left[ 3 \overline{K}_{B_3} - 2 {W} \right]}$ \\
$\mathbf{10}_{1} \cdot \mathbf{5}_{-2}$ & $1 \cdot n_{Y_1}$ & $1 \cdot n_{Y_2}$ & $\cdot$ & $\Mint_{W}{\left[ \overline{K}_{B_3} \right] \wedge \left[  5 \overline{K}_{B_3} - 3 {W} \right]}$ \\ 
\hline
$\mathbf{5}_{3} \cdot \mathbf{5}_{3}$ & $\cdot$ & $\frac{1}{3} \cdot n_{Y_2}$ & $\frac{2}{3} \cdot n_{Y_3}$ & $\Mint_{W}{\left[ 3 \overline{K}_{B_3} - 2 {W} \right] \wedge \left[ 3 \overline{K}_{B_3} - 2 {W} \right]}$ \\
$\mathbf{5}_{3} \cdot \mathbf{5}_{-2}$ & $\cdot$ & $1 \cdot n_{Y_2}$ & $ 1 \cdot n_{Y_3}$ & $\Mint_{W}{\left[ 5 \overline{K}_{B_3} - 3 {W} \right] \wedge \left[ 3 \overline{K}_{B_3} - 2 {W} \right]}$ \\
\hline
$\mathbf{5}_{-2} \cdot \mathbf{5}_{-2}$ & $\frac{1}{2} \cdot n_{Y_1}$ & $2 \cdot n_{Y_2}$ & $\frac{3}{2} \cdot n_{Y_3}$ & $\Mint_{W}{\left[ 5 \overline{K}_{B_3} - 3 {W} \right] \wedge \left[ 5 \overline{K}_{B_3} - 3 {W} \right]}$ \\
\bottomrule
\end{tabular}
\end{center}
\caption{Intersection points and intersection multiplicites (up to linear equivalence) of the matter curves $C_{\mathbf{10}_{1}}$, $C_{\mathbf{5}_{3}}$ and $C_{\mathbf{5}_{-2}}$ in $W$.}
\label{table-8}
\end{table}

\subsection{Vertical and Gauge Invariant Matter Surface Fluxes over Matter Curves} \label{subsec:VerticalAndGaugeInvariantCurveSupportFluxesOverMatterCurves}

We now turn to constructing the matter surface fluxes (\ref{def-mattersurfaceflux})  over the matter curves $C_{\mathbf{10}_{1}}$, $C_{\mathbf{5}_{3}}$, $C_{\mathbf{5}_{-2}}$ and $C_{\mathbf{1}_{5}}$, beginning with the flux associated to one of the matter surfaces $S^a(\mathbf{10}_{1})$. The 2-cycle $A(\mathbf{10}_{1})$ obtained by subtracting suitable correction terms will be a linear combination of rational curves fibred over  $C_{\mathbf{10}_{1}}$. These rational curves arise from the splitting of the fibres of the resolution divisors $E_i \subseteq \hat{Y}_4$, $i=1,\ldots 4,$ over $C_{\mathbf{10}_{1}}$. For example, the fibre of $E_2$ splits into two rational curves $\mathbb{P}_{24}^1$ and $\mathbb{P}_{2B}^1$. The 2-cycles obtained by fibring these over $C_{\mathbf{10}_{1}}$ will be denoted by  $\mathbb{P}_{24}^1 ( \mathbf{10}_{1} )$  and $\mathbb{P}_{2B}^1 ( \mathbf{10}_{1} )$, respectively. An overview of such fibral curves and their associated 2-cycles is given in \autoref{sec:FibreStructure}. More details can be found in \cite{Krause:2011xj, oai:arXiv.org:1202.3138}.

Explicitly, evaluating the general expression (\ref{def-mattersurfaceflux}) following the logic of \cite{Borchmann:2013hta}, we find the complex 2-cycle $A ( \mathbf{10}_{1} ) \in \text{CH}^2 ( \hat{Y}_4 )$ given by
\[ \label{10flux1}
A \left( \mathbf{10}_{1} \right) = S^{a} \left( \mathbf{10}_{1} \right) + {\beta}^a(\mathbf{10})^T \cdot C^{-1} \cdot \left( \begin{array}{c} \mathbb{P}_{14}^1 \left( \mathbf{10}_{1} \right) \\ \mathbb{P}_{24}^1 \left( \mathbf{10}_{1} \right) + \mathbb{P}_{2B}^1 \left( \mathbf{10}_{1} \right) \\ \mathbb{P}_{3C}^1 \left( \mathbf{10}_{1} \right) \\ \mathbb{P}_{14}^1 \left( \mathbf{10}_{1} \right) + \mathbb{P}_{24}^1 \left( \mathbf{10}_{1} \right) + \mathbb{P}_{4D}^1 \left( \mathbf{10}_{1} \right) \end{array} \right) \, . \]
This expression uses the inverse of the $SU(5)$ Cartan matrix 
\[ C = \left( \begin{array}{cccc} 2 & -1 & 0 & 0 \\ -1 & 2 & -1 & 0 \\ 0 & -1 & 2 & -1 \\ 0 & 0 & -1 & 2 \end{array} \right) \, . \]
Note that the column vector appearing in (\ref{10flux1}) contains precisely those $\mathbb{P}^1$-fibrations over $C_{\mathbf{10}_{1}}$ which arise from restriction of $E_1$, $E_2$, $E_3$ and $E_4$ to $C_{\mathbf{10}_{1}}$. Since this procedure gives the same cycle class for all matter surfaces $S^{a} ( \mathbf{10}_{1} )$ we have dropped the superscript `$a$' in $A ( \mathbf{10}_{1})$. The result can be expressed as \footnote{The fractions $\frac{1}{5}$ originate from the \emph{inverse} of the Cartan matrix $C$.}
\[ A \left( \mathbf{10}_{1} \right) \left( \lambda \right) := A \left( \mathbf{10}_{1} \right) \left( 0, 0, \frac{2\lambda}{5}, -\frac{\lambda}{5}, \frac{\lambda}{5}, -\frac{2 \lambda}{5} \right) \in \mathrm{CH}^2(\hat{Y_4}) \, . \label{equ:flux10Final} \]
In this expression we are introducing a rational coefficient $\lambda \in \mathbb{Q}$ and apply the short-hand notation
\begin{align}
\begin{split}
A \left( \mathbf{10}_{1} \right) \left( a_0, a_1, a_2, a_3, a_4, a_5 \right) &= a_0 \cdot \mathbb{P}^1_{0A} \left( \mathbf{10}_{1} \right) + a_1 \cdot \mathbb{P}^1_{14} \left( \mathbf{10}_{1} \right) + a_2 \cdot \mathbb{P}^1_{24} \left( \mathbf{10}_{1} \right) \\
&+ a_3 \cdot \mathbb{P}^1_{2B} \left( \mathbf{10}_{1} \right) + a_4 \cdot \mathbb{P}^1_{3C} \left( \mathbf{10}_{1} \right) + a_5 \cdot \mathbb{P}^1_{4D} \left( \mathbf{10}_{1} \right) \, .
\end{split}
\end{align}
The coefficient $\lambda \in \mathbb{Q}$ has to be chosen subject to the flux quantisation condition (\ref{FW2}).

It is not too hard to repeat this analysis for matter surface fluxes over $C_{\mathbf{5}_{3}}$ and $C_{\mathbf{5}_{-2}}$. We list the relevant $\mathbb{P}^1$-fibrations and the corresponding short-hand notations in \autoref{sec:FibreStructure}. This enables us to state the result as
\begin{align}
\begin{split} \label{5flux32}
A \left( \mathbf{5}_{3} \right) \left( \lambda \right) &= A \left( \mathbf{5}_{3} \right) \left( 0, \frac{\lambda}{5}, \frac{2 \lambda}{5}, \frac{3 \lambda}{5}, -\frac{2 \lambda}{5}, -\frac{\lambda}{5} \right)\in \mathrm{CH}^2(\hat{Y_4}), \\
A \left( \mathbf{5}_{-2} \right) \left( \lambda \right) &= A \left( \mathbf{5}_{-2} \right) \left( 0, \frac{\lambda}{5}, \frac{2 \lambda}{5}, \frac{3 \lambda}{5}, -\frac{2 \lambda}{5}, - \frac{\lambda}{5} \right)\in \mathrm{CH}^2(\hat{Y_4}) \, ,
\end{split}
\end{align}
using analogous notation for the respective fibral curves in the order listed in (\ref{53cuvesplitting}) and  (\ref{52cuvesplitting}).
By similar arguments it can be shown that
\[ A \left( \mathbf{1}_{5} \right) \left( \lambda \right) = \lambda \,  \mathbb{P}^1_A \left( \mathbf{1}_{5} \right) = \lambda \,  V \left( P^\prime, s, a_{3,2}, a_{4,3} \right)\in \mathrm{CH}^2(\hat{Y_4}) \]
is both vertical and gauge invariant.\footnote{The fluxes described above are the ones which exist generically for every choice of $B_3$. 
In addition, there can be extra gauge invariant matter surface fluxes if $B_3$ has special properties, for instance if the matter curves are forced to split.}

As remarked at the very end of \autoref{subsec:NewStrategyForMasslessSpectra} it is sometimes convenient to express these cycles as elements $\mathcal{A} \in \text{CH}^2 ( \hat{X_5} )$ such that $\mathcal{A} |_{\hat{Y}_4} = A$. These take the form
\begin{align}
\begin{split}
\label{equ:MatterSurfaceFluxes-X5}
\mathcal{A} \left( \mathbf{10}_{1} \right) \left( \lambda \right) &= \frac{\lambda}{5} \cdot \left( 2 \mathcal{E}_1 - \mathcal{E}_2 + \mathcal{E}_3 - 2 \mathcal{E}_4 \right) \cdot \overline{\mathcal{K}}_{B_3} + \lambda \cdot \mathcal{E}_2 \cdot \mathcal{E}_4 \, , \\
\mathcal{A} \left( \mathbf{5}_{3} \right) \left( \lambda \right) &= \frac{\lambda}{5} \cdot \left( \mathcal{E}_1 + 2 \mathcal{E}_2 - 2 \mathcal{E}_3 - \mathcal{E}_4 \right) \cdot \left( 3 \overline{\mathcal{K}}_{B_3} - 2 \mathcal{W} \right) + \lambda \cdot \mathcal{E}_3 \cdot \mathcal{X} \, , \\
\mathcal{A} \left( \mathbf{5}_{-2} \right) \left( \lambda \right) &= \frac{\lambda}{5} \cdot \left( \mathcal{E}_1 + 2 \mathcal{E}_2 + 3 \mathcal{E}_3 - \mathcal{E}_4 \right) \cdot \left( 5 \overline{\mathcal{K}}_{B_3} - 3 \mathcal{W} \right) \\
& \qquad \qquad \qquad \qquad \qquad - \lambda \cdot \left( \mathcal{E}_3 \cdot \overline{\mathcal{K}}_{B_3} + \mathcal{E}_3 \cdot \mathcal{Y} - \mathcal{E}_3 \cdot \mathcal{E}_4 \right) \, , \\
\mathcal{A} \left( \mathbf{1}_{5} \right) \left( \lambda \right) &= \lambda \cdot \mathcal{S} \cdot \left( 3 \overline{\mathcal{K}}_{B_3} - 2 \mathcal{W} \right) - \lambda \cdot \mathcal{S} \cdot \mathcal{X} \, .
\end{split}
\end{align}
In writing down the equality for the flux $\mathcal{A} \left( \mathbf{5}_{-2} \right) \left( \lambda \right)$ we have used that 
on $\hat{X_5}$ 
\[ \mathbb{P}^1_{3H} \left( \mathbf{5}_{-2} \right) = V \left( P_T^\prime, e_3, a_{2,1} e_0 x z e_1 e_2 - a_{1,0} y \right) - V \left( P_T^\prime, e_3, e_4 \right) \, . \]
 Also note that the $U \left( 1 \right)_X$-flux -- as introduced in \cite{Krause:2011xj, oai:arXiv.org:1202.3138} -- can be expressed in terms of $\hat{X_5}$ as\footnote{Strictly speaking, we should call this $\mathcal{A}_X \left( \frac{1}{5} \mathcal{F} \right)$. The normalization factor of $\frac{1}{5}$ has been introduced for historical reasons.}  
\[ \mathcal{A}_X \left( \mathcal{F} \right) = - \frac{1}{5} \cdot \mathcal{F} \cdot  \left( 5 \mathcal{S} - 5 \mathcal{Z} - 5 \overline{\mathcal{K}}_{B_3} + 2 \mathcal{E}_1 + 4 \mathcal{E}_2 + 6 \mathcal{E}_3 + 3 \mathcal{E}_4 \right) \, . \label{equ:U1X-Flux-X5} \]
In this expression $F \in \text{CH}^1 ( B_3 )$ and $\mathcal{F} \in \text{CH}^1 ( \hat{X_5} )$ is chosen such that $\mathcal{F} |_{\hat Y_4} = \hat\pi^*F$.

\section{Line Bundles Induced by Matter Surface Fluxes} \label{sec:MasslessSpectraOfCurveSupportFluxesII}

In this section we take the first step towards computing the massless spectra in the presence of gauge backgrounds associated with the matter surface fluxes derived in \autoref{subsec:VerticalAndGaugeInvariantCurveSupportFluxesOverMatterCurves}. Following our general formalism we have to compute the intersection product of the 2-cycle classes defining the gauge background with the relevant matter surfaces, and then project the result onto the matter curves in question. There are two ways to perform these computations in practice. We begin, in \autoref{subsec:ExampleMasslessSpectrumOfUniversalFlux}, with the first approach, which has been outlined in \autoref{subsec:NewStrategyForMasslessSpectra}. Equivalently, though computationally more involved, we can perform the intersection computations with the help of the presentations (\ref{equ:MatterSurfaceFluxes-X5}) of the gauge data as pullbacks from the ambient space. In \autoref{subsec:ExampleOnMasslessSpectrumComputation} we will demonstrate this approach for the $U(1)_X$ gauge background and we present the detailed computation for all matter surface fluxes in \autoref{sec:MasslessSpectraOfCurveSupportFluxesTedious}. The two approaches match precisely.

\subsection{Massless Spectrum of \texorpdfstring{$\mathbf{A ( \mathbf{10}_{1} ) ( \lambda )}$}{A(10)(lambda)} via Transverse Intersections} \label{subsec:ExampleMasslessSpectrumOfUniversalFlux}

To derive the massless spectrum for the gauge background $\mathcal{A} ( \mathbf{10}_{1} ) ( \lambda )$, we first notice that the matter curve $C_{\mathbf{10}_{1}}$ intersects the matter curves $C_{\mathbf{5}_{3}}$, $C_{\mathbf{5}_{-2}}$ and $C_{\mathbf{1}_{5}}$ transversely. Hence to compute the massless spectra induced by ${A} ( \mathbf{10}_{1} ) ( \lambda )$ of states localised on the latter three matter surfaces, we can directly apply (\ref{intersectionMSF}). The transverse intersection numbers in the base have already been computed in (\ref{transverseY}), and it merely remains to determine the intersection numbers in the fibre over these intersection points. These intersection numbers are computed in \autoref{sec:FibreStructure} and specifically listed in \autoref{app_Yukawas}. As explained in more detail therein, due to a seeming $\mathbb{Z}_2$-orbifold singularity in the top over the Yukawa locus $Y_1$ (despite $\hat Y_4$ being smooth) some of the intersection numbers of the rational curves in the top over the locus $Y_1$ are in fact fractional, see (\ref{tab:IntersectionsOfFibralCurvesOverY1}). By use of the information displayed in \autoref{app_Yukawas} the intersection products (\ref{equ:MatterSurfaceFluxes-X5}) eventually take the form \footnote{Since their precise meaning is now clear, we omit the subscripts in the symbols for the projection and the intersection product.}
\begin{align}
\begin{split} \label{transverseintersections10-example}
 \pi_\ast \left(S_{\mathbf{5}_{3}} \cdot  A\left(\mathbf{10}_{1} \right)(\lambda) \right) &= \frac{2 \lambda}{5} \cdot Y_2 \, , \\
 \pi_\ast \left(S_{\mathbf{5}_{-2}} \cdot  A\left(\mathbf{10}_{1} \right)(\lambda) \right)&= \frac{2 \lambda}{5} \cdot Y_2 - \frac{3 \lambda}{5} Y_1 \, ,\\
  \pi_\ast \left(S_{\mathbf{1}_{5}} \cdot  A\left(\mathbf{10}_{1} \right)(\lambda) \right) &= 0 \, .
\end{split}
\end{align}

As an example consider the intersection in the second line. The 2-cycle ${A} ( \mathbf{10}_{1} ) ( \lambda)$ is given explicitly in (\ref{equ:flux10Final}) in terms of $\mathbb P^1$-fibrations over  the curve $C_{{\bf 10}_{1}}$. As for $S_{\mathbf{5}_{-2}}$, since the gauge background respects the $SU(5)$ gauge symmetry, we can pick a matter surface for any of the five states in $\mathbf{5}_{-2}$ as listed in (\ref{app:MatterSurfaces52}) of \autoref{app_Mattersurfaces}. For instance, take $S^{(4)}_{\mathbf{5}_{-2}} = \mathbb P^1_{3H}(C_{\mathbf{5}_{-2X}})$. Now, from (\ref{transverseY}) we read off that $C_{\mathbf{10}_{1}}$ and $C_{\mathbf{5}_{-2X}}$ intersect over the two point sets $Y_1$ and $Y_2$. We hence need to study the splitting of the fibres of ${A} ( \mathbf{10}_{1} ) ( \lambda )$ and of $S^{(4)}_{\mathbf{5}_{-2}}$ over these two loci and compute the intersection numbers in the fibre. The relevant fibre splittings over $Y_1$ are listed in (\ref{tab:Splitting10ToY1}), (\ref{tab:Splitting5M2ToY1}), the ones over $Y_2$ are listed in (\ref{tab:Splitting10ToY2}) and (\ref{tab:Splitting5M2ToY2}). For instance, over $Y_1$
\begin{align}
\begin{split}
\left.{A} \left( \mathbf{10}_{1} \right)\right|_{Y_1} =& \frac{2}{5} \mathbb P^1_{24}(Y_1) -  \frac{1}{5} \mathbb P^1_{23}(Y_1) +  \frac{1}{5} (\mathbb P^1_{34}(Y_1) + \mathbb P^1_{3J}(Y_1)) \\
& - \frac{2}{5}(\mathbb P^1_{24}(Y_1) + \mathbb P^1_{34}(Y_1)) \, ,  \\
\left.S^{(4)}_{\mathbf{5}_{-2}} \right|_{Y_1} =& \mathbb P^1_{3J}(Y_1) \, .
\end{split}
\end{align}
The only non-zero intersection numbers between the fibral curves involved, as tabulated in (\ref{tab:IntersectionsOfFibralCurvesOverY1}), are
\[ \left|\mathbb P^1_{3J}(Y_1) \cdot \mathbb P^1_{3J}(Y_1)\right| = -2, \qquad  \left|\mathbb P^1_{3J}(Y_1) \cdot \mathbb P^1_{34}(Y_1) \right|= 1 \, . \]
These are to be viewed as intersection numbers between two rational curves in the complex two dimensional top over $Y_1$. Summing everything up explains the term proportional to $Y_1$ in the second line of (\ref{transverseintersections10-example}). A similar analysis is to be performed over $Y_2$. 

The element $\frac{2 \lambda}{5} \cdot Y_2 - \frac{3 \lambda}{5} Y_1 \in \mathrm{CH}_0(C_{\mathbf{5}_{-2}}) \simeq \mathrm{Pic}^1(C_{\mathbf{5}_{-2}})$ defines a line bundle on $C_{\mathbf{5}_{-2}}$ given by
\[ L_{\mathbf{5}_{-2}} = {\cal O}_{C_{\mathbf{5}_{-2}}} \left(\frac{2 \lambda}{5} \, Y_2 - \frac{3 \lambda}{5}\, Y_1\right) \, . \]
This line bundle has the property that it cannot be obtained by restriction of another line bundle on $W$ to the curve $C_{\mathbf{5}_{-2}}$. This is equivalent to the statement that the divisor $\frac{2 \lambda}{5} Y_2 - \frac{3 \lambda}{5} Y_1$ does not arise as a complete intersection of $C_{\mathbf{5}_{-2}}$ with another divisor on $W$. This will be important when it comes to evaluating the sheaf cohomology groups associated to this line bundle counting the massless matter states on $C_{\mathbf{5}_{-2}}$.

The computation of $ \pi_\ast (S_{\mathbf{10}_{1}} \cdot  A(\mathbf{10}_{1})(\lambda) )$    is more involved due to the self-intersection of $C_{\mathbf{10}_{1}}$. 
However, as pointed out before, there exist non-trivial linear relations between the 2-cycle representing the gauge backgrounds which allow us to treat non-transverse intersections of this type for a linear combination of transverse ones. In \cite{Bies:2017-2} we explicitly show that in the model at hand these relations take the form 
\begin{align}
\begin{split} \label{relationsforAsinChow}
A \left( \mathbf{10}_{1} \right) \left(  \lambda \right) &= A \left( \mathbf{5}_{3} \right) \left( -\lambda \right) + A \left( \mathbf{5}_{-2} \right) \left( - \lambda \right) \, , \\
A \left( \mathbf{5}_{-2} \right) \left( \lambda \right) &= A_X \left( - \lambda W \right) \, , \\
A \left( \mathbf{1}_{5} \right) \left( \lambda \right) &= A_X \left( - \lambda \left[ 6 \overline{K}_{B_3} - 5 W \right] \right) + A \left( \mathbf{10}_{1} \right) \left( - \lambda \right) \, .
\end{split}
\end{align}
These are the manifestation of a more general set of relations between 2-cycle classes which in fact follow, at a general level, from absence of gauge and gravitational anomalies  in F-theory \cite{Bies:2017-2}. With the help of the first relation, it is readily verified that\footnote{The first relation in (\ref{relationsforAsinChow}) is equivalent to $\mathbb{P}^1_{3x} ( \mathbf{5}_{3} ) + \mathbb{P}^1_{3G} ( \mathbf{5}_{-2} ) = \mathbb{P}^1_{24} ( \mathbf{10}_{1} ) \in \text{CH}^2( \hat{Y}_4)$. Alternatively, this enables us to rewrite the relevant matter surface such that the intersection in question is given as sum of two transverse intersections, leading to the same result.}
\begin{align}
\begin{split}
 \pi_\ast \left(S_{\mathbf{10}_{1}} \cdot  A\left(\mathbf{10}_{1}\right)(\lambda) \right) &= \pi_\ast \left(S_{\mathbf{10}_{1}} \cdot A\left(\mathbf{5}_{3} \right)(-\lambda) \right) + \pi_\ast \left(S_{\mathbf{10}_{1}} \cdot  A\left(\mathbf{5}_{-2} \right)(-\lambda) \right) \\
  & = \frac{3 \lambda}{5} Y_1 - \frac{4 \lambda}{5} Y_2 \, ,
\end{split}
\end{align}
where the two transverse intersections appearing in the first line are computed analogously and the result is tabulated in \autoref{table-9}. This table also contains all other intersections between 2-cycles for matter surface backgrounds and matter surfaces. 

Another straightforward application of this formalism is a derivation of the line bundle induced by a Cartan flux on the matter curves. Such a gauge background can be expressed as
\[ A \left( C \right) := \sum_{i = 1}^{4}{a_i E_i|_C} \in \text{CH}^2 \left( \hat{Y_4} \right) \, , \label{equ:GeneralCartanFlux} \] 
which automatically satisfies the transversality conditions (\ref{TransverseChow1}) and (\ref{TransverseChow2})  but of course violates, by construction, (\ref{condition-gauge-weaker}). Here $C \in \mathrm{CH}_1(W)$ denotes any curve on the 7-brane divisor. For instance, the hypercharge flux takes the form \cite{Donagi:2008kj,Mayrhofer:2013ara,Braun:2014pva}
\[ A_Y \left( C \right) =  \left. \left(2 E_1 + 4 E_2 + 6 E_3 + 3 E_4\right) \right|_{C} \, . \]
The intersections in the fibre of $A \left( C \right) |_{C_{\mathbf{R}}}$ with the matter surfaces $S^{a}_{\mathbf{R}}$ are readily worked out from the results of \autoref{sec:FibreStructure} for the representations present in this model and explicitly confirm the result (\ref{Cartanfluxintersection}).

\begin{landscape}

\begin{table}[h]
\begin{center}
\begin{tabular}{|c||c|c|c|c|c|}
\toprule
& $A^X \left( F \right)$ & $A \left( \mathbf{10}_{1} \right) \left( \lambda \right)$ & $A \left( \mathbf{5}_{3} \right) \left( \lambda \right)$ & $A \left( \mathbf{5}_{-2} \right) \left( \lambda \right)$ & $A \left( \mathbf{1}_{5} \right) \left( \lambda \right)$ \\
\hline \hline
$C_{\mathbf{10}_{1}}$ & $\frac{1}{5} \left. F \right|_{C_{\mathbf{10}_{1}}}$ & $\frac{3\lambda}{5} Y_1 - \frac{4 \lambda}{5} Y_2$ & $\frac{2 \lambda}{5} Y_2$ & $- \frac{3 \lambda}{5} Y_1 + \frac{2 \lambda}{5} Y_2$ & 0 \\
$C_{\mathbf{5}_{3}}$ & $\frac{3}{5} \left. F \right|_{C_{\mathbf{5}_{3}}}$ & $\frac{2\lambda}{5} Y_2$ & $\frac{2 \lambda}{5} Y_2 - \frac{\lambda}{5} Y_3$ & $ \frac{\lambda}{5} Y_3 - \frac{4 \lambda}{5} Y_2$ & $- \lambda Y_3$ \\
$C_{\mathbf{5}_{-2}}$ & $- \frac{2}{5} \left. F \right|_{C_{\mathbf{5}_{-2}}}$ & $- \frac{3 \lambda}{5} Y_1 + \frac{2 \lambda}{5} Y_2$ & $-\frac{4 \lambda}{5} Y_2 + \frac{\lambda}{5} Y_3$ & $\frac{3 \lambda}{5} Y_1 + \frac{2 \lambda}{5} Y_2 - \frac{\lambda}{5} Y_3$ & $+ \lambda Y_3$ \\
$C_{\mathbf{1}_{5}}$ & $\left. F \right|_{C_{\mathbf{1}_{5}}}$ & $0$ & $\lambda Y_3$ & $- \lambda Y_3$ & $- \left. \lambda \left( 6 \overline{K}_{B_3} - 5 W \right) \right|_{C_{\mathbf{1}_{5}}}$ \\
\bottomrule
\end{tabular}
\caption{Line bundles (viewed as divisors on the matter curves) induced by the fluxes $A_X \left( F \right)$, $A \left( \mathbf{10}_{1} \right)$, $A \left( \mathbf{5}_{3} \right)$, $A \left( \mathbf{5}_{-2} \right)$ and $A \left( \mathbf{1}_{5} \right)$. Tensoring with the respective spin bundles gives the line bundles whose cohomologies count the massless matter.}
\label{table-6}
\end{center}
\end{table}

\begin{table}[h]
\begin{center}
{\small
\begin{tabular}{|c||c|c|c|c|c|}
\toprule
& $A^X \left( F \right)$ & $A \left( \mathbf{10}_{1} \right)$ & $A \left( \mathbf{5}_{3} \right)$ & $A \left( \mathbf{5}_{-2} \right)$ & $A \left( \mathbf{1}_{5} \right)$ \\
\hline \hline
$C_{\mathbf{10}_{1}}$ & $\frac{1}{5} \left. F \right|_{S_{\text{GUT}}} \cdot \left. \overline{K}_{B_3} \right|_{S_{\text{GUT}}}$ & $- \frac{\lambda}{5} \overline{K}_{B_3} \left( 6 \overline{K}_{B_3} - 5 W \right)$ & $\frac{2 \lambda}{5} \overline{K}_{B_3} \left( 3 \overline{K}_{B_3} - 2 W \right)$ & $- \frac{\lambda}{5} \overline{K}_{B_3} W$ & 0 \\
\cdashline{1-6}
$C_{\mathbf{5}_{3}}$ & $\frac{3}{5} \left. F \right|_{S_{\text{GUT}}} \cdot \left. \left( 3 \overline{K}_{B_3} - 2 W \right) \right|_{S_{\text{GUT}}}$ & $\frac{2 \lambda}{5} \overline{K}_{B_3} \left( 3 \overline{K}_{B_3} - 2 W \right)$ & $\frac{\lambda}{5} \left( 3 \overline{K}_{B_3} - 2 W \right) \left( 3 W - 2 \overline{K}_{B_3} \right)$ & $- \frac{3 \lambda}{5} \left( 3 \overline{K}_{B_3} - 2 W \right) W$ & \pbox{20cm}{$-\lambda \left( 3 \overline{K}_{B_3} - 2 W \right)$ \\ $\times \left( 4 \overline{K}_{B_3} - 3 W \right)$} \\
\cdashline{1-6}
$C_{\mathbf{5}_{-2}}$ & $- \frac{2}{5} \left. F \right|_{S_{\text{GUT}}} \cdot \left. \left( 5 \overline{K}_{B_3} - 3 W \right) \right|_{S_{\text{GUT}}}$ & $- \frac{\lambda}{5} \overline{K}_{B_3} W$ & $- \frac{3 \lambda}{5} \left( 3 \overline{K}_{B_3} - 2 W \right) W$ & $\frac{2 \lambda}{5} W \cdot \left( 5 \overline{K}_{B_3} - 3 W \right)$ & \pbox{20cm}{$\lambda \left( 3 \overline{K}_{B_3} - 2 W \right)$ \\ $\times \left( 4 \overline{K}_{B_3} - 3 W \right)$} \\
\cdashline{1-6}
$C_{\mathbf{1}_{5}}$ & $F \cdot \left( 3 \overline{K}_{B_3} - 2 W \right) \left( 4 \overline{K}_{B_3} - 3 W \right)$ & 0 & \pbox{20cm}{$\lambda \left( 3 \overline{K}_{B_3} - 2 W \right)$ \\ $\times \left( 4 \overline{K}_{B_3} - 3 W \right)$} & \pbox{20cm}{$-\lambda \left( 3 \overline{K}_{B_3} - 2 W \right)$ \\ $\times \left( 4 \overline{K}_{B_3} - 3 W \right)$} & \pbox{20cm}{$- \lambda \left( 6 \overline{K}_{B_3} - 5 W \right)$ \\ $\times \left( 4 \overline{K}_{B_3} - 3 W \right)$ \\ $\times \left( 3 \overline{K}_{B_3} - 2 W \right)$} \\
\bottomrule
\end{tabular}
} 
\caption{
Chiralities of the massless spectra of the fluxes $A^X \left( F \right)$, $A \left( \mathbf{10}_{1} \right)$, $A \left( \mathbf{5}_{3} \right)$, $A \left( \mathbf{5}_{-2} \right)$ and $A \left( \mathbf{1}_{5} \right)$. These chiralities \emph{include} the contributions from the spin bundle.}
\label{table-9}
\end{center}
\end{table}

\end{landscape}

\subsection{Projection via Restriction from Ambient Space}\label{subsec:ExampleOnMasslessSpectrumComputation}

In this section we exemplify the explicit evaluation of the projection formula (\ref{projectionformula1}) using a different method. It is based on the representation of the 2-cycles describing the gauge background as the pullback of elements of $\mathrm{CH}^2(\hat{X_5})$ with $\hat{X_5}$ the ambient space of $\hat{Y_4}$. Let us make this concrete for the $U ( 1 )_X$-background $A_X( F)$, postponing the analogous computation for the other types of gauge backgrounds to \autoref{sec:MasslessSpectraOfCurveSupportFluxesTedious}. In \autoref{subsec:VerticalAndGaugeInvariantCurveSupportFluxesOverMatterCurves} we pointed out that this background can be described by restriction to $\hat{Y_4}$ of
\[ \mathcal{A}_X \left( F \right) = - \frac{1}{5} \mathcal{F} \cdot \left( 2 \mathcal{E}_1 + 4 \mathcal{E}_2 + 6 \mathcal{E}_3 + 3 \mathcal{E}_4 + 5 \mathcal{S} - 5 \mathcal{Z} - 5 \overline{\mathcal{K}}_{B_3} \right) \in \mathrm{CH}^2 \left( \hat{X_5} \right) \, . \]
Our task is to compute the intersections $S^{a}_{\bfR} \cdot_{\iota_\bfR} {A}_X ( F ) $ for the matter curves $C_{\mathbf{10}_{1}}$, $C_{\mathbf{5}_{3}}$, $C_{\mathbf{5}_{-2}}$ and $C_{\mathbf{1}_{5}}$.  We do so by interpreting also $S^{a}_{\bfR}$ as an element ${\cal S}^{a}_{\bfR} \in \mathrm{CH}^3(\hat{X_5})$ and working entirely on $\hat{X_5}$. By construction $\mathcal{A}_X ( F )$ is gauge invariant. Therefore the result does not depend on which of the matter surfaces listed in (\ref{app_ S10}), (\ref{app_S53}), (\ref{app:MatterSurfaces52}) of \autoref{app_Mattersurfaces} over a given matter curve we pick for each representation. For example, focus on intersections with the following matter surfaces \footnote{In (\ref{app_S53}) $S^{(4)}_{\mathbf{5}_{3}}$ is given as $\mathbb{P}^1_{3F}( \mathbf{5}_{3} )$. Since $\mathbb P^1_3$ splits into $\mathbb P^1_{3x}$ and $\mathbb P^1_{3F}$ over $C_{\mathbf{5}_{3}}$ and since we only consider intersections with 2-cycles representing gauge invariant backgrounds here, we can represent the matter state in this way.}
\begin{align*}
S^{(6)}_{\mathbf{10}_{1}} &= \mathbb{P}^1_{4D} \left( \mathbf{10}_{1} \right) = V \left( a_{10}, e_4, xs e_2 e_3 + a_{21} z^2 e_0 \right), \\
S^{(4)}_{\mathbf{5}_{3}} &= - \mathbb{P}^1_{3x} \left( \mathbf{5}_{3} \right) = - V \left( a_{32}, e_3, x \right), \\
S^{(4)}_{\mathbf{5}_{-2}} &=\mathbb{P}_{3H}^1 \left( \mathbf{5}_{-2} \right) = V \left( a_{3,2} a_{2,1} - a_{4,3} a_{1,0}, e_3, a_{4,3} e_0 x z e_1 e_2 - a_{3,2} y, a_{2,1} e_0 x z e_1 e_2 - a_{1,0} y  \right) \\
&= V \left( e_3, a_{43} e_0 e_1 e_2 x z - a_{32} y, a_{21} e_0 x z e_1 e_2 - a_{10} y \right) \\
S_{\mathbf{1}_{5}} &= \mathbb{P}^1_{A} \left( \mathbf{1}_{5} \right) =  V \left( a_{32}, a_{43}, s \right).
\end{align*}

Let us compute $\mathbb{P}^1_{4D} ( \mathbf{10}_{1} ) \cdot_{\iota_{\mathbf{10}_{1}}} \mathcal{A}_X ( F )$ term by term by first determining the vanishing ideal representing the intersection points in $\hat{X_5}$. In determining this ideal, we are free to use the relations in $\mathrm{CH}(\hat{X_5})$ without changing the result up to rational equivalence. The key point is now that, as far as the toric fibre ambient space is concerned, rational and homological equivalence agree. We are therefore free to use the linear relations (\ref{linearrelationsambient}) and the relations encoded in the SR-ideal (\ref{SRideal}) of the ambient space. Furthermore, in the following $f$ will denote a polynomial in the coordinate ring of $\hat{X_5}$ in the same Chow class as $\mathcal{F}$. With this in mind, we find
\begin{align}
\begin{split}
\mathcal{F} \cdot \mathcal{E}_1 \cdot \mathbb{P}^1_{4D} \left( \mathbf{10}_{1} \right) &= V \left( a_{1,0}, f, e_1, e_4, e_2 + a_{2,1} e_0 \right) \, , \\
\mathcal{F} \cdot \mathcal{E}_2 \cdot \mathbb{P}^1_{4D} \left( \mathbf{10}_{1} \right) &= V \left( a_{1,0}, f, e_2, e_4, a_{2,1} \right) \, , \\
\mathcal{F} \cdot \mathcal{E}_3 \cdot \mathbb{P}^1_{4D} \left( \mathbf{10}_{1} \right) &= V \left( a_{1,0}, f, e_3, e_4, a_{2,1} \right) \, , \\
\mathcal{F} \cdot \mathcal{E}_4 \cdot \mathbb{P}^1_{4D} \left( \mathbf{10}_{1} \right) &= V \left( a_{1,0}, f, e_1, e_4, e_2 + a_{2,1} e_0 \right) - 2 V \left( a_{1,0}, f, y, e_4, e_3 + a_{2,1} e_0 \right), \\\mathcal{F} \cdot \mathcal{S} \cdot \mathbb{P}^1_{4D} \left( \mathbf{10}_{1} \right) &= \emptyset \, , \\
\mathcal{F} \cdot \mathcal{Z} \cdot \mathbb{P}^1_{4D} \left( \mathbf{10}_{1} \right) &= \emptyset \, , \\
\mathcal{F} \cdot \overline{\mathcal{K}}_{B_3} \cdot \mathbb{P}^1_{4D} \left( \mathbf{10}_{1} \right) &= V \left( a_{1,0}, f, e_1, e_4, e_2 + a_{2,1}    
      e_0 \right) - V \left( a_{1,0}, f, y, e_4, e_3 + a_{2,1} e_0 \right) \, .
\end{split}
\end{align}
In the fourth line we used the linear relation $\mathcal{E}_4 = \mathcal{E}_1 + 2 \mathcal{E}_2 + \mathcal{S} + 3 \mathcal{X} - 2 \mathcal{Y}$, and for the last line $\overline{\mathcal{K}}_{B_3} = \mathcal{E}_1 + 2 \mathcal{E}_2 + \mathcal{E}_3 + \mathcal{S} + 2 \mathcal{X} - \mathcal{Y} - \mathcal{Z}$. Concerning the vanishing locus $V ( a_{1,0}, f, e_2, e_4, a_{2,1} )$, note that $e_2 = 0$ implies that this is a sublocus of the fibration over the GUT-surface. Inside $B_3$ this locus is described by the intersection of the divisors $F$ (associated to $f$), $\overline{K}_{B_3}$ and $2 \overline{K}_{B_3} - W$ inside this surface. This intersection is the empty set. Therefore, we can discard this vanishing locus. Along the same lines we can discard $V ( a_{1,0}, f, e_3, e_4, a_{2,1} )$.

Summing up all contributions we obtain
\[ \mathbb{P}^1_{4D} \left( \mathbf{10}_{1} \right) \cdot_{\iota_{\mathbf{10}_{1}}} \mathcal{A}_X \left( F \right) = \frac{1}{5} \cdot V \left( a_{10}, f, y, e_4, e_3 + a_{21} e_0 \right) \in \text{CH}_0 \left(  \hat{Y}_4 |_{C_{\mathbf{10}_{1}}} \right) \, . \]
Upon use of the projection $\pi_{\mathbf{10}_{1}} \colon \hat{Y}_4 |_{C_{\mathbf{10}_{1}}} \twoheadrightarrow C_{\mathbf{10}_{1}}$ this yields
\[ \pi_{\mathbf{10}_{1} \ast} \left( \mathbb{P}^1_{4D} \left( \mathbf{10}_{1} \right) \cdot_{\iota_{\mathbf{10}_{1}}} \mathcal{A}_X \left( F \right) \right) = \frac{1}{5} \cdot \left. \mathcal{F} \right|_{C_{\mathbf{10}_{1}}} \in \text{CH}_0 \left( C_{\mathbf{10}_{1}} \right) \, . \]
It is a simple but teadious exercise to repeat this type of computation for the other matter surfaces with the result
\begin{align}
\begin{split}
\pi_{\mathbf{5}_{3} \ast} \left( - \mathbb{P}^1_{3x} \left( \mathbf{5}_{3} \right) \cdot_{\iota_{\mathbf{5}_{3}}} \mathcal{A}_X \left( F \right) \right) &= \frac{3}{5} \cdot \left. \mathcal{F} \right|_{C_{\mathbf{5}_{3}}} \in \text{CH}_0 \left( C_{\mathbf{5}_{3}} \right) \, , \\
\pi_{\mathbf{5}_{-2} \ast} \left( {\mathbb{P}^1_{3H} \left( \mathbf{5}_{-2} \right)} \cdot_{\iota_{\mathbf{5}_{-2}}} \mathcal{A}_X \left( F \right) \right) &= - \frac{2}{5} \cdot \left. \mathcal{F} \right|_{C_{\mathbf{5}_{-2}}} \in \text{CH}_0 \left( C_{\mathbf{5}_{-2}} \right) \, , \\
\pi_{\mathbf{1}_{5} \ast} \left( \mathbb{P}^1_{A} \left( \mathbf{1}_{5} \right) \cdot_{\iota_{\mathbf{1}_{5}}} \mathcal{A}_X \left( F \right) \right) &= \left. \mathcal{F} \right|_{C_{\mathbf{1}_{5}}} \in \text{CH}_0 \left( C_{\mathbf{1}_{5}} \right) \, .
\end{split}
\end{align}

Consequently the massless spectrum of the flux $\mathcal{A}_X ( F )$ is counted by the sheaf cohomology groups of the line bundles listed in \autoref{table--1}.

\begin{table}
\begin{center}
\begin{tabular}{|c|c|}
\toprule
matter curve & $\mathcal{L} \left( S^{a}_{C_\ast}, \mathcal{A}_X \left( F \right) \right)$ \\
\hline \hline
$C_{\mathbf{10}_{1}}$ & $\mathcal{O}_{C_{\mathbf{10}_{1}}} \left( \frac{1}{5} \cdot \left. F \right|_{C_{\mathbf{10}_{1}}} \otimes \sqrt{K_{C_{\mathbf{10}_{1}}}} \right)$ \\
$C_{\mathbf{5}_{3}}$ & $\mathcal{O}_{C_{\mathbf{5}_{3}}} \left( \frac{3}{5} \cdot \left. F \right|_{C_{\mathbf{5}_{3}}} \otimes \sqrt{K_{C_{\mathbf{5}_{3}}}} \right)$ \\
$C_{\mathbf{5}_{-2}}$ & $\mathcal{O}_{C_{\mathbf{5}_{-2}}} \left( - \frac{2}{5} \cdot \left. F \right|_{C_{\mathbf{5}_{-2}}} \otimes \sqrt{K_{C_{\mathbf{5}_{-2}}}} \right)$ \\
$C_{\mathbf{1}_{5}}$ & $\mathcal{O}_{C_{\mathbf{1}_{5}}} \left( \left. F \right|_{C_{\mathbf{1}_{5}}} \otimes \sqrt{K_{C_{\mathbf{1}_{5}}}} \right)$ \\
\bottomrule
\end{tabular}
\end{center}
\caption{The massless spectrum of states localised on the various matter curves in the presence of the flux $\mathcal{A}_X ( F )$ \cite{Krause:2011xj, 
oai:arXiv.org:1202.3138} with $F \in \text{Pic} ( B_3 )$ is counted by the sheaf cohomologies of the line bundles listed above.}
\label{table--1}
\end{table}

\section{Computing Massless Spectra of Matter Surface Fluxes in an F-theory Model} \label{sec:ComputingTheSpectra}

In this final section we compute the cohomology dimensions of the line bundles on the matter curves deduced previously. If the base space $B_3$ is embedded into a toric variety $X_\Sigma$, we can interpret these line bundles as coherent sheaves on $X_\Sigma$. The computation of the massless spectrum then reduces to the computation of sheaf cohomology groups/dimensions on toric spaces. This mathematical problem has been investigated in great detail by M.\ Barakat and collaborators \cite{2010arXiv1003.1943B, 2012arXiv1202.3337B, 2012arXiv1210.1425B, 2012arXiv1212.4068B, 2014arXiv1409.6100B, BL_GabrielMorphisms}, whose technology we adapt for our purposes. This requires making a concrete choice of a base space $B_3$ and 7-brane divisor $W$ therein, \ie fixing the complex structure moduli defining the fibration.

In \autoref{subsec:ComputingMasslessSpectraWithGAP} we outline the algorithm for the computation of the sheaf cohomology groups/dimensions on $X_\Sigma$. Our algorithms are implemented in \texttt{gap} \cite{GAP4} and phrased in the language of \emph{categorical programming} of \texttt{CAP} \cite{CAP, PosurDoktor, GutscheDoktor}. For the reader's convenience, we provide more introductory material on this rather mathematical topic in \autoref{sec:ATasteOfAlgebraAndAlgebraicGeometry}. We describe our definition of the concrete model on the toric base $B_3 = \mathbb P^3$ in \autoref{sec:Simplifying}.

In \autoref{sec:Spectrum1} we finally demonstrate our computations for different choices of complex structure moduli of the matter curves. Thereby we observe jumps in the cohomology dimensions across the moduli space. Such phenomenon is of relevance when it comes, for instance, to scanning for Standard-Model-like spectra in F-theory compactifications.

\subsection{Computing Sheaf Cohomologies with \texttt{GAP} and \texttt{CAP}} \label{subsec:ComputingMasslessSpectraWithGAP}

Our results so far imply that we need to compute sheaf cohomologies of line bundles 
\[ \mathcal{L}(S_\bfR, A)  = {\cal O}_{C_{\bfR}}\left(  D(S_\bfR, A) \right) \otimes \sqrt{K_{C_\bfR}} \in \text{Pic} \left( C_\bfR \right) \label{calLbundletocompute} \]
on curves $C_\bfR$ which in general cannot be obtained as pullback line bundles from the 7-brane divisor $W$ or any toric ambient space. For instance, this is the case for the line bundle induced by the gauge background $A ( \mathbf{10}_{1} )$ on $C_{\mathbf{5}_{-2}}$, \ie 
\[ \forall L \in \text{Pic} \left( W \right): \qquad   \left. L \right|_{C_{\mathbf{5}_{-2}}} \not \cong \mathcal{L}  \left(S_{\mathbf{5}_{-2}},  A\left(\mathbf{10}_{1} \right) \left( \lambda \right) \right) \, . \]
A similar obstruction exists for the hypercharge gauge background in F-theory GUT-Models \cite{Braun:2014pva, Beasley:2008kw} on the divisor $W$, which cannot be obtained as a pullback bundle from the base. For line bundles on $C_\bfR$ obtained by pullback, the computation of the associated sheaf cohomologies can oftentimes be performed with \emph{cohomCalg} \cite{Blumenhagen:2010pv, cohomCalg:Implementation, 2011JMP....52c3506J, Rahn:2010fm, Blumenhagen:2010ed} or techniques employed in heterotic string compactifications for CICYs \cite{Anderson:2008ex}. Given the above obstruction, these methods are, however, not applicable in our case and we hence need to go beyond this framework.
 
Even though $\mathcal{L}(S_\bfR, A)$ does in general not descend from a line bundle on an ambient space, we can extend this line bundle $\mathcal{L}(S_\bfR, A)$ by zero outside of $C_{\mathbf{R}}$. The so-obtained object is a coherent sheaf $\mathcal{F}(S_\bfR, A)$ on the space into which $C_\bfR$ is embedded. In case $C_{\mathbf{R}}$ is embedded into a toric ambient space $X_\Sigma$, we thus obtain elements in $\mathfrak{Coh} ( X_\Sigma )$, the category of coherent sheaves on the toric ambient space $X_\Sigma$. In this sense, the remaining task is to compute sheaf cohomology groups for such objects.  

Our methods to compute the sheaf cohomologies for all elements of $\mathfrak{Coh} ( X_\Sigma )$ are based on \cite{2010arXiv1003.1943B, 2012arXiv1202.3337B, 2012arXiv1210.1425B, 2012arXiv1212.4068B, 2014arXiv1409.6100B, BL_GabrielMorphisms}. These methods, which we are extending further, apply as long as $X_\Sigma$ is a normal toric variety which is either smooth and complete or simplicial and projective. Note that $\mathfrak{Coh} ( X_\Sigma )$ includes the above mentioned non-pullback line bundles and the hypercharge flux, but is far bigger than that. Also vector bundles which are not direct sums of lines bundles, quotients thereof, T-branes in the language of \cite{Collinucci:2014qfa} or skyscraper sheaves can be modelled by our technology. In particular smoothness of the matter curves is not required.

In \autoref{sec:ATasteOfAlgebraAndAlgebraicGeometry} we briefly review topics from algebra, category theory and algebraic geometry which are necessary to understand our approach to computing sheaf cohomologies of coherent sheaves. Experts may well skip these sections, interested readers can find additional information in \cite{lane1998categories, hartshorne1977algebraic, cox2011toric}.

Let us return to our task of computing the sheaf cohomology groups of the line bundle ${\cal L} ( S_{\bfR}, A )$ defined, as in (\ref{calLbundletocompute}), via a divisor 
\[D  = D(S_\bfR, A) + \sqrt{K_{C_\bfR}} \in \mathrm{CH}^1(C_\bfR) \, . \]
Here $\sqrt{K_{C_{\mathbf{R}}}}$ denotes, by slight abuse of notation, the divisor on $C_{\mathbf{R}}$ associated with the spin bundle induced by the embedding of the curve $C_\bfR$ into $B_3$. It is well-known that a Cartier divisor $D$ on a complex variety $X$ gives rise to a line bundle $\mathcal{O}_X ( D )$, and we take  this opportunity to recall how this line bundle is actually defined. Namely $\mathcal{O}_X ( D )$ can be understood as a sheaf on $X$.\footnote{If the divisor $D$ is Cartier, the sheaf $\mathcal{O}_X ( D )$ is \emph{invertible}, which means that it actually defines a line bundle. Equivalently, there exists another sheaf on $X$ -- in the case at hand $\mathcal{O}_X ( -D )$ -- which satisfies the property $\mathcal{O}_X ( D ) \otimes_{\mathcal{O}_X} \mathcal{O}_X (  - D ) \cong \mathcal{O}_X$, hence the name \emph{invertible}. We will exploit this invertibiliy momentarily. Recall also that on a smooth variety every Weil divisor is Cartier.} A sheaf $\mathcal{F}$ (of Abelian groups) on $X$ assigns to every open subset $U \subseteq X$ an Abelian group $\mathcal{F} ( U )$. For the line bundle $\mathcal{O}_X ( D )$ this Abelian group is given by
\[ \mathcal{O}_X \left( D \right) \left( U \right) = \left\{ f \in \mathbb{C} \left( X \right)^* \; : \; \left( \mathrm{div} \left( f \right) + D \right)_{U} \geq 0 \right\} \cup \{ 0 \} \, , \label{equ:DefinitionOfOX(D)} \]
where $\mathrm{div}(f)$ denotes the divisor associated with the not identically vanishing meromorphic function $f$. The requirement that $(\mathrm{div}(f) + D)|_{U} \geq 0$, \ie that $(\mathrm{div}(f) + D)|_U$ is effective, means that the product $f \, d_U$ has no poles on $U$, where we locally express $D$ on $U$ as the zero-locus of the function $d_U$, \ie $D |_U = V \left( d_U \right)$.

This definition is rather abstract. In addition it is not at all obvious at this stage how we could actually encode this data in a form understandable for computers. To bridge this gap, let us also recall the notion of a so-called ideal sheaf. To this end we first look at the sheaf of holomorphic functions $\mathcal{O}_X$ on a complex variety $X$, which assigns to every open subset $U \subseteq X$ the set $\mathcal{O}_X ( U ) = \{ f \colon U \to \mathbb{C} \; , \; f \text{ holomorphic} \}$. It turns out that $\mathcal{O}_X ( U )$ is a (commutative and unitial) ring. Now let us consider global sections $f_i \in H^0 \left( X, \mathcal{O}_X \left( D_i \right) \right)$ for suitable divisors $D_i \in \text{Cl} ( X )$ and $1 \leq i \leq n$. In addition let $\mathcal{U} = \{ U_j \}_{j \in J}$ be an affine open cover of $X$. Consequently $\left. f_i \right|_{U_j} \in \mathcal{O}_X ( U_j )$. Therefore these global sections $f_i$ cut out an analytic subvariety $Y \subseteq X$ given by
\[ Y \cap U_j = V \left( \left. f_1 \right|_{U_j}, \dots, \left. f_n \right|_{U_j} \right) \, . \]
For every open $W \subseteq U_j$ we have the ideal $\left\langle \left. f_1 \right|_W, \left. f_2 \right|_W, \dots, \left. f_n \right|_W \right\rangle \subseteq \mathcal{O}_X \left( W \right)$. This assignment of ideals forms a sheaf on $U_j$. Finally, these sheaves on the affine patches $U_j$ glue to form a sheaf on $X$ -- \emph{the ideal sheaf $\mathcal{I}_X ( f_1, \dots, f_n )$ of $Y$}.\footnote{More generally, one can use any closed embedding $\iota \colon Y \hookrightarrow X$ to define the morphism of sheaves $\iota^\sharp \colon \mathcal{O}_X \to \iota_\ast \mathcal{O}_Y$. The kernel of $\iota^\sharp$ is then termed the ideal sheaf of $Y$ in $X$.}

Now let us look at a Cartier divisor $D \subseteq C_{\mathbf{R}}$. We assume that $D = V ( f_1, \dots, f_n )$ for global sections $f_1, \dots, f_n$. \footnote{The divisor $D$ need not be a complete intersection. Consequently there is no contradiction between $D$ being of codimension $1$ and $D$ being cut out by more than one global section.} We can then wonder if there is a relation between the ideal sheaf $\mathcal{I}_{C_{\mathbf{R}}} ( f_1, \dots, f_n )$ and the line bundle $\mathcal{O}_{C_{\mathbf{R}}} ( D )$. And indeed, by proposition 6.18 of \cite{hartshorne1977algebraic},
\[ \mathcal{O}_{C_{\mathbf{R}}} ( -D ) \cong \mathcal{I}_{C_{\mathbf{R}}} \left( f_1, \dots, f_n \right) \, . \label{Idealshaefdef} \]
Up to an important $-1$, this brings us close to handling the line bundles $\mathcal{O}_{C_{\mathbf{R}}} ( D )$. 

To overcome this additional $-1$, let us recall that line bundles $\mathcal{O}_{C_{\mathbf{R}}} ( D )$ are invertible sheaves. This means that there exists another sheaf $\mathcal{F}$ on $C_{\mathbf{R}}$ with the property $\mathcal{O}_{C_{\mathbf{R}}} ( D ) \otimes_{\mathcal{O}_{C_{\mathbf{R}}}} \mathcal{F} \cong \mathcal{O}_{C_{\mathbf{R}}}$. We can describe this sheaf quite explicitly. Namely we consider all sheaf homomorphisms from the sheaf $\mathcal{O}_{C_{\mathbf{R}}}$ to the line bundle $\mathcal{O}_{C_{\mathbf{R}}} ( D )$. It is a well-known fact that these homomorphisms form a sheaf, the so-called \emph{sheaf-Hom} $\Hom_{\mathcal{O}_{C_{\mathbf{R}}}} ( \mathcal{O}_{C_{\mathbf{R}}}, \mathcal{O}_{C_{\mathbf{R}}} ( D ) )$. It can be shown that $\Hom_{\mathcal{O}_{C_{\mathbf{R}}}} ( \mathcal{O}_{C_{\mathbf{R}}} ( D ), \mathcal{O}_{C_{\mathbf{R}}} )$ is isomorphic to $\mathcal{O}_{C_{\mathbf{R}}} ( -D )$. Therefore we reach the conclusion
\[ \mathcal{O}_{C_{\mathbf{R}}} \left( D \right) \cong \Hom_{\mathcal{O}_{C_{\mathbf{R}}}} \left( \mathcal{I}_{C_{\mathbf{R}}} \left( f_1, \dots, f_n \right), \mathcal{O}_{C_{\mathbf{R}}} \right) \, . \label{equ:OXDAndIdealSheaf} \]
This formula connects the defining data of the divisor $D  = D(S_\bfR, A) + \sqrt{K_{C_\bfR}} \in \mathrm{CH}^1(C_\bfR)$ to the line bundle ${\cal L} ( S_{\bfR}, A )$ far more explicitly than \eg (\ref{equ:DefinitionOfOX(D)}). But we can still do better.

To achieve an even more explicit description, we now turn our attention to toric varieties $X_\Sigma$ without torus factor. As we are eventually interested in algorithms applicable to computers, we use this opportunity to also let go of \emph{analytic geometry} and model these varieties $X_\Sigma$ over the rational numbers $\mathbb{Q}$. In particular we then employ the language of \emph{algebraic geometry} to describe these toric varieties $X_\Sigma$. A brief review of the topic is given in \autoref{sec:ATasteOfAlgebraAndAlgebraicGeometry}.

Recall that $X_\Sigma$ comes equipped with a coordinate ring $S$ -- typically refered to as the Cox ring -- which is graded by $\text{Cl} ( X_\Sigma )$. The assumption that $X_\Sigma$ has no torus factor ensures the existence of the so-called \emph{sheafification functor}
\[ \widetilde{\phantom{m}} \colon S \mathrm{\textnormal{-}fpgrmod} \to \mathfrak{Coh} X_\Sigma \, . \label{equ:SheafificationFunctorMainText} \]
This functor turns a so-called \emph{finitely presented (f.p.) graded $S$-module} into a coherent sheaf on $X_\Sigma$. Homogeneous ideals $I \subseteq S$ are special examples of \fp graded $S$-modules. Hence for homogeneous polynomials $f_1, \dots, f_n \in S$, we can turn the ideal $I = \langle f_1, \dots, f_n \rangle \subseteq S$ into a coherent sheaf $\tilde{I}$ on $X_\Sigma$. And indeed $\tilde{I} \cong \mathcal{I}_{X_\Sigma} ( f_1, \dots, f_n )$, \ie the sheafification of the ideal $I$ provides nothing but the ideal sheaf on $X_\Sigma$ generated by $f_1, \dots, f_n$. In this sense, our next best model for the sheaf $\mathcal{I}_{X_\Sigma} ( f_1, \dots, f_n )$ is the ideal $\langle f_1, \dots, f_n \rangle$ itself, which provides a very explicit description for this sheaf.

For the representation of the ideal it turns out more practical to specify the relations satisfied by its generators than specifying them. Such relations are conveniently expressed as a linear map $M$ acting on (finite) direct sums of modules over $S$ (respecting the grading, which is explained further in \autoref{subsec:ProjSmodule}). Such a homomorphism $M$ is a \emph{finitely presented (f.p.) graded $S$-module}, in the sense defined in \autoref{subsec:FPGradedSModules}.

The rough idea behind the functor (\ref{equ:SheafificationFunctorMainText}) is then the following. The toric variety $X_\Sigma$ is defined by the combinatorics of a fan $\Sigma$. For every cone $\sigma \in \Sigma$ there is an affine patch $U_\sigma$ and a monomial $x^{\hat{\sigma}} \in S$. Given an \fp graded $S$-module $M$, we can perform a so-called homogeneous localisation of $M$ with respect to $x^{\hat{\sigma}}$. This is explained in \autoref{subsec:LocalisationOfRings}. The result is an \fp $S_{(x^{\hat{\sigma}})}$-module $M_{(x^{\hat{\sigma}})}$, which defines a unique coherent sheaf $\tilde{M_{(x^{\hat{\sigma}})}}$ on $U_\sigma$ with the property \cite{hartshorne1977algebraic}
\[ \tilde{M_{(x^{\hat{\sigma}})}} \left( U_\sigma \right) = M_{(x^{\hat{\sigma}})} \, . \]
We have thus obtained a coherent sheaf on every affine patch $U_\sigma$ of $X_\Sigma$. It turns out that these sheaves glue together to form a sheaf on the entire variety $X_\Sigma$.

For a divisor $D = V ( f_1, \dots, f_n ) \subseteq X_\Sigma$, we see from (\ref{equ:OXDAndIdealSheaf}) that we need to invert the sheaf $\tilde{I}$ associated with $I = \langle f_1, \dots, f_n \rangle$ to describe $\mathcal{O}_X ( D )$. We are thus looking for an analogue of this equation in terms of \fp graded $S$-modules. Motivated by the fact $\tilde{S} \cong \mathcal{O}_X$, which can of course be proven rigorously, the analogue in question is $M = \text{Hom}_S ( I, S )$ (c.f. \autoref{subsec:ExtOfFPModules}). The so-defined \fp graded $S$-module $M$ now satisfies $\tilde{M} \cong \mathcal{O}_{X_\Sigma} ( D )$. We provide more details in \autoref{subsec:SheafCohomologyFromFPGradedSModules}.

Finally note that in general the matter curve $C_{\mathbf{R}}$ is not a divisor in a toric ambient space $X_\Sigma$ but of higher codimension. Suppose therefore that $C_{\mathbf{R}} = V ( g_1, \dots, g_k )$ and $D = V ( f_1, \dots, f_n ) \subseteq C_{\mathbf{R}}$ for homogeneous polynomials $g_i, f_i \in S$. Then we can consider the graded ring $S ( C_{\mathbf{R}} ) = S / \langle g_1, \dots, g_k \rangle$ and construct from $f_1, \dots, f_n$ an \fp graded $S ( C_\bfR )$-module $M_{C_\bfR}$ such that $\tilde{M_{C_\bfR}} \cong \mathcal{O}_{C_{\mathbf{R}}} ( D )$. To make use of the structure of the toric ambient space $X_\Sigma$, where we can for example apply the \emph{cohomCalg}-algorithm \cite{Blumenhagen:2010pv, cohomCalg:Implementation, 2011JMP....52c3506J, Rahn:2010fm, Blumenhagen:2010ed}, we now turn this module $M_{C_\bfR}$ into an \fp graded $S$-module $M$ such that $\tilde{M} \in \mathfrak{Coh} ( X_\Sigma )$ is the coherent sheaf on $X_\Sigma$ which is zero outside of $C_{\mathbf{R}}$ and matches $\mathcal{O}_{C_{\mathbf{R}}} ( D )$ on the matter curve $C_{\mathbf{R}}$. We explain the transition from $M_{C_\bfR}$ to $M$ in \autoref{subsec:IdealSheavesFromModules}.

This now brings us to the final question: Given an \fp graded $S$-module $M$, how do we compute the sheaf cohomology dimension of $\tilde{M}$ from the data defining $M$?
Before we answer this question, let us mention that for any two \fp graded $S$-modules $M$, $N$ one can compute extension groups -- denoted by $\text{Ext}^i_S ( M, N )$ -- which are \fp graded $S$-modules themselves. Note that $\text{Ext}^0_S ( M, N ) \cong \text{Hom}_S ( M, N )$, which we already encountered before. In particular we can truncate $\text{Ext}^i_S ( M, N )$ to any $d \in \text{Cl} ( X_\Sigma )$. Now, to compute the sheaf cohomologies of $\tilde{M}$, we have designed algorithms which compute an ideal $I \subseteq S$ such that 
\[ H^i \left( X_\Sigma, \tilde{M} \right) \cong \text{Ext}^i_S \left( I, M \right)_0 \, . \]
In addition we have implemented performant algorithms for the computation of $\text{Ext}^i_S ( I, M )_0$. They are optimised in that Gr\"obner basis computations are replaced by Gauss eliminations. In addition, the algorithms are parallised. We explain this step in more detail in \autoref{subsec:SheafCohomologyFromFPGradedSModules}. See also \autoref{subsec:ExtOfFPModules} for the computation of $\text{Ext}^i_S \left( I, M \right)_0$.

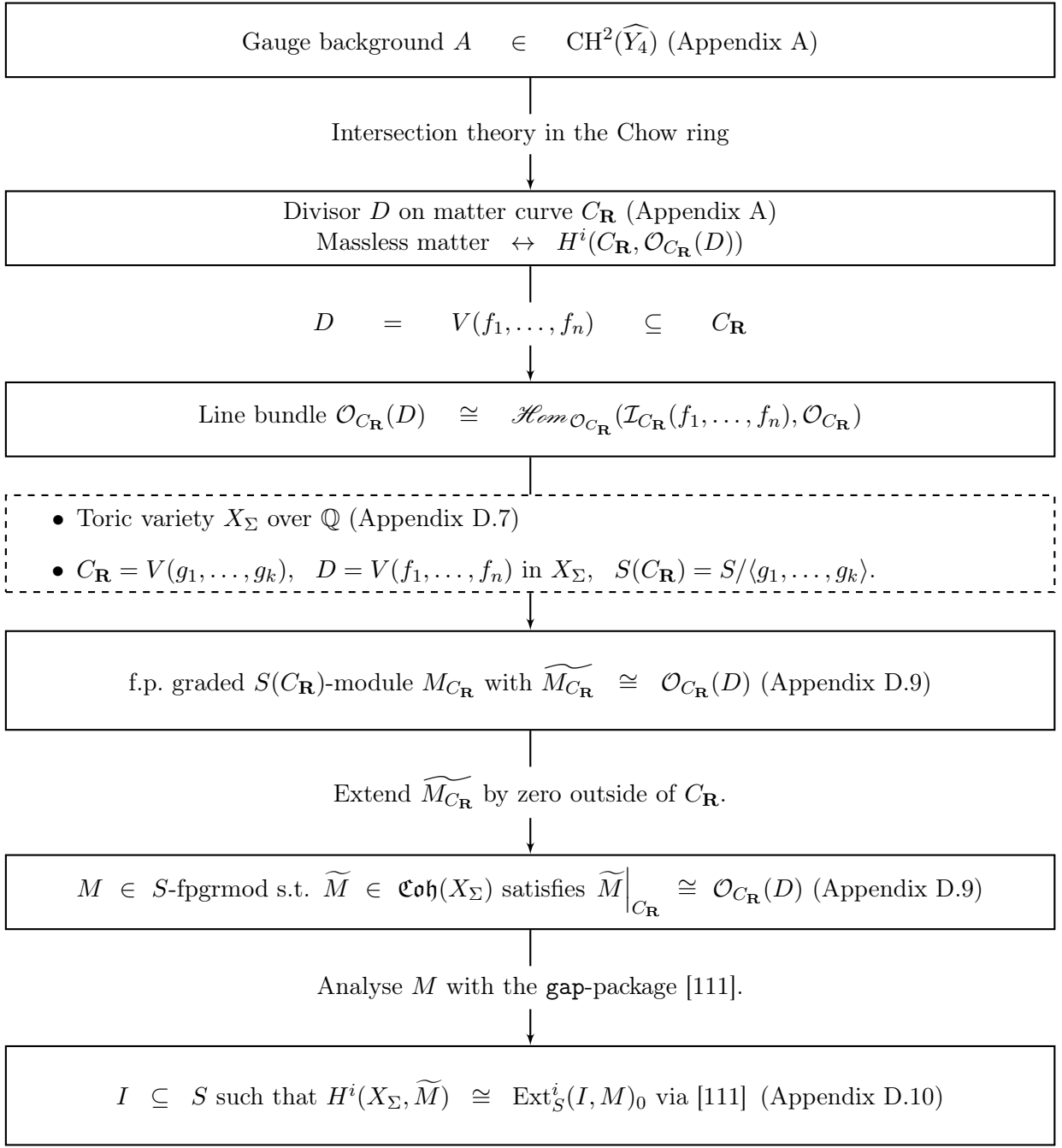
\begin{figure}
\begin{center}
\begin{tikzpicture}

 \matrix [column sep=1em,row sep=1.5em] {
      \node [rectangle, draw=black, thick, fill=white, text width=\textwidth, text centered, minimum height=3em] (Chow) {Gauge background $A \in \text{CH}^2 ( \hat{Y_4} )$  (\autoref{subsec:ReviewChowGroupsAndG4Fluxes}) }; \\
      \node [rectangle, draw=none, thick, fill=none, text width=\textwidth, text centered, minimum height=1em] (Intersect) {Intersection theory in the Chow ring}; \\
      \node [rectangle, draw=black, thick, fill=white, text width=\textwidth, text centered, minimum height=3em] (Divisor) {Divisor $D$ on matter curve $C_{\mathbf{R}}$  
     (\autoref{subsec:ReviewChowGroupsAndG4Fluxes}) \\ Massless matter \, $\leftrightarrow$  \, $H^i(C_\bfR, \mathcal{O}_{C_{\bfR}}(D))$ }; \\
      \node [rectangle, draw=none, thick, fill=none, text width=\textwidth, text centered, minimum height=1em] (DivisorAsZeroLocus) { $D = V ( f_1, \dots, f_n ) \subseteq C_{\mathbf{R}}$ }; \\
      \node [rectangle, draw=black, thick, fill=white, text width=\textwidth, text centered, minimum height=3em] (RelationToIdealSheaf) {Line bundle $\mathcal{O}_{C_{\mathbf{R}}} ( D ) \cong \Hom_{\mathcal{O}_{C_{\mathbf{R}}}} ( \mathcal{I}_{C_{\mathbf{R}}} ( f_1, \dots, f_n ), \mathcal{O}_{C_{\mathbf{R}}} )$}; \\
      \node [rectangle, draw=black, dashed, thick, fill=none, text width=\textwidth, text ragged, minimum height=1em] (SwitchToToric) {
      \vspace{1em}
      \begin{itemize}
       \item Toric variety $X_\Sigma$ over $\mathbb{Q}$ (\autoref{app_sheafification})
       \item $C_{\mathbf{R}} = V ( g_1, \dots, g_k )$, \, $D = V ( f_1, \dots, f_n )$ in $X_\Sigma$, \, $S ( C_{\mathbf{R}} ) = S / \langle g_1, \dots, g_k 
            \rangle$.
      \end{itemize}
      }; \\    
      \node [rectangle, draw=black, thick, fill=white, text width=\textwidth, text centered, minimum height=4em] (SheafifyIdeal) {
       \fp graded $S ( C_{\mathbf{R}} )$-module $M_{C_{\mathbf{R}}}$ with $\tilde{M_{C_{\mathbf{R}}}} \cong \mathcal{O}_{C_{\mathbf{R}}} ( D )$ (\autoref{subsec:IdealSheavesFromModules})}; \\
      \node [rectangle, draw=none, thick, fill=none, text width=\textwidth, text centered, minimum height=1em] (ExtensionByZero) {Extend $\tilde{M_{C_{\mathbf{R}}}}$ by zero outside of $C_{\mathbf{R}}$.}; \\    
      \node [rectangle, draw=black, thick, fill=white, text width=\textwidth, text centered, minimum height=3em] (Module) { $M \in S \mathrm{\textnormal{-}fpgrmod}$ s.t. $\tilde{M} \in \mathfrak{Coh} ( X_\Sigma )$ satisfies $\left. \tilde{M} \right|_{C_{\mathbf{R}}} \cong \mathcal{O}_{C_{\mathbf{R}}} ( D )$  (\autoref{subsec:IdealSheavesFromModules})}; \\
      \node [rectangle, draw=none, thick, fill=none, text width=\textwidth, text centered, minimum height=1em] (IdentifyIdealForSheafCohomologyComputation) {Analyse $M$ with the \texttt{gap}-package \cite{SheafCohomologyOnToricVarieties}.}; \\
      \node [rectangle, draw=black, thick, fill=white, text width=\textwidth, text centered, minimum height=4em] (Compute) {
      $I \subseteq S$ such that $H^i ( X_\Sigma, \tilde{M} ) \cong \text{Ext}^i_S ( I, M )_0$   via \cite{SheafCohomologyOnToricVarieties}  \,(\autoref{subsec:SheafCohomologyFromFPGradedSModules}) }; \\
      };

    \path[draw, thick] (Chow) -- (Intersect);
    \path[draw, thick, -latex'] (Intersect) -- (Divisor);
    \path[draw, thick] (Divisor) -- (DivisorAsZeroLocus);
    \path[draw, thick, -latex'] (DivisorAsZeroLocus) -- (RelationToIdealSheaf);
    \path[draw, thick] (RelationToIdealSheaf) -- (SwitchToToric);
    \path[draw, thick, -latex'] (SwitchToToric) -- (SheafifyIdeal);
    \path[draw, thick] (SheafifyIdeal) -- (ExtensionByZero);
    \path[draw, thick, -latex'] (ExtensionByZero) -- (Module);
    \path[draw, thick] (Module) -- (IdentifyIdealForSheafCohomologyComputation);
    \path[draw, thick, -latex'] (IdentifyIdealForSheafCohomologyComputation) -- (Compute);
    
\end{tikzpicture}
\end{center}
\caption{The sheaf cohomologies $H^i(C_\bfR, \mathcal{O}_{C_{\bfR}}(D)) \simeq H^i ( X_\Sigma, \tilde{M} )$ encode the massless matter. We compute them along the above algorithm by use of the \texttt{gap}-package \cite{SheafCohomologyOnToricVarieties}.}
\label{figure-987654}
\end{figure}

The packages \cite{CAPCategoryOfProjectiveGradedModules, CAPPresentationCategory, PresentationsByProjectiveGradedModules, TruncationsOfPresentationsByProjectiveGradedModules} provide the implementation of the category $S \mathrm{\textnormal{-}fpgrmod}$ in the language of \emph{categorical programming} of \texttt{CAP} \cite{CAP, PosurDoktor, GutscheDoktor}. Basic functionality of toric varieties is provided by the \texttt{gap}-package \texttt{ToricVarieties} of \cite{homalg}. This package is extended by \cite{SheafCohomologyOnToricVarieties}, which provides the algorithms that identify the above-mentioned ideal $I$. Also the specialised algorithms for the computation of $\text{Ext}^i_S ( I, M )_0$ are provided by \cite{SheafCohomologyOnToricVarieties}. The overall procedure is summarised compactly in  \autoref{figure-987654}.

\subsection{Simplifying Assumptions and Geometric Consequences} \label{sec:Simplifying}

To explicitly perform the computation of sheaf cohomologies we return to the fibration defined in \autoref{subsec:SpecialFTheoryGUTModel} and specialise $B_3 = \mathbb{P}_{\mathbb{Q}}^3$, with $[ z_1 \colon z_2 \colon z_3 \colon z_4 ]$ denoting its homogeneous coordinates. The 7-brane divisor is chosen as $W = V ( z_4 ) \cong \mathbb{P}_{\mathbb{Q}}^2$. Consequently the elliptic fibration $\hat{Y}_4$ is a complete intersection in the toric ambient space $\hat{X_\Sigma}$ whose Cox ring $S( \hat{X_\Sigma}) = \mathbb{Q} [ z_1, z_2, z_3, e_0, e_4, e_3, e_2, e_1, x, y, z, s ]$ is graded by $\mathbb{Z}^7$ as summarized in \autoref{table-toricscalingsmodel}.
\begin{table}
\begin{center}
\begin{tabular}{|cccc|cccc|cccc|}
\toprule
$z_1$ & $z_2$ & $z_3$ & $z_4 \equiv e_0$ & $e_1$ & $e_2$ & $e_3$ & $e_4$ & x & y & z & s \\
\hline \hline
1 & 1 & 1 & 1 & 0 & 0 & 0 & 0 & 8 & 12 & 0 & 0 \\
\hline
$\cdot$          & \dots &  $\cdot$    & -1        & 1 & $\cdot$ & $\cdot$ & $\cdot$ & -1 & -1 & $\cdot$ & $\cdot$ \\
$\cdot$          & \dots &  $\cdot$    & -1        & $\cdot$ & 1 & $\cdot$ & $\cdot$ & -2 & -2 & $\cdot$ & $\cdot$ \\
$\cdot$          & \dots &  $\cdot$    & -1        & $\cdot$ & $\cdot$ & 1 & $\cdot$ & -2 & -3 & $\cdot$ & $\cdot$ \\
$\cdot$          & \dots &  $\cdot$    & -1        & $\cdot$ & $\cdot$ & $\cdot$ & 1 & -1 & -2 & $\cdot$ & $\cdot$ \\
\hline
$\cdot$          & \dots &  $\cdot$    & $\cdot$   & $\cdot$ & $\cdot$ & $\cdot$ & $\cdot$ & 2 &  3 & 1 & $\cdot$ \\
$\cdot$          & \dots &  $\cdot$    & $\cdot$   & $\cdot$ & $\cdot$ & $\cdot$ & $\cdot$ & -1 & -1 & $\cdot$ & 1 \\
\bottomrule
\end{tabular}
\end{center} \caption{Scaling relations of the elliptic fibration over $B_3 = \mathbb P^3_{\mathbb{Q}}$. \label{table-toricscalingsmodel}}
\end{table}
The Stanley-Reisner ideal is
\begin{align*}
\begin{split}
I_{\text{SR}} \left( \hat{X_\Sigma} \right) &= \left\langle xy, x e_4, z s, z e_1, z e_2, z e_3, z e_4, s e_0, s e_1, s e_2, s e_4, y e_1, y e_2,
e_0 e_3, \right. \\
& \qquad \qquad \qquad \left. e_1 e_3, e_0 e_2, e_0 z_1 z_2 z_3, e_1 z_1 z_2 z_3, e_2 z_1 z_2 z_3, e_3 z_1 z_2 z_3, e_4 z_1 z_2 z_3 \right\rangle.
\end{split}
\end{align*}
This toric space $\hat{X_\Sigma}$ is not smooth, but rather a toric orbifold. The elliptic fourfold is embedded as $\hat{Y}_4 = V ( P_T^\prime ) \subseteq \hat{X_\Sigma}$ with
\begin{align}
\begin{split}
P_T^\prime &= y^2 s e_3 e_4 + a_1 \left( z_i \right) x y z s + a_{3,2} \left( z_i \right) y z^3 e_0^2 e_1 e_4 - x^3 s^2 e_1 e_2^2 e_3 \\
           &\qquad \qquad - a_{2,1} \left( z_i \right) x^2 z^2 s e_0 e_1 e_2 - a_{4,3} \left( z_i \right) x z^4 e_0^3 e_1^2 e_2 e_4 \, .
\end{split}
\end{align}
The section $a_{i,j}$ are taken according to
\begin{align*}
& a_{1,0} \in H^0 \left( \mathbb{P}_{\mathbb{Q}}^3, \mathcal{O}_{\mathbb{P}_{\mathbb{Q}}^3} \left( 4 \right) \right), & a_{2,1} \in H^0 \left( \mathbb{P}_{\mathbb{Q}}^3, \mathcal{O}_{\mathbb{P}_{\mathbb{Q}}^3} \left( 7 \right) \right), \\
& a_{3,2} \in H^0 \left( \mathbb{P}_{\mathbb{Q}}^3, \mathcal{O}_{\mathbb{P}_{\mathbb{Q}}^3} \left( 10 \right) \right), & a_{4,3} \in H^0 \left( \mathbb{P}_{\mathbb{Q}}^3, \mathcal{O}_{\mathbb{P}_{\mathbb{Q}}^3} \left( 13 \right) \right)
\end{align*}
since $\overline{K}_{B_3} \cong \mathcal{O}_{\mathbb{P}_{\mathbb{Q}}^3} ( 4 )$ and $W = V ( z_4 )$. In particular $\hat{Y}_4$ is smooth for generic such sections, and the matter curves contained in $S_{\text{GUT}}$ can be described in terms of these generic polynomials as
\[ C_{\mathbf{10}_{1}} = V \left( z_4, a_{1,0} \right), \qquad C_{\mathbf{5}_{3}} = V \left( z_4, a_{3,2} \right), \qquad C_{\mathbf{5}_{-2}} = V \left( z_4, a_{1,0} a_{4,3} - a_{2,1} a_{3,2} \right) \, . \]

From the computational point of view, this toy model has a very appealing feature -- the GUT-surface $W$ is itself a toric variety. In such a situation it is always favourable to apply the tools provided by \texttt{gap} \cite{GAP4} to this toric GUT-surface directly, rather than describing it as a subvariety in $B_3$. In particular we can model the matter curves in $W \cong \mathbb{P}_{\mathbb{Q}}^2$ with homogeneous coordinates $[ z_1 \colon z_2 \colon z_3 ]$ by use of the homogeneous polynomials
\begin{equation}
  \begin{aligned}
& \tilde{a_{1,0}} \in H^0 \left( \mathbb{P}_{\mathbb{Q}}^2, \mathcal{O}_{\mathbb{P}_{\mathbb{Q}}^2} \left( 4 \right) \right) \, , \qquad & \tilde{a_{2,1}} \in H^0 \left( \mathbb{P}_{\mathbb{Q}}^2, \mathcal{O}_{\mathbb{P}_{\mathbb{Q}}^2} \left( 7 \right) \right) \, , \\
& \tilde{a_{3,2}} \in H^0 \left( \mathbb{P}_{\mathbb{Q}}^2, \mathcal{O}_{\mathbb{P}_{\mathbb{Q}}^2} \left( 10 \right) \right) \, , \qquad & \tilde{a_{4,3}} \in H^0 \left( \mathbb{P}_{\mathbb{Q}}^2, \mathcal{O}_{\mathbb{P}_{\mathbb{Q}}^2} \left( 13 \right) \right)
  \end{aligned}
\end{equation}
as the loci
\[ C_{\mathbf{10}_{1}} \cong V \left( \tilde{a_{1,0}} \right), \qquad C_{\mathbf{5}_{3}} \cong V \left( \tilde{a_{3,2}} \right), \qquad C_{\mathbf{5}_{-2}} \cong V \left( \tilde{a_{1,0}} \tilde{a_{4,3}} - \tilde{a_{2,1}} \tilde{a_{3,2}} \right) \, . \]
To appreciate to what degree the defining polynomials for the matter curves simplify upon restriction to $W$ we compare the number of polynomials defining $a_{i,j}$ and their restrictions $\tilde a_{i,j}$ displayed in \autoref{table-coefficients}.
\begin{table}
\begin{center}
\begin{tabular}{|c|c|c|}
\hline
$d \in \mathbb{Z}$ & $h^0 \left( \mathbb{P}_{\mathbb{Q}}^3, \mathcal{O}_{\mathbb{P}_{\mathbb{Q}}^3} \left( d \right) \right)$ & $h^0 \left( \mathbb{P}_{\mathbb{Q}}^2, \mathcal{O}_{\mathbb{P}_{\mathbb{Q}}^2} \left( d \right) \right)$ \\
\hline \hline
4 & 35 & 15 \\
7 & 120 & 36 \\
10 & 286 & 66 \\
13 & 560 & 105 \\
\hline
\end{tabular} \caption{Comparision of sections over $B_3 = \mathbb P^3_\mathbb{Q}$ and their restriction to $W = \mathbb P^2_\mathbb{Q}$.  \label{table-coefficients}}
\end{center}
\end{table}
In particular, whilst $a_{4,3} \in H^0( \mathbb{P}_{\mathbb{Q}}^3, \mathcal{O}_{\mathbb{P}_{\mathbb{Q}}^3}( 13 ))$ generically consists of 560 monomials, the corresponding $\tilde{a_{4,3}} \in H^0 ( \mathbb{P}_{\mathbb{Q}}^2, \mathcal{O}_{\mathbb{P}_{\mathbb{Q}}^2} ( 13 ) )$ merely consists of 105 monomials.

With this preparation we now turn to the actual quantities to evaluate. For instance, the massless spectrum on $C_{\mathbf{5}_{-2}}$ induced by the gauge background $A ( \mathbf{10}_{1} ) ( \lambda )$ is counted by the sheaf cohomologies of the line bundle 
\begin{align}
\begin{split}
\mathcal{L} \left( S_{\mathbf{5}_{-2}} , A \left( \mathbf{10}_{1} \right) \left( \lambda \right)\right) &= \mathcal{O}_{C_{\mathbf{5}_{-2}}} \left( - \frac{3 \lambda}{5} Y_1 + \frac{2 \lambda}{5} Y_2 \right) \otimes \left. \mathcal{O}_{\mathbb{P}_{\mathbb{Q}}^2} \left( 7 \right) \right|_{C_{\mathbf{5}_{-2}}} \, , \\
&\cong \mathcal{O}_{C_{\mathbf{5}_{-2}}} \left( - \lambda Y_1 \right) \otimes \left. \mathcal{O}_{\mathbb{P}_{\mathbb{Q}}^2} \left( 7 + \frac{8 \lambda}{5} \right) \right|_{C_{\mathbf{5}_{-2}}} \, .
\end{split}
\end{align}
As outlined in the previous section, with the help of \texttt{gap} \cite{GAP4} we can in principle compute an \fp graded $S( \mathbb{P}_{\mathbb{Q}}^2)$-module which sheafifies to $\mathcal{L} $ on  $C_{\mathbf{5}_{-2}}$. The question whether this also works in practice strongly depends on the complexity of the involved polynomials $a_{i,j}$. Although the restriction from $B_3 = \mathbb{P}_{\mathbb{Q}}^3$ to $W \cong \mathbb{P}_{\mathbb{Q}}^2$ removes many moduli from the polynomials $a_{i,j}$, we are still left with a huge polynomial $\tilde{a_{1,0}} \tilde{a_{4,3}} - \tilde{a_{3,2}} \tilde{a_{2,1}}$. For such a big polynomial, the currently available Gr\"obner basis algorithms come to their limits, which means that for such big polynomials defining the matter curve $C_{\mathbf{5}_{-2}}$ we are in practice unable to compute the \fp graded $S( \mathbb{P}_{\mathbb{Q}}^2)$-module which sheafifies to give the above line bundle.

To overcome this shortcoming, we will compute the massless spectrum for non-generic matter curves instead. In our first example, we pick
\[ a_{1,0} = c_1 \left( x_1 - x_2 \right)^4, \qquad
a_{2,1} = c_2 x_1^7, \qquad
a_{3,2} = c_3 x_2^{10}, \qquad
a_{4,3} = c_4 x_3^{13} \,  \label{aijmodel1} \]
with $c_i \in \mathbb{N}_{>0}$ (pseudo-)random integers. Then the discriminant $\Delta$ of $P_T$ can be expanded in terms of the GUT-coordinate $w$ as
\begin{align}
\begin{split}
\Delta &= 16 a_{1,0}^4 a_{3,2} \left( -a_{2,1} a_{3,2} + a_{1,0} a_{4,3} \right) w^5 \\
& \qquad + 16 a_{1,0}^2 \left( -8 a_{2,1}^2 a_{3,2}^2 + 8 a_{1,0} a_{2,1} a_{3,2} a_{4,3} + a_{1,0} \left( a_{3,2}^3 + a_{1,0} a_{4,3}^2 \right) \right) w^6 + \mathcal{O} \left( w^7 \right) \,,
\end{split}
\end{align}
where for simplicity we have not written out the $a_{i,j}$ explicitly. No further factorisation occurs, and hence this choice of non-generic matter curves still leaves us with a $SU ( 5 ) \times U ( 1 )_X$-gauge theory. The curve $C_{\mathbf{5}_{-2}}$ is given by
\[ C_{\mathbf{5}_{-2}} = V \left( a_{1,0} a_{4,3} - a_{2,1} a_{3,2} \right) = V \left( \left( x_1 - x_2 \right)^4 x_3^{13} - \frac{c_2 c_3}{c_1 c_4} x_2^7 x_3^{10} \right) \, . \]
This curve $C_{\mathbf{5}_{-2}}$ is \emph{not} smooth. Let us therefore emphasize again that the techniques implemented in \texttt{gap} \cite{GAP4} are not limited to generic or smooth matter curves. In fact we are able to handle just about any subvariety of smooth and complete toric varieties, provided its defining polynomials are of reasonable size so that the currently available Gr\"obner basis algorithms terminate in a timely fashion. We will have far more to say about this in \cite{Bies:2017-3}.

\subsection{Massless Spectrum on \emph{Non}-Generic Matter Curves} \label{sec:Spectrum1}

As an example consider the gauge background
\bea \label{totalfluxchoice}
A \equiv A \left( \mathbf{10}_{1} \right) \left( 5 \right) + A_X \left( \frac{5}{2} \cdot H \right)
\eea
with $H \in \text{Cl} \left( \mathbb{P}_{\mathbb{Q}}^3 \right)$ the hyperplane class on $\mathbb{P}_{\mathbb{Q}}^3$. This gauge background can be checked to satisfy the quantisation condition (\ref{FW2}). This follows already from the analysis in \cite{oai:arXiv.org:1202.3138} around equ. (3.18) therein. The massless spectrum is counted by the sheaf cohomologies of the line bundles
\begin{align}
\begin{split}
\mathcal{L} \left( A, C_{\mathbf{10}_{1}} \right) = \left. \mathcal{O}_{\mathbb{P}_{\mathbb{Q}}^2} \left( - 18 \right) \right|_{C_{\mathbf{10}_{1}}} \,,
\qquad \mathcal{L} \left( A, C_{\mathbf{5}_{3}} \right) &= \left. \mathcal{O}_{\mathbb{P}_{\mathbb{Q}}^2} \left( 13 \right) \right|_{C_{\mathbf{5}_{3}}}  \,, \cr
\mathcal{L} \left( A, C_{\mathbf{5}_{-2}} \right) \cong \mathcal{O}_{C_{\mathbf{5}_{-2}}} \left( - 5 Y_1 \right) \otimes \left. \mathcal{O}_{\mathbb{P}_{\mathbb{Q}}^2} \left( 14 \right) \right|_{C_{\mathbf{5}_{-2}}} \,,
\qquad \mathcal{L} \left( A, C_{\mathbf{1}_{5}} \right) &= \left. \mathcal{O}_{\mathbb{P}_{\mathbb{Q}}^3} \left( 12 \right) \right|_{C_{\mathbf{1}_{5}}} \,.
\end{split}
\end{align}
The first two and the fourth line bundle manifestly arise by pullback of a line bundle on the toric base ${\mathbb{P}_{\mathbb{Q}}^3}$. Therefore we can resolve these bundles by Koszul resolutions, formed from vector bundles on ${\mathbb{P}_{\mathbb{Q}}^3}$. For all of these vector bundles it is possible to compute the cohomology dimensions \eg via \emph{cohomCalg} \cite{Blumenhagen:2010pv, cohomCalg:Implementation, 2011JMP....52c3506J, Rahn:2010fm, Blumenhagen:2010ed}. 

In general this information alone does not suffice to determine the cohomology dimensions of a pullback line bundle uniquely, rather the maps in the resolution need to be taken into account. However, in fortunate cases the exact sequences describing the resolution involve a sufficient number of zeroes which allow one to predict the cohomology dimensions of the pullback line bundle without any knowledge about the involved mappings in the resolution. Indeed, the bundles on $C_{\mathbf{10}_{1}}$, $C_{\mathbf{5}_{3}}$ and $C_{\mathbf{1}_{5}}$ are such fortunate instances. Therefore we are able to determine their cohomology dimension along the algorithms implemented in the \emph{Koszul extension of cohomCalg} \cite{Blumenhagen:2010pv, cohomCalg:Implementation, 2011JMP....52c3506J, Rahn:2010fm, Blumenhagen:2010ed}. The results are
\begin{align}
\begin{split}
h^i \left( C_{\mathbf{10}_{1}}, \mathcal{L} \left( A, C_{\mathbf{10}_{1}} \right) \right) &= \left( 0, 74 \right) \, , \qquad
h^i \left( C_{\mathbf{5}_{3}}, \mathcal{L} \left( A, C_{\mathbf{5}_{3}} \right) \right) = \left( 95, 0 \right) \, , \\
h^i \left( C_{\mathbf{1}_{5}}, \mathcal{L} \left( A, C_{\mathbf{1}_{5}} \right) \right) &= \left( 445, 120 \right) \, .
\end{split}
\end{align}
Note that as a consequence of the zeroes in the resolution, these values are independent of the complex structure moduli of the matter curves. In fact, if the matter curves in question were smooth, the above results for the cohomology groups on $C_{\mathbf{10}_{1}}$ and $C_{\mathbf{5}_{3}}$  would follow already from the Kodaira vanishing theorem and the  Riemann-Roch index theorem.\footnote{For a line bundle $\mathcal{L}$ on a smooth genus $g$ curve $C$ this theorem states that if $\mathrm{deg}(\mathcal{L})  < 0$, $h^0(C, \mathcal{L}) = 0$. By Serre duality, this implies that if $\mathrm{deg}(\mathcal{L})   >  \mathrm{deg}(K_C)  = g-1$, $h^1(C, \mathcal{L}) = 0$.}

By contrast, to determine the cohomology dimensions of the line bundle $\mathcal{L}( A, C_{\mathbf{5}_{-2}})$ we have to invoke the machinery described in  \autoref{subsec:ComputingMasslessSpectraWithGAP} as this line bundle does not descend from a line bundle on $\mathbb{P}^3_\mathbb{Q}$. As it turns out, the result is 
sensitive to the actual choice of Tate polynomials $\tilde{a_{ij}}$, \ie of complex structure moduli defining the elliptic fibration. 

For a number of choices, we compute an \fp graded $S$-module $M$ and then deduce the cohomology dimension of $\tilde{M}$ by use of the technologies described in \autoref{subsec:SheafCohomologyFromFPGradedSModules}. The relevant technical details of the modules involved are displayed in \autoref{sec:Data}. We summarise our findings in the following table:

\begin{center}
\begin{tabular}{|c|c|c|c|c|c|}
\toprule
Module & $\tilde{a_{1,0}}$ & $\tilde{a_{2,1}}$ & $\tilde{a_{3,2}}$ & $\tilde{a_{4,3}}$ & $h^i \left( C_{\mathbf{5}_{-2}}, \mathcal{L} \left( A, C_{\mathbf{5}_{-2}} \right) \right)$ \\
\hline \hline
$M_1$ & $\left( x_1 - x_2 \right)^4$ & $x_1^7$ & $x_2^{10}$ & $x_3^{13}$ & $\left( 22, 43 \right)$ \\
$M_2$ & $\left( x_1 - x_2 \right) \cdot x_3^3$ & $x_1^7$ & $x_2^{10}$ & $x_3^{13}$ & $\left( 21, 42 \right)$ \\
$M_3$ & $x_3^4$ & $x_1^7$ & $x_2^7 \cdot \left( x_1 + x_2 \right)^3$ & $x_3^{12} \cdot \left( x_1 - x_2 \right)$ & $\left( 11, 32 \right)$ \\
$M_4$ & $\left( x_1 - x_2 \right)^3 \cdot x_3$ & $x_1^7$ & $x_2^{10}$ & $x_3^{13}$ & $\left( 9, 30 \right)$ \\
$M_5$ & $x_3^4$ & $x_1^7$ & $x_2^{8} \left( x_1 + x_2 \right)^2$ & $x_3^{11} \cdot \left( x_1 - x_2 \right)^2$ & $\left( 7, 28 \right)$ \\
$M_6$ & $x_3^4$ & $x_1^7$ & $x_2^{10}$ & $x_3^{8} \cdot \left( x_1 - x_2 \right)^5$ & $\left( 6, 27 \right)$ \\
$M_7$ & $x_3^4$ & $x_1^7$ & $x_2^9 \cdot \left( x_1 + x_2 \right)$ & $x_3^{10} \cdot \left( x_1 - x_2 \right)^3$ & $\left( 5, 26 \right)$ \\
\bottomrule
\end{tabular}
\end{center}

In particular we observe jumps in the cohomology dimensions of the line bundle on $C_{\mathbf{5}_{-2}}$ as we wander in the moduli space of the elliptic fibration $\hat{\pi} \colon \hat{Y_4} \twoheadrightarrow B_3$. E.g.\ moving from the first line to the second, we observe that a pair of a chiral and anti-chiral (super)-field becomes massive, and is therefore no longer accounted for by the massless spectrum. In moving to the last line, another 16 such pairs become massive.

\section{Conclusion and Outlook} \label{sec:Conclusion}

In this work we have taken what we believe is an important step forward in our understanding of F-theory vacua beyond the computation of topologically protected quantities such as chiral indices or gauge anomalies. By extending  the framework developed in \cite{Bies:2014sra} we have computed the exact massless charged spectrum in F-theory compactifications to four dimensions. Our first main result is to extract the gauge bundles on matter curves induced in presence of all types of gauge backgrounds underlying gauge fluxes in $H^{2,2}_{\mathrm{vert}}(\hat{Y_4})$. This includes both the gauge backgrounds in the presence of $U(1)$ gauge group factors, studied already in \cite{Bies:2014sra}, and all additional types of vertical gauge backgrounds, which we have called matter surface fluxes. The gauge bundle induced on the matter curves by this second type of backgrounds  pushes forward to a coherent sheaf - as opposed to a line bundle - on the ambient space of the curve. Our second main result is to apply and further develop methods from constructive algebraic geometry to calculate the associated sheaf cohomology groups. This technique has allowed us to determine the exact massless charged matter spectrum in an F-theory vacuum with gauge group $SU(5) \times U(1)$. In particular we have explicitly observed an explicit dependence on the number of chiral-anti-chiral pairs of massless matter on the complex structure moduli. 

The framework developed here opens up many new directions both of conceptual interest and of practical relevance. In \cite{Bies:2017-3} we will apply similar techniques to evaluate also the cohomology groups associated with non-vertical gauge backgrounds, in particular of the type which in the F-theory GUT literature goes by the name of `hypercharge flux.' Again these backgrounds have the property that they do not descend from line bundles on the base and hence the full power of the machinery to compute sheaf cohomology groups will be at work.  

As a spin-off of our investigation of the Chow groups describing the gauge backgrounds we will present in \cite{Bies:2017-2} an intriguing set of  relations between cohomology classes of rational 2-cycles. These will be proven to hold on every smooth elliptically fibred Calabi-Yau 4-fold as a consequence of anomaly cancellation in the associated F-theory vacuum. This generalizes and extends observations made in \cite{Lin:2016vus}. We conjecture that these relations hold even in the Chow ring, as we verify in non-trivial examples. In fact, said relations in the Chow ring have been used in the present work in order to simplify the intersection theoretic operations which extract the gauge bundles on the matter curves. They are yet another manifestation of the close interrelations between the consistency conditions of effective field  theories obtained from string theory and the geometry of the compactifications spaces. 

More generally it would be desirable to advance our understanding of the second Chow group on elliptic 4-folds further. 
There are two aspects to this, one depending on the fibration and one depending on the explicit choice of base.
Concerning the first, the example fibration studied in this work has the property that $h^{2,1}(\hat{Y_4}) = 0$ so that  the intermediate Jacobian in (\ref{SESDeligne})  is trivial. As a result, Deligne cohomology and ordinary cohomology coincide. Despite this simplification it is important to perform all computations within the Chow ring, as done in this work, if we want to extract the exact matter spectrum and not only the chiral index. Nevertheless it would be exciting to explore gauge backgrounds associated with non-trivial, but flat configurations of $C_3$ as encoded in a non-trivial intermediate Jacobian. A generalized Abel-Jacobi map relates these data to the kernel of the cycle map from $\mathrm{CH}^2(\hat{Y_4}) \rightarrow H^{2,2}_{\mathrm{alg}}(\hat{Y_4})$. 
This way, both continuous flat $C_3$ connections and 
discrete $C_3$ backgrounds from torsional $H^{2,1}(\hat Y_4)$ can arise.
Various aspects of the intermediate Jacobian in F-theory compactified on elliptic 4-folds have been studied recently in \cite{Greiner:2015mdm,Greiner:2017ery}.
Concerning the base $B_3$, the explicit example we have studied is manifestly torsion free, and we have therefore not encountered any effects from torsional 4-cycles on the base.
It would be interesting to detect such effects by modifying our computations.

An interesting outcome of our investigations is the aforementioned jump in the number of massless states as we vary the complex structure moduli. From general field theory reasoning such jumps in moduli space are clearly expected. They are the manifestation of the lifting of vectorlike pairs as we vary the vacuum expectation value of some of the chiral fields of the model. Analogous effects have been studied intensively for heterotic compactifications such as \cite{Braun:2005xp,Bouchard:2005ag,Anderson:2013qca} and references therein. It would be exciting to determine the minimal number of vectorlike pairs for a given topological type of F-theory model as we vary the complex structure moduli and to interpret this result from an effective field theory point of view. 

In fact the explicit computations in this work have been performed at highly non-generic points in moduli space. The practical reason behind this was the need to reduce the complexity of the involved polynomials. Only this reduction allowed \texttt{gap} \cite{GAP4} to model the line bundle in question by an \fp graded $S$-module $M$. This limitation in turn is caused by the involved Gr\"obner basis algorithms. 
Recall that the module $M$ sheafifes to give a coherent sheaf $\tilde{M}$. The computation of the sheaf cohomologies of $\tilde{M}$ involves Gr\"obner basis algorithms as well. If we restrict ourselves to the computation of only the cohomology dimensions, then it is indeed possible to apply algorithms in which the Gr\"obner basis computations are replaced by Gauss eliminations. For the latter far more performant algorithms exist, \eg in \texttt{MAGMA} \cite{MR1484478}. Consequently, this approach increased the performance of our algorithms a lot. However, both computing models $M$ for line bundles at more general points of the moduli space and subsequently identifying an explicit basis of the cohomology groups of $\tilde{M}$ hing on more efficient Gr\"obner basis algorithms.  It is therefore desirable to find improvements to such algorithms.

We have focused, in this article, on the computation of the massless matter spectrum in F-theory compactified to four dimensions. As stressed already in the introduction, in the recently explored F-theory compactifications to two dimensions \cite{Schafer-Nameki:2016cfr,Apruzzi:2016iac} the massless matter spectrum likewise depends on the gauge background, which can be described in very similar terms. It will be interesting to extend our formalism to F-theory compactifications on Calabi-Yau 5-folds.

Finally, we have insisted throughout this work that the elliptic fibration $\hat{Y_4}$ be smooth in order to avoid dealing with singularities. Since resolving an elliptic fibration amounts to moving along the Coulomb branch of the dual three-dimensional M-theory vacuum this limits ourselves to studying Abelian gauge backgrounds. 
Similar challenges arise when it comes to describing certain T-brane configurations which obstruct a resolution of the 4-fold on singular spaces and force us to work on singular spaces instead \cite{Anderson:2013rka,Collinucci:2014qfa,Collinucci:2014taa,Collinucci:2016hpz,Anderson:2017rpr}. We are optimistic that these are no unsurpassable obstacles. In particular, a generalisation of intersection theory within the Chow ring on non-smooth varieties exists, and it is hence a natural question how far one can push the present formalism concerning more general, non-Abelian backgrounds. We look forward to coming back to these questions in the near future.

\paragraph{Acknowledgements}
We thank Sebastian Gutsche, Jim Halverson, Craig Lawrie, Ling Lin, Sebastian Posur, Sakura Sch\"afer-Nameki, Wati Taylor, and especially Mohamed Barakat for valuable discussions, and \texttt{plesken} for his/her reliable computations.
M.B.\ thanks the University of Siegen and the University of Aachen, C.M.\ thanks Heidelberg University, M.B.\ and C.M.\ thank CERN, and  T.W.\ thanks MIT, Oxford University as well as MPI and LMU Munich for hospitality. This work was supported in part by Transregio TR 33 `The Dark Universe' and GK `Particle Physics Beyond the Standard Model'. The work of M.B.\ and C.M.\ are supported by the Studienstiftung des Deutschen Volkes and  by the Munich Excellence Cluster for Fundamental Physics `Origin and
the Structure of the Universe', respectively.

\newpage
\appendix

\section{From Chow Groups to \texorpdfstring{$\mathbf{C_3}$}{C3}-Backgrounds} \label{subsec:ReviewChowGroupsAndG4Fluxes}

\subsection{Rational Equivalence and the Refined Cycle Map}

In this appendix we describe the parametrisation of elements in $H^4_D ( \hat{Y}_4, \mathbb{Z} ( 2 ))$ representing the 3-form gauge background in F/M-theory on a smooth elliptic 4-fold $\hat{Y_4}$ in terms of the Chow group $\mathrm{CH}^2(\hat{Y_4})$ of complex 2-cycles modulo rational equivalence. For further details we also refer to \cite{Bies:2014sra}.

The group of algebraic cycles $Z^p ( \hat{Y}_4 )$ of complex codimension $p$ in $\hat{Y}_4$ is formed from elements
\[ C = \sum_{i = 1}^{N}{n_i C_i} \]
for suitable $N \in \mathbb{N}$, $n_i \in \mathbb{Z}$ and $C_i$ \emph{not necessarily smooth} but irreducible subvarieties of $\hat{Y}_4$.\footnote{We use the symbol $Z^p ( \hat{Y}_4 )$ to denote the group of algebraic cycle of complex \textbf{co}dimension $p$ in $\hat{Y}_4$. In contrast, $Z_p ( \hat{Y}_4 )$ is to denote the group of algebraic cycle of complex dimension $p$ in $\hat{Y}_4$. We adopt this notation also for the Chow groups, \ie $\text{CH}^2 ( \hat{Y}_4 )$ represents classes of algebraic cycles of codimension $2$ in $\hat{Y}_4$, whilst \eg $\text{CH}_1 ( \hat{Y}_4 )$ is for classes of algebraic cycle of dimension $1$ in $\hat{Y}_4$.} An algebraic cycle $C \in Z_p ( \hat{Y}_4)$ is rationally equivalent to zero, $C \sim 0$, if and only if
\[ C = \sum_{i = 1}^{N}{\left[ \text{div} \left( r_i \right) \right]} \]
for suitable $N \in \mathbb{N}$ and invertible rational functions $r_i \in \mathbb{C} ( W )^\ast$ on some $( p + 1 )$-dimensional subvarieties $W_i$ of $\hat{Y}_4$. Let $\text{Rat}_p ( \hat{Y}_4 )$ be the subgroup of $Z_p ( \hat{Y}_4 )$ formed from all algebraic cycles which are rationally equivalent to $0$. Then the Chow group $\text{CH}_p ( \hat{Y}_4 )$ is defined as the quotient
\[ \text{CH}_p \left( \hat{Y}_4 \right) := Z_p \left( \hat{Y}_4 \right) / \text{Rat}_p \left( \hat{Y}_4 \right) \, . \]
To any algebraic cycle in $Z^p ( \hat{Y}_4 )$ one can associated a cocycle in $H_{\mathbb{Z}}^{p,p} ( \hat{Y}_4 )$ via a group homomorphism $\gamma_{\hat{Y}_4, p} \colon Z^p ( \hat{Y}_4 ) \to H^{p,p}_{\mathbb{Z}} ( \hat{Y}_4 )$ which is termed the \emph{cycle map}.

Next let us explain how we specify 3-form data, given by elements of $H^4_D ( \hat{Y}_4, \mathbb{Z} ( 2 ) )$, by a class of algebraic cycles in $\text{CH}^2 ( \hat{Y}_4 )$. The key insight is that there exists a so-called \emph{refined cycle map} (see \eg p.~123 in \cite{GreenMurreVoisin})
\[ \hat{\gamma}_{\hat{Y}_4, p} \colon Z^p \left( \hat{Y}_4 \right) \to H^4_D \left( \hat{Y}_4, \mathbb{Z} \left( 2 \right) \right) \, . \]
This morphism is a group homomorphism and respects rational equivalence. Hence given algebraic cycles $C_1, C_2 \in Z^p ( \hat{Y}_4 )$ it holds
\[ C_1 \sim C_2 \; \Rightarrow \; \hat{\gamma}_{\hat{Y}_4, p} \left( C_1 \right) = \hat{\gamma}_{\hat{Y}_4, p} \left( C_2 \right) \, . \]
Therefore the refined cycle map extends to a map $\text{CH}^2 ( \hat{Y}_4 ) \to H^4_D ( \hat{Y}_4, \mathbb{Z} \left( 2 \right) )$. 

In concrete applications, it is oftentimes possible to express an algebraic cycle $A \in Z^2 ( \hat{Y}_4) $ in terms of data of a toric ambient space. This is possible whenever $\hat{Y}_4$ can be embedded into a smooth toric variety $X_\Sigma$. Let $j \colon \hat{Y}_4 \hookrightarrow X_\Sigma$ denote the corresponding embedding. Given that $X_\Sigma$ is smooth it is known that this map indeed induces pullback maps of the Chow groups 
\[ j^\ast \colon \text{CH}^2 ( X_\Sigma ) \to \text{CH}^2 ( \hat{Y}_4 ) \, . \] 
Let $S$ be the Cox ring of $X_\Sigma$ (over $\mathbb{Q}$), $I_{\text{SR}} \leq S$ its Stanley-Reisner ideal and $I_{\text{LR}} \leq S$ the ideal of linear relations. Then by smoothness of $X_\Sigma$ it even holds \cite{cox2011toric}
\[ \text{CH}^{\bullet} \left( X_\Sigma \right) \cong H^{\bullet} \left( X_\Sigma, \mathbb{Z} \right) \cong S / \left( I_{\text{SR}} + I_{\text{LR}} \right) \equiv R_{\mathbb{Q}} \left( \Sigma \right) \, . \]
Consequently, in such situations modifications of the pre-image of $A$ on $X_\Sigma$ which leave its class in $H^{\bullet} ( X_\Sigma, \mathbb{Z} )$ unchanged do not alter the gauge background described by $A$ via the refined cycle map. Indeed, we will make use of this freedom to simplify our computations later.

By use of the commutative diagram\footnote{The maps $\pi$ are not to be confused with the projection map in the elliptic fibration $\pi \colon Y_4 \twoheadrightarrow B_3$ that is used frequently in the main text.} in \autoref{fig:C3BackgroundFromChow}  we can now summarise our strategy as follows:
\begin{enumerate}
 \item Specify $\mathcal{A} \in Z^2 ( X_\Sigma )$. Use manipulations respecting the homology class associated with ${\cal A}$ to simplify this algebraic cycle or 
      represent it differently whenever necessary.
 \item $\mathcal{A}$ induces $\mathcal{\alpha} = \pi ( \mathcal{A} ) \in \text{CH}^2 ( X_\Sigma )$ and $a \in \text{CH} ( \hat{Y}_4 )$. 
 \item $\hat{\gamma} ( a ) \in H^4_D ( \hat{Y}_4, \mathbb{Z} ( 2 ) )$ is the 3-form data that we are after in the first place.
 \item $\gamma ( a ) = \hat{c_2} \circ \hat{\gamma} ( a ) = G_4$ is the (class of) differential forms usually referred to as $G_4$-flux.
\end{enumerate}

\begin{figure}
\begin{align*}
\xymatrix{
         & & & \hat{\gamma} \left( a \right) \ar@{}[d]|-*[@]{\in} \ar@{|->}[r] & G_4 \ar@{}[d]|-*[@]{\in} \\
0 \ar[r] & J^2 \left( \hat{Y}_4 \right) \ar[rr] & & H^4_D \left( \hat{Y}_4, \mathbb{Z} \left( 2 \right) \right) \ar[r]^-{\hat{c}_2} & H^{2,2}_{\mathbb{Z}} \left( 
                                                 \hat{Y}_4 \right) \ar[r] & 0 \\
         &                                     & a \ar@{}[r]|-*[@]{\in} \ar@/^2pc/@{|->}[uur] & \text{CH}^2 \left( \hat{Y}_4 \right) \ar[u]^-{\hat{\gamma}} \ar[ru]^-{\gamma} & Z^2 \left( \hat{Y}_4 \right) \ar[l]_-{\pi} \ar@{-->}[u] \ar@{}[r]|-*[@]{\ni} & A \\
         & & \mathcal{\alpha} \ar@{}[r]|-*[@]{\in} \ar[u] & \text{CH}^2 \left( X_\Sigma \right) \ar[u]^-{j^\ast} \\
         & & \mathcal{A} \ar@{}[r]|-*[@]{\in} \ar[u] & \text{Z}^2 \left( X_\Sigma \right) \ar[u]^-{\pi}
         }
\end{align*}
\caption{Summary on how (classes of) algebraic cycles $A \in \text{CH}^2 \left( \hat{Y_4} \right)$ encode gauge backgrounds in $H^4_D \left( \hat{Y_4}, \mathbb{Z} \left( 2 \right) \right)$.}
\label{fig:C3BackgroundFromChow}
\end{figure}

Note that in pratical applications, including the model presented in \ref{sec:Simplifying}), it can happen that the toric ambient space of $\hat{Y}_4$ is not smooth, but rather a complete toric orbifold. Since such a toric variety is simplicial it follows from theorem 12.5.3 in \cite{cox2011toric} that
\[ \text{CH}^{\bullet} \left( X_\Sigma \right)_{\mathbb{Q}} \cong H^\bullet \left( X_\Sigma, \mathbb{Q} \right) \cong R_{\mathbb{Q}} \left( \Sigma \right)_{\mathbb{Q}} \, . \]
Therefore we should wonder if even in such a geometric setup, the pullback $j^\ast \colon \text{CH} ( X_\Sigma ) \to \text{CH} ( \hat{Y}_4 )$ is well-defined. Hence note that this very pullback amounts to computing intersection products of elements of $\text{CH} ( X_\Sigma )$ and $\hat{Y}_4$, and this intersection is well-defined as long as the embedding $\iota \colon \hat{Y}_4 \hookrightarrow X_\Sigma$ is a closed embedding and $\hat{Y}_4$ is a \emph{local} complete intersection \cite{Bies:2014sra}, as these conditions guarantee a well-defined Gysin-homomorphism. Hence, under these assumptions even for a complete toric orbifold $X_\Sigma$ we can start with elements $\mathcal{A} \in Z^2 ( X_\Sigma )$, modify them by manipulations which respect the homology class of $\mathcal{A}$ in $X_\Sigma$ to simplify computations, and use the pullback of this cycle to $\hat{Y}_4$ to model $G_4$-fluxes on $\hat{Y}_4$.

Finally, let us mention that in this article we will not distinguish algebraic cycles $A \in Z^2 ( \hat{Y}_4 )$ from their classes in $\text{CH}^2 ( \hat{Y}_4 )$. Rather we use capital and upright letters to denote both the relevant class and (if necessary) an explicit representant of this class. Similarly $\mathcal{A}$ will denote elements of $\text{CH}^2 ( X_\Sigma )$ and $Z^2 ( X_\Sigma)$ depending on the context. Whenever it is not explicitly necessary to assume a toric ambient space $X_\Sigma$ we will use the label $\hat X_5$ for a (non-toric) 5 complex-dimensional ambient space of $\hat{Y}_4$.

\subsection{Intersection Product and Extraction of Line Bundles} \label{app:intersection}

The Chow ring is endowed with a natural intersection product. To introduce this intersection product, let $f \colon X \to Y$ be a morphism from a variety $X$ (not necessarily smooth) to a smooth variety $Y$ of dimension $n$. Now the graph morphism $\gamma_f \colon X \to X \times Y \; , \; x \mapsto ( x, f ( x ) )$ induces a Gysin morphism $\gamma_f^\ast$ such that
\[ \text{CH}_k \left( X \right) \otimes \text{CH}_l \left( Y \right) \stackrel{\times}{\rightarrow} \text{CH}_{k+l} \left( X \times Y \right) \stackrel{\gamma^\ast_f}{\rightarrow} \text{CH}_{k+l-n} \left( X \right) \]
is a well-defined mapping. Given $\alpha \in \text{CH}_k ( X )$, $\beta \in \text{CH}_l ( Y )$ we denote the image of $\alpha \otimes \beta$ under the above map as $\alpha \cdot_f \beta$. We term this the intersection product of $\alpha$, $\beta$ under the morphism $f$. We will oftentimes omit $f$.

Next recall that we have matter curves $C_\bfR \subseteq B_3$ over which the elliptic fibre degenerates. Therefore the generic fibre over these matter curves in the resolved 4-fold $\hat{Y}_4 \stackrel{\pi}{\rightarrow} B_3$ is such that $\pi^{-1} ( C_\bfR ) \subseteq \hat{Y}_4$ is a sum of $\mathbb{P}^1$'s (the rational curves into which the $T^2$ splits along these loci) fibred over $C_\bfR$. Formal sums of $\mathbb{P}^1$'s fibred over $C_\bfR$ constitute the matter surfaces $S_\bfR^{a}$ - each of which represents a state of localised matter in representation $\bfR$ with weight $\beta^a(\bfR)$ over the matter curve $C_\bfR$.

Let $\iota_{\bfR,a} \colon S^a_{\bfR} \hookrightarrow \hat{Y}_4$ be the canonical embedding of this matter surface into $\hat{Y}_4$. In a slight abuse of notation we also denote the class of this matter surface by $S^a_{\bfR} \in \text{CH}_2 ( S^a_{\bf R} )$. Moreoever let $\pi_{\bfR,a}$ be the projection
\[ \pi_{\bfR,a} \colon S^a_{\bfR} \twoheadrightarrow C_{\bfR} \, . \]
Then we can compute the intersection product $S^a_{\bfR} \cdot_{\iota_{\bfR,a}} A \in \text{CH}_0 ( S^a_\bfR )$, where $A \in \text{CH}^2 ( \hat{Y}_4 )$ represents a $C_3$ background. Consequently
\[ L_{\bfR,a} := \pi_{\bfR,a \ast} \left( S^a_\bfR \cdot_{\iota_{\bfR,a}} A \right) \in \text{CH}_0 \left( C_\bfR \right) \cong \text{Pic} \left( C_\bfR \right) \]
defines a line bundle on the matter curve $C_\bfR$. 

In explicit examples, one of the obstacles to overcome is the computation of the relevant intersection products. 
We have presented two ways to perform this computation. Either, we proceed as detailed in \autoref{subsec:NewStrategyForMasslessSpectra} and \autoref{subsec:ExampleMasslessSpectrumOfUniversalFlux}.
Alternatively, in suitable models we have an embedding $j \colon \hat{Y}_4 \hookrightarrow X_\Sigma$. For the fluxes of our interest it is then possible to describe $A \in \text{CH}^2 ( \hat{Y}_4 )$ as $\mathcal{A} \in \text{CH}^2 ( X_\Sigma )$. We then pick an explicit algebraic cycle to represent this class, and manipulate this cycle by use of $I_{\text{SR}} ( X_\Sigma )$ and $I_{\text{LR}} ( X_\Sigma )$. These manipulations allow us to ensure for transverse intersections, in which case it suffices to merely compute the set-theoretic intersections.

A similar computation allows us to extract the line bundles on the non-Abelian brane stacks along $\Delta_I$, as described in \autoref{bulk-global}:
Here we consider the resolution divisor $E_{i_I}$ with projection $\pi_{i_I}: E_{i_I} \, {\twoheadrightarrow} \, \Delta_I$ and canonical embedding  $\iota_{i_I}: E_{i_I} \hookrightarrow \hat{Y_4}$, giving rise to
\[ L_{i_I} := \pi_{i_I \ast} \left( E_{i_I} \cdot_{\iota_{i_I}}  A  \right) \in \mathrm{CH}_1 (\Delta_I) \simeq \mathrm{CH}^1 (\Delta_I) \simeq \mathrm{Pic}(\Delta_I) \, . \]

\section{Fibre Structure of a Resolved Tate Model} \label{sec:FibreStructure}

In this appendix we analyse the fibre structure of the resolved 4-fold $\hat{\pi} \colon \hat{Y}_4 \twoheadrightarrow B_3$ of the F-theory GUT models described in \autoref{subsec:SpecialFTheoryGUTModel}, which are based on \cite{Krause:2011xj}. 

The general philosophy is to identify $\mathbb{P}^1$-fibrations -- of which some will turn out to be present only over 'special' subloci of $\text{S}_{\text{GUT}}$, namely the matter curves and the Yukawa points -- and then to work out their intersection numbers in the fibre. The discussion of the massless spectrum of matter surface fluxes, \eg in \autoref{sec:MasslessSpectraOfCurveSupportFluxesII}, relies heavily on this information. This in turn is the main reason why we wish to present the details of the fibre structure here. 

The fibre structure of the $SU ( 5 ) \times U ( 1 )_X$-top in question was originally worked out in \cite{Krause:2011xj}. We refine this analysis in two important respects:
\begin{itemize}
 \item A $\mathbb{P}^1$-fibration present over generic points of $S_{\text{GUT}}$ -- that is points which are not contained in any matter curve -- in general splits into 
      a formal linear combination of $\mathbb{P}^1$-fibrations over the matter curves. Similarly a $\mathbb{P}^1$-fibration present over generic points of the matter curves -- \ie non-Yukawa points of the matter curves -- in general splits into a formal linear combination of $\mathbb{P}^1$-fibrations. In order to deduce these splittings we use primary decompositions of the relevant ideals, which differs from the approach in \cite{Krause:2011xj}. \\
      In consequence we find different defining equations for $\mathbb{P}^1_{3G} ( \mathbf{5}_{-2} )$, $\mathbb{P}^1_{3H} ( \mathbf{5}_{-2} )$ and different splittings for these fibrations once restricted to the Yukawa locus $Y_1$, namely
      \[ \mathbb{P}^1_{3G} \left( \mathbf{5}_{-2} \right) \to \mathbb{P}^1_{34} \left( Y_1 \right), \qquad \mathbb{P}^1_{3H} \left( \mathbf{5}_{-2} \right) \to \mathbb{P}^1_{3J} \left( Y_1 \right) \, . \]
 \item In \cite{Krause:2011xj} the intersections of the $\mathbb{P}^1$-fibrations were identified to have the \emph{structure} of certain Dynkin diagrams. However, the 
      intersection numbers -- in particular the self-intersection numbers -- were not stated. Here we work out these numbers. \\
      To describe our findings let us introduce the notation $T^2 ( C_{\bfR} )$ to indicate the total elliptic fibre over a matter curve $C_{\bfR}$. Similarly $T^2 ( Y_i )$ is to indicate the total elliptic fibre over the Yukawa locus $Y_i$. Furthermore let $\mathbb{P}_i ( C_{\bfR} )$ and $\mathbb{P}_i ( Y_j )$ denote a $\mathbb{P}^1$-fibration over a matter curve $C_{\bfR}$ and a Yukawa locus $Y_j$ respectively. Then for all matter curves $C_{\bfR}$ and all Yukawa loci $Y_j$ described by the $SU ( 5 ) \times U ( 1 )_X$-top in question, the following holds true:
      \[ T^2 \left( Y_j \right) \cdot \mathbb{P}^1_i \left( Y_j \right) = T^2 \left( C_{\bfR} \right) \cdot \mathbb{P}^1_i \left( C_{\bfR} \right) = 0 \, . \]
      Next suppose $Y_j \subseteq C_{\bfR}$ and that the splitting of $\mathbb{P}^1_i ( C_{\bfR} )$ onto $Y_j$ takes the form
      \[ \mathbb{P}^1_i \left( C_{\bfR} \right) \to \sum_{k = 1}^{N \left( i \right)}{\alpha_{k}^{(i)} \mathbb{P}^1_{k} \left( Y_j \right)} \, . \]
      Naively we might expect
      \[ \mathbb{P}^1_i \left( C_{\bfR} \right) \cdot \mathbb{P}^1_j \left( C_{\bfR} \right) = \left( \sum_{k = 1}^{N \left( i \right)}{\alpha_{k}^{(i)} \mathbb{P}^1_{k} \left( Y_j \right)} \right) \cdot \left( \sum_{l = 1}^{N \left( j \right) }{\alpha_{l}^{(j)} \mathbb{P}^1_{l} \left( Y_j \right)} \right) \, . \label{equ:IntersectionsStableWRTIntersections} \]
      As it turns out, in the geometry described by the $SU ( 5 ) \times U ( 1 )_X$ this need not be the case. Rather it fails 
      precisely for the restriction of $C_{\mathbf{10}_1}$ to $Y_1$ which involves either $\mathbb{P}^1_{24} ( Y_1 )$ or $\mathbb{P}^1_{2B} ( Y_1 )$. The reason for this failure is that these two $\mathbb{P}^1$-fibrations encounter a $\mathbb{Z}_2$-singularity over their intersection point. This parallels the situation studied for the enhancement $A_5 \hookrightarrow E_6$ in \cite{Morrison:2011mb}. As a consequence we find half-integer intersection numbers
      \[ \mathbb{P}^1_{24} \left( Y_1 \right)^2 = \mathbb{P}^1_{2B} \left( Y_1 \right)^2 = - \frac{3}{2}, \qquad \mathbb{P}^1_{24} \left( Y_1 \right) \cdot \mathbb{P}^1_{2B} \left( Y_1 \right) = \frac{1}{2} \, . \]
   A particular example where (\ref{equ:IntersectionsStableWRTIntersections}) fails is follows:
      \[ \mathbb{P}_{24} \left( \mathbf{10}_{1} \right)^2 = -2 \neq - \frac{3}{2} = \mathbb{P}^1_{24} \left( Y_1 \right)^2 \, . \]
\end{itemize}

Before we proceed, let us mention that we work with the triangulation $T_{11}$ of \cite{Krause:2011xj} throughout this entire article.

\subsection{Intersection Structure away from Matter Curves}

We start by looking at the five divisors $E_i := V ( P_T^\prime, e_i ) \subseteq \hat{Y}_4$ for $0 \leq i \leq 4$. Note that $e_i = 0$ automatically implies that the 'new' GUT-coordinate $e_0 e_1 e_2 e_3 e_4$ vanishes. Therefore $E_i$ indeed is a subset of $\hat{Y}_4$. These subsets can be understood as fibration of the i-th exceptional divisor over the GUT-surface $S_{\text{GUT}}$.

Now let $p \in S_{\text{GUT}}$ a point which is not contained in any matter curve. By means of the projection map $\hat{\pi} \colon \hat{Y}_4 \twoheadrightarrow B_3$ we can describe the fibre over the point $p$ as $\hat{\pi}^{-1} ( p )$. We now wish to work out the intersection structure of the divisor classes $E_i$ in $\hat{\pi}^{-1} ( p )$. For simplicity we merely focus on the set-theoretic intersection $E_0 \cap E_1$. By use of the Stanley-Rei{\ss}ner ideal of the top -- see \autoref{subsec:SpecialFTheoryGUTModel} -- it is readily confirmed that
\begin{align}
\begin{split}
E_0 \cap E_1 \cap \hat{\pi}^{-1} \left( p \right) &= V \left( e_0, e_1, e_3 e_4 s y^2 + a_1 \left( z_i \right) s x y z \right) \cap \pi^{-1} \left( p \right) \\
           &= \left\{ p \right\} \times \underbrace{\left\{ \left[ 0 : 1 : 1 : - a_1 \left( p \right) : 1 : 1 : 1 : 1 \right] \right\}}_{\text{ fibre coordinates }} \, ,
\end{split}
\end{align}
where $p = [ p_1 : p_2 : \dots : p_{n-1} : e_0 = 0 ]$ are inhomogeneous coordinates of the point $p$. Hence we have found a single intersection point in $\hat{\pi}^{-1} ( p )$. This finding can be made precise to state that in $\hat{\pi}^{-1} ( p )$ the divisor classes $E_0$ and $E_1$ intersect with intersection number 1 \cite{Krause:2011xj}. Moreover this analysis is easily repeated for all intersections of the divisor classes $E_i$, $0 \leq i \leq 4$. 

To compute $U ( 1 )_X$-charges, intersection numbers with the $U ( 1 )_X$-generator $U_X$ are required. These intersections involve \cite{Krause:2011xj}
\[ E_5 = V \left( P_T^\prime, s \right) - V \left( P_T^\prime, z \right) - V \left( P_T^\prime, k_{B_3} \right) \] 
where $k_{B_3}$ is a polynomial in the coordinate ring of $X_5$ such that its degree matches $\overline{\mathcal{K}}_{B_3}$. To simplify notation we set $\alpha = V ( P_T^\prime, s )$, $\beta = V ( P_T^\prime, z )$ and $\gamma = V ( P_T^\prime, k_{B_3} )$.

The intersection numbers are then as follows:
\begin{align}
\begin{tabular}{|c||c|c|c|c|c||c|c|c|c|}
\toprule
$E_i \cdot E_j \cdot \hat{\pi}^{-1} \left( p \right)$ & $E_0$ & $E_1$ & $E_2$ & $E_3$ & $E_4$ & $E_5$ & $\alpha$ & $\beta$ & $\gamma$ \\
\hline \hline
$E_0$ & -2 & 1  & 0  & 0  & 1 & -1 & 0 & 1 & 0 \\
$E_1$ &  1 & -2 & 1  & 0  & 0 & 0 & 0 & 0 & 0 \\
$E_2$ &  0 &  1 & -2 & 1  & 0 & 0 & 0 & 0 & 0 \\
$E_3$ &  0 &  0 & 1  & -2 & 1 & 1 & 1 & 0 & 0 \\
$E_4$ &  1 &  0 & 0  & 1  & -2 & 0 & 0 & 0 & 0 \\
\bottomrule
\end{tabular}
\end{align}

\subsection{Matter Curves}\label{app_Mattersurfaces}

\subsubsection*{Intersection Structure over \texorpdfstring{$\mathbf{C_{\mathbf{10}_{1}}}$}{C10} away from Yukawa Loci}

Over the matter curves singularity enhancements occur. This expresses itself geometrically in the presence of new $\mathbb{P}^1$-fibrations, of which linear combinations eventually serve as matter surfaces. Over $C_{\mathbf{10}_{1}}$ the following six $\mathbb{P}^1$-fibrations are present:
\begin{align} 
\begin{split}
\mathbb{P}_{0,A}^1 \left( \mathbf{10}_{1} \right) &= V \left( a_{1,0}, e_0, y^2 e_4 - x^3 s e_1 e_2^2 \right) \, , \\
\mathbb{P}_{14}^1 \left( \mathbf{10}_{1} \right) &= V \left( a_{1,0}, e_1, e_4 \right) \, , \\
\mathbb{P}_{24}^1 \left( \mathbf{10}_{1} \right) &= V \left( a_{1,0}, e_2, e_4 \right) \, , \\
\mathbb{P}_{2B}^1 \left( \mathbf{10}_{1} \right) &= V \left( a_{1,0}, e_2, y s e_3 + a_{3,2} z^3 e_0^2 e_1 \right) \, , \\
\mathbb{P}_{3C}^1 \left( \mathbf{10}_{1} \right) &= V \left( a_{1,0}, e_3, a_{3,2} y z e_0 e_4 - a_{2,1} x^2 s e_2 - a_{4,3} x z^2 e_0^2 e_1 e_2 e_4 \right) \, , \\
\mathbb{P}^1_{4D} \left( \mathbf{10}_{1} \right) &= V \left( a_{1,0}, e_4, x s e_2 e_3 + a_{2,1} z^2 e_0 \right) \, .
\end{split}
\end{align}
The total elliptic fibre over $C_{\mathbf{10}_{1}}$ is given by
\begin{align}
\begin{split}
T^2 \left( \mathbf{10}_{1} \right) &= \mathbb{P}_{0A}^1 \left( \mathbf{10}_{1} \right) + 2 \mathbb{P}_{14}^1 \left( \mathbf{10}_{1} \right) +  2 \mathbb{P}_{24}^1 \left( \mathbf{10}_{1} \right) +  \mathbb{P}_{2B}^1 \left( \mathbf{10}_{1} \right) \\
& \qquad +  \mathbb{P}_{3C}^1 \left( \mathbf{10}_{1} \right) +  \mathbb{P}_{4D}^1 \left( \mathbf{10}_{1} \right) \, .
\end{split}
\end{align}
They above $\mathbb{P}^1$-fibration originate from the divisors $E_i$ according to the following splitting:
\begin{align}
\begin{tabular}{|c|c|}
\toprule
Original & Split components over $C_{10_{1}}$ \\
\hline \hline
$E_0$ & $\mathbb{P}_{0A}^1 \left( \mathbf{10}_{1} \right)$ \\
$E_1$ & $\mathbb{P}_{14}^1 \left( \mathbf{10}_{1} \right)$ \\
$E_2$ & $\mathbb{P}_{24}^1\left( \mathbf{10}_{1} \right)$ + $\mathbb{P}_{2B}^1\left( \mathbf{10}_1 \right)$ \\
$E_3$ & $\mathbb{P}_{3C}^1\left( \mathbf{10}_1 \right)$ \\
$E_4$ & $\mathbb{P}_{14}^1\left( \mathbf{10}_1 \right)$ + $\mathbb{P}_{24}^1\left( \mathbf{10}_1 \right)$ + $\mathbb{P}_{4D}^1\left( \mathbf{10}_1 \right)$ \\
\bottomrule
\end{tabular}
\end{align}
Over $p \in C_{\mathbf{10}_1}$ which is not a Yukawa point, these $\mathbb{P}^1$-fibrations intersect in $\hat{\pi}^{-1} \left( p \right)$ as follows:
\begin{align}
\begin{tabular}{|c||c|c|c|}
\toprule
 & $\mathbb{P}_{0A}^1 \left( \mathbf{10}_1 \right)$ & $\mathbb{P}_{14}^1 \left( \mathbf{10}_1 \right)$ & $\mathbb{P}_{24}^1 \left( \mathbf{10}_1 \right)$ \\
\hline \hline
$\mathbb{P}_{0A}^1 \left( \mathbf{10}_1 \right)$ & -2 & 1  & 0 \\
$\mathbb{P}_{14}^1 \left( \mathbf{10}_1 \right)$  &  1 & -2 & 1 \\
$\mathbb{P}_{24}^1 \left( \mathbf{10}_1 \right)$  &  0 &  1 & -2 \\
$\mathbb{P}_{2B}^1 \left( \mathbf{10}_1 \right)$ &  0 &  0 & 1 \\
$\mathbb{P}_{3C}^1 \left( \mathbf{10}_1 \right)$ &  0 &  0 & 1 \\
$\mathbb{P}_{4D}^1 \left( \mathbf{10}_1 \right)$ &  0 &  1 & 0 \\
\hline \hline
 & $\mathbb{P}_{2B}^1 \left( \mathbf{10}_1 \right)$ & $\mathbb{P}_{3C}^1 \left( \mathbf{10}_1 \right)$ & $\mathbb{P}^1_{4D} \left( \mathbf{10}_1 \right)$ \\
\hline \hline
$\mathbb{P}_{0A}^1 \left( \mathbf{10}_1 \right)$ & 0  & 0  & 0 \\
$\mathbb{P}_{14}^1 \left( \mathbf{10}_1 \right)$ & 0  & 0  & 1 \\
$\mathbb{P}_{24}^1 \left( \mathbf{10}_1 \right)$ & 1  & 1  & 0 \\
$\mathbb{P}_{2B}^1 \left( \mathbf{10}_1 \right)$ & -2 & 0  & 0 \\
$\mathbb{P}_{3C}^1 \left( \mathbf{10}_1 \right)$ & 0  & -2 & 0 \\
$\mathbb{P}_{4D}^1 \left( \mathbf{10}_1 \right)$ & 0  & 0  & -2 \\
\bottomrule
\end{tabular}
\end{align}
The intersection numbers between the $\mathbb{P}^1$-fibrations over $C_{\mathbf{10}_1}$ and the pullbacks of the divisors $E_i$ -- including $E_5$ -- onto the fibration over $C_{\mathbf{10}_1}$ are readily computed. The results are as follows:
\begin{align}
\begin{tabular}{|c||c|c|c|c|c||c|c|c|c|}
\toprule
 & $E_0$ & $E_1$ & $E_2$ & $E_3$ & $E_4$ & $E_5 = \alpha - \beta - \gamma$ & $\alpha$ & $\beta$ & $\gamma$ \\
\hline \hline
$\mathbb{P}_{0A}^1 \left( \mathbf{10}_1 \right)$ & -2 & 1  & 0  & 0  & 1  & -1 & 0 & 1 & 0 \\
$\mathbb{P}_{14}^1 \left( \mathbf{10}_1 \right)$ &  1 & -2 & 1  & 0  & 0  & 0  & 0 & 0 & 0 \\
$\mathbb{P}_{24}^1 \left( \mathbf{10}_1 \right)$ &  0 &  1 & -1 & 1  & -1 & 0  & 0 & 0 & 0 \\
$\mathbb{P}_{2B}^1 \left( \mathbf{10}_1 \right)$ &  0 &  0 & -1 & 0  & 1  & 0  & 0 & 0 & 0 \\
$\mathbb{P}_{3C}^1 \left( \mathbf{10}_1 \right)$ &  0 &  0 & 1  & -2 & 1  & 1  & 1 & 0 & 0 \\
$\mathbb{P}_{4D}^1 \left( \mathbf{10}_1 \right)$ &  0 &  1 & 0  & 0  & -1 & 0  & 0 & 0 & 0 \\
\bottomrule
\end{tabular}
\end{align}
The matter surfaces $S_{\mathbf{10}_1}^{(a)}$ over $C_{\mathbf{10}_1}$ are linear combinations of the above $\mathbb{P}^1$-fibrations. We use $\mathbb{P}^1$ to denote such a linear combination compactly. To this end $\vec{P}$ is a list of the multiplicites with which these $\mathbb{P}^1$-fibrations appear in the above order. Hence
\[ \vec{P} = \left( 0, 1, 0, 4, 0, 0 \right) \quad \leftrightarrow \quad 1 \cdot \mathbb{P}^1_{14} \left( \mathbf{10}_1 \right) + 4 \cdot \mathbb{P}^1_{2B} \left( \mathbf{10}_1 \right) \, . \]
$\vec{\beta}$ indicates the Cartan charges of such a linear combination, \ie lists the intersection numbers with the resolution divisors $E_i$, $1 \leq i \leq 4$. We will adopt these notations also for the other matter curves. All that said, the matter surfaces over $C_{\mathbf{10}_1}$ take the following form:
{\small
\begin{align}
\begin{tabular}{|c||c|c||c||c|c|}
\toprule
Label & $\vec{P}$ & $\vect{\beta}$ & Label & $\vec{P}$ & $\vect{\beta}$ \\
\hline \hline
$S_{\mathbf{10}_1}^{(1)}$ & $\left( 0, -1, -2, -1, -1, 0 \right)$ & $\left( 0, 1, 0, 0 \right)$ & $S_{\mathbf{10}_1}^{(6)}$ & $\left( 0, 0, 0, 0, 0, 1 \right)$ & $\left( 1, 0, 0, -1 \right)$ \\
$S_{\mathbf{10}_1}^{(2)}$ & $\left( 0, -1, -1, 0, -1, 0 \right)$ & $\left( 1, -1, 1, 0 \right)$ & $S_{\mathbf{10}_1}^{(7)}$ & $\left( 0, 0, 0, 1, 0, 0 \right)$ & $\left( 0, -1, 0, 1 \right)$ \\
$S_{\mathbf{10}_1}^{(3)}$ & $\left( 0, 0, -1, 0, -1, 0 \right)$ & $\left( -1, 0, 1, 0 \right)$ & $S_{\mathbf{10}_1}^{(8)}$ & $\left( 0, 1, 0, 0, 0, 1 \right)$ & $\left( -1, 1, 0, -1 \right)$ \\
$S_{\mathbf{10}_1}^{(4)}$ & $\left( 0, -1, -1, 0, 0, 0 \right)$ & $\left( 1, 0, -1, 1 \right)$ & $S_{\mathbf{10}_1}^{(9)}$ & $\left( 0, 1, 1, 1, 0, 1 \right)$ & $\left( 0, -1, 1, -1 \right)$ \\
$S_{\mathbf{10}_1}^{(5)}$ & $\left( 0, 0, -1, 0, 0, 0 \right)$ & $\left( -1, 1, -1, 1 \right)$ & $S_{\mathbf{10}_1}^{(10)}$ & $\left( 0, 1, 1, 1, 1, 1 \right)$ & $\left( 0, 0, -1, 0 \right)$ \\
\bottomrule
\end{tabular} \label{app_ S10}
\end{align}
}

\subsubsection*{Intersection Structure over \texorpdfstring{$\mathbf{C_{\mathbf{5}_3}}$}{C53} away from Yukawa Loci}

Over $C_{\mathbf{5}_{3}}$ the following six $\mathbb{P}^1$-fibrations are present:
\begin{align}
\begin{split}\label{53cuvesplitting}
 \mathbb{P}_{0}^1 \left( \mathbf{5}_3 \right) &= V \left( a_{3,2}, e_0, a_{1,0} x y z - e_1 e_2^2 e_3 s x^3 + e_3 e_4 y^2 \right) \, , \\
\mathbb{P}_{1}^1 \left( \mathbf{5}_3 \right) &= V \left( a_{3,2}, e_1, e_3 e_4 y + a_{1,0} x z \right) \, , \\
\mathbb{P}_{2E}^1 \left( \mathbf{5}_3 \right) &= V \left( a_{3,2}, e_2, e_3 e_4 y + a_{1,0} x z \right) \, , \\
\mathbb{P}_{3x}^1 \left( \mathbf{5}_3 \right) &= V \left( a_{3,2}, e_3, x \right) \, , \\
\mathbb{P}_{3F}^1 \left( \mathbf{5}_3 \right) &= V \left( a_{3,2}, e_3, a_{1,0} s y - a_{2,1} e_0 e_1 e_2 s x z - a_{4,3} e_0^3 e_1^2 e_2 e_4 z^3 \right) \, , \\ 
\mathbb{P}_{4}^1 \left( \mathbf{5}_3 \right) &= V \left( a_{3,2}, e_4, a_{1,0} y z - e_1 e_2^2 e_3 s x^2 - a_{2,1} e_0 e_1 e_2 x z^2 \right) \, .
\end{split}
\end{align}
The total elliptic fibre over $C_{\mathbf{5}_3}$ is given by
\[ T^2 \left( \mathbf{5}_3 \right) = \mathbb{P}_{0}^1 \left( \mathbf{5}_3 \right) + \mathbb{P}_{1}^1 \left( \mathbf{5}_3 \right) +  \mathbb{P}_{2E}^1 \left( \mathbf{5}_3 \right) +  \mathbb{P}_{3x}^1 \left( \mathbf{5}_3 \right) +  \mathbb{P}_{3F}^1 \left( \mathbf{5}_3 \right) +  \mathbb{P}_{4}^1 \left( \mathbf{5}_3 \right) \, . \]
They above $\mathbb{P}^1$-fibrations emerge from the $E_i$ according to the following table.
\begin{align}
\begin{tabular}{|c|c|}
\toprule
Original & Split components over $C_{5_{3_X}}$ \\
\hline \hline
$E_0$ & $\mathbb{P}_{0}^1 \left( \mathbf{5}_3 \right)$ \\
$E_1$ & $\mathbb{P}_{1}^1 \left( \mathbf{5}_3 \right)$ \\
$E_2$ & $\mathbb{P}_{2E}^1 \left( \mathbf{5}_3 \right)$ \\
$E_3$ & $\mathbb{P}_{3x}^1 \left( \mathbf{5}_3 \right)$ + $\mathbb{P}_{3F}^1 \left( \mathbf{5}_3 \right)$ \\
$E_4$ & $\mathbb{P}_{4}^1 \left( \mathbf{5}_3 \right)$ \\
\bottomrule
\end{tabular}
\end{align}
Over $p \in C_{\mathbf{5}_3}$ which is not a Yukawa point, these $\mathbb{P}^1$-fibrations intersect in $\hat{\pi}^{-1} \left( p \right)$ as follows:
\begin{align}
\begin{tabular}{|c||c|c|c|c|c|c|}
\toprule
 & $\mathbb{P}_{0}^1 \left( \mathbf{5}_3 \right)$ & $\mathbb{P}_{1}^1 \left( \mathbf{5}_3 \right)$ & $\mathbb{P}_{2E}^1 \left( \mathbf{5}_3 \right)$ & $\mathbb{P}_{3x}^1 \left( \mathbf{5}_3 \right)$ & $\mathbb{P}_{3F}^1 \left( \mathbf{5}_3 \right)$ & $\mathbb{P}^1_{4} \left( \mathbf{5}_3 \right)$ \\
\hline \hline
$\mathbb{P}_{0}^1 \left( \mathbf{5}_3 \right)$  & -2 &  1 & 0  & 0  & 0  & 1  \\
$\mathbb{P}_{1}^1 \left( \mathbf{5}_3 \right)$  &  1 & -2 & 1  & 0  & 0  & 0  \\
$\mathbb{P}_{2E}^1 \left( \mathbf{5}_3 \right)$ &  0 &  1 & -2 & 1  & 0  & 0  \\
$\mathbb{P}_{3x}^1 \left( \mathbf{5}_3 \right)$ &  0 &  0 & 1  & -2 & 1  & 0  \\
$\mathbb{P}_{3F}^1 \left( \mathbf{5}_3 \right)$ &  0 &  0 & 0  & 1  & -2 & 1  \\
$\mathbb{P}^1_{4} \left( \mathbf{5}_3 \right)$  &  1 &  0 & 0  & 0  & 1  & -2 \\
\bottomrule
\end{tabular}
\end{align}
The intersections with the pullbacks of the divisors $E_i$ onto the fibre over $C_{\mathbf{5}_3}$ are as follows.
\begin{align}
\begin{tabular}{|c||c|c|c|c|c||c|c|c|c|}
\toprule
 & $E_0$ & $E_1$ & $E_2$ & $E_3$ & $E_4$ & $E_5 = \alpha - \beta - \gamma$ & $\alpha$ & $\beta$ & $\gamma$ \\
\hline \hline
$\mathbb{P}_{0}^1 \left( \mathbf{5}_3 \right)$  & -2 & 1  & 0  & 0  & 1  & -1 & 0 & 1 & 0 \\
$\mathbb{P}_{1}^1 \left( \mathbf{5}_3 \right)$  &  1 & -2 & 1  & 0  & 0  & 0  & 0 & 0 & 0 \\
$\mathbb{P}_{2E}^1 \left( \mathbf{5}_3 \right)$ &  0 &  1 & -2 & 1  & 0  & 0  & 0 & 0 & 0 \\
$\mathbb{P}_{3x}^1 \left( \mathbf{5}_3 \right)$ &  0 &  0 & 1  & -1 & 0  & 1  & 1 & 0 & 0 \\
$\mathbb{P}_{3F}^1 \left( \mathbf{5}_3 \right)$ &  0 &  0 & 0  & -1 & 1  & 0  & 0 & 0 & 0 \\
$\mathbb{P}_{4}^1 \left( \mathbf{5}_3 \right)$  &  1 &  0 & 0  & 1  & -2 & 0  & 0 & 0 & 0 \\
\bottomrule
\end{tabular}
\end{align}
The matter surfaces over $C_{\mathbf{5}_3}$ are:
\begin{align}
\begin{tabular}{|c||c|c||c||c|c|}
\toprule
Label & $\vec{P}^1$ & $\vect{\beta}$ & Label & $\vec{P}^1$ & $\vect{\beta}$ \\
\hline \hline
$S_{\mathbf{5}_3}^{(1)}$ & $\left( 0, -1, -1, -1, 0, 0 \right)$ & $\left( 1, 0, 0, 0 \right)$ & $S_{\mathbf{5}_3}^{(4)}$ & $\left( 0, 0, 0, 0, 1, 0 \right)$ & $\left( 0, 0, -1, 1 \right)$ \\
$S_{\mathbf{5}_3}^{(2)}$ & $\left( 0, 0, -1, -1, 0, 0 \right)$ & $\left( -1, 1, 0, 0 \right)$ & $S_{\mathbf{5}_3}^{(5)}$ & $\left( 0, 0, 0, 0, 1, 1 \right)$ & $\left( 0, 0, 0, -1 \right)$ \\
$S_{\mathbf{5}_3}^{(3)}$ & $\left( 0, 0, 0, -1, 0, 0 \right)$ & $\left( 0, -1, 1, 0 \right)$ & & & \\
\bottomrule
\end{tabular}  \label{app_S53}
\end{align}

\subsubsection*{Intersection Structure over \texorpdfstring{$\mathbf{C_{\mathbf{5}_{-2}}}$}{C5-2} away from Yukawa Loci}

By primary decompositions it is readily confirmed that over $C_{\mathbf{5}_{-2}}$ the following $\mathbb{P}^1$-fibrations are present:
\begin{align}
\begin{split} \label{52cuvesplitting}
\mathbb{P}_{0}^1 \left( \mathbf{5}_{-2} \right) &= V \left( a_{3,2} a_{2,1} - a_{4,3} a_{1,0}, e_0, e_3 e_4 y^2 + a_{1,0} x y z - e_1 e_2^2 e_3 s x^3 \right) \, , \\
\mathbb{P}_{1}^1 \left( \mathbf{5}_{-2} \right) &= V \left( a_{3,2} a_{2,1} - a_{4,3} a_{1,0}, e_1, e_3 e_4 y + a_{1,0} x z \right) \, , \\
\mathbb{P}_{2}^1 \left( \mathbf{5}_{-2} \right) &= V \left( a_{3,2} a_{2,1} - a_{4,3} a_{1,0}, e_2, e_0^2 z^3 e_1 e_4 a_{3,2} + y s e_3 e_4 + a_{1,0} x z s, \right. \\
      & \hspace{10em} \left. a_{1,0} a_{4,3} e_0^2 z^3 e_1 e_4 + a_{2,1} y s e_3 e_4 + a_{1,0} a_{2,1} x z s \right) \, , \\
\mathbb{P}_{3G}^1 \left( \mathbf{5}_{-2} \right) &= V \left( a_{3,2} a_{2,1} - a_{4,3} a_{1,0}, e_3, a_{4,3} e_0^2 z^2 e_1 e_4 + a_{2,1} x s, a_{3,2} e_0^2 z^2 e_1 e_4 + a_{1,0} x s \right) \, , \\
\mathbb{P}_{3H}^1 \left( \mathbf{5}_{-2} \right) &= V \left( a_{3,2} a_{2,1} - a_{4,3} a_{1,0}, e_3, a_{4,3} e_0 x z e_1 e_2 - a_{3,2} y, a_{2,1} e_0 x z e_1 e_2 - a_{1,0} y  \right) \, , \\
\mathbb{P}_{4}^1 \left( \mathbf{5}_{-2} \right) &= V \left( a_{3,2} a_{2,1} - a_{4,3} a_{1,0}, e_4, a_{1,0} y z - a_{2,1} e_0 e_1 e_2 x z^2 - e_1 e_2^2 e_3 s x^2 \right) \, .
\end{split}
\end{align}
Note that these results differ from \cite{Krause:2011xj}, where primary decomposition was not applied. The total elliptic fibre over $C_{\mathbf{5}_{-2}}$ is given by
\[ T^2 \left( \mathbf{5}_{-2} \right) = \mathbb{P}_{0}^1 \left( \mathbf{5}_{-2} \right) + \mathbb{P}_{1}^1 \left( \mathbf{5}_{-2} \right) +  \mathbb{P}_{2}^1 \left( \mathbf{5}_{-2} \right) +  \mathbb{P}_{3G}^1 \left( \mathbf{5}_{-2} \right) +  \mathbb{P}_{3H}^1 \left( \mathbf{5}_{-2} \right) +  \mathbb{P}_{4}^1 \left( \mathbf{5}_{-2} \right) \, . \]
The above $\mathbb{P}^1$-fibrations emerge from the $E_i$ according to the following table.
\begin{align}
\begin{tabular}{|c|c|}
\toprule
Original & Split components over $C_{5_{-2_X}}$ \\
\hline \hline
$E_0$ & $\mathbb{P}_{0}^1 \left( \mathbf{5}_{-2} \right)$ \\
$E_1$ & $\mathbb{P}_{1}^1 \left( \mathbf{5}_{-2} \right)$ \\
$E_2$ & $\mathbb{P}_{2}^1 \left( \mathbf{5}_{-2} \right)$ \\
$E_3$ & $\mathbb{P}_{3G}^1\left( \mathbf{5}_{-2} \right)$ + $\mathbb{P}_{3H}^1 \left( \mathbf{5}_{-2} \right)$ \\
$E_4$ & $\mathbb{P}_{4}^1 \left( \mathbf{5}_{-2} \right)$ \\
\bottomrule
\end{tabular}
\end{align}
Over $p \in C_{\mathbf{5}_3}$ which is not a Yukawa point, these $\mathbb{P}^1$-fibrations intersect in $\hat{\pi}^{-1} \left( p \right)$ as follows:
\begin{align}
\begin{tabular}{|c||c|c|c|}
\toprule
 & $\mathbb{P}_{0}^1 \left( \mathbf{5}_{-2} \right)$ & $\mathbb{P}_{1}^1 \left( \mathbf{5}_{-2} \right)$ & $\mathbb{P}_{2}^1 \left( \mathbf{5}_{-2} \right)$ \\
\hline \hline
$\mathbb{P}_{0}^1 \left( \mathbf{5}_{-2} \right)$  & -2 &  1 & 0 \\
$\mathbb{P}_{1}^1 \left( \mathbf{5}_{-2} \right)$  &  1 & -2 & 1 \\
$\mathbb{P}_{2}^1 \left( \mathbf{5}_{-2} \right)$  &  0 &  1 & -2 \\
$\mathbb{P}_{3G}^1 \left( \mathbf{5}_{-2} \right)$ &  0 &  0 & 1 \\
$\mathbb{P}_{3H}^1 \left( \mathbf{5}_{-2} \right)$ &  0 &  0 & 0 \\
$\mathbb{P}^1_{4} \left( \mathbf{5}_{-2} \right)$  &  1 &  0 & 0 \\
\hline \hline
& $\mathbb{P}_{3G}^1 \left( \mathbf{5}_{-2} \right)$ & $\mathbb{P}_{3H}^1 \left( \mathbf{5}_{-2} \right)$ & $\mathbb{P}^1_{4} \left( \mathbf{5}_{-2} \right)$ \\
\hline \hline
$\mathbb{P}_{0}^1 \left( \mathbf{5}_{-2} \right)$  & 0  & 0  & 1  \\
$\mathbb{P}_{1}^1 \left( \mathbf{5}_{-2} \right)$  & 0  & 0  & 0  \\
$\mathbb{P}_{2}^1 \left( \mathbf{5}_{-2} \right)$  & 1  & 0  & 0  \\
$\mathbb{P}_{3G}^1 \left( \mathbf{5}_{-2} \right)$ & -2 & 1  & 0  \\
$\mathbb{P}_{3H}^1 \left( \mathbf{5}_{-2} \right)$ & 1  & -2 & 1  \\
$\mathbb{P}^1_{4} \left( \mathbf{5}_{-2} \right)$  & 0  & 1  & -2 \\
\bottomrule
\end{tabular}
\end{align}
The intersections with the pullbacks of the divisors $E_i$ onto the fibre over $C_{\mathbf{5}_{-2}}$ are as follows.
\begin{align}
\begin{tabular}{|c||c|c|c|c|c||c|c|c|c|} 
\toprule
 & $E_0$ & $E_1$ & $E_2$ & $E_3$ & $E_4$ & $E_5 = \alpha - \beta - \gamma$ & $\alpha$ & $\beta$ & $\gamma$ \\
\hline \hline
$\mathbb{P}_{0}^1 \left( \mathbf{5}_{-2} \right)$  & -2 & 1  & 0  & 0  & 1  & -1 & 0 & 1 & 0 \\
$\mathbb{P}_{1}^1 \left( \mathbf{5}_{-2} \right)$  &  1 & -2 & 1  & 0  & 0  & 0  & 0 & 0 & 0 \\
$\mathbb{P}_{2}^1 \left( \mathbf{5}_{-2} \right)$ &  0 &  1 & -2 & 1  & 0  & 0  & 0 & 0 & 0 \\
$\mathbb{P}_{3G}^1 \left( \mathbf{5}_{-2} \right)$ &  0 &  0 & 1  & -1 & 0  & 0  & 0 & 0 & 0 \\
$\mathbb{P}_{3H}^1 \left( \mathbf{5}_{-2} \right)$ &  0 &  0 & 0  & -1 & 1  & 1  & 1 & 0 & 0 \\
$\mathbb{P}_{4}^1 \left( \mathbf{5}_{-2} \right)$  &  1 &  0 & 0  & 1  & -2 & 0  & 0 & 0 & 0 \\
\bottomrule
\end{tabular}
\end{align}
The matter surfaces over $C_{\mathbf{5}_{-2}}$ are:
\begin{align}
\begin{tabular}{|c||c|c||c||c|c|}
\toprule
Label & $\vec{P}^1$ & $\vect{\beta}$ & Label & $\vec{P}^1$ & $\vect{\beta}$ \\
\hline \hline
$S_{\mathbf{5}_{-2}}^{(1)}$ & $\left( 0, -1, -1, -1, 0, 0 \right)$ & $\left( 1, 0, 0, 0 \right)$ & $S_{\mathbf{5}_{-2}}^{(4)}$ & $\left( 0, 0, 0, 0, 1, 0 \right)$ & $\left( 0, 0, -1, 1 \right)$ \\
$S_{\mathbf{5}_{-2}}^{(2)}$ & $\left( 0, 0, -1, -1, 0, 0 \right)$ & $\left( -1, 1, 0, 0 \right)$ & $S_{\mathbf{5}_{-2}}^{(5)}$ & $\left( 0, 0, 0, 0, 1, 1 \right)$ & $\left( 0, 0, 0, -1 \right)$ \\
$S_{\mathbf{5}_{-2}}^{(3)}$ & $\left( 0, 0, 0, -1, 0, 0 \right)$ & $\left( 0, -1, 1, 0 \right)$ & & & \\
\bottomrule 
\end{tabular} \label{app:MatterSurfaces52}
\end{align}

\subsubsection*{Intersection Structure over \texorpdfstring{$\mathbf{C_{\mathbf{1}_{5}}}$}{C15} away from Yukawa Loci}

Over the singlet curve $C_{\mathbf{1}_{5}} = V \left( P^\prime, a_{3,2}, a_{4,3} \right)$ the following two $\mathbb{P}^1$-fibrations are present:
\begin{align}
\begin{split}
\mathbb{P}_{A}^1 \left( \mathbf{1}_{5} \right) &= V \left( a_{3,2}, a_{4,3}, s \right) \, , \\
\mathbb{P}_{B}^1 \left( \mathbf{1}_{5} \right) &= V \left( a_{3,2}, a_{4,3}, y^2 e_3 e_4 + a_{1,0} x y z - x^3 s e_1 e_2^2 e_3 - a_{2,1} x^2 z^2 e_0 e_1 e_2 \right) \, .
\end{split}
\end{align}
These fibrations intersect in $\hat{\pi}^{-1} \left( p \right)$ as follows:
\begin{align}
\begin{tabular}{|c||c|c|}
\toprule
 & $\mathbb{P}_{A}^1 \left( \mathbf{1}_{5} \right)$ & $\mathbb{P}_{B}^1 \left( \mathbf{1}_{5} \right)$ \\
\hline \hline
$\mathbb{P}_{A}^1 \left( \mathbf{1}_{5} \right)$ & -2 &  2 \\
$\mathbb{P}_{B}^1 \left( \mathbf{1}_{5} \right)$ &  2 & -2 \\
\bottomrule
\end{tabular}
\end{align}
The intersection numbers with the divisors $E_i$ are given by:\footnote{Recall that $E_5 = \alpha - \beta - \gamma$.}
\begin{align}
\begin{tabular}{|c||c|c|c|c|c||c|c|c|c||c|}
\toprule
 & $E_0$ & $E_1$ & $E_2$ & $E_3$ & $E_4$ & $E_5$ & $\alpha$ & $\beta$ & $\gamma$ & $q$ \\
\hline \hline
$\mathbb{P}_{A}^1 \left( \mathbf{1}_{5} \right) = S_{\mathbf{1}_{5}}^{(1)}$ & 0 & 0 & 0 & 0 & 0 & -1 & -1 & 0 & 0 & $5$ \\
$\mathbb{P}_{B}^1 \left( \mathbf{1}_{5} \right) = S_{\mathbf{1}_{5}}^{(2)}$ & 0 & 0 & 0 & 0 & 0 & 1  & 2  & 1 & 0 & $-5$ \\
\bottomrule
\end{tabular}
\end{align}
Note that only the $\mathbb{P}^1$-fibration $\mathbb{P}^1_A \left( \mathbf{1}_5 \right)$ has vanishing intersection with the zero-section $\beta = V \left( z \right)$
and hence defines a  viable matter surface. As this fibration satisfies $q = 5$, we denote the singlet curve as $C_{\mathbf{1}_5}$.

\subsection{Intersection Structure over Yukawa Loci} \label{app_Yukawas}

\subsubsection*{Intersection Structure over Yukawa Locus \texorpdfstring{$\mathbf{Y_1}$}{Y1}}

Over the Yukawa point $Y_1 = V ( w, a_{1,0}, a_{2,1} )$ the following six $\mathbb{P}^1$-fibrations are present:
\begin{align}
\begin{split}
 \mathbb{P}_{0A}^1 \left( Y_1 \right) &= V \left( a_{1,0}, a_{2,1}, e_0, y^2 e_4 - x^3 s e_1 e_2^2 \right) \, , \\
 \mathbb{P}_{14}^1 \left( Y_1 \right) &= V \left( a_{1,0}, a_{2,1}, e_1, e_4 \right) \, , \\
 \mathbb{P}_{24}^1 \left( Y_1 \right) &= V \left( a_{1,0}, a_{2,1}, e_2, e_4 \right) \, , \\
 \mathbb{P}_{2B}^1 \left( Y_1 \right) &= V \left( a_{1,0}, a_{2,1}, e_2, y s e_3 + a_{3,2} z^3 e_0^2 e_1 \right) \, , \\
 \mathbb{P}_{34}^1 \left( Y_1 \right) &= V \left( a_{1,0}, a_{2,1}, e_3, e_4 \right) \, , \\
 \mathbb{P}_{3J}^1 \left( Y_1 \right) &= V \left( a_{1,0}, a_{2,1}, e_3, a_{3,2} y - a_{4,3} x z e_0 e_1 e_2 \right) \, .
\end{split}
\end{align}
The total elliptic fibre over $Y_1$ is given by
\[ T^2 \left( Y_1 \right) = \mathbb{P}_{0A}^1 \left( Y_1 \right) + 2 \mathbb{P}_{14}^1 \left( Y_1 \right) + 3 \mathbb{P}_{24}^1 \left( Y_1 \right)
+ \mathbb{P}_{2B}^1 \left( Y_1 \right) + 2 \mathbb{P}_{34}^1 \left( Y_1 \right) + \mathbb{P}_{3J}^1 \left( Y_1 \right) \, . \]
Starting from $C_{\mathbf{10}_1}$, the above fibrations emerge from the following splitting process:
\begin{align}
\begin{tabular}{|c|c|}
\toprule
Split components over $C_{10_{1_X}}$ & Split components over $Y_1$ \\
\hline \hline
$\mathbb{P}_{0A}^1 \left( \mathbf{10}_1 \right)$ & $\mathbb{P}_{0A}^1 \left( Y_1 \right)$ \\
$\mathbb{P}_{14}^1 \left( \mathbf{10}_1 \right)$ & $\mathbb{P}_{14}^1 \left( Y_1 \right)$ \\
$\mathbb{P}_{24}^1\left( \mathbf{10}_1 \right)$ & $\mathbb{P}_{24}^1\left( Y_1 \right)$ \\
$\mathbb{P}_{2B}^1\left( \mathbf{10}_1 \right)$ & $\mathbb{P}_{2B}^1\left( Y_1 \right)$ \\
$\mathbb{P}_{3C}^1\left( \mathbf{10}_1 \right)$ & $\mathbb{P}_{34}^1\left( Y_1 \right)$ + $\mathbb{P}_{3J}^1\left( Y_1 \right)$ \\
$\mathbb{P}_{4D}^1\left( \mathbf{10}_1 \right)$ & $\mathbb{P}_{24}^1\left( Y_1 \right)$ + $\mathbb{P}_{34}^1\left( Y_1 \right)$ \\
\bottomrule
\end{tabular}
\label{tab:Splitting10ToY1}
\end{align}
The splitting behaviour, when approached from $C_{\mathbf{5}_{-2}}$, is different:
\begin{align}
\begin{tabular}{|c|c|c|}
\toprule
Split components over $C_{5_{-2_X}}$ & Split components over $Y_1$ \\
\hline \hline
$\mathbb{P}_{0}^1 \left( \mathbf{5}_{-2} \right)$ & $\mathbb{P}_{0A}^1 \left( Y_1 \right)$ \\
$\mathbb{P}_{1}^1 \left( \mathbf{5}_{-2} \right)$ & $\mathbb{P}_{14}^1 \left( Y_1 \right)$ \\
$\mathbb{P}_{2}^1 \left( \mathbf{5}_{-2} \right)$ & $\mathbb{P}^{1}_{24} \left( Y_1 \right)$ + $\mathbb{P}^{1}_{2B} \left( Y_1 \right)$ \\
$\mathbb{P}_{3G}^1\left( \mathbf{5}_{-2} \right)$ & $\mathbb{P}^{1}_{34} \left( Y_1 \right)$ \\
$\mathbb{P}_{3H}^1\left( \mathbf{5}_{-2} \right)$ & $\mathbb{P}^{1}_{3J} \left( Y_1 \right)$ \\
$\mathbb{P}_{4}^1 \left( \mathbf{5}_{-2} \right)$ & $\mathbb{P}_{14}^1\left( Y_1 \right)$ + $2 \mathbb{P}_{24}^1\left( Y_1 \right)$ + $\mathbb{P}_{34}^1\left( Y_1 \right)$ \\
\bottomrule
\end{tabular}
\label{tab:Splitting5M2ToY1}
\end{align}
The intersection numbers in the fibre over $Y_1$ are as follows:
\begin{align}
\begin{tabular}{|c||c|c|c|c|c|c|}
\toprule
 & $\mathbb{P}_{0A}^1 \left( Y_1 \right)$ & $\mathbb{P}_{14}^1 \left( Y_1 \right)$ & $\mathbb{P}_{24}^1 \left( Y_1 \right)$ & $\mathbb{P}_{2B}^1 \left( Y_1 \right)$ & $\mathbb{P}_{34}^1 \left( Y_1 \right)$ & $\mathbb{P}^1_{3J} \left( Y_1 \right)$ \\
\hline \hline
$\mathbb{P}_{0A}^1 \left( Y_1 \right)$  & -2 & 1  & 0  &  0 & 0  & 0  \\
$\mathbb{P}_{14}^1 \left( Y_1 \right)$   & 1  & -2 & 1  &  0 & 0  & 0  \\
$\mathbb{P}_{24}^1 \left( Y_1 \right)$   & 0  & 1 & $-\frac{3}{2}$ & $\frac{1}{2}$ & 1  & 0  \\
$\mathbb{P}_{2B}^1 \left( Y_1 \right)$  & 0  & 0  & $\frac{1}{2}$ & $-\frac{3}{2}$ & 0  & 0  \\
$\mathbb{P}_{34}^1 \left( Y_1 \right)$  & 0  & 0  & 1  & 0  & -2 & 1  \\
$\mathbb{P}_{3J}^1 \left( Y_1 \right)$  & 0  & 0  & 0  & 0  & 1  & -2 \\
\bottomrule
\end{tabular}
\label{tab:IntersectionsOfFibralCurvesOverY1}
\end{align}

\subsubsection*{Intersection Structure over Yukawa Locus \texorpdfstring{$\mathbf{Y_2}$}{Y2}}

Over the Yukawa locus $Y_2 = V ( w, a_{1,0}, a_{3,2} )$ the following seven $\mathbb{P}^1$-fibrations are present:
\begin{align}
\begin{split}
 \mathbb{P}_{0A}^1 \left( Y_2 \right) &= V \left( a_{1,0}, a_{3,2}, e_0, y^2 e_4 - x^3 s e_1 e_2^2 \right) \, , \\
 \mathbb{P}_{14}^1 \left( Y_2 \right) &= V \left( a_{1,0}, a_{3,2}, e_1, e_4 \right) \, , \\
 \mathbb{P}_{24}^1 \left( Y_2 \right) &= V \left( a_{1,0}, a_{3,2}, e_2, e_4 \right) \, , \\
 \mathbb{P}_{23}^1 \left( Y_2 \right) &= V \left( a_{1,0}, a_{3,2}, e_2, e_3 \right) \, , \\
 \mathbb{P}_{3x}^1 \left( Y_2 \right) &= V \left( a_{1,0}, a_{3,2}, e_3, x \right) \, , \\
 \mathbb{P}_{3K}^1 \left( Y_2 \right) &= V \left( a_{1,0}, a_{3,2}, e_3, a_{2,1} x s + a_{4,3} z^2 e_0^2 e_1 e_4 \right) \, , \\
 \mathbb{P}_{4D}^1 \left( Y_2 \right) &= V \left( a_{1,0}, a_{3,2}, e_4, x s e_2 e_3 + a_{2,1} z^2 e_0 \right) \, .
\end{split}
\end{align}
The total elliptic fibre over $Y_2$ is given by
\[ T^2 \left( Y_2 \right) = \mathbb{P}_{0A}^1 \left( Y_2 \right) + 2 \mathbb{P}_{14}^1 \left( Y_2 \right) + 2 \mathbb{P}_{24}^1 \left( Y_2 \right)
+ 2 \mathbb{P}_{23}^1 \left( Y_2 \right) + \mathbb{P}_{3x}^1 \left( Y_2 \right) + \mathbb{P}_{3K}^1 \left( Y_2 \right) + \mathbb{P}_{4D}^1 \left( Y_2 \right) \, . \]
The above $\mathbb{P}^1$-fibrations result from splittings the corresponding fibrations over $C_{\mathbf{10}_1}$ according to the following table.
\begin{align}
\begin{tabular}{|c|c|}
\toprule
Split components over $C_{10_{1_X}}$ & Split components over $Y_2$ \\
\hline \hline
$\mathbb{P}_{0A}^1 \left( \mathbf{10}_1 \right)$ & $\mathbb{P}_{0A}^1 \left( Y_2 \right)$ \\
$\mathbb{P}_{14}^1 \left( \mathbf{10}_1 \right)$ & $\mathbb{P}_{14}^1 \left( Y_2 \right)$ \\
$\mathbb{P}_{24}^1\left( \mathbf{10}_1 \right)$ & $\mathbb{P}_{24}^1\left( Y_2 \right)$ \\
$\mathbb{P}_{2B}^1\left( \mathbf{10}_1 \right)$ & $\mathbb{P}_{23}^1\left( Y_2 \right)$ \\
$\mathbb{P}_{3C}^1\left( \mathbf{10}_1 \right)$ & $\mathbb{P}_{3x}^1\left( Y_2 \right)$ + $\mathbb{P}_{23}^1 \left( Y_2 \right)$ + $\mathbb{P}_{3K}^1 \left( Y_2 
\right)$ \\
$\mathbb{P}_{4D}^1\left( \mathbf{10}_1 \right)$ & $\mathbb{P}_{4D}^1\left( Y_2 \right)$ \\
\bottomrule
\end{tabular}
\label{tab:Splitting10ToY2}
\end{align}
When approached from $C_{\mathbf{5}_3}$ the spittings are different, namely
\begin{align}
\begin{tabular}{|c|c|c|}
\toprule
Split components over $C_{5_{3_X}}$ & Split components over $Y_2$ \\
\hline \hline
$\mathbb{P}_{0}^1 \left( \mathbf{5}_3 \right)$ & $\mathbb{P}_{0A}^1 \left( Y_2 \right)$ \\
$\mathbb{P}_{1}^1 \left( \mathbf{5}_3 \right)$ & $\mathbb{P}_{14}^1 \left( Y_2 \right)$ \\
$\mathbb{P}_{2E}^1 \left( \mathbf{5}_3 \right)$ & $\mathbb{P}^{1}_{23} \left( Y_2 \right)$ + $\mathbb{P}^{1}_{24} \left( Y_2 \right)$ \\
$\mathbb{P}_{3x}^1\left( \mathbf{5}_3 \right)$ & $\mathbb{P}^{1}_{3x} \left( Y_2 \right)$ \\
$\mathbb{P}_{3F}^1\left( \mathbf{5}_3 \right)$ & $\mathbb{P}^{1}_{23} \left( Y_2 \right)$ + $\mathbb{P}^{1}_{3K} \left( Y_2 \right)$ \\
$\mathbb{P}_{4}^1 \left( \mathbf{5}_3 \right)$ & $\mathbb{P}_{14}^1\left( Y_2 \right)$ + $\mathbb{P}_{24}^1\left( Y_2 \right)$ + $\mathbb{P}_{4D}^1\left( Y_2 \right)$ \\
\bottomrule
\end{tabular}
\end{align}
Finally, the splitting as seen from $C_{\mathbf{5}_{-2}}$, is yet again different:
\begin{align}
\begin{tabular}{|c|c|c|}
\toprule
Split components over $C_{5_{-2_X}}$ & Split components over $Y_2$ \\
\hline \hline
$\mathbb{P}_{0}^1 \left( \mathbf{5}_{-2} \right)$ & $\mathbb{P}_{0A}^1 \left( Y_2 \right)$ \\
$\mathbb{P}_{1}^1 \left( \mathbf{5}_{-2} \right)$ & $\mathbb{P}_{14}^1 \left( Y_2 \right)$ \\
$\mathbb{P}_{2}^1 \left( \mathbf{5}_{-2} \right)$ & $\mathbb{P}^{1}_{23} \left( Y_2 \right)$ + $\mathbb{P}^{1}_{24} \left( Y_2 \right)$ \\
$\mathbb{P}_{3G}^1\left( \mathbf{5}_{-2} \right)$ & $\mathbb{P}_{3K}^1\left( Y_2 \right)$ \\
$\mathbb{P}_{3H}^1\left( \mathbf{5}_{-2} \right)$ & $\mathbb{P}_{23}^1\left( Y_2 \right)$ + $\mathbb{P}_{3x}^1\left( Y_2 \right)$ \\
$\mathbb{P}_{4}^1 \left( \mathbf{5}_{-2} \right)$ & $\mathbb{P}_{14}^1\left( Y_2 \right)$ + $\mathbb{P}_{24}^1\left( Y_2 \right)$ + $\mathbb{P}_{4D}^1\left( Y_2 \right)$ \\
\bottomrule
\end{tabular}
\label{tab:Splitting5M2ToY2}
\end{align}
The intersection numbers in the fibre over $Y_2$ are as follows:
{\small{
\begin{align}
\begin{tabular}{|c||c|c|c|c|c|c|c|}
\toprule
 & $\mathbb{P}_{0A}^1 \left( Y_2 \right)$ & $\mathbb{P}_{14}^1 \left( Y_2 \right)$ & $\mathbb{P}_{24}^1 \left( Y_2 \right)$ & $\mathbb{P}_{23}^1 \left( Y_2 \right)$ & $\mathbb{P}_{3x}^1 \left( Y_2 \right)$ & $\mathbb{P}^1_{3K} \left( Y_2 \right)$ & $\mathbb{P}^1_{4D} \left( Y_2 \right)$ \\
\hline \hline
$\mathbb{P}_{0A}^1 \left( Y_2 \right)$  & -2 & 1  & 0  &  0 & 0  & 0  & 0  \\
$\mathbb{P}_{14}^1 \left( Y_2 \right)$  & 1  & -2 & 1  &  0 & 0  & 0  & 1  \\
$\mathbb{P}_{24}^1 \left( Y_2 \right)$  & 0  & 1  & -2 &  1 & 0  & 0  & 0  \\
$\mathbb{P}_{23}^1 \left( Y_2 \right)$  & 0  & 0  & 1  & -2 & 1  & 1  & 0  \\
$\mathbb{P}_{3x}^1 \left( Y_2 \right)$  & 0  & 0  & 0  & 1  & -2 & 0  & 0  \\
$\mathbb{P}_{3K}^1 \left( Y_2 \right)$  & 0  & 0   & 0 & 1  & 0  & -2 & 0  \\
$\mathbb{P}_{4D}^1 \left( Y_2 \right)$  & 0  & 1   & 0 & 0  & 0  & 0  & -2 \\
\bottomrule
\end{tabular}
\end{align}
}}

\subsubsection*{Intersection Structure over Yukawa Locus \texorpdfstring{$\mathbf{Y_3}$}{Y3}}

Over the Yukawa locus $Y_3 = V ( w, a_{32}, a_{43} )$. There the following seven $\mathbb{P}^1$-fibrations are present:
\begin{align}
\begin{split}
 \mathbb{P}_{0A}^1 \left( Y_3 \right) &= V \left( a_{3,2}, a_{4,3}, e_0, a_{1,0} x y z - e_1 e_2^2 e_3 s x^3 + e_3 e_4 y^2 \right) \, , \\
 \mathbb{P}_{1}^1 \left( Y_3 \right) &= V \left( a_{3,2}, a_{4,3}, e_1, e_3 e_4 y + a_{1,0} x z \right) \, , \\
 \mathbb{P}_{2E}^1 \left( Y_3 \right) &= V \left( a_{3,2}, a_{4,3}, e_2, e_3 e_4 y + a_{1,0} x z \right) \, , \\
 \mathbb{P}_{3x}^1 \left( Y_3 \right) &= V \left( a_{3,2}, a_{4,3}, e_3, x \right) \, , \\
 \mathbb{P}_{3s}^1 \left( Y_3 \right) &= V \left( a_{3,2}, a_{4,3}, e_3, s \right) \, , \\
 \mathbb{P}_{3L}^1 \left( Y_3 \right) &= V \left( a_{3,2}, a_{4,3}, e_3, a_{1,0} y - a_{2,1} e_0 e_1 e_2 x z \right) \, , \\
 \mathbb{P}_{4}^1 \left( Y_3 \right) &= V \left( a_{3,2}, a_{4,3}, e_4, a_{1,0} y z - e_1 e_2^2 e_3 s x^2 - a_{2,1} e_0 e_1 e_2 x z^2 \right) \, .
\end{split}
\end{align}
The total elliptic fibre over $Y_3$ is given by
\[ T^2 \left( Y_3 \right) = \mathbb{P}_{0A}^1 \left( Y_3 \right) + \mathbb{P}_{1}^1 \left( Y_3 \right) + \mathbb{P}_{2E}^1 \left( Y_3 \right) + \mathbb{P}_{3x}^1 \left( Y_3 \right) + \mathbb{P}_{3s}^1 \left( Y_3 \right) + \mathbb{P}_{3L}^1 \left( Y_3 \right)  + \mathbb{P}_{4}^1 \left( Y_3 \right) \, . \]
The individual $\mathbb{P}^1$-fibrations appear from the split components over $C_{\mathbf{5}_3}$ as follows:
\begin{align}
\begin{tabular}{|c|c|c|}
\toprule
Split components over $C_{5_{3_X}}$ & Split components over $Y_3$ \\
\hline \hline
$\mathbb{P}_{0}^1 \left( \mathbf{5}_3 \right)$ & $\mathbb{P}_{0A}^1 \left( Y_3 \right)$ \\
$\mathbb{P}_{1}^1 \left( \mathbf{5}_3 \right)$ & $\mathbb{P}_{1}^1 \left( Y_3 \right)$ \\
$\mathbb{P}_{2E}^1 \left( \mathbf{5}_3 \right)$ & $\mathbb{P}^{1}_{2E} \left( Y_3 \right)$ \\
$\mathbb{P}_{3x}^1\left( \mathbf{5}_3 \right)$ & $\mathbb{P}^{1}_{3x} \left( Y_3 \right)$ \\
$\mathbb{P}_{3F}^1\left( \mathbf{5}_3 \right)$ & $\mathbb{P}^{1}_{3s} \left( Y_3 \right)$ + $\mathbb{P}^{1}_{3L} \left( Y_3 \right)$ \\
$\mathbb{P}_{4}^1 \left( \mathbf{5}_3 \right)$ & $\mathbb{P}_{4}^1\left( Y_3 \right)$ \\
\bottomrule
\end{tabular}
\end{align}
However, if we approach $Y_3$ from $C_{\mathbf{5}_{-2}}$ we have the following behaviour.
\begin{align}
\begin{tabular}{|c|c|c|}
\toprule
Split components over $C_{5_{-2_X}}$ & Split components over $Y_3$ \\
\hline \hline
$\mathbb{P}_{0}^1 \left( \mathbf{5}_{-2} \right)$ & $\mathbb{P}_{0A}^1 \left( Y_3 \right)$ \\
$\mathbb{P}_{1}^1 \left( \mathbf{5}_{-2} \right)$ & $\mathbb{P}_1^1 \left( Y_3 \right)$ \\
$\mathbb{P}_{2}^1 \left( \mathbf{5}_{-2} \right)$ & $\mathbb{P}^{1}_{2E} \left( Y_3 \right)$ \\
$\mathbb{P}_{3G}^1\left( \mathbf{5}_{-2} \right)$ & $\mathbb{P}^{1}_{3x} \left( Y_3 \right)$ + $\mathbb{P}^{1}_{3s} \left( Y_3 \right)$ \\
$\mathbb{P}_{3H}^1\left( \mathbf{5}_{-2} \right)$ & $\mathbb{P}^{1}_{3L} \left( Y_3 \right)$ \\
$\mathbb{P}_{4}^1 \left( \mathbf{5}_{-2} \right)$ & $\mathbb{P}_{4}^1\left( Y_3 \right)$ \\
\bottomrule
\end{tabular}
\end{align}
Finally, when approached from the singlet curve $C_{\mathbf{1}_{5}}$ we have the following splitting:
\begin{align}
\begin{tabular}{|c|c|} 
\toprule
Split components over $C_{1_{5_X}}$ & Split components over $Y_3$ \\
\hline \hline
$\mathbb{P}_{A}^1 \left( \mathbf{1}_{5} \right)$ & $\mathbb{P}_{3s}^1 \left( Y_3 \right)$ \\
$\mathbb{P}_{B}^1 \left( \mathbf{1}_{5} \right)$ & \pbox{20cm}{$\mathbb{P}_{0A}^1 \left( Y_3 \right) + \mathbb{P}_{1}^1 \left( Y_3 \right) + \mathbb{P}_{2E}^1 \left( Y_3 \right)$ \\ $+\mathbb{P}_{3x}^1 \left( Y_3 \right) + \mathbb{P}_{3L}^1 \left( Y_3 \right) + \mathbb{P}_{4}^1 \left( Y_3 \right)$} \\
\bottomrule
\end{tabular}
\end{align}
The intersection numbers in the fibre over $Y_3$ are:
\begin{align}
\begin{tabular}{|c||c|c|c|c|c|c|c|}
\toprule
 & $\mathbb{P}_{0A}^1 \left( Y_3 \right)$ & $\mathbb{P}_{1}^1 \left( Y_3 \right)$ & $\mathbb{P}_{2E}^1 \left( Y_3 \right)$ & $\mathbb{P}_{3x}^1 \left( Y_3 \right)$ & $\mathbb{P}_{3s}^1 \left( Y_3 \right)$ & $\mathbb{P}^1_{3L} \left( Y_3 \right)$ & $\mathbb{P}^1_{4} \left( Y_3 \right)$ \\
\hline \hline
$\mathbb{P}_{0A}^1 \left( Y_3 \right)$  & -2 & 1  & 0  &  0 & 0  & 0  & 1  \\
$\mathbb{P}_{1}^1 \left( Y_3 \right)$   & 1  & -2 & 1  &  0 & 0  & 0  & 0  \\
$\mathbb{P}_{2E}^1 \left( Y_3 \right)$  & 0  & 1  & -2 &  1 & 0  & 0  & 0  \\
$\mathbb{P}_{3x}^1 \left( Y_3 \right)$  & 0  & 0  & 1  & -2 & 1  & 0  & 0  \\
$\mathbb{P}_{3s}^1 \left( Y_3 \right)$  & 0  & 0  & 0  & 1  & -2 & 1  & 0  \\
$\mathbb{P}_{3L}^1 \left( Y_3 \right)$  & 0  & 0  & 0  & 0  & 1  & -2 & 1  \\
$\mathbb{P}_{4}^1 \left( Y_3 \right)$   & 1  & 0  & 0  & 0  & 0  & 1  & -2 \\
\bottomrule
\end{tabular}
\end{align}

\section{Line Bundles Induced by Matter Surface Fluxes from Ambient Space Intersections} \label{sec:MasslessSpectraOfCurveSupportFluxesTedious}

In this appendix we compute the line bundles induced by the matter surface fluxes. As in \autoref{subsec:ExampleOnMasslessSpectrumComputation} we perform the following steps:
\begin{itemize}
 \item The $SU ( 5 ) \times U ( 1 )_X$-top induces the linear relations (\ref{linearrelationsambient}) on the ambient space $\hat{X_5}$. Gauge backgrounds represented by 
      elements of $\text{CH}^2 ( \hat{X_5} )$ can thus be altered upon use of these linear relations.
 \item In doing so, we ensure that the intersections between gauge backgrounds $\mathcal{A} \in \text{CH}^2 ( \hat{X_5} )$ and matter surfaces 
      ${\cal S}^a_{C_\mathbf{R}} \in \text{CH}^3 ( \hat{X}_5 )$ are transverse. Consequently the relevant intersections can then be worked out merely from the corresponding set-theoretic intersections.
 \item Given a gauge invariant background, we compute these intersections for one matter surface over each matter curve and project the result to the matter curve 
      $C_\mathbf{R}$. Tensoring the associated line bundle with the spin bundle $\sqrt{K_{C_{\mathbf R}}}$ induced by the holomorphic embedding of the matter curve  gives the final result for the line bundle $\mathcal{L}(S_{\mathbf{R}},\mathcal{A})$ such that $H^i(C_{\mathbf{R}},\mathcal{L}(S_{\mathbf{R}},\mathcal{A}))$ counts the massless matter in representation $\mathbf{R}$.
\end{itemize}
Let us mention that the following computations make us of rational equivalence on $B_3$ to (re)express cycles in the classes $\overline{K}_{B_3}$ and $W$ whenever this simplifies the computations. Furthermore, we stress again that we are assuming that $B_3$ is torsion-free. Together with the requirement that $H^1(B_3,\mathbb Q) = 0$, which follows from the fact that  $\hat Y_4$ is a `proper' Calabi-Yau, this implies that $H^1(B_3,\mathbb Z) = 0$ and thus $\mathrm{CH}^1(B_3)= H^{1,1}(B_3)$.

\subsection{Line Bundle Induced by \texorpdfstring{$\mathbf{\Delta \mathcal{A} ( \lambda )}$}{Delta A(lambda)}}
\label{sec:MasslessSpectrumOfDeltaAG}

As a preparation we show that the gauge background
\begin{align}
\begin{split}
\Delta \mathcal{A} \left( \lambda \right) &= \lambda \left[ \mathcal{E}_2 \mathcal{E}_4 + \overline{\mathcal{K}}_{B_3} \left( \mathcal{E}_1 + \mathcal{E}_2 - \mathcal{E}_3 - \mathcal{E}_4 \right) \right. \\
&\qquad \qquad \left. + 2 \mathcal{E}_3 \mathcal{W} + \mathcal{E}_4 \mathcal{W} - \overline{\mathcal{K}}_{B_3} \mathcal{W} + \mathcal{S} \mathcal{W} + \mathcal{E}_3 \mathcal{X} - \mathcal{W} \mathcal{Z} \right] \in \text{CH}^2 \left( \hat{X_5} \right)
\end{split}
\end{align}
projects to the trivial line bundle on all matter curves. This result will be used in the next section.

\subsubsection*{Line Bundle Induced on \texorpdfstring{$\mathbf{C_{\mathbf{10}_1}}$}{C10}}

By use of the linear relations we have
\begin{align}
\begin{split}
\Delta \mathcal{A} \left( \lambda \right) &= \lambda \left[ \mathcal{E}_0 \mathcal{E}_3 + \mathcal{E}_1 \mathcal{E}_3 + \mathcal{E}_0 \mathcal{S} + \mathcal{E}_1 \mathcal{S} + \mathcal{E}_2 \mathcal{S} + 2 \mathcal{E}_0 \mathcal{X} + \mathcal{E}_1 \mathcal{X} + \mathcal{E}_2 \mathcal{X} \right. \\
& \left. \qquad \qquad + \mathcal{E}_3 \mathcal{X} - \mathcal{W} \mathcal{X} - \mathcal{E}_0 \mathcal{Y} - \mathcal{E}_0 \mathcal{Z} - 2 \mathcal{E}_1 \mathcal{Z} - 2 \mathcal{E}_1 \mathcal{Z} - 2 \mathcal{E}_2 \mathcal{Z} + \mathcal{W} \mathcal{Z} \right] \, .
\end{split}
\end{align}
By use of the Stanley-Reisner ideal it then follows
\[ \mathbb{P}^1_{4D} \left( \mathbf{10}_1 \right) \cdot \Delta \mathcal{A} \left( \lambda \right) = - V \left( a_{10}, e_0, e_4, y, x s e_2 e_3 + a_{21} z^2 e_0 \right) = \emptyset \, . \]

\subsubsection*{Line Bundle Induced on \texorpdfstring{$\mathbf{C_{\mathbf{5}_3}}$}{C53}}

By use of the linear relations we can write
\begin{align}
\begin{split}
\Delta \mathcal{A} \left( \lambda \right) &= \lambda \left[ - \frac{1}{6} \mathcal{E}_0 \mathcal{E}_1 - \frac{4}{3} \mathcal{E}_0 \mathcal{E}_2 - \frac{1}{2} \mathcal{E}_1 \mathcal{E}_2 + \frac{2}{3} \mathcal{E}_0 \mathcal{E}_4 + \frac{1}{2} \mathcal{E}_1 \mathcal{E}_4 + \mathcal{E}_2 \mathcal{E}_4 + \frac{1}{3} \mathcal{E}_0 \mathcal{S} + \frac{1}{2} \mathcal{E}_1 \mathcal{S} \right. \\
& \qquad \qquad \left. + \mathcal{E}_0 \overline{\mathcal{K}}_{B_3} + \frac{3}{2} \mathcal{E}_1 \overline{\mathcal{K}}_{B_3} + \mathcal{E}_4 \overline{\mathcal{K}}_{B_3} + \mathcal{E}_2 \mathcal{S} - \frac{1}{2} \mathcal{E}_1 \mathcal{W} - \mathcal{E}_4 \mathcal{W} - \frac{2}{3} \mathcal{E}_0 \mathcal{Y} - \frac{1}{2} \mathcal{E}_1 \mathcal{Y} \right. \\
& \qquad \qquad \left. - \mathcal{E}_4 \mathcal{Y} - \frac{1}{2} \mathcal{E}_1 \mathcal{Z} - 2 \mathcal{E}_2 \mathcal{Z} + \mathcal{E}_4 \mathcal{Z} + \mathcal{W} \mathcal{Z} \right] \, .
\end{split}
\end{align}
By use of the Stanley-Reisner ideal it then follows $-1 \mathbb{P}^1_{3x} ( \mathbf{5}_3 ) \cdot \Delta \mathcal{A} ( \lambda )  = \emptyset$.

\subsubsection*{Line Bundle Induced on \texorpdfstring{$\mathbf{C_{\mathbf{5}_{-2}}}$}{C5-2}}

By use of the linear relations we can write
\begin{align}
\begin{split}
\Delta \mathcal{A} \left( \lambda \right) &= \lambda \left[ 
\mathcal{E}_2 \mathcal{E}_4 + \mathcal{E}_0 \overline{\mathcal{K}}_{B_3} + 2 \mathcal{E}_1 \overline{\mathcal{K}}_{B_3} + 2 \mathcal{E}_2 \overline{\mathcal{K}}_{B_3}- 2 \mathcal{E}_1 \mathcal{W} - 4 \mathcal{E}_2 \mathcal{W} + \mathcal{E}_4 \mathcal{W} - \mathcal{S} \mathcal{W} \right. \\
& \qquad \left. + 2 \mathcal{E}_0 \mathcal{X} + \mathcal{E}_1 \mathcal{X} + 3 \overline{\mathcal{K}}_{B_3} \mathcal{X} - \mathcal{S} \mathcal{X} - 6 \mathcal{W} \mathcal{X} + 2 \mathcal{W} \mathcal{Y} - \mathcal{X} \mathcal{Y} + \mathcal{W} \mathcal{Z} + 3 \mathcal{X} \mathcal{Z} \right] \, .
\end{split}
\end{align}
By use of the Stanley-Reisner ideal it then follows 
\[\mathbb{P}^1_{3H} \left( \mathbf{5}_{-2} \right) \cdot  \Delta \mathcal{A} \left( \lambda \right) = V \left( a_{1,0}, a_{3,2}, e_2, e_3, e_4 \right) - V \left(  a_{1,0}, a_{3,2}, s, e_3, s \right) \, . \]
Now we project this locus down onto $C_{\mathbf{5}_{-2}}$. Thereby we find
\[ \pi_{C_{\mathbf{5}_{-2}} \ast} \left(  \mathbb{P}^1_{3H} \left( \mathbf{5}_{-2} \right) \cdot \Delta \mathcal{A} \left( \lambda \right) \right) = Y_2 - Y_2 = 0 \in \text{CH}^1 \left( C_{\mathbf{5}_{-2}} \right) \, . \]

\subsubsection*{Line Bundle Induced on \texorpdfstring{$\mathbf{C_{\mathbf{1}_{5}}}$}{C15}}

By use of the linear relations we can write
\begin{align}
\begin{split}
\Delta \mathcal{A} \left( \lambda \right) &= \lambda \left[ \mathcal{E}_2 \mathcal{E}_4 + \mathcal{E}_1 \overline{\mathcal{K}}_{B_3} + \mathcal{E}_2 \overline{\mathcal{K}}_{B_3} - \mathcal{E}_3 \overline{\mathcal{K}}_{B_3} - \mathcal{E}_4 \overline{\mathcal{K}}_{B_3} - \mathcal{E}_1 \mathcal{W} - 2 \mathcal{E}_2 \mathcal{W} \right. \\
& \qquad \qquad \left. + 2 \mathcal{E}_3 \mathcal{W} + 2 \mathcal{E}_4 \mathcal{W} - \mathcal{W} \overline{\mathcal{K}}_{B_3} + \mathcal{E}_3  \mathcal{X} - 3 \mathcal{W} \mathcal{X} + 2 \mathcal{W} \mathcal{Y} - \mathcal{W} \mathcal{Z} \right] \, .
\end{split}
\end{align}
Let $k_{B_3}$ and $w$ be polynomials in the coordinate ring of $\hat{X_5}$ with degree of $\overline{\mathcal{K}}_{B_3}$ and $\mathcal{W}$ respectively. Then it follows by use of the Stanley-Reisner ideal that
\begin{align}
\begin{split}
\mathbb{P}^1_{A} \left( \mathbf{1}_{5} \right) \cdot \Delta \mathcal{A} \left( \lambda \right) &= - V \left( k_{B_3}, w, a_{3,2}, a_{4,3}, s \right) + V \left( e_3, x, s, a_{3,2}, a_{4,3} \right) \\
& \qquad - 3 V \left( w, x, s, a_{3,2}, a_{4,3} \right) + 2 V \left( w, y, s, a_{3,2}, a_{4,3} \right) \, .
\end{split}
\end{align}
Next we project down this locus onto $C_{\mathbf{1}_{5}}$. Thereby we find
\[ \pi_{C_{\mathbf{1}_{5}} \ast} \left( \mathbb{P}^1_{A} \left( \mathbf{1}_{5} \right) \Delta \mathcal{A} \left( \lambda \right) \right) = Y_3 - 3 Y_3 + 2 Y_3 = 0 \in \text{CH}_0 \left( C_{\mathbf{1}_{5}} \right) \, . \]

\subsection{Line Bundles Induced by \texorpdfstring{$\mathbf{\mathcal{A} ( \mathbf{10}_1 ) ( \lambda )}$}{A(10)(lambda)} and \texorpdfstring{$\mathbf{\mathcal{A} ( \mathbf{5}_3 ) ( \lambda )}$}{A(53)(lambda)}}

In this subsection we will work out the massless spectrum for the following two matter surface fluxes:
\begin{itemize}
 \item $\mathcal{A} \left( \mathbf{10}_1 \right) \left( \lambda \right) = \frac{\lambda}{5} \cdot \left( 2 \mathcal{E}_1 - \mathcal{E}_2 + \mathcal{E}_3 - 2 
      \mathcal{E}_4 \right) \cdot \overline{\mathcal{K}}_{B_3} + \lambda \cdot \mathcal{E}_2 \cdot \mathcal{E}_4$
 \item $\mathcal{A} \left( \mathbf{5}_3 \right) \left( \lambda \right) = \frac{\lambda}{5} \cdot \left( \mathcal{E}_1 + 2 \mathcal{E}_2 - 2 \mathcal{E}_3 - 
      \mathcal{E}_4 \right) \cdot \left( 3 \overline{\mathcal{K}}_{B_3} - 2 \mathcal{W} \right) + \lambda \cdot \mathcal{E}_3 \cdot \mathcal{X}$
\end{itemize}

\subsubsection*{Line Bundle Induced by \texorpdfstring{$\mathbf{\mathcal{A} ( \mathbf{10}_1 ) ( \lambda )}$}{A(10)(lambda)} on \texorpdfstring{$\mathbf{C_{\mathbf{5}_3}}$}{C53}}

The relevant matter surface is $- \mathbb{P}^1_{3x} ( \mathbf{5}_3 ) = - V ( a_{3,2}, e_3, x )$. Upon use of the linear relations induced from the $SU ( 5 ) \times U ( 1 )_X$-top, we can write
\[ \mathcal{A} \left( \mathbf{10}_1 \right) \left( \lambda \right) = \frac{\lambda}{5} \left[ 5 \mathcal{E}_2 \mathcal{E}_4 + \left( - \mathcal{E}_0 + \mathcal{E}_1 - 2 \mathcal{E}_2 - 3 \mathcal{E}_4 + \mathcal{W} \right) \overline{\mathcal{K}}_{B_3} \right] \, . \]
From this it follows
\[  - \mathbb{P}^1_{3x} \left( \mathbf{5}_3 \right) \cdot  \mathcal{A} \left( \mathbf{10}_1 \right) \left( \lambda \right) = \frac{2 \lambda}{5} \cdot V \left( k_{B_3}, a_{3,2}, e_2, e_3, x \right) \, . \]
In this expression $k_{B_3}$ is a polynomial in the coordinate ring of $\hat{X_5}$ whose degree matches $\overline{\mathcal{K}}_{B_3}$. Consequently
$\pi_{C_{\mathbf{5}_3} \ast} ( - \mathbb{P}^1_{3x} ( \mathbf{5}_3 ) \cdot  \mathcal{A} ( \mathbf{10}_1 ) ( \lambda ) ) = \frac{2 \lambda}{5} Y_2 \in \text{CH}^1 ( C_{\mathbf{5}_3} )$, which implies
\[ \mathcal{L} \left(  S_{\mathbf{5}_3},\mathcal{A} \left( \mathbf{10}_1 \right) \right) = \mathcal{O}_{C_{\mathbf{5}_3}} \left( \frac{2 \lambda}{5} Y_2 \right) \otimes \sqrt{K_{C_{\mathbf{5}_3}}} \, . \]

\subsubsection*{Line Bundle Induced by \texorpdfstring{$\mathbf{\mathcal{A} ( \mathbf{10}_1 ) ( \lambda )}$}{A(10)(lambda)} on \texorpdfstring{$\mathbf{C_{\mathbf{5}_{-2}}}$}{C5-2}}

The relevant matter surface is $\mathbb{P}^1_{3H} ( \mathbf{5}_{-2} ) = V ( e_3, a_{21} e_0 x z e_1 e_2 - a_{10} y, a_{43} e_0 x z e_1 e_2 - a_{32} y )$. Upon use of the linear relations induced from the $SU ( 5 ) \times U ( 1 )_X$-top, we can write
\[ \mathcal{A} \left( \mathbf{10}_1 \right) \left( \lambda \right) = \frac{\lambda}{5} \left[ 5 \mathcal{E}_2 \mathcal{E}_4 + \left( - \mathcal{E}_0 + \mathcal{E}_1 - 2 \mathcal{E}_2 - 3 \mathcal{E}_4 + \mathcal{W} \right) \overline{\mathcal{K}}_{B_3} \right] \, . \]
From this it follows
\begin{align}
\begin{split}
\mathcal{A} \left( \mathbf{10}_1 \right) \left( \lambda \right) & \cdot \mathbb{P}^1_{3H} \left( \mathbf{5}_{-2} \right) = 
\lambda \cdot V \left( a_{1,0}, a_{3,2}, e_2, e_3, e_4 \right) \\
& - \frac{3 \lambda}{5} \cdot V \left( k_{B_3}, e_3, e_4, a_{2,1} e_0 x z e_1 e_2 - a_{1,0} y, a_{4,3} e_0 x z e_1 e_2 - a_{3,2} y  \right) \\
& + \frac{\lambda}{5} \cdot V \left( w, k_{B_3}, e_3, a_{21} e_0 x z e_1 e_2 - a_{1,0} y, a_{43} e_0 x z e_1 e_2 - a_{3,2} y \right) \, . 
\end{split}
\end{align}
In this expression $k_{B_3}$ and $w$ are polynomials in the coordinate ring of $\hat{X_5}$ whose degree match $\overline{\mathcal{K}}_{B_3}$ and $\mathcal{W}$ respectively.
Consequently 
\begin{align}
\begin{split}
\pi_{C_{\mathbf{5}_{-2}} \ast} \left( \mathbb{P}^1_{3H}  \left(\mathbf{5}_{-2} \right) \cdot  \mathcal{A} \left( \mathbf{10}_1 \right) \left( \lambda \right) \right) &= \lambda Y_2 - \frac{3 \lambda}{5} \left( Y_1 + Y_2 \right) \\
&= - \frac{3 \lambda}{5} Y_1 + \frac{2 \lambda}{5} Y_2 \in \text{CH}^1 \left( C_{\mathbf{5}_{-2}} \right) \, .
\end{split}
\end{align}
This leads us to conclude
\[ \mathcal{L}\left( S_{\mathbf{5}_{-2}}, \mathcal{A} \left( \mathbf{10}_1 \right) \right) = \mathcal{O}_{C_{\mathbf{5}_{-2}}} \left( - \frac{3 \lambda}{5} Y_1 + \frac{2 \lambda}{5} Y_2 \right) \otimes \sqrt{K_{C_{\mathbf{5}_{-2}}}} \, . \]

\subsubsection*{Line Bundle Induced by \texorpdfstring{$\mathbf{\mathcal{A} ( \mathbf{10}_1 ) ( \lambda )}$}{A(10)(lambda)} on \texorpdfstring{$\mathbf{C_{\mathbf{1}_{5}}}$}{C15}}

The relevant matter surface is $\mathbb{P}^1_{A} ( \mathbf{1}_{5} ) = V ( s, a_{3,2}, a_{4,3} )$. Upon use of the linear relations  induced from the $SU ( 5 ) \times U ( 1 )_X$-top we can write
\[ \mathcal{A} \left( \mathbf{10}_1 \right) \left( \lambda \right) = \frac{\lambda}{5} \left[ 5 \mathcal{E}_2 \cdot \mathcal{E}_4 + \left( - \mathcal{E}_0 + \mathcal{E}_1 - 2 \mathcal{E}_2 - 3 \mathcal{E}_4 + \mathcal{W} \right) \cdot \overline{\mathcal{K}}_{B_3} \right] \, . \]
From this it follows
\[ \mathbb{P}^1_{A} \left( \mathbf{1}_{5} \cdot \mathcal{A} \left( \mathbf{10}_1 \right) \left( \lambda \right)  \right) = \frac{\lambda}{5} \cdot V \left( k_{B_3}, w, s, a_{3,2}, a_{4,3} \right) = \emptyset \, , \]
where the polynomials $k_{B_3}$ and $w$ are picked as before. Consequently 
\[ \pi_{C_{\mathbf{1}_{5}} \ast} \left( \mathbb{P}^1_{A} \left( \mathbf{1}_{5} \right) \cdot  \mathcal{A} \left( \mathbf{10}_1 \right) \left( \lambda \right) \right) = 0 \in \text{CH}^1 \left( C_{\mathbf{1}_{5}} \right) \, . \]
This implies $\mathcal{L} (  S_{\mathbf{1}_{5}}, \mathcal{A} ( \mathbf{10}_1 ) ) = \sqrt{K_{C_{\mathbf{1}_{5}}}}$.

\subsubsection*{Line Bundle Induced by \texorpdfstring{$\mathbf{\mathcal{A} ( \mathbf{10}_1 ) ( \lambda )}$}{A(10)(lambda)} on \texorpdfstring{$\mathbf{C_{\mathbf{10}_1}}$}{C10}}

To compute the massless spectrum of $\mathcal{A} ( \mathbf{10}_1 ) ( \lambda )$ on $C_{\mathbf{10}_1}$ we note that 
\[ \mathcal{A} \left( \mathbf{10}_1 \right) \left( \lambda \right) = \mathcal{A} \left( \mathbf{5}_3 \right) \left( - \lambda \right) - \mathcal{A}_X \left( - \lambda W \right) + \Delta \mathcal{A} \left( \lambda \right) \, . \]
We found in \autoref{sec:MasslessSpectrumOfDeltaAG} that the line bundle induced by $\mathcal{A} ( \lambda )$ on the matter curves is trivial. The massless spectrum of $\mathcal{A}_X ( - \lambda W )$ follows from \autoref{subsec:ExampleOnMasslessSpectrumComputation}. In particular we have
\[ D \left(  S_{\mathbf{10}_1},\mathcal{A}_X \left( - \lambda W \right) \right) = - \frac{\lambda}{5} \left. W \right|_{C_{\mathbf{10}_1}} \, . \]
Note that $W = 3 ( 2 \overline{K}_{B_3} - W ) + ( -2 ) ( 3 \overline{K}_{B_3} - 2 W )$. Therefore
\[ D \left(  S_{\mathbf{10}_1},\mathcal{A}_X \left( - \lambda W \right) \right)  = - \frac{3 \lambda}{5} Y_1 + \frac{2 \lambda}{5} Y_2 \in \text{Div} \left( C_{\mathbf{10}_1} \right) \, . \]
The line bundle induced by $\mathcal{A} ( \mathbf{10}_1 ) ( \lambda )$ on $C_{\mathbf{10}_1}$ follows from $D ( S_{\mathbf{10}_1},\mathcal{A} ( \mathbf{5}_3 ) ( \lambda ) )$. So let us compute this divisor now. We first recall that the relevant matter surface is given by $\mathbb{P}^1_{4D} ( \mathbf{10}_1 ) = V ( a_{1,0}, e_4, x s e_2 e_3 + a_{2,1} z^2 e_0 )$. Upon use of the linear relations we can write
\[ \mathcal{A} \left( \mathbf{5}_3 \right) \left( \lambda \right) = \frac{\lambda}{5} \cdot \left( \mathcal{E}_0 + 2 \mathcal{E}_1 + 3 \mathcal{E}_2 - \mathcal{E}_3 - \mathcal{W} \right) \cdot \left( 3 \overline{\mathcal{K}}_{B_3} - 2 \mathcal{W} \right) + \lambda \mathcal{E}_3 \cdot \mathcal{X} \, . \]
From this it follows
\[ \mathcal{A} \left( \mathbf{5}_3 \right) \left( \lambda \right) \cdot \mathbb{P}^1_{4D} \left( \mathbf{10}_1 \right) = \frac{2 \lambda}{5} \cdot V \left(  a_{1,0}, a_{3,2}, e_0, e_4, x s e_2 e_3 + a_{2,1} z^2 e_0 \right) \, , \]
which implies
\[ \mathcal{L} \left( S_{\mathbf{10}_1},\mathcal{A} \left( \mathbf{5}_3 \right) \left( \lambda \right) \right) = \mathcal{O}_{C_{\mathbf{10}_1}} \left( \frac{2 \lambda}{5} Y_2 \right) \otimes \sqrt{K_{C_{\mathbf{10}_1}}} \, . \]
This finally enables us conclude that $D ( S_{\mathbf{10}_1},\mathcal{A} ( \mathbf{10}_1 ) ( \lambda ) ) = \frac{3 \lambda}{5} Y_1 - \frac{4 \lambda}{5} Y_2$.

\subsubsection*{Line Bundles Induced by \texorpdfstring{$\mathbf{\mathcal{A} ( \mathbf{5}_3 ) ( \lambda )}$}{A(53)(lambda)}}

We can now turn the logic around and use the relation
\[ \mathcal{A} \left( \mathbf{10}_1 \right) \left( \lambda \right) = \mathcal{A} \left( \mathbf{5}_3 \right) \left( - \lambda \right) - \mathcal{A}_X \left( - \lambda W \right) + \Delta \mathcal{A} \left( \lambda \right) \]
to compute the massless spectrum of $\mathcal{A} ( \mathbf{5}_3 ) ( \lambda )$ from knowledge of the massless spectra of all other fluxes. This leads to the results summarised in \autoref{table-6}.

\subsection{Line Bundles Induced by \texorpdfstring{$\mathbf{\mathcal{A} ( \mathbf{5}_{-2} ) ( \lambda )}$}{A(5-2)(lambda)}}

We now turn to the massless spectrum of the flux
\[ \mathcal{A} \left( \mathbf{5}_{-2} \right) \left( \lambda \right) = \frac{\lambda}{5} \left[ \left( \mathcal{E}_1 + 2 \mathcal{E}_2 + 3 \mathcal{E}_3 - \mathcal{E}_4 \right) \left( 5 \overline{\mathcal{K}}_{B_3} - 3 \mathcal{W} \right) + 5 \mathcal{E}_3 \mathcal{E}_4 - 5 \mathcal{E}_3 \overline{\mathcal{K}}_{B_3} - 5 \mathcal{E}_3 \mathcal{Y} \right] \, . \]

\subsubsection*{Line Bundle Induced by \texorpdfstring{$\mathbf{\mathcal{A} ( \mathbf{5}_{-2} ) ( \lambda )}$}{A(5-2)(lambda)} on \texorpdfstring{$\mathbf{C_{\mathbf{10}_1}}$}{C10}}

The relevant matter surface is $\mathbb{P}^1_{4D} ( \mathbf{10}_1 ) = V ( a_{1,0}, e_4, x s e_2 e_3 + a_{2,1} z^2 e_0 )$. Upon use of the linear relations induced from the $SU ( 5 ) \times U ( 1 )_X$-top we can write
\[ \mathcal{A} \left( \mathbf{5}_{-2} \right) \left( \lambda \right) = \frac{\lambda}{5} \left( 5 \overline{\mathcal{K}}_{B_3} - 3 \mathcal{W} \right) \left( 3 \mathcal{E}_3 + 2 \mathcal{Y} \right) + \lambda \left( 3 \mathcal{E}_2 \mathcal{E}_3 - \mathcal{W} \mathcal{E}_3 - 4 \mathcal{E}_3 \mathcal{Y} \right) \, . \]
It is now readily confirmed that this implies
\[ \mathcal{L} \left( S_{\mathbf{10}_1}, \mathcal{A} \left( \mathbf{5}_{-2} \right) \left( \lambda \right) \right) = \mathcal{O}_{C_{\mathbf{5}_{-2}}} \left( - \frac{3 \lambda}{5} Y_1 + \frac{2 \lambda}{5} Y_2 \right) \otimes \sqrt{K_{C_{\mathbf{10}_1}}} \, . \]

\subsubsection*{Line Bundle Induced by \texorpdfstring{$\mathbf{A_G ( \mathbf{5}_{-2} ) ( \lambda )}$}{A_G(5-2)(lambda)} on \texorpdfstring{$\mathbf{C_{\mathbf{5}_3}}$}{C53}}

The relevant matter surface is $- \mathbb{P}^1_{3x} ( \mathbf{5}_3 ) = - V ( a_{3,2}, e_3, x )$. Upon use of the linear relations induced from the $SU ( 5 ) \times U ( 1 )_X$-top, we can write
\begin{align}
\begin{split}
\mathcal{A} \left( \mathbf{5}_{-2} \right) \left( \lambda \right) &= \frac{\lambda}{5} \left( 5 \overline{\mathcal{K}}_{B_3} - 3 \mathcal{W} \right)  \left( - 3 \mathcal{E}_0 - 2 \mathcal{E}_1 - \mathcal{E}_2 - 4 \mathcal{E}_4 + 3 \mathcal{W} \right) \\
& \qquad + \lambda \left( 2 \mathcal{E}_0 \mathcal{E}_4 + \mathcal{E}_1 \mathcal{E}_4 + \mathcal{E}_0 \overline{\mathcal{K}}_{B_3} + \mathcal{E}_1 \overline{\mathcal{K}}_{B_3} + \mathcal{E}_2 \overline{\mathcal{K}}_{B_3} + 4 \mathcal{E}_4 \overline{\mathcal{K}}_{B_3} - \mathcal{E}_4 s \right. \\
& \qquad \qquad \qquad \qquad \left. - 2 \mathcal{E}_4 \mathcal{W} - \overline{\mathcal{K}}_{B_3} \mathcal{W}  + \mathcal{E}_0 \mathcal{Y} + \mathcal{E}_1 \mathcal{Y} + \mathcal{E}_2 \mathcal{Y} - \mathcal{W} \mathcal{Y} + 3 \mathcal{E}_4 \mathcal{Z} \right) \, .
\end{split}
\end{align}
From this it is readily confirmed that
\[ \mathcal{L} \left(S_{\mathbf{5}_3}, \mathcal{A} \left( \mathbf{5}_{-2} \right) \left( \lambda \right) \right) = \mathcal{O}_{C_{\mathbf{5}_3}} \left( \frac{\lambda}{5} Y_3 - \frac{4 \lambda}{5} Y_2 \right) \otimes \sqrt{K_{C_{\mathbf{10}_1}}} \, . \]

\subsubsection*{Line Bundle Induced by \texorpdfstring{$\mathbf{\mathcal{A} ( \mathbf{5}_{-2} ) ( \lambda )}$}{A(5-2)(lambda)} on \texorpdfstring{$\mathbf{C_{\mathbf{1}_{5}}}$}{C15}}

The relevant matter surface is $\tilde{\mathbb{P}^1_{A} ( \mathbf{1}_{5} )} = V ( s, a_{3,2}, a_{4,3} )$. Therefore all intersections are already transverse and we find
\[ \mathcal{L} \left(S_{\mathbf{1}_{5}}, \mathcal{A} \left( \mathbf{5}_{-2} \right) \left( \lambda \right) \right) = \mathcal{O}_{C_{\mathbf{1}_{5}}} \left( - \lambda Y_3 \right) \otimes \sqrt{K_{C_{\mathbf{1}_{5}}}} \, . \]

\subsubsection*{Line Bundle Induced by \texorpdfstring{$\mathbf{\mathcal{A} ( \mathbf{5}_{-2} ) ( \lambda )}$}{A(5-2)(lambda)} on \texorpdfstring{$\mathbf{C_{\mathbf{5}_{-2}}}$}{C5-2}}

To compute this spectrum we introduce the difference
\[ \Delta \mathcal{A}^{(2)} \left( \lambda \right) := \mathcal{A} \left( \mathbf{5}_{-2} \right) \left( \lambda \right) - \mathcal{A}^X \left( - \lambda W \right) \, . \]
Upon use of the linear relations induced from the $SU ( 5 ) \times U ( 1 )_X$-top, we can write
\begin{align}
\begin{split}
\Delta \mathcal{A}^{(2)} \left( \lambda \right) &= \lambda \left[ - \mathcal{E}_1 \mathcal{E}_4 - 2 \mathcal{E}_2 \mathcal{E}_4 - 2 \mathcal{E}_0 \overline{\mathcal{K}}_{B_3} - 3 \mathcal{E}_1 \overline{\mathcal{K}}_{B_3} - 4 \mathcal{E}_2 \overline{\mathcal{K}}_{B_3} - \mathcal{E}_4 \mathcal{S} + \mathcal{E}_4 \mathcal{Z} - 2 \mathcal{W} \mathcal{Z} \right. \\
& \qquad \qquad \left. -2 \overline{\mathcal{K}}_{B_3} \mathcal{S} + 2 \mathcal{E}_1 \mathcal{W} + 4 \mathcal{E}_2 \mathcal{W} + 2 \mathcal{S} \mathcal{W} -2 \mathcal{E}_4 \mathcal{X} - 6 \overline{\mathcal{K}}_{B_3} \mathcal{X} - 2 \mathcal{Y} \mathcal{Z} \right. \\
& \qquad \qquad \left. + 6 \mathcal{W} \mathcal{X} - \mathcal{E}_0 \mathcal{Y} + \mathcal{E}_2 \mathcal{Y} + \mathcal{E}_4 \mathcal{Y} + 2 \overline{\mathcal{K}}_{B_3} \mathcal{Y} + \mathcal{S} \mathcal{Y} - 2 \mathcal{W} \mathcal{Y} + \mathcal{X} \mathcal{Y} \right] \, .
\end{split}
\end{align}
Then it is readily confirmed that
\begin{align*}
\Delta \mathcal{A}^{(2)} \left( \lambda \right) \cdot \mathbb{P}^1_{3H} \left( \mathbf{5}_{-2} \right) &= \lambda \left[ - 2 \mathcal{E}_2 \mathcal{E}_4 + 2 \left( \mathcal{W} - \overline{\mathcal{K}}_{B_3} \right) \mathcal{S} + \mathcal{E}_4 \mathcal{Y} + \mathcal{S} \mathcal{Y} \right] \cdot \mathbb{P}^1_{3H} \left( \mathbf{5}_{-2} \right).
\end{align*}
Upon projecting onto $C_{\mathbf{5}_{-2}}$ we then find $\pi_{C_{\mathbf{5}_{-2}} \ast} ( \mathbb{P}^1_{3H} ( \mathbf{5}_{-2} ) \cdot \Delta \mathcal{A}^{(2)} ( \lambda ) ) = 0$. Therefore
\[ \mathcal{L} \left(S_{\mathbf{5}_{-2}}, \mathcal{A} \left( \mathbf{5}_{-2} \right) \left( \lambda \right) \right) = \mathcal{O}_{C_{\mathbf{5}_{-2}}} \left( \frac{3 \lambda}{5} Y_1 + \frac{2 \lambda}{5} Y_2 - \frac{\lambda}{5} Y_3 \right) \otimes \sqrt{K_{C_{\mathbf{5}_{-2}}}} \, . \]

\subsection{Line Bundles Induced by \texorpdfstring{$\mathbf{\mathcal{A} ( \mathbf{1}_{5} ) ( \lambda )}$}{A(15)(lambda)}}

We express $A ( \mathbf{1}_{5} ) ( \lambda ) = \lambda \mathbb{P}^1_A ( \mathbf{1}_{5} )$ as $\mathcal{A} ( \mathbf{1}_{5} ) ( \lambda ) = \lambda [ V ( s, a_{3,2} ) - V ( s, x ) ] \in \text{CH}^2 ( \hat{X_5} )$.

\subsubsection*{Line Bundle Induced by \texorpdfstring{$\mathbf{\mathcal{A} ( \mathbf{1}_{5} ) ( \lambda )}$}{A(15)(lambda)} on \texorpdfstring{$\mathbf{C_{\mathbf{10}_1}}$}{C10}}

The relevant matter surface is $\mathbb{P}^1_{4D} ( \mathbf{10}_1 ) = V ( a_{1,0}, e_4, x s e_2 e_3 + a_{2,1} z^2 e_0 )$. From this it follows $\mathcal{A} ( \mathbf{1}_{5} ) ( \lambda ) \cdot \mathbb{P}^1_{4D} ( \mathbf{10}_1 ) = \emptyset$. Consequently $\mathcal{L} ( S_{\mathbf{10}_1},  \mathcal{A} ( \mathbf{1}_{5} ) ( \lambda ) ) = \sqrt{K_{C_{\mathbf{10}_1}}}$.

\subsubsection*{Line Bundle Induced by \texorpdfstring{$\mathbf{A_G ( \mathbf{1}_{5} ) ( \lambda )}$}{A_G(15)(lambda)} on \texorpdfstring{$\mathbf{C_{\mathbf{5}_3}}$}{C53}}

The relevant matter surface is $S = - \mathbb{P}^1_{3x} ( \mathbf{5}_3 ) = - V ( a_{3,2}, e_3, x )$. Upon use of the linear relations induced from the $SU ( 5 ) \times U ( 1 )_X$-top, we can write
\[ \mathcal{A} \left( \mathbf{1}_{5} \right) \left( \lambda \right) = \lambda \left[ 4 \mathcal{S} \overline{\mathcal{K}}_{B_3} - 3 \mathcal{S} \mathcal{W} + \mathcal{S} \mathcal{E}_0 + \mathcal{S} \mathcal{E}_1 + \mathcal{S} \mathcal{E}_2 - \mathcal{S} \mathcal{Y} + \mathcal{S} \mathcal{Z} \right] \, . \]
From this it follows
\[ S \cdot \mathcal{A} \left( \mathbf{1}_{5} \right) \left( \lambda \right)  = - \lambda \left[ 4 V \left( k_{B_3}, a_{3,2}, s, e_3, x \right) - 3 V \left( w, a_{3,2}, s, e_3, x \right) \right] \, . \]
To work out the projection onto $C_{\mathbf{5}_3}$ first write $W = \frac{4}{3} \overline{K}_{B_3} - \frac{1}{3} ( 4 \overline{K}_{B_3} - 3 W )$, to see that this implies $\left. W \right|_{C_{\mathbf{5}_3}} = \frac{4}{3} Y_2 - \frac{1}{3} Y_3$. Therefore $\pi_{C_{\mathbf{5}_3} \ast} (  S \cdot \mathcal{A} ( \mathbf{1}_{5} ) ( \lambda ) ) = - \lambda Y_3$, this shows
\[ \mathcal{L} \left(S_{\mathbf{5}_3} , \mathcal{A} \left( \mathbf{1}_{5} \right) \left( \lambda \right) \right) = \mathcal{O}_{C_{\mathbf{5}_3}} \left( - \lambda Y_3 \right) \otimes \sqrt{K_{C_{\mathbf{5}_3}}} \, . \]

\subsubsection*{Line Bundle Induced by \texorpdfstring{$\mathbf{\mathcal{A} ( \mathbf{1}_{5} ) ( \lambda )}$}{A(15)(lambda)} on \texorpdfstring{$\mathbf{C_{\mathbf{5}_{-2}}}$}{C5-2}}

The relevant matter surface is $S = \mathbb{P}^1_{3H} ( \mathbf{5}_{-2} ) = V ( a_{21} e_0 x z e_1 e_2 - a_{10} y, a_{32} y - a_{43} e_0 e_1 e_2 x z )$. The relevant intersections are therefore automatically transverse and we find
\begin{align}
\begin{split}
S \cdot \mathcal{A} \left( \mathbf{1}_{5} \right) \left( \lambda \right)  &= \lambda V \left( a_{3,2}, s, e_3, a_{2,1} x - a_{1,0} y, a_{4,3} x - a_{3,2} y \right) - \lambda V \left( a_{1,0}, a_{3,2}, e_3, s, x \right) \, .
\end{split}
\end{align}
Therefore $\pi_{C_{\mathbf{5}_{-2}} \ast} (S \cdot  \mathcal{A} ( \mathbf{1}_{5} ) ( \lambda ) ) = \lambda Y_3$. Consequently
\[ \mathcal{L} \left( S_{\mathbf{5}_{-2}} \cdot \mathcal{A} \left( \mathbf{1}_{5} \right) \left( \lambda \right)\right) = \mathcal{O}_{C_{\mathbf{5}_{-2}}} \left( \lambda Y_3 \right) \otimes \sqrt{K_{C_{\mathbf{5}_{-2}}}} \, . \]

\subsubsection*{Line bundle induced by \texorpdfstring{$\mathbf{A_G ( \mathbf{1}_{5} ) ( \lambda )}$}{A_G(15)(lambda)} on \texorpdfstring{$\mathbf{C_{\mathbf{1}_{5}}}$}{C15}}

To compute this spectrum we introduce
\[ \Delta \mathcal{A}^{(3)} \left( \lambda \right) := \mathcal{A} \left( \mathbf{1}_{5} \right) \left( \lambda \right) - \mathcal{A}^X \left( - \lambda \left[ 6 \overline{K}_{B_3} - 5 W \right] \right) - \mathcal{A} \left( \mathbf{10}_1 \right) \left( - \lambda \right) \, . \label{equ:SpectrumOfMSFOnSingletCurveOntoSingletCurve} \]
Upon use of the linear relations induced from the $SU ( 5 ) \times U ( 1 )_X$-top we can write
\begin{align}
\begin{split}
\Delta \mathcal{A}^{(3)} \left( \lambda \right) &= \frac{\lambda}{5} \left[ -31 \mathcal{E}_0 \overline{\mathcal{K}}_{B_3} -26 \mathcal{E}_1 \overline{\mathcal{K}}_{B_3} -36 \mathcal{E}_3 \overline{\mathcal{K}}_{B_3} -36 \mathcal{E}_4 \overline{\mathcal{K}}_{B_3} -5 \mathcal{E}_1 \mathcal{W} + 30 \mathcal{E}_3 \mathcal{W} \right. \\
& \qquad \qquad \left. + 30 \mathcal{E}_4 \mathcal{W} + 6 \overline{\mathcal{K}}_{B_3} \mathcal{W} + 5 \mathcal{E}_1 \mathcal{X} + 15 \mathcal{E}_3 \mathcal{X} + 10 \mathcal{E}_4 \mathcal{X} - 45 \mathcal{W} \mathcal{X} + 30 \mathcal{W} \mathcal{Y} \right. \\
& \qquad \qquad \left. + \mathcal{E}_2 \left( 5 \mathcal{E}_4 - 26 \overline{\mathcal{K}}_{B_3} - 10 \mathcal{W} + 10 \mathcal{X} \right) + 5 \mathcal{X} \mathcal{Y} - 5 \mathcal{Z} \left( 5 \mathcal{W} + 3 \mathcal{X} \right) \right] \, .
\end{split}
\end{align}
By use of the Stanley-Reisner ideal it is readily verified that
\begin{align}
\begin{split}
\mathbb{P}^1_A \left( \mathbf{1}_{5} \right)    \cdot \Delta \mathcal{A}^{(3)} \left( \lambda \right)  &= \frac{\lambda}{5} \left[ 15 V \left( s, x, e_3, a_{3,2}, a_{4,3} \right) - 45 V \left( s, x, w, a_{3,2}, a_{4,3} \right) \right. \\
& \hspace{15em} \left. + 30 V \left( s, y, w, a_{3,2}, a_{4,3} \right) \right] \, .
\end{split}
\end{align}
By projecting this quantity onto $C_{\mathbf{1}_{5}}$ we obtain 
\[ \pi_{C_{\mathbf{1}_{5}} \ast} \left( \Delta \mathcal{A}^{(3)} \left( \lambda \right) \cdot \mathbb{P}^1_A \left( \mathbf{1}_{5} \right) \right) = 0 \in \text{Pic} \left( C_{\mathbf{1}_{5}} \right) \, . \]

That said, we can now use (\ref{equ:SpectrumOfMSFOnSingletCurveOntoSingletCurve}) to compute the massless spectrum of $\mathcal{A} ( \mathbf{1}_{5} ) ( \lambda )$ on $C_{\mathbf{1}_{5}}$. This yields
\[ \mathcal{L} \left( \mathcal{A} \left( \mathbf{1}_{5} \right) \left( \lambda \right), C_{\mathbf{1}_{5}} \right) = \left. \mathcal{O}_{B_3} \left( - \lambda \left( 6 \overline{K}_{B_3} - 5 W \right) \right) \right|_{C_{\mathbf{1}_{5}}} \otimes \sqrt{K_{C_{\mathbf{1}_{5}}}} \, . \]

\section{From Modules to Coherent Sheaves and Sheaf Cohomology Groups} \label{sec:ATasteOfAlgebraAndAlgebraicGeometry}

This appendix provides some of the mathematical background underlying the computations outlined in \autoref{sec:ComputingTheSpectra}. Our task which arises in the main text is the following: Given a toric variety $X_\Sigma$ with Cox ring $S$, a matter curve $C \subseteq X_\Sigma$ and a divisor $D \in \text{Div} ( C )$, construct \fp $S$-modules $M_{\pm}$ such that $\tilde{M_{\pm}} \in \textit{Coh} \left( X_\Sigma \right)$ are supported on $C$ only and satisfy $\left. \tilde{M_{\pm}} \right|_C \cong \mathcal{O}_{C} ( \pm D )$. This construction makes use of the sheafification functor
\[ \widetilde{\phantom{m}} \colon S \mathrm{\textnormal{-}fpgrmod} \to \mathfrak{Coh} X_\Sigma \, , \]
which turns an \fp graded $S$-module $M$ into a coherent sheaf $\tilde{M}$ on $X_\Sigma$. See \cite{fulton1993introduction, cox2011toric, hartshorne1977algebraic} and references therein for further information. 

We give details on this construction in \autoref{subsec:IdealSheavesFromModules}. As a preparation we begin with a general review of the category of \fp graded modules in \autoref{subsec:ProjSmodule} and \autoref{subsec:FPGradedSModules} and explain the computation of extension modules in \autoref{subsec:ExtOfFPModules}. The connection to sheaves (\autoref{subsec:ABriefingOnSheaves}) is described in \autoref{subsec:LocalisationOfRings} to \autoref{subsec:CoherentSheavesOnAbstractVarieties} in general terms, and specialised to toric varieties in \autoref{app_sheafification} to \autoref{subsec:IdealSheavesFromModules}. Finally, \autoref{subsec:SheafCohomologyFromFPGradedSModules} describes how to extract the sheaf cohomologies of $\mathcal{O}_{C} ( \pm D )$ from $M_{\pm}$. 

Some words of caution before we start:
In a supersymmetric context, it is common pratice to work with \emph{analytic} geometry in physics. Unfortunately since the complex numbers $\mathbb{C}$ are not suitable to be modelled in a computer, it is necessary for us to switch to a finite field extension of the rational numbers $\mathbb{Q}$. The same limitations hold for holomorphic functions, \ie absolutely convergent power series. Hence computer applications require us to switch to \emph{algebraic} geometry over the rational numbers $\mathbb{Q}$ (or finite field extensions thereof such as $\mathbb{Q} [ i ] = \mathbb{Q} + i \mathbb{Q}$). Therefore, unless stated explicitly, this appendix works with (toric) varieties over $\mathbb{Q}$ with Cox ring $S$. We will provide a basic introduction to these spaces from the point of view of algebraic geometry. For simplicity, this appendix is formulated in the language of varieties. However a scheme-theoretic approach is indeed possible. The interested reader may consult \cite{RohrerDissertation, 2011arXiv1107.2483R, 2012arXiv1212.3956R, 2011arXiv1107.2713R} and references therein.

\subsection{The Category of Projective Graded \texorpdfstring{$\mathbf{S}$}{S}-Modules} \label{subsec:ProjSmodule}

We assume that $X_\Sigma$ is a normal toric variety over $\mathbb{Q}$ which is either smooth and complete or simplicial and projective. For such varieties $X_\Sigma$ the coordinate ring $S$ -- termed the Cox ring -- is a polynomial ring $S = \mathbb{Q} [ x_1, \dots, x_m ]$ which is graded by $\text{Cl} ( X_\Sigma ) \cong \mathbb{Z}^n$. This means that there is a homomorphism of monoids
\[ \text{deg} \colon \text{Mon} ( S ) \to \mathbb{Z}^n \]
such that the images of the monomials $x_1, \dots, x_m$ generate $\mathbb{Z}^n$ as a group. Here $\text{Mon} ( S )$  denotes the set of monomials in $S$. For a monomial $f \in \text{Mon} ( S )$ we term $\text{deg} ( f )$ the \emph{degree of $f$}. A polynomial $P \in S$ for which all its monomials have identical degree $d \in \mathbb{Z}^n$ is termed homogenous polynomial (of degree $d$). The homogeneous elements of degree $d$ form a subgroup $S_d$ of $S$. As a group, the ring $S$ therefore admits a direct sum decomposition
\[ S = \bigoplus_{d \in \mathbb{Z}^n}{S_d} \]
such that the multiplication in $S$ satisfies $S_d \cdot S_e \subseteq S_{d+e}$ for all $d,e \in \mathbb{Z}^n$. We term the group $S_d$ the \emph{degree $d$ layer of $S$}.

Given a $\mathbb{Z}^n$-graded Cox ring $S = \mathbb{Q} [ x_1, \dots, x_m ]$, we can define for every $d \in \mathbb{Z}^n$ a degree-shift of this ring. Namely $S ( d )$ is the $\mathbb{Z}^n$-graded ring with $S ( d )_e = S ( 0 )_{e+d} \equiv S_{e+d}$.

As an example consider $\mathbb{P}^2_{\mathbb{Q}}$. This toric variety has $\text{Cl} ( \mathbb{P}^2_{\mathbb{Q}} ) = \mathbb{Z}$. Its Cox ring $\mathbb{Q} [ x_1, x_2, x_3 ]$ is $\mathbb{Z}$-graded upon $\text{deg} ( x_1 ) = \text{deg} ( x_2 ) = \text{deg} ( x_3 ) = 1$. In particular $1 \in S ( 0 )$ satisfies $\text{deg} ( 1 ) = 0$. Now consider the ring $S ( -1 )$. By definition it satisfies
\[ S \left( -1 \right)_1 = S \left( 0 \right)_{1 + \left( -1 \right)} = S \left( 0 \right)_0 \, . \]
Consequently those $x \in S ( 0 )$ which have degree $0$ are considered elements of degree $1$ in $S ( -1 )$. For example $1 \in S ( - 1 )$ therefore satisfies $\text{deg} ( 1 ) = +1$.

For a $\mathbb{Z}^n$-graded ring $S = \mathbb{Q} [ x_1, \dots, x_m ]$, we now wish to give a brief introduction to the \emph{category of projective graded $S$-modules}. To this end first recall that a (left) $S$-module $M$ is an Abelian group $( M,+ )$ together with a scalar multiplication
\[ S \times M \to M \; , \; \left( s, m \right) \mapsto s \cdot m \]
such that for all $s_1, s_2 \in S$ and all $m_1, m_2 \in M$ it holds
\begin{itemize}
 \item $s_1 \cdot \left( s_2 \cdot m_1 \right) = \left( s_1 \cdot s_2 \right) \cdot m_1$,
 \item $\left( s_1 + s_2 \right) \cdot m_1 = s_1 \cdot m_1 + s_2 \cdot m_1$,
 \item $s_1 \cdot \left( m_1 + m_2 \right) = s_1 \cdot m_1 + s_1 \cdot m_2$,
 \item $1 \cdot m_1 = m_1$.
\end{itemize}
Consequently a (left) $S$-module $M$ looks very much like a `vector space over $S$', except for the fact that $S$ need not be a field. An $S$-module
\[ M = \bigoplus_{d \in I}{S \left( d \right)}, \qquad I \subseteq \mathbb{Z}^n \]
is called \emph{graded} precisely if $S_i M_j \subseteq M_{i+j}$. Note that the indexing set $I$ need not be finite. However, if $I$ is finite, then $| I |$ is termed the \emph{rank of $M$}. For reasons that will become clear eventually, we will refer to such modules as \emph{projective graded (left) $S$-modules}. For ease of notation we will oftentimes drop the term `left'. So unless stated explicitly, we always mean left-modules.

The morphisms in the \emph{category of projective graded (left) $S$-modules} are module homomorphisms which respect the grading. For example $\varphi \colon S ( -1 ) \xrightarrow{( x_1 )} S ( 0 )$ is such a morphism since 
\[ \underbrace{S \left( -1 \right) \ni 1}_{\text{degree 1}} \mapsto \varphi \left( 1 \right) = \underbrace{x_1 \in S \left( 0 \right)}_{\text{degree 1}} \, . \]

Given the projective graded $S$-modules $M$ and $N$ of finite rank $m$ and $n$, a morphism $M \to N$ is given by a matrix $A$ with entries from $S$. It is now a pure matter of convention to express elements of $M$, $N$ either as columns or rows of polynomials from $S$. Suppose that we express $e \in M$, $f \in N$ as rows of polynomials, then $A$ has to be a matrix with $m$ rows and $n$ columns. In particular, we multiply $e \in M$ from the left to the matrix $A$ to obtain its image $e \cdot A \in N$. As this multiplication is performed from the left, this convention applies to \emph{left}-modules.

Of course, one can also choose to represent elements of $M$ and $N$ as columns. In this case $A$ must be a matrix with $n$ rows and $m$ columns and we multiply $e \in M$ from the right to obtain its image $A \cdot e \in N$. In this case one deals with \emph{projective graded right $S$-modules} and \emph{projective graded right $S$-module homomorphisms}.

For historical reasons, it is tradition in algebra to use left-modules in papers, and we follow this tradition here. Hence elements of projective graded $S$-modules are always expressed as rows of polynomials in $S$.

Given a $\mathbb{Z}^n$-graded ring $S$, the \emph{category of projective graded $S$-modules} happens to be an additive monoidal category, which is both strict and rigid symmetric closed \cite{CAP, PosurDoktor, GutscheDoktor}. We provide an implementation of this category in the language of \emph{CAP} \cite{CAP, PosurDoktor, GutscheDoktor} in the software package \cite{CAPCategoryOfProjectiveGradedModules}.

\subsection{The Category \texorpdfstring{$\mathbf{S\mathrm{\textnormal{-}fpgrmod}}$}{fpgrmod}} \label{subsec:FPGradedSModules}

Based on a $\mathbb{Z}^n$-graded ring $S = \mathbb{Q} [ x_1, \dots, x_m ]$ and its associated \emph{category of projective graded $S$-modules}, we now wish to build a new category -- the category of \fp graded $S$-modules ($S \mathrm{\textnormal{-}fpgrmod}$). 
The importance of this construction for us is that it allows us to describe ideals (or vanishing loci) via the relations enjoyed by the generators of the ideal.
In order to understand this, we need a bit of preparation.

The basic idea is very simple:
\ebox{Objects in $S \mathrm{\textnormal{-}fpgrmod}$ are presented by morphisms of projective graded $S$-modules of finite rank.}

For an example let us pick $\mathbb{P}^2_{\mathbb{Q}}$ again and look at the following two morphisms of projective graded $S$-modules (of finite rank):
\begin{itemize}
 \item $\varphi \colon 0 \to S \left( 0 \right)$. 
 \item $\psi \colon S \left( -2 \right)^{\oplus 3} \xrightarrow{R} S \left( -1 \right)^{\oplus 3}$, where $R = \left( \begin{array}{ccc} 0 & -x_3 & x_2 \\  x_3 & 0 & -x_1 \\ - x_2 & x_1 & 0 \end{array} \right)$,
\end{itemize}
Abstractly we intend to describe the modules
\[ M_\varphi \equiv \text{coker} \left( \varphi \right) = \text{codomain} \left( \varphi \right) / \text{im} \left( \varphi \right), \qquad M_{\psi} \equiv \text{coker} \left( \psi \right) = \text{codomain} \left( \psi \right) / \text{im} \left( \psi \right) \, . \]
A means to present these modules $M_\varphi$, $M_\psi$ is indeed provided by the morphisms $\varphi$, $\psi$. Therefore we term the codomain of $\varphi, \psi$ the \emph{generators} of $M_{\varphi}$ and $M_\psi$ respectively. Similarly the domain of $\varphi, \psi$ is given the name \emph{relations} of $M_{\varphi}$ and $M_\psi$ respectively.

In the following we will make use of commutative diagrams of projective graded $S$-modules. In these diagrams we box morphisms of projective graded $S$-modules (of finite rank) in blue colour if they are to present an \fp graded $S$-module. Consequently we depicture $M_{\psi}$, $M_{\varphi}$ as follows:
\[ 
\begin{tikzpicture}[baseline=(current  bounding  box.center)]

  \matrix (m) [matrix of math nodes,row sep=3em,column sep=10em,minimum width=2em]
  {
     S \left( -2 \right)^{\oplus 3} & 0 \\
     S \left( -1 \right)^{\oplus 3} & S \left( 0 \right) \\};

  \path[->] (m-1-1) edge node [left] {$R$} (m-2-1);
  \path[->] (m-1-2) edge node [right] {$0$} (m-2-2);

  \node[draw=blue,inner sep=8pt, thick,rounded corners, fit= (m-1-2) (m-2-2) (m-2-2) (m-1-2) ]{};
  \node[draw=blue,inner sep=8pt, thick,rounded corners, fit= (m-1-1) (m-2-1) (m-2-1) (m-1-1) ]{};
  
\end{tikzpicture}
\]

Obviously the \emph{relations} of $M_\varphi$ are $0$. Therefore $M_\varphi$ is canonically isomorphic to the projective graded $S$-module $S ( 0 )$. $M_\psi$ however is not quite so simple -- its generators have to satisfy 3 relations. Let us work out these relations in detail. To this end we first identify generating sets:
\begin{itemize}
 \item $S \left( -2 \right)^{\oplus 3}$ is (freely) generated as $S$-module by $\mathcal{R} = \left\{ \left( 1,0,0 \right), \left( 0,1,0 \right), \left( 0,0,1 \right) 
      \right\} \equiv \left\{ r_1, r_2, r_3 \right\}$.
 \item $S \left( -1 \right)^{\oplus 3}$ is (freely) generated as $S$-module by $\mathcal{G} = \left\{ \left( 1,0,0 \right), \left( 0,1,0 \right), \left( 0,0,1 \right) 
      \right\} \equiv \left\{ g_1, g_2, g_3 \right\}$. 
\end{itemize}
Consequently we have
\[\label{eq:relation_1} \mathcal{R} \ni r_1 = \left( 1,0,0 \right) \mapsto \left( 1,0,0 \right) \cdot R = \left( 0, -x_3, x_2 \right) = - x_3 g_2 + x_2 g_3 \, . \]
Now let us think of the cokernel of $\psi$ in terms of classes. Then the representants of these classes are not unique, but can be chosen up to addition of elements of the form $-x_3 g_2 + x_2 g_3$. In fact, there are two more such redundancies, which follow from the images of $r_2$ and $r_3$. Namely
\[\label{eq:relation_2_and_3} r_2 \mapsto  x_3 g_1 - x_1 g_3, \qquad r_3 \mapsto -x_2 g_1 + x_1 g_2 \, . \]
In this sense there are the 3 relations \eqref{eq:relation_1} and \eqref{eq:relation_2_and_3} for the generators $g_1, g_2, g_3$ of $M_\psi$.

Let us now turn to the morphisms in $S\mathrm{\textnormal{-}fpgrmod}$. A morphism $M_\psi \to M_\varphi$ of \fp graded $S$-modules is a commutative diagram of the following form:
\[
\begin{tikzpicture}[baseline=(current  bounding  box.center)]

  \matrix (m) [matrix of math nodes,row sep=3em,column sep=10em,minimum width=2em]
  {
     S \left( -2 \right)^{\oplus 3} & 0 \\
     S \left( -1 \right)^{\oplus 3} & S \left( 0 \right) \\};

  \path[->] (m-1-1) edge node [left] {$R$} (m-2-1);
  \path[->] (m-1-1.east |- m-1-2) edge node [above] {$A$} (m-1-2);
  \path[->] (m-2-1.east |- m-2-2) edge node [above] {$B$} (m-2-2);
  \path[->] (m-1-2) edge node [right] {$0$} (m-2-2);

  \node[draw=blue,inner sep=8pt, thick,rounded corners, fit= (m-1-2) (m-2-2) (m-2-2) (m-1-2) ]{};
  \node[draw=blue,inner sep=8pt, thick,rounded corners, fit= (m-1-1) (m-2-1) (m-2-1) (m-1-1) ]{};
  
\end{tikzpicture}
\]
Recall that we are working with projective graded \emph{left} $S$-module homomorphisms. Therefore the commutativity is to say that $A \cdot 0 = R \cdot B$. We say that the above morphism is congruent to the zero morphism \footnote{Note that there is a slight difference between the notion of a \emph{classical category} and a \emph{CAP-category}. The latter comes equipped with the additional datum of congruence of morphisms. Upon factorisation of this congruence, a \emph{CAP-category} turns into the corresponding \emph{classical category}. For ease of computer implementations, the congruences are added as additional datum. See \cite{CAP, PosurDoktor, GutscheDoktor} for more information.} precisely if there exists a morphism of projective graded $S$-modules $\gamma \colon S ( -1 )^{\oplus 3} \xrightarrow{D} 0$ such that the following diagram commutes:
\[
\begin{tikzpicture}[baseline=(current bounding box.center)]

  \matrix (m) [matrix of math nodes,row sep=4em,column sep=10em,minimum width=2em]
  {
     S \left( -2 \right)^{\oplus 3} & 0 \\
     S \left( -1 \right)^{\oplus 3} & S \left( 0 \right) \\};

  \path[->] (m-1-1) edge node [left] {$R$} (m-2-1);
  \path[->] (m-1-1.east |- m-1-2) edge node [above] {$A$} (m-1-2);
  \path[dashed, ->] (m-2-1) edge node [above] {$D$} (m-1-2);
  \path[->] (m-2-1.east |- m-2-2) edge node [above] {$B$} (m-2-2);
  \path[->] (m-1-2) edge node [right] {$0$} (m-2-2);

  \node[draw=blue,inner sep=8pt, thick,rounded corners, fit= (m-1-2) (m-2-2) (m-2-2) (m-1-2) ]{};
  \node[draw=blue,inner sep=8pt, thick,rounded corners, fit= (m-1-1) (m-2-1) (m-2-1) (m-1-1) ]{};
  
\end{tikzpicture}
\]
Intuitively, the existence of such a morphism implies that all generators of $M_\psi$ can be thought of as relations of $M_\varphi$. A particular example of such a morphism of \fp graded $S$-modules is as follows:
\[ 
\begin{tikzpicture}[baseline=(current  bounding  box.center)]

  \matrix (m) [matrix of math nodes,row sep=4em,column sep=10em,minimum width=2em]
  {
     S \left( -2 \right)^{\oplus 3} & 0 \\
     S \left( -1 \right)^{\oplus 3} & S \left( 0 \right) \\};

  \path[->] (m-1-1) edge node [left] {$R$} (m-2-1);
  \path[->] (m-1-1.east |- m-1-2) edge node [above] {0} (m-1-2);
  \path[->] (m-2-1.east |- m-2-2) edge node [above] {$\left( \begin{array}{c} x_1 \\ x_2 \\ x_3 \end{array} \right)$} (m-2-2);
  \path[->] (m-1-2) edge node [right] {0} (m-2-2);

  \node[draw=blue,inner sep=8pt, thick,rounded corners, fit= (m-1-2) (m-2-2) (m-2-2) (m-1-2) ]{};
  \node[draw=blue,inner sep=8pt, thick,rounded corners, fit= (m-1-1) (m-2-1) (m-2-1) (m-1-1) ]{};
  
\end{tikzpicture}
\label{equ:IdealEmbedding}
\]
It is readily seen that this morphism is not congruent to the zero morphism. To gain some intuition, let us investigate this morphism in more detail. To this end recall the generating sets $\mathcal{R}$ and $\mathcal{G}$ of domain and codomain of $\psi$ as introduced above. In particular we can apply the displayed mapping of projective graded modules to the elements $g_i \in S ( -1 )^{\oplus 3}$. Thereby we find
\[ \mathcal{G} \ni g_1 = \left( 1,0,0 \right) \mapsto \left( 1,0,0 \right) \cdot \left( \begin{array}{c} x_1 \\ x_2 \\ x_3 \end{array} \right) = x_1 \, . \]
Similarly $g_2$ maps to $x_2$ and $g_3$ to $x_3$. We denote these images by $\mathcal{H} = \{ h_1, h_2, h_3 \}$. Note that the images of the relations map to zero, \eg $- x_3 g_2 + x_2 g_3$ turns into
\[ - x_3 h_2 + x_2 h_3 = - x_3 x_2 + x_2 x_3 = 0 \, . \]
In fact it turns out that the map $S ( -2 )^{\oplus 3} \stackrel{R}{\rightarrow} S ( -1 )^{\oplus 3}$ is the kernel embedding of the following morphism of projective graded $S$-modules
\[ S \left( -1 \right)^{\oplus 3} \xrightarrow{\left( \begin{array}{c} x_1 \\ x_2 \\ x_3 \end{array} \right)} S \left( 0 \right) \, . \]
For this very reason, the morphism in (\ref{equ:IdealEmbedding}) is a monomorphism of \fp graded $S$-modules. Consequently it describes the embedding of an ideal into $S ( 0 )$, and this very ideal is the one generated by $x_1, x_2, x_3$. Therefore $M_\psi$ is nothing but a presentation of the irrelevant ideal $B_\Sigma = \langle x_1, x_2, x_3 \rangle \subseteq S$ of $\mathbb{P}_{\mathbb{Q}}^2$ and the morphism in (\ref{equ:IdealEmbedding}) is its standard embedding $B_\Sigma \hookrightarrow S$. To emphasize this finding we extend the diagram to take the following shape:
\[
\begin{tikzpicture}[baseline=(current  bounding  box.center)]

  \matrix (m) [matrix of math nodes,row sep=4em,column sep=10em,minimum width=2em]
  {
     S \left( -2 \right)^{\oplus 3} & 0 \\
     S \left( -1 \right)^{\oplus 3} & S \left( 0 \right) \\
     B \left( \Sigma \right) & S \left( \mathbb{P}_{\mathbb{Q}}^2 \right) \\ };

  \path[->] (m-1-1) edge node [left] {$R$} (m-2-1);
  \path[->] (m-1-1.east |- m-1-2) edge node [above] {0} (m-1-2);
  \path[->] (m-2-1.east |- m-2-2) edge node [above] {$\left( \begin{array}{c} x_1 \\ x_2 \\ x_3 \end{array} \right)$} (m-2-2);
  \path[->] (m-1-2) edge node [right] {0} (m-2-2);
  \path[->, red] (m-2-1) edge node [right] {$\sim$} (m-3-1);
  \path[red, right hook-latex] (m-3-1.east |- m-3-2) edge node [above] {$\iota$} (m-3-2);
  \path[->, red] (m-2-2) edge node [right] {$\sim$} (m-3-2);
  
  \node[draw=blue,inner sep=8pt, thick,rounded corners, fit= (m-1-2) (m-2-2) (m-2-2) (m-1-2) ]{};
  \node[draw=blue,inner sep=8pt, thick,rounded corners, fit= (m-1-1) (m-2-1) (m-2-1) (m-1-1) ]{};
  
\end{tikzpicture}
\]
It is of crucial importance to distinguish the black and red arrows in this diagram. The black ones are morphisms of projective graded $S$-modules, whilst the red ones  mediate between \fp graded $S$-modules.

We have therefore arrived at the representation of an ideal - here the ideal with generators $B_\Sigma = \langle x_1, x_2, x_3 \rangle$  - via its relations in form of an \fp graded S-module. This is precisely the connection promised at the beginning of this section.

Let us use this opportunity to point out that for a given \fp graded $S$-module, there exist numerous presentations. For example the following is an isomorphism of two presentations that are both canonically isomorphic to the projective graded $S$-module $S ( 0 )$:
\[
\begin{tikzpicture}[baseline=(current  bounding  box.center)]

  \matrix (m) [matrix of math nodes,row sep=3em,column sep=10em,minimum width=2em]
  {
     0 & S \left( 0 \right) \\
     S \left( 0 \right) & S \left( 0 \right)^{\oplus 2} \\};

  \path[->] (m-1-1) edge node [left] {$0$} (m-2-1);
  \path[->] (m-1-1.east |- m-1-2) edge node [above] {$0$} (m-1-2);
  \path[->] (m-1-2) edge node [right] {$\left( 1, 0 \right)$} (m-2-2);
  \path[->] (m-2-1.east |- m-2-2) edge node [above] {$\left( 0, 1 \right)$} (m-2-2);

  \node[draw=blue,inner sep=8pt, thick,rounded corners, fit= (m-1-2) (m-2-2) (m-2-2) (m-1-2) ]{};
  \node[draw=blue,inner sep=8pt, thick,rounded corners, fit= (m-1-1) (m-2-1) (m-2-1) (m-1-1) ]{};
  
\end{tikzpicture}
\]

Note also that the rank of generators and the rank of relations are unrelated. We have already given examples of \fp graded $S$-modules for which the rank of the relations is either smaller than or identical to the rank of the generators. Finally consider the ideal $\langle x_1^2, x_1 x_2, x_1 x_3, x_2^2, x_2 x_3 \rangle$ and let 
\[ R^\prime = \left( \begin{array}{ccccc} 0 & -x_3 & x_2 & 0 & 0 \\ 0 & 0 & 0 & -x_3 & x_2 \\ -x_2 & x_1 & 0 & 0 & 0 \\ -x_3 & 0 & x_1 & 0 & 0 \\ 0 & -x_2 & 0 & x_1 & 0 \\ 0 & -x_3 & 0 & 0 & x_1 \end{array} \right) \, . \]
Then the standard embedding of this ideal takes the following form:
\[ 
\begin{tikzpicture}[baseline=(current  bounding  box.center)]

  \matrix (m) [matrix of math nodes,row sep=6em,column sep=10em,minimum width=2em]
  {
     S \left( -3 \right)^{\oplus 6} & 0 \\
     S \left( -2 \right)^{\oplus 5} & S \left( 0 \right) \\};

  \path[->] (m-1-1) edge node [left] {$R^\prime$} (m-2-1);
  \path[->] (m-1-1.east |- m-1-2) edge node [above] {0} (m-1-2);
  \path[->] (m-2-1.east |- m-2-2) edge node [above] {$\left( \begin{array}{c} x_1^2 \\ x_1 x_2 \\ x_1 x_3 \\ x_2^2 \\ x_2 x_3 \end{array} \right)$} (m-2-2);
  \path[->] (m-1-2) edge node [right] {0} (m-2-2);

  \node[draw=blue,inner sep=8pt, thick,rounded corners, fit= (m-1-2) (m-2-2) (m-2-2) (m-1-2) ]{};
  \node[draw=blue,inner sep=8pt, thick,rounded corners, fit= (m-1-1) (m-2-1) (m-2-1) (m-1-1) ]{};
  
\end{tikzpicture}
\]
Hence the 5 generators of this ideal satisfy 6 relations.

It can be proven that the category $S \mathrm{\textnormal{-}fpgrmod}$ is an Abelian monoidal category which is both strict and symmetric closed. See \cite{CAP, PosurDoktor, GutscheDoktor} for further details. This category being Abelian, kernel and cokernel exist for all morphisms in $S \mathrm{\textnormal{-}fpgrmod}$. Let us use this opportunity to display the kernel and cokernel of $\iota \colon B_\Sigma \hookrightarrow S$ in the following diagram:
\[
\begin{tikzpicture}[baseline=(current  bounding  box.center)]

  \matrix (m) [matrix of math nodes,row sep=5em,column sep=6em]
  {
     0 & S \left( -2 \right)^{\oplus 3} & 0 & S \left( -1 \right)^{\oplus 3} \\
     0 & S \left( -1 \right)^{\oplus 3} & S \left( 0 \right) & S \left( 0 \right) \\
     0 & B \left( \Sigma \right) & S \left( \mathbb{P}_{\mathbb{Q}}^2 \right) & S \left( \mathbb{P}_{\mathbb{Q}}^2 \right) / B_\Sigma \\ };

  \path[->] (m-1-1) edge node [above] {0} (m-1-2);
  \path[->] (m-1-1) edge node [left] {0} (m-2-1);
  \path[->] (m-2-1) edge node [above] {0} (m-2-2);
  \path[->] (m-1-2) edge node [left] {$R$} (m-2-2);
  \path[->] (m-1-2) edge node [above] {0} (m-1-3);
  \path[->] (m-2-2.east |- m-2-3) edge node [above] {$\left( \begin{array}{c} x_1 \\ x_2 \\ x_3 \end{array} \right)$} (m-2-3);
  \path[->] (m-1-3) edge node [right] {0} (m-2-3);
  \path[->] (m-1-3) edge node [above] {0} (m-1-4);
  \path[->] (m-2-3) edge node [above] {$\text{id}$} (m-2-4);
  \path[->] (m-1-4) edge node [right] {$\left( \begin{array}{c} x_1 \\ x_2 \\ x_3 \end{array} \right)$} (m-2-4);
  
  \path[->, red] (m-2-1) edge node [right] {$\sim$} (m-3-1);
  \path[red, right hook-latex] (m-3-1.east |- m-3-2) edge node [above] {$\text{ker} \left( \iota \right)$} (m-3-2);
  \path[->, red] (m-2-2) edge node [right] {$\sim$} (m-3-2);
  \path[red, right hook-latex] (m-3-2.east |- m-3-3) edge node [above] {$\iota$} (m-3-3);
  \path[->, red] (m-2-3) edge node [right] {$\sim$} (m-3-3);
  \path[red, ->>] (m-3-3.east |- m-3-4) edge node [above] {$\text{coker} \left( \iota \right)$} (m-3-4);
  \path[->, red] (m-2-4) edge node [right] {$\sim$} (m-3-4);
  
  \node[draw=blue,inner sep=8pt, thick,rounded corners, fit= (m-1-1) (m-2-1) (m-2-1) (m-1-1) ]{};
  \node[draw=blue,inner sep=8pt, thick,rounded corners, fit= (m-1-2) (m-2-2) (m-2-2) (m-1-2) ]{};
  \node[draw=blue,inner sep=8pt, thick,rounded corners, fit= (m-1-3) (m-2-3) (m-2-3) (m-1-3) ]{};
  \node[draw=blue,inner sep=20pt, thick,rounded corners, fit= (m-1-4) (m-2-4) (m-2-4) (m-1-4) ]{};

\end{tikzpicture}
\]
$S \mathrm{\textnormal{-}fpgrmod}$ being Abelian, a morphism is a monomorphism precisely if its kernel object is the zero object. The trivial box on the left hence reflects the fact that $\iota \colon B_\Sigma \hookrightarrow S ( 0 )$ is a monomorphism. The object boxed on the very right is a factor object. Such factor objects will later serve as models for structure sheaves of subloci of $X_\Sigma$. We will make use of this in \autoref{subsec:IdealSheavesFromModules}.

Quite generally, it is possible to associate to a given category $\mathcal{C}$ its category of morphisms. An implementation of this mechanism is provided in the \texttt{gap}-package \cite{CAPPresentationCategory}. Applying this technique to the category of projective graded $S$-modules, as introduced in this subsection, provides the category $S \mathrm{\textnormal{-}fpgrmod}$. Along this philosophy, this very category is implemented in the language of \texttt{CAP} \cite{CAP, PosurDoktor, GutscheDoktor} in the software package \cite{PresentationsByProjectiveGradedModules}.

Finally a word on the terminology of the \emph{projective} graded $S$-modules (of finite rank). These modules are canonically embedded into the category $S \mathrm{\textnormal{-}fpgrmod}$. In the latter they constitute the projective objects. Hence their name.

\subsection{A Briefing on Sheaves} \label{subsec:ABriefingOnSheaves}

Since our next goal is to understand the sheafifcation of the modules constructed so far, we now include a brief review of the definition of a sheaf. Experts may safely skip this standard exposition.
Let $( X, \tau )$ be a topological space. A presheaf $\mathcal{F}$ of Abelian groups on $( X, \tau )$ consists of Abelian groups $\mathcal{F} ( U )$ for all open $U \subseteq X$ and group homomorphisms -- termed \emph{restriction maps}
\[ \text{res}^U_V \colon \mathcal{F} \left( U \right) \to \mathcal{F} \left( V \right) \]
for all open $V \subseteq U$, such that the following conditions hold true:
\begin{enumerate}
 \item $\mathcal{F} \left( \emptyset \right) = 0$ - the trivial group,
 \item $\text{res}^U_U = \text{id}_{\mathcal{F} \left( U \right)}$ and
 \item for $W \subseteq V \subseteq U$ we have $\text{res}^U_W = \text{res}^V_W \circ \text{res}^U_V$.
\end{enumerate}
The elements of $\mathcal{F} ( U )$ are termed (local) sections of $\mathcal{F}$ over $U$. The restriction maps are typically denoted as $\text{res}^U_V ( s ) = s |_V$ for $s \in F ( U )$.

A presheaf $\mathcal{F}$ of Abelian groups on a topological space $( X, \tau )$ is a sheaf precisely if for every open $U \subseteq X$ and every open cover $\mathcal{U} = \{ U_i \}_{i \in I}$ of $U$ the following diagram is exact:
\begin{displaymath}
\xymatrix{\mathcal{F} \left( U \right) \ar[r] & \prod \limits_{i \in I}{\mathcal{F} \left( U_i \right)} \ar@<0.5ex>[r]\ar@<-0.5ex>[r] & \prod \limits_{i,j \in I}{\mathcal{F} \left( U_i \cap U_j \right)}}
\end{displaymath}
By this we mean the following:
\begin{itemize}
 \item The map $s \mapsto ( \left. s \right|_{U_i} )_{i \in I}$ is injective, \ie given that $s |_{U_i} = 0$ for all $i \in I$ it holds $s = 0$.
 \item The image of the first map is the kernel of the double arrow. So for a family $\left\{ s_i \in \mathcal{F} \left( U_i \right) \right\}_{i \in I}$ with
      $ s_i |_{U_i \cap U_j} = s_j |_{U_i \cap U_j}$, there exists $s \in \mathcal{F} ( U )$ with the property $ s |_{U_i} = s_i$ for all $i \in I$.
\end{itemize}

Next we pick a point $p \in X$. Given a sheaf $\mathcal{F}$ on the topological space $( X, \tau )$, we consider pairs $( U, s )_p$ where $U$ is an open neighbourhood of $p$ and $s \in \mathcal{F} ( U )$. For such pairs we can define an equivalence relation
\[ \left( U, s \right)_p \sim \left( V, t \right)_p \; \Leftrightarrow \; \exists \text{ an open set $W$ such that } p \in W \subseteq U \cap V \text{ and } \left. s \right|_W = \left. t \right|_W \, . \]
The equivalence class $[ U, s ]_p := \{ ( V, t )_p \; , \; ( V, t )_p \sim ( U, s )_p \}$ is the germ of the section $s$ at the point $p$. The union of all such germs
\[ \mathcal{F}_a := \left\{ \left. \left[ U, s \right]_p \; \right| \; p \in U \subseteq X \text{ open and } s \in \mathcal{F} \left( U \right) \right\} \]
is the stalk of the sheaf $\mathcal{F}$ in the point $p$. $\mathcal{F}_p$ is an Abelian group.

By replacing \emph{Abelian group} in the above lines by \emph{ring}, \emph{module}, \emph{algebra}, \dots one defines along the very same lines the concept of sheaves of \emph{rings}, \emph{modules}, \emph{algebras} \dots on a topological space.

Given a topological manifold $X$, for every open $U \subseteq X$ the set of continuous functions $f \colon U \to \mathbb{R}$ forms an Abelian group. Let $\text{res}^U_V$ be the ordinary restriction of functions. Then we obtain from this data a sheaf on the topological manifold $X$. This sheaf of continuous functions $\mathcal{O}_X$ is conventionally denoted by $\mathcal{O}_X$. By inspection, $\mathcal{O}_X$ is even a sheaf of rings, \ie $\{ f \colon U \to \mathbb{R} \; , \; f \text{ continuous} \}$ possesses the structure of a ring, and the ordinary restriction of functions respects this ring structure.

Likewise, on a smooth manifold the smooth (real valued) functions give rise to the sheaf of smooth (real valued) functions. On a complex manifold the sheaf of holomorphic functions can be considered. As continuous/ smooth/ holomorphic functions are characteristic to the structure of topological/ smooth/ complex manifolds the above sheaves are referred to as the \emph{structure sheaf of the manifold}.

All of the above structure sheaves are sheaves of rings. This observation leads to the concept of a \emph{ringed space}. A ringed space is a pair $( X, \mathcal{O}_X )$ consisting of a topological space $X$ and a sheaf of rings $\mathcal{O}_X$ on $X$. On a ringed space it is possible to consider sheaves of $\mathcal{O}_X$-modules. Such a sheaf $\mathcal{F}$ on $X$ assigns to every open $U \subseteq X$ an $\mathcal{O}_X ( U )$-module $\mathcal{F} ( U )$. In addition the restriction maps of $\mathcal{F}$ respect this module structure. Coherent sheaves are special such sheaves of $\mathcal{O}_X$-modules, as we will point out in \autoref{subsec:CoherentSheavesOnAbstractVarieties}.

\subsection{Localisation of Rings} \label{subsec:LocalisationOfRings}

Sheafification of modules on affine varieties (and eventually also toric varieties) makes use of \emph{localisation of rings}. Therefore, let us use this subsection to recall the basics behind this procedure. 

We consider a commutative unitial ring $R$ and a multiplicatively closed subset $1 \in S \subseteq R$. We now construct a new ring $R_S$ from this data. To this end define the following (equivalence) relation on $R \times S$
\[ \left( r_1, s_1 \right) \sim \left( r_2, s_2 \right) \; \Leftrightarrow \; \exists t \in S \colon t \cdot \left( r_1 s_2 - r_2 s_1 \right) = 0 \]
and use it to consider the equivalence classes
\[ \frac{r_1}{s_1} := \left[ \left( r_1, s_1 \right) \right] := \left\{ \left. \left( r_2, s_2 \right) \in R \times S \; \right| \; \left( r_1, s_1 \right) \sim \left( r_2, s_2 \right) \right\} \]
We denote the collection of all of these equivalence classes by $S^{-1} R = R_S$. It is readily verified that the binary compositions
\begin{align}
\begin{split}
+ \colon R_S \times R_S \to R_S \; , \; & \left( \frac{r_1}{s_1}, \frac{r_2}{s_2} \right) \mapsto \frac{r_1 s_2 + r_2 s_1}{s_1 s_2}\, , \\
\cdot \colon R_S \times R_S \to R_S \; , \; & \left( \frac{r_1}{s_1}, \frac{r_2}{s_2} \right) \mapsto \frac{r_1 r_2}{s_1 s_2}
\end{split}
\end{align}
turn this set $R_S$ into a ring. This ring $R_S$ is termed the \emph{localisation of $R$ at $S$}.

Oftentimes one localises rings at prime ideals $\mathfrak{p} \subseteq R$. Recall that a prime ideal $\mathfrak{p} \subseteq R$ in a commutative ring is a proper ideal (\ie $\mathfrak{p} \neq R$) such that for all $a,b \in R$ with $a b \in \mathfrak{p}$ it holds $a \in \mathfrak{p}$ or $b \in \mathfrak{p}$. In particular $1 \notin \mathfrak{p}$ for otherwise $\mathfrak{p} = R$ in contradiction to $\mathfrak{p}$ being proper. However, for the localisation $R_S$ we assumed $1 \in S$! Therefore, by convention, localisation at a prime ideal $\mathfrak{p} \subseteq R$ means to localise at the multiplicatively closed set $R - \mathfrak{p}$, \ie
\[ R_{\mathfrak{p}} := \left( R - \mathfrak{p} \right)^{-1} R = R_{R - \mathfrak{p}} \, . \]

Another common type of localisation is at an element $0 \neq f \in R$. By this we mean to form the set $\{ 1, f, f^2, f^3, \dots \}$ and then to localise at this set $S$. The localisation at this set $S$, induced from $0 \neq f \in R$, is denoted by $R_f$.

Yet another common situation is to start with a graded ring $R$. Given a multiplicately closed set $1 \in S$ of homogeneous elements, one can perform the so-called \emph{homogeneous localisation}. To this end we first define the degree of the equivalence class $\frac{r_1}{r_2}$ as
\[ \text{deg} \left( \frac{r_1}{r_2}\right) = \text{deg} \left( r_1 \right) - \text{deg} \left( r_2 \right) \, . \]
Thereby one realises that the localisation $R_S$ is a graded ring. The homogenous localisation $R_{(S)}$ is now defined by
\[ R_{(S)} := \left( R_S \right)_{0} \, , \]
\ie $R_{(S)}$ consists of all elements of $R_S$ that have vanishing degree. This is easily generalised to homogeneous elements $0 \neq f \in R$ and homogeneous prime ideals $\mathfrak{p} \subseteq R$.

\subsection{Sheafification of Modules on Affine Varieties} \label{subsec:SheafifcationOnAffineVarieties}

The general idea of sheafification on affine varieties is very simple: Given an affine variety $X$ with coordinate ring $R$, we define a sheaf $\tilde{M}$ associated to an $R$-module $M$ by stating its (local) sections on (special) open subsets of $X$, and subsequently check that these assignments glue together to form a sheaf on $X$. The open sets in question are of the form $D(f) = X - V(\langle f \rangle)$ with $f \in R - 0$, \ie the total space $X$ minus the vanishing locus of $f$. On these open set, the sheaf $\tilde{M}$ has the (local) sections given by
\[ \tilde{M} \left( D \left( f \right) \right) = M \otimes_R R_f \, , \]
where $R_f$ denotes the localisation of $R$ at $f$, as introduced in \autoref{subsec:LocalisationOfRings}. Intuitively, $R_f$ consists of all (rational) functions on $D(f)$. In the remainder of this section we make these statements more precise. The reader not interested in all the technical details may safely jump directly to \autoref{app_sheafification}.

Let us first  recall the notion of an affine variety with a view towards algebraic geometry. To this end let $R$ be a commutative unitial ring. Recall that an ideal $\mathfrak{m} \subseteq R$ is termed a maximal ideal precisely if for every proper ideal $\mathfrak{a} \subseteq R$ with $\mathfrak{m} \subseteq \mathfrak{a}$ it holds $\mathfrak{m} = \mathfrak{a}$. 

Whilst every maximal ideal is a prime ideal, the converse is not quite true. To see this consider the ring $k [ x ]$ where $k$ is an algebraically closed field. For every $a \in k$ the ideal
\[ \mathfrak{m}_{a} := \left\langle x - a \right\rangle \subseteq k \left[ x \right] \]
is a maximal ideal, and -- provided that $k$ is algebraically closed -- all maximal ideals of $k [ x ]$ are of this form. However, since $k$ is free of zero divisors \footnote{This means that for every $a,b \in R - 0$ it holds $a b \neq 0$.} the trivial ideal $\langle 0 \rangle$ is a prime ideal as well. Clearly this ideal is not maximal since $\langle 0 \rangle \subsetneq \langle x \rangle$!

Let us construct a topological space from $R$. A scheme-theoretic approach would consider
\[ \text{Spec} \left( R \right) = \left\{ \left. \mathfrak{p} \subseteq R \; \right| \; \mathfrak{p} \text{ a prime ideal } \right\} \]
and equip it with the Zariski topology. As just pointed out, for an algebraically closed field $k$ this means 
\[ \text{Spec} \left( R \right) = \left\{ \left. \left\langle x - a \right\rangle \; \right| \; a \in k \right\} \cup \left\{ \left\langle 0 \right\rangle \right\} \cong k \cup \left\{ \left\langle 0 \right\rangle \right\} \, . \]
So besides containing all `points' of $k$, this affine scheme contains also the so-called \emph{generic point} $\langle 0 \rangle$. Such points distinguish the scheme-theoretic approach from a variety-theoretic approach. Here we present our results in the language of (toric) varieties over $\mathbb{Q}$. However it is possible to formulate these findings also in the language of (toric) schemes \cite{RohrerDissertation, 2011arXiv1107.2483R, 2012arXiv1212.3956R, 2011arXiv1107.2713R}.

In following \cite{cox2011toric} we therefore consider the set of maximal ideals
\[ \text{Specm} \left( R \right) = \left\{ \left. \mathfrak{p} \subseteq R \; \right| \; \mathfrak{p} \text{ a maximal ideal} \right\} \, . \]
The Zariski topology on $\text{Specm} ( R )$ is defined by saying that for every ideal $\mathfrak{a} \subseteq R$ the following set is closed
\[ V \left( \mathfrak{a} \right) = \left\{ \left. \mathfrak{p} \in \text{Specm} \left( R \right) \; \right| \; \mathfrak{a} \subseteq \mathfrak{p} \right\} \subseteq \text{Specm} \left( R \right) \, . \]

As an example let us consider the ring $\mathbb{C} [ x ]$. Since $\mathbb{C}$ is algebraically closed we have an isomorphism of sets $\text{Specm} ( \mathbb{C} [ x ] ) \cong \mathbb{C}$. Next recall that every $f \in \mathbb{C} [ x ]$ has finitely many zeros. In addition, by the Hilbert Nullstellensatz, every ideal $\mathfrak{a} \subseteq \mathbb{C} [ x ]$ is finitely generated. Consequently every (Zariski) closed subset of $\text{Specm} ( \mathbb{C} [ x ] )$ is a finite subset. Or conversely, an open subset of $\text{Specm} ( \mathbb{C} [ x ] )$ consists of the entire space $\text{Specm} ( \mathbb{C} [ x ] )$ up to finitely many exceptions. Therefore, for any two $\mathfrak{p}, \mathfrak{q} \in \text{Specm} ( \mathbb{C} [ x ] )$, any open neighbourhood $U$ of $\mathfrak{p}$ and any open neighbourhood $V$ of $\mathfrak{q}$ it holds $U \cap V \neq \emptyset$. This shows that the Zariski topology need not be Hausdorff. For this reason it defies intuition from Euclidean or complex geometry.

For every $f \in R - 0 $ we can consider the open set
\[ D \left( f \right) := \text{Specm} \left( R \right) - V \left( \left\langle f \right\rangle \right) = \left\{ \left. \mathfrak{p} \in \text{Specm} \left( A \right) \; \right| \; f \notin \mathfrak{p} \right\} \, . \]
$\mathcal{Z} := \{ D ( f ) \}_{f \in R - 0}$ is a basis of the Zariski topology. Moreover $D ( f ) \cong \text{Specm} ( R_f )$ where $R_f$ denotes the localisation of the ring $R$ at $f \in R - 0$.

All that said, we finally turn to the sheafification of an $R$-module $M$. This means that we wish to turn this module $M$ into a sheaf $\tilde{M}$ on the affine variety $\text{Specm} ( R )$ \cite{hartshorne1977algebraic}. We describe this sheaf by merely stating its local sections over the basis of topology $\mathcal{Z}$. Namely it holds
\[ \tilde{M} \left( D \left( f \right) \right) = M \otimes_R R_f \, . \]
Of course one has to check that these local assignments satisfy all conditions of a sheaf as stated in \autoref{subsec:ABriefingOnSheaves}. For further details see \cite{hartshorne1977algebraic}.

In particular the ring $R$ is an $R$-module. It sheafifies to form the structure sheaf $\mathcal{O}_{\text{Specm} ( R )} \cong \tilde{R}$. Therefore, since $M \otimes_R R_f$ is an $R_f$-module, the sheaf $\tilde{M}$ is a sheaf of $\mathcal{O}_{\text{Specm} ( R )}$-modules.

\subsection{Coherent Sheaves on (Abstract) Varieties} \label{subsec:CoherentSheavesOnAbstractVarieties}

Let $X$ be a topological space and $\mathcal{O}_X$ a sheaf of rings on $X$. The pair $( X, \mathcal{O}_X )$ is termed a \emph{locally ringed space} if for every $p \in X$ the stalk $\mathcal{O}_{X,p}$ is a local ring.\footnote{This means that the ring $\mathcal{O}_{X,p}$ has a unique maximal ideal.} An abstract variety is a locally ringed space $( X, \mathcal{O}_X )$ such that for every $p \in X$ there exists an open neighbourhood $p \in U \subseteq X$ such that $( U, \left. \mathcal{O}_X \right|_U )$ is isomorphic (as locally ringed space) to $( \text{Specm} ( R ), \tilde{R})$ for a suitable commutative unitial ring $R$.

For a sheaf $\mathcal{F}$ of $\mathcal{O}_X$-modules on a variety $( X, \mathcal{O}_X )$ we define the following notions:
\begin{itemize}
 \item For an open subset $U \subseteq X$, the restriction $\mathcal{F} |_U$ of $\mathcal{F}$ to $U$ is the sheaf of $\mathcal{O}_U$-modules 
      obtained from $ \mathcal{F} |_U ( V ) = \mathcal{F} ( V )$ for every $V \subseteq U$ open.
 \item $\mathcal{F}$ is \emph{quasicoherent} precisely if $X$ admits an open affine cover \footnote{A variety $( X, \mathcal{O}_X )$ may admit various open affine 
      covers!} $\mathcal{U} = \{ U_\alpha \}_{\alpha \in I}$, \ie
      \[ \left( U_\alpha, \left. \mathcal{O}_X \right|_{U_\alpha} \right) \cong \left( \text{Specm} \left( R_{\alpha} \right), \tilde{R_\alpha} \right) \, , \]
      such that for every $\alpha \in I$ there exists an $R_\alpha$-module $M_\alpha$ with the property $\tilde{M_\alpha} \cong \mathcal{F} |_{U_\alpha}$.
 \item $\mathcal{F}$ is \emph{coherent} if in addition for every $\alpha \in I$ the module $M_\alpha$ is finitely presented.
\end{itemize}

\subsection{Sheafification of F.P.\ Graded \texorpdfstring{$\mathbf{S}$}{S}-Modules on Toric Varieties} \label{app_sheafification}

We will assume that $X_\Sigma$ is a toric variety over $\mathbb{Q}$ without torus factor. Let $S = \mathbb{Q} [ x_1, \dots, x_m ]$ be its Cox ring. As we already mentioned, this ring is graded by $\text{Cl} ( X_\Sigma )$. Very much along the same lines as for affine varieties, an \fp graded $S$-module $M$ can be turned into a coherent sheaf $\tilde{M}$ on $X_\Sigma$. Here we briefly review this sheafification. Further details can be found in \cite{cox2011toric}.

Recall that $X_\Sigma$ is defined by the combinatorics of a fan $\Sigma$. There are precisely $m$ rays in the fan $\Sigma$. We denote the associated ray generators by $\rho_1, \dots \rho_m$. To each of these ray generators there is precisely one indeterminate $x_i \in S$ associated. We may assume that $x_1 \leftrightarrow \rho_1$, $x_2 \leftrightarrow \rho_2$ \dots. Given a cone $\sigma \in \Sigma$ we can now form the monomial
\[ x^{\hat{\sigma}} := \prod_{\rho \notin \sigma \left( 1 \right)}{x_\rho} \in S \, . \]
The affine variety associated to this cone $\sigma$ is given by $U_\sigma \cong \text{Specm} ( \mathbb{Q} [ \sigma^\vee \cap M ] )$.\footnote{In this expression $M$ is the character lattice of the toric variety $X_\Sigma$. It is not to be confused with the \fp graded $S$-module $M$ which we intend to sheafify.} In addition there is an isomorphism of graded rings $\pi_\sigma^\ast \colon \mathbb{Q} [ \sigma^\vee \cap M ] \xrightarrow{\sim} ( S_{\hat{\sigma}} )_{0}$. Consequently we can also understand the affine variety associated to the cone $\sigma$ as $U_\sigma = \text{Specm} ( ( S_{\hat{\sigma}} )_{0} )$. Of course these affine varieties have to glue together in a meaningful fashion. The key observation is that for a face $\tau = \sigma \cap m^\bot$ of $\sigma$ it holds $( S_{x^{\hat{\tau}}} )_0 = ( ( S_{x^{\hat{\sigma}}} )_0 )_{\pi_\sigma^\ast ( \chi^m )}$. The commutativity of the following diagram then establishes the desired gluing:
\begin{displaymath}
\begin{split}
\label{equ:Gluing}
\xymatrix{ \left( S_{x^{\hat{\sigma}}} \right)_0 \ar[r] \ar[d] & \left( \left( S_{x^{\hat{\sigma}}} \right)_0 \right)_{\pi_\sigma^\ast \left( \chi^m \right)} \ar[d] \\
\mathbb{Q} \left[ \sigma^\vee \cap M \right] \ar[r] & \mathbb{Q} \left[ \tau^\vee \cap M \right]_{\chi^m}.
}
\end{split}
\end{displaymath}
With this background, we can now turn an \fp graded $S$-module $M$ into a coherent sheaf $\tilde{M}$ on $X_\Sigma$. This works as follows:
\begin{itemize}
 \item The localisation $M_{x^{\hat{\sigma}}}$ turns out to be an \fp graded $S_{x^{\hat{\sigma}}}$-module. Therefore $( M_{x^{\hat{\sigma}}} )_0$ is an \fp graded 
      $( S_{x^{\hat{\sigma}}} )_0$-module. Since $U_\sigma = \text{Specm} ( ( S_{x^{\hat{\sigma}}} )_0 )$, we can now follow the steps outlined in \autoref{subsec:SheafifcationOnAffineVarieties}, to turn $( M_{x^{\hat{\sigma}}} )_0$ into the coherent sheaf $\tilde{ ( M_{x^{\hat{\sigma}}} )_0}$ on $U_\sigma$.
 \item It is guaranteed by (\ref{equ:Gluing}) that these sheaves $\tilde{ ( M_{x^{\hat{\sigma}}} )_0}$ on $U_\sigma$ glue to form a sheaf on $X_\Sigma$.
\end{itemize}

Let us illustrate these abstract words in an example. We look at the toric variety $\mathbb{P}^1_{\mathbb{Q}}$ with Cox ring $S = \mathbb{Q} [ x_1, x_2 ]$. Its fan $\Sigma \subseteq \mathbb{R}$ consists of the trivial cone and the following two maximal cones:
\begin{center}
\begin{tabular}{|c||c|c|c|c|}
\hline
cone & generator & indeterminate & affine variety & $x^{\hat{\sigma}}$ \\
\hline \hline
$\sigma_1$ & $1$ & $x_1$ & $U_1$ & $x_2$ \\
$\sigma_2$ & $-1$ & $x_2$ & $U_2$ & $x_1$ \\
\hline
\end{tabular}
\end{center}
The affine open cover $\mathcal{U} = \{ U_1, U_2 \}$ is given by $U_i \cong \text{Specm} ( ( S_{\hat{\sigma}} )_0 )$. As sets we have
\[ S ( n )_{(x_i)} \equiv \left( S \left( n \right)_{x_i} \right)_0 = \left\{ \left. \frac{f}{x_i^k} \; \right| \; f \in S \left( 0 \right) \text{ homogeneous } \; , \; \text{deg} \left( f \right) = n + k \right\} \, . \]
Therefore
\[ U_1 \cong \text{Specm} \left( \mathbb{Q} \left[ \frac{x_1}{x_2} \right] \right), \qquad U_2 \cong \text{Specm} \left( \mathbb{Q} \left[ \frac{x_2}{x_1} \right] \right) \, . \]
Upon homogenisation of $X_\Sigma$ we thus understand $U_1$ as the locus $x_2 \neq 0$, and similarly $U_2$ as $x_1 \neq 0$.

Now consider the \fp graded $S$-Module $M$ encoded by the morphism $\varphi \colon 0 \xrightarrow{0} S$ of projective graded $S$-modules of finite rank. On the affine patches $U_1$, $U_2$ we assign to this module $M$ the following coherent sheaves:
\[ \left( M_{x^{\hat{\sigma_1}}} \right)_0 = \tilde{\mathbb{Q} \left[ \frac{x_1}{x_2}\right]}, \qquad \left( M_{x^{\hat{\sigma_2}}} \right)_0 = \tilde{\mathbb{Q} \left[ \frac{x_2}{x_1}\right]} \, . \]
Note that we can also pick the trivial cone and localise at its monomial $x^{\hat{0}} = x_0 x_1$. This corresponds to the restriction to $U_1 \cap U_2$. On this intersection we therefore associate to the module $M$ the coherent sheaf $( M_{x^{\hat{0}}} )_0 = \tilde{\mathbb{Q} [ \frac{x_1}{x_2}, \frac{x_2}{x_1} ]}$. Overall we obtain the following commutative diagram of coherent sheaves:
\begin{displaymath}
\begin{split}
\label{equ:SheafifyS}
 \xymatrix{ & \tilde{M} \ar[dl]_-{\left. \right|_{U_1}} \ar[dr]^-{\left. \right|_{U_2}} & \\ \tilde{\mathbb{Q} \left[ \frac{x_1}{x_2}\right]} \ar[dr]_-{\left. \right|_{U_1 \cap U_2}} & & \tilde{\mathbb{Q} \left[ \frac{x_2}{x_1}\right]} \ar[dl]^-{\left. \right|_{U_1 \cap U_2}} \\ &  \tilde{\mathbb{Q} \left[ \frac{x_1}{x_2}, \frac{x_2}{x_1} \right]} }
\end{split}
\end{displaymath}
In this diagram we use the symbol $|_{U}$ to denote the restriction of a sheaf to an open subset $U \subseteq X_\Sigma$. Recall that we have described this process (briefly) in \autoref{subsec:CoherentSheavesOnAbstractVarieties}.

Since the module $M$ described by the cokernel of $\varphi \colon 0 \xrightarrow{0} S$ is isomorphic to $S$ itself, this gives by definition the structure sheaf $\tilde M = \mathcal{O}_{\mathbb{P}^1_\mathbb{Q}}$. 
The global sections of the coherent sheaves $\tilde{\mathbb{Q} [ \frac{x_1}{x_2}]}$, $\tilde{\mathbb{Q} [ \frac{x_2}{x_1}]}$ and $\tilde{\mathbb{Q} [ \frac{x_1}{x_2}, \frac{x_2}{x_1} ]}$ are given by $\mathbb{Q} [ \frac{x_1}{x_2}]$, $\mathbb{Q} [ \frac{x_2}{x_1}]$ and $\mathbb{Q} [ \frac{x_1}{x_2}, \frac{x_2}{x_1} ]$. Hence, if we set $t \equiv \frac{x_1}{x_2}$, we obtain from (\ref{equ:SheafifyS}) the commutative diagram of sections:
\begin{displaymath}
\begin{split}
 \xymatrix{ & H^0 \left( \mathbb{P}^1_{\mathbb{Q}}, \tilde{M} \right) \ar@{_{(}->}[dl]_-{\text{res}^{\mathbb{P}^1_{\mathbb{Q}}}_{U_1}} 
 \ar@{^{(}->}[dr]^-{\text{res}^{\mathbb{P}^1_{\mathbb{Q}}}_{U_2}} & \\ \mathbb{Q} \left[ t \right] \ar@{^{(}->}[dr]_-{\text{res}^{U_1}_{U_1 \cap U_2}} & & \mathbb{Q} \left[ t^{-1} \right] \ar@{_{(}->}[dl]^-{\text{res}^{U_2}_{U_1 \cap U_2}} \\ &  \mathbb{Q} \left[ t, t^{-1} \right] }
\end{split}
\end{displaymath}
In this diagram, the maps $\text{res}^{U}_V$ are the restriction maps of the sheaf $\tilde{M}$. For example $\text{res}^{U_1}_{U_1 \cap U_2}$ is the canonical inclusion map $\mathbb{Q} [ t ] \hookrightarrow \mathbb{Q} [ t, t^{-1} ]$. A global section of $\tilde{M}$, \ie $s \in H^0 ( \mathbb{P}^1_{\mathbb{Q}}, \tilde{M} )$, can be regarded as a pair
\[ \left( \alpha, \beta \right) \in \tilde{M} \left( U_1 \right) \times \tilde{M} \left( U_2 \right) \cong \mathbb{Q} \left[ t \right] \times \mathbb{Q} \left[ t^{-1} \right] \]
with the property $\text{res}^{U_1}_{U_1 \cap U_2} ( \alpha ) = \text{res}^{U_2}_{U_1 \cap U_2} ( \beta )$. It is readily verified that the only such pairs are the diagonal elements $( \alpha, \alpha ) \in \mathbb{Q} \times \mathbb{Q}$. Consequently $H^0 ( \mathbb{P}^1_{\mathbb{Q}}, \tilde{M} ) \cong \mathbb{Q}$, in agreement with the expected result for $\tilde{M} \cong \mathcal{O}_{\mathbb{P}^1_{\mathbb{Q}}}$.

Similarly the twist $S ( n )$ -- considered as \fp graded $S$-modules with trivial relation module -- sheafifies to give the twisted structure sheaf $\mathcal{O}_{\mathbb{P}^1_{\mathbb{Q}}} ( n )$. Let us exemplify this statement. For $n = 1$, the commutative diagram of (local) sections takes the following shape:
\begin{displaymath}
\begin{split}
 \xymatrix{ & H^0 \left( \mathbb{P}^1_{\mathbb{Q}}, \tilde{S \left( 1 \right)} \right) \ar@{_{(}->}[ddl]_-{\text{res}^{\mathbb{P}^1_{\mathbb{Q}}}_{U_1}} 
 \ar@{^{(}->}[ddr]^-{\text{res}^{\mathbb{P}^1_{\mathbb{Q}}}_{U_2}} & \\ \\ 
 x_2 \cdot \mathbb{Q} \left[ \frac{x_1}{x_2}\right] \ar@{^{(}->}[ddr]_-{\text{res}^{U_1}_{U_1 \cap U_2}} & & x_1 \cdot \mathbb{Q} \left[ \frac{x_2}{x_1}\right] \ar@{_{(}->}[ddl]^-{\text{res}^{U_2}_{U_1 \cap U_2}} \\ \\ 
 & \text{span}_{\mathbb{Q}} \left\{ x_1, \frac{x_1^2}{x_2}, \frac{x_1^3}{x_2^2}, \dots, x_2, \frac{x_2^2}{x_1}, \frac{x_2^3}{x_1^2}, \dots \right\} }
\end{split}
\end{displaymath}
From this we see $H^0 ( \mathbb{P}^1_{\mathbb{Q}}, \tilde{ S ( 1 )} ) \cong \langle x_1, x_2 \rangle_{\mathbb{Q}}$. Similarly we find for $n = -1$ the commutative diagram:
\begin{displaymath}
\begin{split}
 \xymatrix{ & H^0 \left( \mathbb{P}^1_{\mathbb{Q}}, \tilde{S \left( -1 \right)} \right) \ar@{_{(}->}[ddl]_-{\text{res}^{\mathbb{P}^1_{\mathbb{Q}}}_{U_1}} 
 \ar@{^{(}->}[ddr]^-{\text{res}^{\mathbb{P}^1_{\mathbb{Q}}}_{U_2}} & \\ \\ 
 \frac{1}{x_2} \cdot \mathbb{Q} \left[ \frac{x_1}{x_2}\right] \ar@{^{(}->}[ddr]_-{\text{res}^{U_1}_{U_1 \cap U_2}} & & \frac{1}{x_1} \cdot \mathbb{Q} \left[ \frac{x_2}{x_1}\right] \ar@{_{(}->}[ddl]^-{\text{res}^{U_2}_{U_1 \cap U_2}} \\ \\ 
 & \text{span}_{\mathbb{Q}} \left\{ \frac{1}{x_1}, \frac{x_2}{x_1^2}, \frac{x_2^2}{x_1^3}, \dots, \frac{1}{x_2}, \frac{x_1}{x_2^2}, \frac{x_1^2}{x_2^3}, \dots \right\} }
\end{split}
\end{displaymath}
This shows $H^0 ( \mathbb{P}^1_{\mathbb{Q}}, \tilde{S ( -1 )} ) \cong \{ 0 \}$.

\subsection{More Properties of the Sheafification Functor}

What we have described thus far is the functor
\[ \widetilde{\phantom{m}} \colon S \mathrm{\textnormal{-}fpgrmod} \to \mathfrak{Coh} X_\Sigma \label{redundant_description_II} \]
which turns an \fp graded $S$-module into a coherent sheaf on the toric variety $X_\Sigma$. It is this functor which enables us to model coherent sheaves in a way suitable for computer manipulations  as objects of the category $S \mathrm{\textnormal{-}fpgrmod}$. An important question is whether this description is unique. As it turns out, it is not.

Assume for a moment that $X_\Sigma$ were a smooth toric variety with irrelevant ideal $B ( \Sigma ) \subseteq S$. Then an \fp graded $S$-module $M$ satisifes
by proposition 5.3.10 in \cite{cox2011toric} \footnote{Proposition 5.3.10 of \cite{cox2011toric} fails for $\mathbb{P}_{\mathbb{C}} ( 2,3,1 )$, \ie is not applicable if $X_\Sigma$ is not smooth. See \cite{cox2011toric} for generalisations of this proposition.}
\[ \tilde{M} = 0 \quad \Leftrightarrow \quad B \left( \Sigma \right)^l M = 0 \text{ for } l \gg 0 \, . \]
Consequently, even in the smooth case the functor in \autoref{redundant_description_II} is no equivalence of categories. Let $S \mathrm{\textnormal{-}fpgrmod}^0$ be the thick subcategory of $S \mathrm{\textnormal{-}fpgrmod}$ of \fp graded $S$-modules which are supported on $V ( B_\Sigma )$. Then it can be proven that
\[ S \mathrm{\textnormal{-}fpgrmod} / S \mathrm{\textnormal{-}fpgrmod}^0 \stackrel{\sim}{\rightarrow} \mathfrak{Coh} X_\Sigma \, . \] 
is an equivalence of categories, which therefore allows for an ideal parametrisation of $\mathfrak{Coh} ( X_\Sigma )$ \cite{2012arXiv1210.1425B, 2011arXiv1110.0323P}.

In summary, any coherent sheaf $\mathcal{F} \in \mathfrak{Coh} X_\Sigma$ can be modelled by $M \in S \mathrm{\textnormal{-}fpgrmod}$ - however not uniquely, \ie in general there exist many non-isomorphic modules $M$ with the property $\tilde{M} \cong \mathcal{F}$. 

We give an example on $\mathbb{P}^2_{\mathbb{Q}}$. In \autoref{subsec:FPGradedSModules} we discussed the \fp graded $S$-module $B_\Sigma$. Also recall also that $S ( 0 )$ can canonically be considered a \fp graded $S$-module and that $B_\Sigma \not \cong S ( 0 )$ (as \fp graded $S$-modules). In addition we pointed out in \autoref{app_sheafification} that $\tilde{S ( 0 )} \cong \mathcal{O}_{\mathbb{P}^2_{\mathbb{Q}}}$. As it turns out $\tilde{S} \cong \tilde{B_\Sigma} \cong \mathcal{O}_{\mathbb{P}^2_{\mathbb{Q}}}$. 

In \autoref{subsec:SheafCohomologyFromFPGradedSModules}, we will describe our strategy to extract from an \fp graded $S$-module $M$ the sheaf cohomologies of $\tilde{M}$. In so doing we will point out why $S$ is a perfect model for the structure sheaf of $\mathbb{P}^2_{\mathbb{Q}}$, whilst $B_\Sigma$ is not quite as good a choice.

\subsection{Line Bundles from F.P.\ Graded \texorpdfstring{$\mathbf{S}$}{S}-Modules} \label{subsec:IdealSheavesFromModules}

Before we discuss the computation of sheaf cohomologies from \fp graded modules, we first describe how we actually obtain modules that sheafify to the coherent sheaves in question. The major task from the main text is as follows:

\ebox{Be $X_\Sigma$ a normal toric variety that is either smooth, complete or simplicial, projective. Its Cox ring be $S$. Given a subvariety $C = V ( g_1, \dots, g_k ) \subseteq X_\Sigma$ and $D = V ( f_1, \dots, f_n ) \in \text{Div} ( C )$ -- both not necessarily complete intersections -- we want to construct an \fp graded $S$-module $M$ such that $\tilde{M} \in \mathfrak{Coh} \left( X_\Sigma \right)$ is supported only over $C$ and satisfies $\tilde{M} |_{C} \cong \mathcal{O}_{C} ( \pm D )$.}

To find a module $M$ such that $\tilde{M}$ is supported only over $C$ and satisfies $\tilde{M} |_C \cong \mathcal{O}_{C} ( - D )$ we proceed as follows:
\begin{enumerate}
 \item The polynomials $f_i$ and $g_j$ are homogeneous. In consequence the ring $S ( C ) := S / \langle g_1, \dots, g_k \rangle$ is graded by $\text{Cl} ( X_\Sigma )$ 
      and the canonical projection map $\pi \colon S \twoheadrightarrow S ( C )$ allows us to consider the matrix
      \[ M = \left( \pi \left( f_1 \right), \dots, \pi \left( f_n \right) \right) \in M \left( 1 \times n, S \left( C \right) \right) \, . \]
      Let $\text{ker} ( M )$ denote the kernel matrix of $M$. Then the following diagram describes a monomorphism of \fp graded $S ( C )$-modules.\footnote{Note that $I, J \subseteq \text{Cl} ( X_\Sigma )$ are \emph{finite} indexing sets.}
      \begin{center}
      \begin{tikzpicture}

         \matrix (m) [matrix of math nodes,row sep=3em,column sep=9em,minimum width=2em]
         {
         \bigoplus_{j \in J}{S \left( C \right) \left( j \right)} & 0 \\
         \bigoplus_{i \in I}{S \left( C \right) \left( i \right)} & S \left( C \right) \\
         A_C & S \left( C \right) \\ };

         \path[->] (m-1-1) edge node [left] {$\text{ker} \left( M \right)$} (m-2-1);
         \path[->] (m-1-1.east |- m-1-2) edge node [above] {0} (m-1-2);
         \path[->] (m-2-1.east |- m-2-2) edge node [above] {$M$} (m-2-2);
         \path[->] (m-1-2) edge node [right] {0} (m-2-2);
         \path[->, red] (m-2-1) edge node [right] {$\sim$} (m-3-1);
         \path[red, right hook-latex] (m-3-1.east |- m-3-2) edge node [above] {$\iota$} (m-3-2);
         \path[->, red] (m-2-2) edge node [right] {$\sim$} (m-3-2);
  
         \node[draw=blue,inner sep=8pt, thick,rounded corners, fit= (m-1-2) (m-2-2) (m-2-2) (m-1-2) ]{};
         \node[draw=blue,inner sep=8pt, thick,rounded corners, fit= (m-1-1) (m-2-1) (m-2-1) (m-1-1) ]{};
  
      \end{tikzpicture}
      \end{center}
      By proposition 6.18 of \cite{hartshorne1977algebraic} it holds $\tilde{A_C} \cong \mathcal{O}_{C} \left( -D \right)$.
 \item As a next step we extend $\tilde{A_C}$ to become a (proper) coherent sheaf on $X_\Sigma$. To this end note that the entries of the matrix $\text{ker} ( M )$ are 
      elements of $S ( C )$, \ie are equivalence classes of elements of the ring $S$. For each entry of $\text{ker} \left( M \right)$ pick one representant in $S$. Thereby we obtain a matrix $\text{ker} ( M )^\prime$ with entries in $S$.\footnote{Of course this matrix $\text{ker} ( M )^\prime$ is not unique. Among others, this ambiguity is taken care of by tensoring with the structure sheaf $\mathcal{O}_C$ of $C$, as explained in step 3.} This enables us to construct the following \fp graded $S$-module $A$:
      \begin{center}
      \begin{tikzpicture}

         \matrix (m) [matrix of math nodes,row sep=4em,column sep=9em,minimum width=2em]
         {
           \bigoplus_{j \in J}{S \left( j \right)} \\
           \bigoplus_{i \in I}{S \left( i \right)} \\
           A \\ };

         \path[->] (m-1-1) edge node [left] {$\text{ker} \left( M \right)^\prime$} (m-2-1);
         \path[->, red] (m-2-1) edge node [right] {$\sim$} (m-3-1);
  
         \node[draw=blue,inner sep=15pt, thick,rounded corners, fit= (m-1-1) (m-2-1) (m-2-1) (m-1-1) ]{};
  
      \end{tikzpicture}
      \end{center}
 \item The coherent sheaf $\tilde{A}$ could have support outside of $C$. To ensure that $\tilde{A_C}$ has been extended by zero outside of $C$, we tensor $\tilde{A}$ 
      with the structure sheaf $\mathcal{O}_C$ of $C$. Given the matrix $N = ( g_1, g_2, \dots, g_k )^T$, the following \fp graded $S$-module $B$ sheafifes to give this structure sheaf $\mathcal{O}_C$: 
      \begin{center}
      \begin{tikzpicture}

        \matrix (m) [matrix of math nodes,row sep=3em,column sep=9em,minimum width=2em]
        {
          \bigoplus_{k \in K}{S \left( k \right)} \\
          S \left( C \right) \\
          B \\ };

        \path[->] (m-1-1) edge node [left] {$N$} (m-2-1);
        \path[->, red] (m-2-1) edge node [right] {$\sim$} (m-3-1);
  
        \node[draw=blue,inner sep=8pt, thick,rounded corners, fit= (m-1-1) (m-2-1) (m-2-1) (m-1-1) ]{};
  
      \end{tikzpicture}
      \end{center}
      Consequently consider $I_D = A \otimes_S B \in S ( C ) \mathrm{\textnormal{-}fpgrmod}$. It defines $\tilde{I_D} \in \mathfrak{Coh} ( X_\Sigma )$ which is supported only on $C$ and satisfies $\tilde{I_D} |_C \cong \mathcal{O}_C \left( -D \right)$.
\end{enumerate}

To obtain an \fp graded $S$-module $M$ such that $\tilde{M}$ is zero outside of $C$ and satisfies $\tilde{M} |_C \cong \mathcal{O}_C ( +D )$, we perform step 1. This yields the module $A_C$ described above. Now we dualise this module in the category $S ( C ) \mathrm{\textnormal{-}fpgrmod}$, which means (c.f. \autoref{subsec:ExtOfFPModules} for more details on the dualisation)
\[ A_C^\vee := \text{Hom}_{S \left( C \right)} \left( S \left( C \right), A_C \right) \, . \]
With this \fp graded $S \left( C \right)$-module we now proceed just as before, \ie we turn $A_C^\vee$ into an $S$-module $A^\vee$ and then compute the tensor product $O_D = A^\vee \otimes_S B$. This module $O_D$ defines a coherent sheaf $\tilde{O_D} \in \mathfrak{Coh} \left( X_\Sigma \right)$ which is supported only on $C$ and satisfies $\tilde{O_D} |_C \cong \mathcal{O}_C ( +D )$.

\subsection{Sheaf Cohomologies from F.P.\ Graded \texorpdfstring{$\mathbf{S}$}{S}-Modules} \label{subsec:SheafCohomologyFromFPGradedSModules}

Let $X_\Sigma$ be a normal toric variety which is either smooth, complete or simplicial, projective. Its Cox ring be $S$. By now we have described means to parametrise coherent sheaves $\mathcal{F} \in \mathfrak{Coh} ( X_\Sigma )$ by \fp graded $S$-modules. Given such an \fp graded $S$-module $M$, we can wonder how we extract the sheaf cohomologies of $\tilde{M}$ from the data defining $M$. Our algorithm is as follows:
\begin{enumerate}
 \item Given that $X_\Sigma$ is smooth and complete or simplicial and projective, the \emph{cohomCalg}-algorithm applies to it \cite{Blumenhagen:2010pv, 
      cohomCalg:Implementation, 2011JMP....52c3506J, Rahn:2010fm, Blumenhagen:2010ed}. This enables us to compute the vanishing sets
      \[ V^i \left( X_\Sigma \right) := \left\{ \left. D \in \text{Cl} \left( X_\Sigma \right) \; \right| \; h^i \left( X_\Sigma, \mathcal{O}_{X_\Sigma} \left( D \right) \right) = 0 \right\} \]
      fairly rapidly. Note that the so-obtained vanishing sets serve as a properly refined version of the semigroup $\mathbb{\mathcal{K}}^{\text{sat}}$ introduced in \cite{Maclagan03multigradedcastelnuovo-mumford} which was used in \cite{Oberwolfach} to propose a means to compute sheaf cohomology of coherent sheaves.
 \item Our algorithm now determines an ideal $I \subseteq S$ and $e \in \mathbb{Z}_{\geq 0}$ such that the pair $( I^{(e)}, M )$ \footnote{$I^{(e)}$ denotes 
      the e-th Frobenius power of the ideal $I$. If $I = \left\langle f_1, f_2, \dots, f_n \right\rangle$ then $I^{(e)} = \langle f_1^e, f_2^e, \dots, f_n^e \rangle$.} satisfies a number of conditions. These conditions are phrased in terms of the sets $V^i ( X_\Sigma )$ and designed such that
      \[ H^i \left( X_\Sigma, \tilde{M} \right) \cong \left[ \text{Ext}^i_{S} \left( I, M \right) \right]_{0} \, . \label{equ:CohoIso} \]
      See \cite{Bies-Thesis2017} for details on this isomorphism. 
 \item The $i$-th (global) extension module $\text{Ext}^i_{S} ( I, M )$ of the \fp graded $S$-modules $I, M$ happens to be an \fp graded $S$-module by itself. Its definition and properties are discussed in detail in \autoref{subsec:ExtOfFPModules}.
      We truncate it to degree $0 \in \text{Cl} ( X_\Sigma )$. Since $\tilde{M}$ is a \emph{coherent} sheaf, this degree-$0$-layer happens to be a finite-dimensional $\mathbb{Q}$-vector space. Our algorithm returns its $\mathbb{Q}$-dimension.
\end{enumerate}

The packages \cite{CAPCategoryOfProjectiveGradedModules, CAPPresentationCategory, PresentationsByProjectiveGradedModules, TruncationsOfPresentationsByProjectiveGradedModules} provide the implementation of the category $S \mathrm{\textnormal{-}fpgrmod}$ in the language of \emph{categorical programming} of \texttt{CAP} \cite{CAP, PosurDoktor, GutscheDoktor}. In addition, basic functionality of toric varieties is provided by the \texttt{gap}-package \texttt{ToricVarieties} of \cite{homalg}. The package \cite{SheafCohomologyOnToricVarieties} extends this package and provides routines to find an ideal $I$ and $e \in \mathbb{Z}_{\geq 0}$ which fit the criteria in the second step. In addition this package provides implementations of algorithms which allow for a quick computation of $\text{Ext}^i_{S} ( I, M )$, as explained in \autoref{subsec:ExtOfFPModules}.

Let us finally give an example on how these computations work in practise. To this end we pick $\mathbb{P}^2_{\mathbb{Q}}$ again and consider the \fp graded $S$-module $B_\Sigma$. We already stated that this module sheafifies to give $\tilde{B_\Sigma} \cong \mathcal{O}_{\mathbb{P}^2_{\mathbb{Q}}}$. Let us justify this statement by computing the sheaf cohomology dimensions of $\tilde{B_\Sigma}$ to see that they match up with this assertation. The first step of our algorithm computes the following vanishing sets:
\[ V^0 \left( \mathbb{P}^2_{\mathbb{Q}} \right) = \left\{ -1, -2, -3, -4, \dots \right\}, \quad V^1 \left( \mathbb{P}^2_{\mathbb{Q}} \right) = \mathbb{Z}, \quad
V^2 \left( \mathbb{P}^2_{\mathbb{Q}} \right) = \left\{ -2, -1, 0, 1, 2, \dots \right\} \, . \]
Based on these we compute the sheaf cohomology dimension of $\tilde{B_\Sigma}$:
\begin{itemize}
 \item $H^0 \left( \mathbb{P}^2_{\mathbb{Q}}, \tilde{B_\Sigma} \right)$: \\
      The algorithm finds $I = \langle x_1, x_2, x_3 \rangle$, $e = 1$.\footnote{It is a coincidence that $I$ happens to be the irrelevant ideal for projective spaces. For sufficiently more involved toric spaces, this is no longer true. Rather we pick an ideal associated to an ample divisor in $X_\Sigma$, as this guarantees the existence of a \emph{finite} $e$ for which the isomorphism in (\ref{equ:CohoIso}) holds true.} So $H^0 \left( \mathbb{P}^2_{\mathbb{Q}}, \tilde{B_\Sigma} \right) \cong \text{Hom}_S \left( I^{(1)}, B_\Sigma \right)_0 \cong \mathbb{Q}$.
 \item $H^1 ( \mathbb{P}^2_{\mathbb{Q}}, \tilde{B_\Sigma} )$: \\
      Now $I = \langle x_1, x_2, x_3 \rangle$, $e = 1$ and so $H^1 \left( \mathbb{P}^2_{\mathbb{Q}}, \tilde{B_\Sigma} \right) \cong \text{Ext}^1_S \left( I^{(1)}, B_\Sigma \right)_0 \cong \mathbb{Q}^0$.
 \item $H^2 ( \mathbb{P}^2_{\mathbb{Q}}, \tilde{B_\Sigma} )$: \\
      The algorithm gives $I = \langle x_1, x_2, x_3 \rangle$, $e = 0$. So $H^2 \left( \mathbb{P}^2_{\mathbb{Q}}, \tilde{B_\Sigma} \right) \cong \text{Ext}^2_S \left( I^{(0)}, B_\Sigma \right)_0 \cong \mathbb{Q}^0$.
\end{itemize}

This shows that indeed $\tilde{B_\Sigma}$ has the cohomology dimensions of $\mathcal{O}_{\mathbb{P}^2_{\mathbb{Q}}}$. To tell how good or bad a model for $\mathcal{O}_{\mathbb{P}^2_{\mathbb{Q}}}$ this module $B_\Sigma$ really is, let us consider the sequence
\[ \mathfrak{h}^i \left( e \right) = \text{dim}_{\mathbb{Q}} \left( \text{Ext}^{(i)}_S \left( I^{(e)}, M \right)_{0} \right) \, . \]
With our choice of ideal $I$ it is guaranteed that this sequence becomes constant for sufficiently large $e$. As the computations become harder and harder with increasing $e$, the art is to find the minimal integer $e$ such that the isomorphism (\ref{equ:CohoIso}) holds true. The quality of both our estimate of this minimal integer $e$ and the \fp graded $S$-module $B_\Sigma$ to serve as model for $\mathcal{O}_{\mathbb{P}^2_{\mathbb{Q}}}$ can therefore be estimated from looking at this sequence. In this particular example we have
\[ \mathfrak{h}^0 \left( e \right) = \left( 0, 1, 1, 1, \dots \right), \qquad \mathfrak{h}^1 \left( e \right) = \left( 0, 0, 0, 0, \dots \right), \qquad \mathfrak{h}^2 \left( e \right) = \left( 0, 0, 0, 0, \dots \right) \, . \]
So the perfect minimal integers are $\check{e}_0 = 1$, $\check{e}_1 = 0$ and $\check{e}_2 = 0$. Our estimates for $e_0$ and $e_2$ are therefore the ideal choices, but we overestimated $\check{e}_1$ and chose $e_1 = 1$ whilst $e_1 = 0$ would have worked already. 

As for the quality of $B_\Sigma$ - a perfect model of $\mathcal{O}_{\mathbb{P}^2_{\mathbb{Q}}}$ has $\check{e}_0 = \check{e}_1 = \check{e}_2 = 0$. For $B_\Sigma$ however $\check{e}_0 = \check{e}_1 = 1$. Therefore it is not an ideal model for $\mathcal{O}_{\mathbb{P}^2_{\mathbb{Q}}}$. $S ( 0 )$ however indeed satisfies $\check{e}_0 = \check{e}_1 = \check{e}_2 = 0$. Therefore this \fp $S$-module is \emph{the} ideal choice to model $\mathcal{O}_{\mathbb{P}^2_{\mathbb{Q}}}$.

\subsection{Computing Extension Modules in \texorpdfstring{$\mathbf{S\mathrm{\textnormal{-}fpgrmod}}$}{S-fpgrmod}} \label{subsec:ExtOfFPModules}

Given a normal toric variety $X_\Sigma$ without torus factor, we have explained in \autoref{app_sheafification} that an \fp graded $S$-module $M$ serves as model for the coherent sheaf $\tilde{M}$ on $X_\Sigma$. In \autoref{subsec:SheafCohomologyFromFPGradedSModules} we have outlined how one can extract the cohomology dimensions of $\tilde{M}$ from the module $M$ itself. This process involves the computation of the extension modules $\text{Ext}^i_S ( M, N )$ for \fp graded $S$-modules $M, N$. As the extension modules are \fp graded $S$-modules by themselves, they can be truncated to any $d \in \text{Cl} ( X_\Sigma )$. In particular we have seen that the truncations $\text{Ext}^i_S ( M, N )_0$ are important for the computation of the cohomology dimensions. Let us therefore, in this section, provide details on the computation of $\text{Ext}^i_S ( M, N )_0$. For the sake of simplicity, we assume that $S = \mathbb{Q} \left[ x_0, \dots, x_n \right]$ is a $\mathbb{Z}^n$-graded ring and start by investigating $\text{Hom}_S ( M, N)_0$ first. 

First recall that $\text{Hom}_S \left( M, N \right)$ is the module formed from all morphisms $\varphi \colon M \to N$ of \fp graded $S$-modules. To gain some intuition for this module and its truncations, we first consider the category of \emph{projective graded $S$-modules}. So pick two projective graded $S$-modules $M$, $N$ given by
\[ M \cong \bigoplus_{i \in I}{S( i )}, \qquad N \cong \bigoplus_{j \in J}{S( j )} \, . \]
Then the module of all homomorphism $\varphi \colon M \to N$ of projective graded $S$-modules is given by
\[ \text{Hom}_S \left( M, N \right) \cong \bigoplus_{i \in I, j \in J}{S \left( -i + j \right)} \, . \]
So indeed $\text{Hom}_S ( M, N )$ is a projective $\mathbb{Z}^n$-graded $S$-module. Therefore $\text{Hom}_S ( M, N )$ can be truncated to any $d \in \mathbb{Z}^n$. The dual module of the projective graded $S$-module $M = \bigoplus_{i \in I}{S \left( i \right)}$ is defined as
\[ M^\vee := \text{Hom}_S \left( M, S \right) \cong \bigoplus_{i \in I}{S \left( - i \right)} \, . \]
Intuitively, we write $\text{Hom}_S \left( M, N \right) \cong M^\vee \otimes_S N$.

Let us now turn back to $S\mathrm{\textnormal{-}fpgrmod}$ in order to generalise the above procedure to compute $\text{Hom}_S ( M, N )_0$ for (proper) \fp graded $S$-modules $M$ and $N$. We assume that $M$, $N$ are presented as follows:
\[ 
\begin{tikzpicture}[baseline=(current  bounding  box.center)]

  \matrix (m) [matrix of math nodes,row sep=3em,column sep=10em,minimum width=2em]
  {
     R_M & R_N \\
     G_M & G_N \\};

  \path[->] (m-1-1) edge node [left] {$\rho_M$} (m-2-1);
  \path[->] (m-1-2) edge node [right] {$\rho_N$} (m-2-2);

  \node[draw=blue,inner sep=8pt, thick,rounded corners, fit= (m-1-2) (m-2-2) (m-2-2) (m-1-2) ]{};
  \node[draw=blue,inner sep=8pt, thick,rounded corners, fit= (m-1-1) (m-2-1) (m-2-1) (m-1-1) ]{};
  
\end{tikzpicture}
\]
Then $\text{Hom}_S ( M, N )$ is the kernel object of the following morphism of \fp graded $S$-modules:
\[ \label{equ:HomEmbedding}
\begin{tikzpicture}[baseline=(current  bounding  box.center)]

  \matrix (m) [matrix of math nodes,row sep=6em,column sep=9em,minimum width=2em]
  {
     R & G_M^\vee \otimes R_N & R_M^\vee \otimes R_N \\
     G & G_M^\vee \otimes G_N & R_M^\vee \otimes G_N \\
     \text{Hom}_S \left( M, N \right) & G_M^\vee \otimes N & R_M^\vee \otimes N \\};

  \path[->] (m-1-1) edge node [left] {$\rho$} (m-2-1);
  \path[->] (m-1-2) edge node [below, rotate = -90] {$\text{id}_{G_M^\vee} \otimes \rho_N$} (m-2-2);
  \path[->] (m-1-3) edge node [above, rotate = -90] {$\text{id}_{R_M^\vee} \otimes \rho_N$} (m-2-3);
  \path[->] (m-2-2.east |- m-2-3) edge node [above] {$\rho_M^\vee \otimes \text{id}_{G_N}$} (m-2-3);
  \path[->] (m-1-2.east |- m-1-3) edge node [above] {$\rho_M^\vee \otimes \text{id}_{R_N}$} (m-1-3);
  \path[->] (m-1-1.east |- m-1-2) edge node [above] {$\beta$} (m-1-2);
  \path[->] (m-2-1.east |- m-2-2) edge node [above] {$\alpha$} (m-2-2);
  
  \path[->, red] (m-2-1) edge node [right] {$\sim$} (m-3-1);
  \path[->, red] (m-2-2) edge node [right] {$\sim$} (m-3-2);
  \path[->, red] (m-2-3) edge node [right] {$\sim$} (m-3-3);
  \path[red, right hook-latex] (m-3-1.east |- m-3-2) edge node [above] {$\iota$} (m-3-2);  
  \path[->, red] (m-3-2.east |- m-3-3) edge node [above] {$\rho_M^\vee \otimes \text{id}_N$} (m-3-3);
  
  \node[draw=blue,inner sep=12pt, thick,rounded corners, fit= (m-1-1) (m-2-1) (m-2-1) (m-1-1) ]{};
  \node[draw=blue,inner sep=12pt, thick,rounded corners, fit= (m-1-2) (m-2-2) (m-2-2) (m-1-2) ]{};
  \node[draw=blue,inner sep=12pt, thick,rounded corners, fit= (m-1-3) (m-2-3) (m-2-3) (m-1-3) ]{};
  
\end{tikzpicture}
\]
The monomorphism $\iota$ is termed the \text{Hom-embedding}.

One way to compute $\text{Hom}_S ( M, N )_0$ is therefore to first identify $\text{Hom}_S ( M, N )$ as the domain of the Hom-embedding $\iota$ and then truncate it to degree $0$. This process proceeds in the following 3 steps:
\begin{enumerate}
 \item First compute the following pullback diagram in the category of projective graded $S$-modules:
      \[ \label{equ:pullback1}
      \begin{tikzpicture}[baseline=(current  bounding  box.center)]

      \matrix (m) [matrix of math nodes,row sep=5em,column sep=10em,minimum width=2em]
      {
        G                    & R_M^\vee \otimes R_N \\
        G_M^\vee \otimes G_N & R_M^\vee \otimes G_N \\};

        \path[->] (m-1-1) edge node [left] {$\alpha$} (m-2-1);
        \path[dashed, ->] (m-1-1.east |- m-1-2) edge (m-1-2);
        \path[->] (m-1-2) edge node [above, rotate = -90] {$\text{id}_{R_M^\vee} \otimes \rho_N$} (m-2-2);
        \path[->] (m-2-1.east |- m-2-2) edge node [above] {$\rho_M^\vee \otimes \text{id}_{G_N}$} (m-2-2);
      
      \end{tikzpicture}
      \]
      Hence $G$ is the pullback object of the morphisms $( \rho_M^\vee \otimes \text{id}_{G_N}, \text{id}_{R_M^\vee} \otimes \rho_N )$ and $\alpha$ is its canonical projection onto $G_M^\vee \otimes G_N$.\footnote{$G$ and $\alpha$ are unique up to isomorphism by general category theory. The identification of $\alpha$ and $G$ in the computer involves Gr\"obner basis algorithms to be applied to the matrices representing the morphisms $\rho_M^\vee \otimes \text{id}_{G_N}$ and $\text{id}_{R_M^\vee} \otimes \rho_N$.}
 \item Next compute the following pullback diagram in the category of projective graded $S$-modules:
      \[ \label{equ:pullback2}
      \begin{tikzpicture}[baseline=(current  bounding  box.center)]

      \matrix (m) [matrix of math nodes,row sep=5em,column sep=10em,minimum width=2em]
      {
        R & G_M^\vee \otimes R_N \\
        G & G_M^\vee \otimes G_N \\};

        \path[->] (m-2-1.east |- m-2-2) edge node [above] {$\alpha$} (m-2-2);
        \path[->] (m-1-1) edge node [left] {$\rho$} (m-2-1);
        \path[dashed, ->] (m-1-1.east |- m-1-2) edge (m-1-2);
        \path[->] (m-1-2) edge node [above, rotate = -90] {$\text{id}_{G_M^\vee} \otimes \rho_N$} (m-2-2);
    
      \end{tikzpicture}
      \]
      So $R$ is the pullback object of $( \alpha, \text{id}_{G_M^\vee} \otimes \rho_N )$ and $\rho$ is its projection onto $G$.\footnote{The computation of $R$ and $\rho$ in the computer involves Gr\"obner basis algorithms.} \\
      Let us mention that the computation of kernel embeddings in the category $S \mathrm{\textnormal{-}fpgrmod}$ always proceeds along these first two steps. The interested reader might find it illustrative to use this knowledge to check that all kernel embeddings presented in \autoref{subsec:FPGradedSModules} are indeed obtained along this strategy.
 \item Finally truncate the morphism $R \xrightarrow{\rho} G$ of projective graded $S$-modules to degree $0$. Thereby we obtain a morphism of finite-dimensional 
      vector spaces over $\mathbb{Q}$. The so-obtained object
      \[ \label{equ:presentationOfHomS(I,M)_0}
      \begin{tikzpicture}[baseline=(current  bounding  box.center)]

      \matrix (m) [matrix of math nodes,row sep=2em,column sep=10em,minimum width=2em]
      {
        R_0 \\
        G_0 \\};

      \path[->] (m-1-1) edge node [left] {$\rho_0$} (m-2-1);  

      \node[draw=blue,inner sep=8pt, thick,rounded corners, fit= (m-1-1) (m-2-1) (m-2-1) (m-1-1) ]{};
  
      \end{tikzpicture}
      \]
      is an \fp $\mathbb{Q}$-vector space with the property $\text{Hom}_S ( M, N )_0 \cong \text{coker} ( \rho_0 )$. Therefore it is now a simple matter to determine the $\mathbb{Q}$-dimension of $\text{Hom}_S ( M, N )_0$ from $\text{rk} ( \rho_0 )$.
\end{enumerate}

This algorithm requires a lot of computational resources and time, due to the use of Gr\"obner basis algorithms for the computation of the pullback diagrams (\ref{equ:pullback1}) and (\ref{equ:pullback2}) in the first two steps. To overcome this shortcoming we have developed an alternative algorithm. This algorithm applies the truncation functor $\text{tr} \colon S \mathrm{\textnormal{-}fpgrmod} \to \mathbb{Q} \mathrm{\textnormal{-}}fpvec$ to (\ref{equ:HomEmbedding}) directly. Thereby we obtain the following diagram:
\[ \label{equ:HomEmbeddingTruncated}
\begin{tikzpicture}[baseline=(current  bounding  box.center)]

  \matrix (m) [matrix of math nodes,row sep=6.5em,column sep=8em,minimum width=2em]
  {
     R_0 & \left( G_M^\vee \otimes R_N \right)_0 & \left( R_M^\vee \otimes R_N \right)_0 \\
     G_0 & \left( G_M^\vee \otimes G_N \right)_0 & \left( R_M^\vee \otimes G_N \right)_0 \\
     \text{Hom}_S \left( M, N \right)_0 & \left( G_M^\vee \otimes N \right)_0 & \left( R_M^\vee \otimes N \right)_0 \\};

  \path[->] (m-1-1) edge node [left] {$\rho_0$} (m-2-1);
  \path[->] (m-1-2) edge node [below, rotate = -90] {$\left( \text{id}_{G_M^\vee} \otimes \rho_N \right)_0$} (m-2-2);
  \path[->] (m-1-3) edge node [above, rotate = -90] {$\left( \text{id}_{R_M^\vee} \otimes \rho_N \right)_0$} (m-2-3);
  \path[->] (m-2-2.east |- m-2-3) edge node [above] {$\left( \rho_M^\vee \otimes \text{id}_{G_N} \right)_0$} (m-2-3);
  \path[->] (m-1-2.east |- m-1-3) edge node [above] {$\left( \rho_M^\vee \otimes \text{id}_{R_N} \right)_0$} (m-1-3);
  \path[->] (m-1-1.east |- m-1-2) edge node [above] {$\beta_0$} (m-1-2);
  \path[->] (m-2-1.east |- m-2-2) edge node [above] {$\alpha_0$} (m-2-2);
  
  \path[->, red] (m-2-1) edge node [right] {$\sim$} (m-3-1);
  \path[->, red] (m-2-2) edge node [right] {$\sim$} (m-3-2);
  \path[->, red] (m-2-3) edge node [right] {$\sim$} (m-3-3);
  \path[red, right hook-latex] (m-3-1.east |- m-3-2) edge node [above] {$\text{ker}$} (m-3-2);  
  \path[->, red] (m-3-2.east |- m-3-3) edge node [above] {$\left( \rho_M^\vee \otimes \text{id}_N \right)_0$} (m-3-3);
  
  \node[draw=blue,inner sep=7pt, thick,rounded corners, fit= (m-1-1) (m-2-1) (m-2-1) (m-1-1) ]{};
  \node[draw=blue,inner sep=7pt, thick,rounded corners, fit= (m-1-2) (m-2-2) (m-2-2) (m-1-2) ]{};
  \node[draw=blue,inner sep=7pt, thick,rounded corners, fit= (m-1-3) (m-2-3) (m-2-3) (m-1-3) ]{};
  
\end{tikzpicture}
\]
This is a diagram in the category $\mathbb{Q} \mathrm{\textnormal{-}}fpvec$ of \fp $\mathbb{Q}$-vector spaces -- the black arrows are morphisms of $\mathbb{Q}$-vector spaces and the red ones mediate between \fp $\mathbb{Q}$-vector spaces. The crucial observation is that $\text{tr} \colon S \mathrm{\textnormal{-}fpgrmod} \to \mathbb{Q} \mathrm{\textnormal{-}}fpvec$ is an exact functor. Therefore $\text{Hom}_S ( M, N )_0$ is the kernel object of the map $( \rho_M^\vee \otimes \text{id}_N )_0$ of \fp $\mathbb{Q}$-vector spaces. This now comes with two advantages:
\begin{itemize}
 \item For once, the application of the truncation to (\ref{equ:HomEmbedding}) is easily parallelised by applying the functor $\text{tr} \colon S 
      \mathrm{\textnormal{-}fpgrmod} \to \mathbb{Q} \mathrm{\textnormal{-}}fpvec$ to the morphisms $\text{id}_{G_M^\vee} \otimes \rho_N$, $\text{id}_{R_M^\vee} \otimes \rho_N$ and $\rho_M^\vee \otimes \text{id}_{G_N}$ in parallel.
 \item Subsequently we compute the following two pullback diagrams:
      \begin{center}
      \begin{tikzpicture}[baseline=(current  bounding  box.center)]

      \matrix (m) [matrix of math nodes,row sep=7em,column sep=5.5em,minimum width=2em]
      {
        G_0                                   & \left( R_M^\vee \otimes R_N \right)_0 & R_0 & \left( G_M^\vee \otimes R_N \right)_0 \\
        \left( G_I^\vee \otimes G_M \right)_0 & \left( R_M^\vee \otimes G_N \right)_0 & G_0 & \left( G_M^\vee \otimes G_N \right)_0 \\};

      \path[->] (m-1-1) edge node [left] {$\alpha_0$} (m-2-1);
      \path[dashed, ->] (m-1-1.east |- m-1-2) edge (m-1-2);
      \path[->] (m-1-2) edge node [above, rotate = -90] {$\left( \text{id}_{R_M^\vee} \otimes \rho_N \right)_0$} (m-2-2);
      \path[->] (m-2-1.east |- m-2-2) edge node [above] {$\left( \rho_M^\vee \otimes \text{id}_{G_N} \right)_0$} (m-2-2);
      \path[->] (m-2-3.east |- m-2-4) edge node [above] {$\alpha_0$} (m-2-4);
      \path[->] (m-1-3) edge node [left] {$\rho_0$} (m-2-3);
      \path[dashed, ->] (m-1-3.east |- m-1-4) edge (m-1-4);
      \path[->] (m-1-4) edge node [above, rotate = -90] {$\left( \text{id}_{G_M^\vee} \otimes \rho_N \right)_0$} (m-2-4);

      \end{tikzpicture}
      \end{center}
      Crucially though, this time the involved matrices are $\mathbb{Q}$-valued. Therefore the pullbacks can be computed from Gau{\ss} eliminations. For the latter there exist highly performant computer implementations \eg in \texttt{MAGMA} \cite{MR1484478}. Such parallelised algorithms are not (yet) available for Gr\"obner basis computations.
\end{itemize}
For these two reasons, this alternative algorithm allows us to identify the object (\ref{equ:presentationOfHomS(I,M)_0}) faster. In consequence, we can also identify $\text{dim}_{\mathbb{Q}} ( \text{coker} ( \rho_0 ) )$ more efficiently along these lines.

The examples in the main text are indeed computed by applying $\text{tr} \colon S \mathrm{\textnormal{-}fpgrmod} \to \mathbb{Q} \mathrm{\textnormal{-}}fpvec$ to (\ref{equ:HomEmbedding}) first and use \texttt{MAGMA} \cite{MR1484478} to perform the subsequent Gau{\ss} eliminations for huge matrices over $\mathbb{Q}$ -- they easily reach sizes of more than 10.000 x 10.000. The overall algorithm is provided by the \texttt{gap}-package \cite{SheafCohomologyOnToricVarieties}.

Finally, let us turn to the computation of the extension modules $\text{Ext}^i_S ( M, N )$ and their truncations. To this end we first recall the theoretical foundations of this bivariate functor $\text{Ext}_S^n ( -, - )$. To this end let us pick a \fp graded $S$-module $B$ and use it to define an endofunctor $G_B$ of $S\mathrm{\textnormal{-}fpgrmod}$ as follows:
\begin{itemize}
 \item $A \mapsto \text{Hom}_S \left( A,B \right)$
 \item $\left[ \varphi \colon A_1 \to A_2 \right] \mapsto \tilde{\varphi} \equiv \left[ \text{Hom}_S \left( A_2 , B \right) \to \text{Hom}_S \left( A_1, B \right) \; , \; \alpha \mapsto \alpha 
      \circ \varphi \right]$
\end{itemize}
It can then be verified that $G_B$ is a contravariant left-exact endofunctor of $S\mathrm{\textnormal{-}fpgrmod}$. Now the abstract mathematical definition of 
$\text{Ext}_S^n ( A,B )$ is to say
\[ \text{Ext}_S^n \left( A,B \right) := \left( R^n G_B \right) \left( A \right) \, , \]
which means that $\text{Ext}_S^n ( - , - )$ is the n-th right-derived functor of the functor $G_B$. This abstract statement can be made far more explicit. Namely, to compute $\text{Ext}_S^n ( A , B )$ ($n \geq 0$) for two \fp graded $S$-modules $A,B$ we perform the following steps:
\begin{enumerate}
 \item Consider a minimal free resolution $F^\bullet ( A )$ of $A$ by projective objects in $S\mathrm{\textnormal{-}fpgrmod}$. Use this opportunity to recall 
      that the projective objects in $S\mathrm{\textnormal{-}fpgrmod}$ are the projective graded $S$-modules of finite rank. In addition we state it as a fact that such a resolution exists for every $A \in S\mathrm{\textnormal{-}fpgrmod}$. The resolution $F^\bullet ( A )$ will be denoted as follows:
      \[ \dots \xrightarrow{\alpha_4} P^3 \xrightarrow{\alpha_3} P^2 \xrightarrow{\alpha_2} P^1 \xrightarrow{\alpha_1} P^0 \to 0 \, . \label{equ:MinimalFreeResolution}\]
 \item Now apply the functor $G_B$ to (\ref{equ:MinimalFreeResolution}). Recall that this functor is a contravariant left-exact endofunctor of 
      $S\mathrm{\textnormal{-}fpgrmod}$. Therefore we obtain the following complex in $S\mathrm{\textnormal{-}fpgrmod}$:
      \[ \dots \xleftarrow{\tilde{\alpha_3}} \text{Hom}_S \left( P^2, B \right) \xleftarrow{\tilde{\alpha_2}} \text{Hom}_S \left( P^1, B \right) \xleftarrow{\tilde{\alpha_1}} \text{Hom}_S \left( P^0, B \right) \leftarrow 0 \, . \label{equ:complex} \]
 \item Finally, recall that $S\mathrm{\textnormal{-}fpgrmod}$ is an Abelian category. Therefore we can compute the homology of this complex at position $n$ and the 
      result will be a \fp graded $S$-module. For $n > 0$ we say that $( R^n G_B ) ( A )$ is the homology of this complex (\ref{equ:complex}) at position $n$. For $n = 0$ we set
     \[ \left( R^0 G_B \right) \left( A \right) = G_B \left( A \right) = \text{Hom}_S \left( A,B \right) \, . \]
\end{enumerate}

That all said, we are finally in the position to explain how we compute $\text{Ext}_S^n ( A,B )$ in the package \cite{SheafCohomologyOnToricVarieties}. The computation of $\text{Ext}_S^0 ( A,B )$ of course makes use of the fact
\[ \text{Ext}_S^0 \left( A,B \right) = \text{Hom}_S \left( A, B \right) \]
so that we simply have to compute $\text{Hom}_S ( A, B )$ as outlined above. The interesting case is therefore $\text{Ext}_S^n ( A,B )$ with $n > 0$. In this case we compute a minimal free resolution of $A$. It takes the following form:
\[
\begin{tikzpicture}[baseline=(current  bounding  box.center)]

  \matrix (m) [matrix of math nodes,row sep=2em,column sep=7em,minimum width=2em]
  {
     \dots & 0 & 0 & 0 \\
     \dots & P^2 & P^1 & P^0 \\};

  \path[->] (m-1-2) edge node [left] {$0$} (m-2-2);
  \path[->] (m-1-3) edge node [left] {$0$} (m-2-3);
  \path[->] (m-1-4) edge node [left] {$0$} (m-2-4);
  
  \path[->] (m-1-1.east |- m-1-2) edge node [above] {$0$} (m-1-2);
  \path[->] (m-1-2.east |- m-1-3) edge node [above] {$0$} (m-1-3);
  \path[->] (m-1-3.east |- m-1-4) edge node [above] {$0$} (m-1-4);
  
  \path[->] (m-2-1.east |- m-2-2) edge node [above] {$\alpha_3$} (m-2-2);
  \path[->] (m-2-2.east |- m-2-3) edge node [above] {$\alpha_2$} (m-2-3);
  \path[->] (m-2-3.east |- m-2-4) edge node [above] {$\alpha_1$} (m-2-4);
  
  \node[draw=blue,inner sep=12pt, thick,rounded corners, fit= (m-1-2) (m-2-2) (m-2-2) (m-1-2) ]{};
  \node[draw=blue,inner sep=12pt, thick,rounded corners, fit= (m-1-3) (m-2-3) (m-2-3) (m-1-3) ]{};
  \node[draw=blue,inner sep=12pt, thick,rounded corners, fit= (m-1-4) (m-2-4) (m-2-4) (m-1-4) ]{};
  
\end{tikzpicture}
\]
Recall that we want to compute the homology of the complex (\ref{equ:complex}) at position $n$. Therefore we need to take into account both $\alpha_n$ and $\alpha_{n+1}$. To this end we compute the kernel embedding of the cokernel projection of the $n$-th morphism in the above resolution, \ie of the morphism:
\[
\begin{tikzpicture}[baseline=(current  bounding  box.center)]

  \matrix (m) [matrix of math nodes,row sep=2em,column sep=7em,minimum width=2em]
  {
     0 & 0 \\
     P^{n} & P^{n-1} \\};

  \path[->] (m-1-1) edge node [left] {$0$} (m-2-1);
  \path[->] (m-1-2) edge node [left] {$0$} (m-2-2);
  
  \path[->] (m-1-1.east |- m-1-2) edge node [above] {$0$} (m-1-2);
  \path[->] (m-2-1.east |- m-2-2) edge node [above] {$\alpha_n$} (m-2-2);
  
  \node[draw=blue,inner sep=12pt, thick,rounded corners, fit= (m-1-1) (m-2-1) (m-2-1) (m-1-1) ]{};
  \node[draw=blue,inner sep=12pt, thick,rounded corners, fit= (m-1-2) (m-2-2) (m-2-2) (m-1-2) ]{};

\end{tikzpicture}
\]
It is readily verified that the so-obtained morphism $\mu \colon X \to Y$ is given by
\[
\begin{tikzpicture}[baseline=(current  bounding  box.center)]

  \matrix (m) [matrix of math nodes,row sep=3em,column sep=9em,minimum width=2em]
  {
     P^{n+1} & 0 \\
     P^{n} & P^{n-1} \\};

  \path[->] (m-1-1) edge node [left] {$\alpha_{n+1}$} (m-2-1);
  \path[->] (m-1-2) edge node [left] {$0$} (m-2-2);
  
  \path[->] (m-1-1.east |- m-1-2) edge node [above] {$0$} (m-1-2);
  \path[->] (m-2-1.east |- m-2-2) edge node [above] {$\alpha_n$} (m-2-2);
  
  \node[draw=blue,inner sep=12pt, thick,rounded corners, fit= (m-1-1) (m-2-1) (m-2-1) (m-1-1) ]{};
  \node[draw=blue,inner sep=12pt, thick,rounded corners, fit= (m-1-2) (m-2-2) (m-2-2) (m-1-2) ]{};

\end{tikzpicture}
\]
The functor $G_B$ now induces the morphism $\tilde{\mu} \colon \text{Hom}_S \left( Y, B \right) \to \text{Hom}_S \left( X, B \right)$. It turns out that the cokernel object of this morphism is isomorphic to the n-th homology of the complex (\ref{equ:complex}) at position $n$. So $\text{coker} ( \tilde{\mu} ) \cong \text{Ext}^n_S ( A, B )$. We compute $\tilde{\mu}$ based on the commutative diagram in \autoref{DiagramForComputationOfGradedHomOnMorphisms}.

\begin{figure}
\begin{tikzpicture}[baseline=(current bounding box.center)]

  \tikzset{font=\footnotesize}
  
  \matrix (m) [matrix of math nodes,row sep=6.1em,column sep=8.5em,minimum width=2em]
  {
     \text{Hom}_S \left( Y, B \right) & \left( P^{n-1} \right)^\vee \otimes B & 0^\vee \otimes B \\
                                      & \left( P^{n-1} \right)^\vee \otimes R_B & 0 \\
                                      & \left( P^{n-1} \right)^\vee \otimes G_B & 0 \\ \\
                                      & \left( P^{n} \right)^\vee \otimes R_B & \left( P^{n+1} \right)^\vee \otimes R_B \\
                                      & \left( P^{n} \right)^\vee \otimes G_B & \left( P^{n+1} \right)^\vee \otimes G_B \\
     \text{Hom}_S \left( X, B \right) & \left( P^{n} \right)^\vee \otimes B & \left( P^{n+1} \right)^\vee \otimes B \\
     };

  \path[->] (m-2-2) edge node [below, rotate = -90] {$\text{id}_{\left( P^{n-1} \right)^\vee} \otimes \rho_B$} (m-3-2);
  \path[->] (m-2-3) edge node [left] {$0$} (m-3-3);
  \path[->] (m-2-2.east |- m-2-3) edge node [above] {$0$} (m-2-3);
  \path[->] (m-3-2.east |- m-3-3) edge node [above] {$0$} (m-3-3);

  \path[->] (m-5-2) edge node [below, rotate = -90] {$\text{id}_{\left( P^{n} \right)^\vee} \otimes \rho_B$} (m-6-2);
  \path[->] (m-5-3) edge node [below, rotate = -90] {$\text{id}_{\left( P^{n+1} \right)^\vee} \otimes \rho_B$} (m-6-3);
  \path[->] (m-5-2) edge node [above] {$\alpha_{n+1}^\vee \otimes \text{id}_{R_B}$} (m-5-3);
  \path[->] (m-6-2) edge node [above] {$\alpha_{n+1}^\vee \otimes \text{id}_{G_B}$} (m-6-3);
  
  \path[->, red] (m-2-2) edge node [right] {$\sim$} (m-1-2);
  \path[->, red] (m-2-3) edge node [right] {$\sim$} (m-1-3);
  \path[red, right hook-latex] (m-1-1.east |- m-1-2) edge node [above] {$\iota_1$} (m-1-2);  
  \path[->, red] (m-1-2.east |- m-1-3) edge node [above] {$0$} (m-1-3);

  \path[->, red] (m-6-2) edge node [right] {$\sim$} (m-7-2);
  \path[->, red] (m-6-3) edge node [right] {$\sim$} (m-7-3);
  \path[red, right hook-latex] (m-7-1.east |- m-7-2) edge node [above] {$\iota_2$} (m-7-2);
  \path[->, red] (m-7-2.east |- m-7-3) edge node [above] {$\alpha_{n+1}^\vee \otimes \text{id}_B$} (m-7-3);
  
  \path[->] (m-3-2.south west) edge [bend right=50,looseness=0.5] node [below, rotate = -90] {$\alpha_n^\vee \otimes \text{id}_{G_B}$} (m-6-2.north west);
  \path[->] (m-2-2.south east) edge [bend left=50,looseness=0.5] node [above, rotate = -90] {$\alpha_n^\vee \otimes \text{id}_{R_B}$} (m-5-2.north east);
  \path[->, red] (m-1-2.south west) edge [bend right=50,looseness=0.5] node [below, rotate = -90] {$\alpha_n^\vee \otimes \text{id}_{B}$} (m-7-2.north west);
  
  \path[->, red, dashed] (m-1-1) edge node [left] {$\tilde{\mu}$} (m-7-1);
  
  \node[draw=blue,inner sep=12pt, thick,rounded corners, fit= (m-2-2) (m-3-2) (m-3-2) (m-2-2) ]{};
  \node[draw=blue,inner sep=12pt, thick,rounded corners, fit= (m-2-3) (m-3-3) (m-3-3) (m-2-3) ]{};
  \node[draw=blue,inner sep=12pt, thick,rounded corners, fit= (m-5-2) (m-6-2) (m-6-2) (m-5-2) ]{};
  \node[draw=blue,inner sep=12pt, thick,rounded corners, fit= (m-5-3) (m-6-3) (m-6-3) (m-5-3) ]{};
  
  \node[draw=darkgreen,inner sep=15pt, thick,rounded corners, fit= (m-1-1) (m-1-3) (m-3-3) (m-3-2) (m-1-1) ]{};
  \node[draw=darkgreen,inner sep=15pt, thick,rounded corners, fit= (m-7-1) (m-7-3) (m-5-3) (m-5-2) (m-7-1) ]{};  
  
\end{tikzpicture}
\caption{Given $\mu \colon X \to Y$ we can compute the morphism $\tilde{\mu} \colon \text{Hom}_S \left( Y, B \right) \to \text{Hom}_S \left( X, B \right)$ based on the above diagram.}
\label{DiagramForComputationOfGradedHomOnMorphisms}
\end{figure}

The green boxes denote the Hom-embeddings of $\text{Hom}_S ( Y, B )$ and $\text{Hom}_S ( X, B )$ (c.f. (\ref{equ:HomEmbedding})). As outlined above we can therefore compute the monomorphisms $\iota_1$ and $\iota_2$. The ranges of these maps are connected by $\alpha_n^\vee \otimes \text{id}_B$. The map $\tilde{\mu}$ is now the lift of the pair of morphisms $\left( \alpha_n^\vee \otimes \text{id}_B \circ \iota_1, \iota_2 \right)$.

The computation of cohomology dimension will involve $\text{Ext}^n_S \left( I, M \right)_0$ for ideals $I \subseteq S$ and `special' modules $M$. Instead of computing $\tilde{\mu}$ along \autoref{DiagramForComputationOfGradedHomOnMorphisms}, determining its cokernel module and then truncate the latter to degree $0$, it turns out to be much more efficient computational-wise to truncate the diagram in \autoref{DiagramForComputationOfGradedHomOnMorphisms} to degree $0$, compute $\tilde{\mu}_0$ and then determine its cokernel vector space (presentation). This parallels the philosophy employed for $\text{Hom}_S ( A,B )_0$, which allows us to replace Gr\"obner basis computations by Gau{\ss} eliminations. The corresponding algorithms are implemented in the \texttt{gap}-package \cite{SheafCohomologyOnToricVarieties}.

\section{Data Computed with \texttt{plesken}} \label{sec:Data}

In this appendix we  detail the structure of the modules $M_i$, $i=1, \ldots, 7$ describing the line bundle $\mathcal{L}(A, C_{\mathbf{5}_{-2}})$ for the family of models presented in \autoref{sec:Spectrum1} and apply the algorithm of \autoref{subsec:SheafCohomologyFromFPGradedSModules} for the computation of its sheaf cohomology groups. The computations are performed with the help of the computer \texttt{plesken} at Universitaet Siegen.

\subsection{The Module \texorpdfstring{$\mathbf{M_1}$}{M1}}

The first step is compute a minimal free resolution of $M_1$. Indeed, every coherent sheaf possesses a locally free resolution, \ie it fits into a long exact sequence where the objects other than $M_1$ are locally free sheaves. In the present case,
the module $M_1$ has a minimal free resolution given by
\begin{align}
\begin{split}
0 & \to S \left( -36 \right) \oplus S \left( -47 \right) \oplus S \left( -48 \right) \\
  &\to S \left( -34 \right) \oplus S \left( -35 \right) \oplus S \left( -45 \right) \oplus S \left( -23 \right) \oplus S \left( -38 \right) \\
  &\to S \left( -6 \right) \oplus S \left( -21 \right) \twoheadrightarrow M_1 \to 0 \, .
\end{split}
\end{align}
From this resolution \texttt{plesken} first read off the Betti numbers of $M_1$, and then used them to compute $e_0 = e_1 = 46$. For the definition of the numbers $e_i$, see again  \autoref{subsec:SheafCohomologyFromFPGradedSModules}. From this we conclude $h^i ( C_{\mathbf{5}_{-2}}, \tilde{M_1} ) = ( 22, 43 )$. Note that we could have applied the so-called \emph{linear regularity} as well \cite{2014arXiv1409.6100B}. For this particular module $M_1$ this method predicts $e_0 = 45$, which is slightly better. Explicitly we computed with \texttt{plesken} the values listed in \autoref{table1001}, which are visualised in \autoref{fig1003}. This shows that for this module the ideal values are $\check{e}_0 = 39$ and $\check{e}_1 = 46$.

\begin{table}
\begin{center}
{\footnotesize
\begin{tabular}{|c|c|c||c|c|c|}
\hline
$e$ & $\text{Hom}_S \left( B_\Sigma^{(e)}, M_1 \right)_{0}$ & $\text{Ext}^1_S \left( B_\Sigma^{(e)}, M_1 \right)_{0}$ & $e$ & $\text{Hom}_S \left( B_\Sigma^{(e)}, M_1 \right)_{0}$ & $\text{Ext}^1_S \left( B_\Sigma^{(e)}, M_1 \right)_{0}$ \\
\hline \hline
0 & 0 & 0 & 25 & 0 & 576 \\
1 & 0 & 0 & 26 & 0 & 558 \\
2 & 0 & 0 & 27 & 0 & 537 \\
3 & 0 & 0 & 28 & 0 & 513 \\
4 & 0 & 0 & 29 & 0 & 486 \\
\hline
5 & 0 & 0 & 30 & 0 & 456 \\
6 & 0 & 0 & 31 & 0 & 423 \\
7 & 0 & 0 & 32 & 0 & 387 \\
8 & 0 & 3 & 33 & 0 & 348 \\
9 & 0 & 15 & 34 & 0 & 309 \\
\hline
10 & 0 & 36 & 35 & 0 & 273 \\
11 & 0 & 62 & 36 & 2 & 239\\
12 & 0 & 89 & 37 & 8 & 209 \\
13 & 0 & 116 & 38 & 15 & 183 \\
14 & 0 & 150 & 39 & 22 & 160 \\
\hline
15 & 0 & 195 & 40 & 22 & 133 \\
16 & 0 & 256 & 41 & 22 & 109 \\
17 & 0 & 325 & 42 & 22 & 88 \\
18 & 0 & 394 & 43 & 22 & 70 \\
19 & 0 & 467 & 44 & 22 & 56 \\
\hline
20 & 0 & 528 & 45 & 22 & 46 \\
21 & 0 & 573 & 46 & 22 & 43 \\
22 & 0 & 600 & 47 & 22 & 43 \\
23 & 0 & 603 & 48 & 22 & 43 \\
24 & 0 & 591 & 49 & 22 & 43 \\
   &   &     & 50 & 22 & 43 \\
\hline
\end{tabular}}
\end{center}
\caption{$\text{dim}_{\mathbb{Q}} \left( \text{Hom}_S \left( B_\Sigma^{(e)}, M_1 \right)_{0} \right)$ and $\text{dim}_{\mathbb{Q}} \left( \text{Ext}^1_S \left( B_\Sigma^{(e)}, M_1 \right)_{0} \right)$ for $0 \leq e \leq 50$.}
\label{table1001}
\end{table}

\begin{figure}
\begin{center}

  \subfloat[$\text{dim}_{\mathbb{Q}} \left( \text{Hom}_S \left( B_\Sigma^{(e)}, M_1 \right)_{0} \right)$ for $0 \leq e \leq 50$.]{\label{figure 1001}

    \begin{tikzpicture}[scale = 0.45, x=.8cm,font=\sffamily]

      \draw[-stealth] (0,0) -- coordinate (x axis mid) (18,0) node [below] {$e$};
      \draw[-stealth] (0,0) -- coordinate (y axis mid) (0,12);

      \foreach \x in {0,6,12,...,48}
      \draw (\x/3,1pt) -- (\x/3,-3pt)
      node[anchor=north] {\x};

      \foreach \y in {0,4,8,...,20}
      \draw (1pt,\y/2) -- (-3pt,\y/2)
      node[anchor=east] {\y};

      \draw plot[mark=+,only marks] coordinates{ 
      (0/3, 0)
      (1/3, 0)
      (2/3, 0)
      (3/3, 0)
      (4/3, 0)
      (5/3, 0)
      (6/3, 0)
      (7/3, 0)
      (8/3, 0)
      (9/3, 0)
      (10/3, 0)
      (11/3, 0)
      (12/3, 0)
      (13/3, 0)
      (14/3, 0)
      (15/3, 0)
      (16/3, 0)
      (17/3, 0)
      (18/3, 0)
      (19/3, 0)
      (20/3, 0)
      (21/3, 0)
      (22/3, 0)
      (23/3, 0)
      (24/3, 0)
      (25/3, 0)
      (26/3, 0)
      (27/3, 0)
      (28/3, 0)
      (29/3, 0)
      (30/3, 0)
      (31/3, 0)
      (32/3, 0)
      (33/3, 0)
      (34/3, 0)
      (35/3, 0)
      (36/3, 2/2)
      (37/3, 8/2)
      (38/3, 15/2)
      (39/3, 22/2)
      (40/3, 22/2)
      (41/3, 22/2)
      (42/3, 22/2)
      (43/3, 22/2)
      (44/3, 22/2)
      (45/3, 22/2)
      (46/3, 22/2)
      (47/3, 22/2)
      (48/3, 22/2)
      (49/3, 22/2) 
      (50/3, 22/2)
      };
      
    \end{tikzpicture}
  }  \subfloat[$\text{dim}_{\mathbb{Q}} \left( \text{Ext}^1_S \left( B_\Sigma^{(e)}, M_1 \right)_{0} \right)$ for $0 \leq e \leq 50$.]{\label{figure 1002}

    \begin{tikzpicture}[scale = 0.45, x=.8cm,font=\sffamily]

      \draw[-stealth] (0,0) -- coordinate (x axis mid) (18,0) node [below] {$e$};
      \draw[-stealth] (0,0) -- coordinate (y axis mid) (0,12);

      \foreach \x in {0,6,12,...,48}
      \draw (\x/3,1pt) -- (\x/3,-3pt)
      node[anchor=north] {\x};

      \foreach \y in {0,100,...,600}
      \draw (1pt,\y/55) -- (-3pt,\y/55)
      node[anchor=east] {\y};

      \draw plot[mark=+,only marks] coordinates {
      (0/3, 0)
      (1/3, 0)
      (2/3, 0)
      (3/3, 0)
      (4/3, 0)
      (5/3, 0)
      (6/3, 0)
      (7/3, 0)
      (8/3, 3/55)
      (9/3, 15/55)
      (10/3, 36/55)
      (11/3, 62/55)
      (12/3, 89/55)
      (13/3, 116/55)
      (14/3, 150/55)
      (15/3, 195/55)
      (16/3, 256/55)
      (17/3, 325/55)
      (18/3, 394/55)
      (19/3, 467/55)
      (20/3, 528/55)
      (21/3, 573/55)
      (22/3, 600/55)
      (23/3, 603/55)
      (24/3, 591/55)
      (25/3, 576/55)
      (26/3, 558/55)
      (27/3, 537/55)
      (28/3, 513/55)
      (29/3, 486/55)
      (30/3, 456/55)
      (31/3, 423/55)
      (32/3, 387/55)
      (33/3, 348/55)
      (34/3, 309/55)
      (35/3, 273/55)
      (36/3, 239/55)
      (37/3, 209/55)
      (38/3, 183/55)
      (39/3, 160/55)
      (40/3, 133/55)
      (41/3, 109/55)
      (42/3, 88/55)
      (43/3, 70/55)
      (44/3, 56/55)
      (45/3, 46/55)
      (46/3, 43/55)
      (47/3, 43/55)
      (48/3, 43/55)
      (49/3, 43/55)
      (50/3, 43/55)
      };

    \end{tikzpicture}
  }
  \end{center}
  \caption{Computing the sheaf cohomologies of $\tilde{M_1}$ with \texttt{plesken}.}
  \label{fig1003}
\end{figure}

\subsection{The Module \texorpdfstring{$\mathbf{M_2}$}{M2}}

For this module \texttt{plesken} computed the following minimal free resolution:
\begin{align}
\begin{split}
0 & \to S \left( -40 \right)^{\oplus 2} \oplus S \left( -49 \right) \\
  &\to S \left( -34 \right) \oplus S \left( -39 \right)^{\oplus 2} \oplus S \left( -23 \right) \oplus S \left( -38 \right) \oplus S \left( -38 \right) \\
  &\to S \left( -6 \right) \oplus S \left( -21 \right) \twoheadrightarrow M_2 \to 0 \, .
\end{split}
\end{align}
From it we read off the Betti numbers of $M_2$. Thereby \texttt{plesken} found $e_0 = e_1 = 47$. Consequently $h^i ( C_{\mathbf{5}_{-2}}, \tilde{M_2} ) = ( 21, 42 )$. Explicitly we computed with \texttt{plesken} the values listed in \autoref{table1002}, which are visualised in \autoref{fig1004}. So in this case, our algorithm picked the ideal values $\check{e}_0 = \check{e}_1 = 47$.

\begin{table}
\begin{center}
{\footnotesize
\begin{tabular}{|c|c|c||c|c|c|}
\hline
$e$ & $\text{Hom}_S \left( B_\Sigma^{(e)}, M_2 \right)_{0}$ & $\text{Ext}^1_S \left( B_\Sigma^{(e)}, M_2 \right)_{0}$ & $e$ & $\text{Hom}_S \left( B_\Sigma^{(e)}, M_2 \right)_{0}$ & $\text{Ext}^1_S \left( B_\Sigma^{(e)}, M_2 \right)_{0}$ \\
\hline \hline
0 & 0 & 0 & 25 & 0 & 576 \\
1 & 0 & 0 & 26 & 1 & 559 \\
2 & 0 & 0 & 27 & 3 & 540 \\
3 & 0 & 0 & 28 & 6 & 519 \\
4 & 0 & 0 & 29 & 6 & 492 \\
\hline
5 & 0 & 0 & 30 & 6 & 462 \\
6 & 0 & 0 & 31 & 6 & 429 \\
7 & 0 & 0 & 32 & 7 & 394 \\
8 & 0 & 3 & 33 & 8 & 356 \\
9 & 0 & 15 & 34 & 9 & 318 \\
\hline
10 & 0 & 36 & 35 & 9 & 279 \\
11 & 0 & 62 & 36 & 9 & 240 \\
12 & 0 & 90 & 37 & 9 & 201 \\
13 & 0 & 125 & 38 & 9 & 165 \\
14 & 0 & 175 & 39 & 9 & 138 \\
\hline
15 & 0 & 236 & 40 & 9 & 114 \\
16 & 0 & 299 & 41 & 9 & 93 \\
17 & 0 & 366 & 42 & 10 & 76 \\
18 & 0 & 430 & 43 & 12 & 63 \\
19 & 0 & 498 & 44 & 15 & 54 \\
\hline
20 & 0 & 544 & 45 & 18 & 48 \\
21 & 0 & 576 & 46 & 20 & 44 \\
22 & 0 & 594 & 47 & 21 & 42 \\
23 & 0 & 600 & 48 & 21 & 42 \\
24 & 0 & 591 & 49 & 21 & 42 \\
   &   &     & 50 & 21 & 42 \\
\hline
\end{tabular}}
\end{center}
\caption{$\text{dim}_{\mathbb{Q}} \left( \text{Hom}_S \left( B_\Sigma^{(e)}, M_2 \right)_{0} \right)$ and $\text{dim}_{\mathbb{Q}} \left( \text{Ext}^1_S \left( B_\Sigma^{(e)}, M_2 \right)_{0} \right)$ for $0 \leq e \leq 50$.}
\label{table1002}
\end{table}

\begin{figure}
\begin{center}

  \subfloat[$\text{dim}_{\mathbb{Q}} \left( \text{Hom}_S \left( B_\Sigma^{(e)}, M_2 \right)_{0} \right)$ for $0 \leq e \leq 50$.]{\label{figure 10001}

    \begin{tikzpicture}[scale = 0.45, x=.8cm,font=\sffamily]

      \draw[-stealth] (0,0) -- coordinate (x axis mid) (18,0) node [below] {$e$};
      \draw[-stealth] (0,0) -- coordinate (y axis mid) (0,12);

      \foreach \x in {0,6,12,...,48}
      \draw (\x/3,1pt) -- (\x/3,-3pt)
      node[anchor=north] {\x};

      \foreach \y in {0,4,8,...,20}
      \draw (1pt,\y/2) -- (-3pt,\y/2)
      node[anchor=east] {\y};

      \draw plot[mark=+,only marks] coordinates { 
      (0/3, 0)
      (1/3, 0)
      (2/3, 0)
      (3/3, 0)
      (4/3, 0)
      (5/3, 0)
      (6/3, 0)
      (7/3, 0)
      (8/3, 0)
      (9/3, 0)
      (10/3, 0)
      (11/3, 0)
      (12/3, 0)
      (13/3, 0)
      (14/3, 0)
      (15/3, 0)
      (16/3, 0)
      (17/3, 0)
      (18/3, 0)
      (19/3, 0)
      (20/3, 0)
      (21/3, 0)
      (22/3, 0)
      (23/3, 0)
      (24/3, 0)
      (25/3, 0)
      (26/3, 1/2)
      (27/3, 3/2)
      (28/3, 6/2)
      (29/3, 6/2)
      (30/3, 6/2)
      (31/3, 6/2)
      (32/3, 7/2)
      (33/3, 8/2)
      (34/3, 9/2)
      (35/3, 9/2)
      (36/3, 9/2)
      (37/3, 9/2)
      (38/3, 9/2)
      (39/3, 9/2)
      (40/3, 9/2)
      (41/3, 9/2)
      (42/3, 10/2)
      (43/3, 12/2)
      (44/3, 15/2)
      (45/3, 18/2)
      (46/3, 20/2)
      (47/3, 21/2)
      (48/3, 21/2)
      (49/3, 21/2)
      (50/3, 21/2)
      };

    \end{tikzpicture}
  }
  \subfloat[$\text{dim}_{\mathbb{Q}} \left( \text{Ext}^1_S \left( B_\Sigma^{(e)}, M_2 \right)_{0} \right)$ for $0 \leq e \leq 50$.]{\label{figure 10002}

    \begin{tikzpicture}[scale = 0.45, x=.8cm,font=\sffamily]

      \draw[-stealth] (0,0) -- coordinate (x axis mid) (18,0) node [below] {$e$};
      \draw[-stealth] (0,0) -- coordinate (y axis mid) (0,12);

      \foreach \x in {0,6,12,...,48}
      \draw (\x/3,1pt) -- (\x/3,-3pt)
      node[anchor=north] {\x};

      \foreach \y in {0,100,...,600}
      \draw (1pt,\y/55) -- (-3pt,\y/55)
      node[anchor=east] {\y};

      \draw plot[mark=+,only marks] coordinates {
       (0/3, 0)
       (1/3, 0)
       (2/3, 0)
       (3/3, 0)
       (4/3, 0)
       (5/3, 0)
       (6/3, 0)
       (7/3, 0)
       (8/3, 3/55)
       (9/3, 15/55)
       (10/3, 36/55)
       (11/3, 62/55)
       (12/3, 90/55)
       (13/3, 125/55)
       (14/3, 175/55)
       (15/3, 236/55)
       (16/3, 299/55)
       (17/3, 366/55)
       (18/3, 430/55)
       (19/3, 498/55)
       (20/3, 544/55)
       (21/3, 576/55)
       (22/3, 594/55)
       (23/3, 600/55)
       (24/3, 591/55)
       (25/3, 576/55)
       (26/3, 559/55)
       (27/3, 540/55)
       (28/3, 519/55)
       (29/3, 492/55)
       (30/3, 462/55)
       (31/3, 429/55)
       (32/3, 394/55)
       (33/3, 356/55)
       (34/3, 318/55)
       (35/3, 279/55)
       (36/3, 240/55)
       (37/3, 201/55)
       (38/3, 165/55)
       (39/3, 138/55)
       (40/3, 114/55)
       (41/3, 93/55)
       (42/3, 76/55)
       (43/3, 63/55)
       (44/3, 54/55)
       (45/3, 48/55)
       (46/3, 44/55)
       (47/3, 42/55)
       (48/3, 42/55)
       (49/3, 42/55)
       (50/3, 42/55)
       };
       
    \end{tikzpicture}
  }
  \end{center}
  \caption{Computing the sheaf cohomologies of $\tilde{M_2}$ with \texttt{plesken}.}
  \label{fig1004}
\end{figure}

\subsection{The Module \texorpdfstring{$\mathbf{M_3}$}{M3}}

For the module $M_3$ \texttt{plesken} computed the following minimal free resolution:
\begin{align}
\begin{split}
0 & \to S \left( -36 \right) \oplus S \left( -47 \right) \oplus S \left( -48 \right) \\
  &\to S \left( -34 \right)^{\oplus 2} \oplus S \left( -35 \right) \oplus S \left( -45 \right) \oplus S \left( -23 \right) \oplus S \left( -38 \right) \\
  &\to S \left( -6 \right) \oplus S \left( -21 \right) \twoheadrightarrow M_3 \to 0 \, .
\end{split}
\end{align}
From it we deduce the Betti numbers of $M_3$ and thereby computed $e_0 = e_1 = 46$ by the help of \texttt{plesken}. Consequently $h^i ( C_{\mathbf{5}_{-2}}, \tilde{M_3} ) = ( 11, 32 )$. Explicitly we computed with \texttt{plesken} the values listed in \autoref{table1003}, which are visualised in \autoref{fig1016}. Hence the ideal values are $\check{e}_0 = 39$ and $\check{e}_1 = 45$.

\begin{table}
\begin{center}
{\footnotesize
\begin{tabular}{|c|c|c||c|c|c|}
\hline
$e$ & $\text{Hom}_S \left( B_\Sigma^{(e)}, M_3 \right)_{0}$ & $\text{Ext}^1_S \left( B_\Sigma^{(e)}, M_3 \right)_{0}$ & $e$ & $\text{Hom}_S \left( B_\Sigma^{(e)}, M_3 \right)_{0}$ & $\text{Ext}^1_S \left( B_\Sigma^{(e)}, M_3 \right)_{0}$ \\
\hline \hline
0 & 0 & 0 & 25 & 0 & 576 \\
1 & 0 & 0 & 26 & 0 & 558 \\
2 & 0 & 0 & 27 & 0 & 537 \\
3 & 0 & 0 & 28 & 1 & 514 \\
4 & 0 & 0 & 29 & 1 & 487 \\
\hline
5 & 0 & 0 & 30 & 1 & 457 \\
6 & 0 & 0 & 31 & 1 & 424 \\
7 & 0 & 0 & 32 & 2 & 389 \\
8 & 0 & 3 & 33 & 4 & 352 \\
9 & 0 & 15 & 34 & 6 & 312 \\
\hline
10 & 0 & 34 & 35 & 8 & 272 \\
11 & 0 & 60 & 36 & 9 & 231 \\
12 & 0 & 87 & 37 & 10 & 190 \\
13 & 0 & 113 & 38 & 10 & 151 \\
14 & 0 & 144 & 39 & 11 & 116 \\
\hline
15 & 0 & 198 & 40 & 11 & 86 \\
16 & 0 & 274 & 41 & 11 & 65 \\
17 & 0 & 350 & 42 & 11 & 50 \\
18 & 0 & 426 & 43 & 11 & 41 \\
19 & 0 & 502 & 44 & 11 & 35 \\
\hline
20 & 0 & 567 & 45 & 11 & 32 \\
21 & 0 & 600 & 46 & 11 & 32 \\
22 & 0 & 609 & 47 & 11 & 32 \\
23 & 0 & 603 & 48 & 11 & 32 \\
24 & 0 & 591 & 49 & 11 & 32 \\
   &   &     & 50 & 11 & 32 \\
\hline
\end{tabular}}
\end{center}
\caption{$\text{dim}_{\mathbb{Q}} \left( \text{Hom}_S \left( B_\Sigma^{(e)}, M_3 \right)_{0} \right)$ and $\text{dim}_{\mathbb{Q}} \left( \text{Ext}^1_S \left( B_\Sigma^{(e)}, M_3 \right)_{0} \right)$ for $0 \leq e \leq 50$.}
\label{table1003}
\end{table}

\begin{figure}
\begin{center}

  \subfloat[$\text{dim}_{\mathbb{Q}} \left( \text{Hom}_S \left( B_\Sigma^{(e)}, M_3 \right)_{0} \right)$ for $0 \leq e \leq 50$.]{\label{figure 19001}

    \begin{tikzpicture}[scale = 0.45, x=.8cm,font=\sffamily]

      \draw[-stealth] (0,0) -- coordinate (x axis mid) (18,0) node [below] {$e$};
      \draw[-stealth] (0,0) -- coordinate (y axis mid) (0,12);

      \foreach \x in {0,6,12,...,48}
      \draw (\x/3,1pt) -- (\x/3,-3pt)
      node[anchor=north] {\x};

      \foreach \y in {0,4,8,...,20}
      \draw (1pt,\y/2) -- (-3pt,\y/2)
      node[anchor=east] {\y};

      \draw plot[mark=+,only marks] coordinates{ 
       (0/3, 0)
       (1/3, 0)
       (2/3, 0)
       (3/3, 0)
       (4/3, 0)
       (5/3, 0)
       (6/3, 0)
       (7/3, 0)
       (8/3, 0)
       (9/3, 0)
       (10/3, 0)
       (11/3, 0)
       (12/3, 0)
       (13/3, 0)
       (14/3, 0)
       (15/3, 0)
       (16/3, 0)
       (17/3, 0)
       (18/3, 0)
       (19/3, 0)
       (20/3, 0)
       (21/3, 0)
       (22/3, 0)
       (23/3, 0)
       (24/3, 0)
       (25/3, 0)
       (26/3, 0)
       (27/3, 0)
       (28/3, 1/2)
       (29/3, 1/2)
       (30/3, 1/2)
       (31/3, 1/2)
       (32/3, 2/2)
       (33/3, 4/2)
       (34/3, 6/2)
       (35/3, 8/2)
       (36/3, 9/2)
       (37/3, 10/2)
       (38/3, 10/2)
       (39/3, 11/2)
       (40/3, 11/2)
       (41/3, 11/2)
       (42/3, 11/2)
       (43/3, 11/2)
       (44/3, 11/2)
       (45/3, 11/2)
       (46/3, 11/2)
       (47/3, 11/2)
       (48/3, 11/2)
       (49/3, 11/2)
       (50/3, 11/2)
      };
       
    \end{tikzpicture}
  }
  \subfloat[$\text{dim}_{\mathbb{Q}} \left( \text{Ext}^1_S \left( B_\Sigma^{(e)}, M_3 \right)_{0} \right)$ for $0 \leq e \leq 50$.]{\label{figure 19002}

    \begin{tikzpicture}[scale = 0.45, x=.8cm,font=\sffamily]

      \draw[-stealth] (0,0) -- coordinate (x axis mid) (18,0) node [below] {$e$};
      \draw[-stealth] (0,0) -- coordinate (y axis mid) (0,12);

      \foreach \x in {0,6,12,...,48}
      \draw (\x/3,1pt) -- (\x/3,-3pt)
      node[anchor=north] {\x};

      \foreach \y in {0,100,...,600}
      \draw (1pt,\y/55) -- (-3pt,\y/55)
      node[anchor=east] {\y};

      \draw plot[mark=+,only marks] coordinates{ 
       (0/3, 0)
       (1/3, 0)
       (2/3, 0)
       (3/3, 0)
       (4/3, 0)
       (5/3, 0)
       (6/3, 0)
       (7/3, 0)
       (8/3, 3/55)
       (9/3, 15/55)
       (10/3, 34/55)
       (11/3, 60/55)
       (12/3, 87/55)
       (13/3, 113/55)
       (14/3, 144/55)
       (15/3, 198/55)
       (16/3, 274/55)
       (17/3, 350/55)
       (18/3, 426/55)
       (19/3, 502/55)
       (20/3, 567/55)
       (21/3, 600/55)
       (22/3, 609/55)
       (23/3, 603/55)
       (24/3, 591/55)
       (25/3, 576/55)
       (26/3, 558/55)
       (27/3, 537/55)
       (28/3, 514/55)
       (29/3, 487/55)
       (30/3, 457/55)
       (31/3, 424/55)
       (32/3, 389/55)
       (33/3, 352/55)
       (34/3, 312/55)
       (35/3, 272/55)
       (36/3, 231/55)
       (37/3, 190/55)
       (38/3, 151/55)
       (39/3, 116/55)
       (40/3, 86/55)
       (41/3, 65/55)
       (42/3, 50/55)
       (43/3, 41/55)
       (44/3, 35/55)
       (45/3, 32/55)
       (46/3, 32/55)
       (47/3, 32/55)
       (48/3, 32/55)
       (49/3, 32/55)
       (50/3, 32/55)
      };

    \end{tikzpicture}
  }
  \end{center}
  \caption{Computing the sheaf cohomologies of $\tilde{M_3}$ with \texttt{plesken}.}
  \label{fig1016}
\end{figure}

\subsection{The Module \texorpdfstring{$\mathbf{M_4}$}{M4}}

For the module $M_4$ a minimal free resolution is given by
\begin{align}
\begin{split}
0 & \to S \left( -37 \right) \oplus S \left( -45 \right)^{\oplus 3} \\
  &\to S \left( -36 \right)^{\oplus 2} \oplus S \left( -40 \right) \oplus S \left( -43 \right) \oplus S \left( -23 \right) \oplus S \left( -38 \right) \\
  &\to S \left( -6 \right) \oplus S \left( -21 \right) \twoheadrightarrow M_4 \to 0 \, .
\end{split}
\end{align}
The Betti numbers of $M_4$ therefore let \texttt{plesken} to compute $e_0 = e_1 = 43$. This in turn implies $h^i ( C_{\mathbf{5}_{-2}}, \tilde{M_4} ) = ( 9, 30 )$. Explicitly we computed with \texttt{plesken} the values listed in \autoref{table1004}, which are visualised in \autoref{fig1005}. Hence, for this module our algorithm picked the ideal values $\check{e}_0 = \check{e}_1 = 43$.

\begin{table}
\begin{center}
{\footnotesize
\begin{tabular}{|c|c|c||c|c|c|}
\hline
$e$ & $\text{Hom}_S \left( B_\Sigma^{(e)}, M_4 \right)_{0}$ & $\text{Ext}^1_S \left( B_\Sigma^{(e)}, M_4 \right)_{0}$ & $e$ & $\text{Hom}_S \left( B_\Sigma^{(e)}, M_4 \right)_{0}$ & $\text{Ext}^1_S \left( B_\Sigma^{(e)}, M_4 \right)_{0}$ \\
\hline \hline
0 & 0 & 0 & 25 & 0 & 576 \\
1 & 0 & 0 & 26 & 0 & 558 \\
2 & 0 & 0 & 27 & 0 & 537 \\
3 & 0 & 0 & 28 & 0 & 513 \\
4 & 0 & 0 & 29 & 0 & 486 \\
\hline
5 & 0 & 0 & 30 & 0 & 456 \\
6 & 0 & 0 & 31 & 0 & 423 \\
7 & 0 & 0 & 32 & 0 & 387 \\
8 & 0 & 3 & 33 & 0 & 348 \\
9 & 0 & 15 & 34 & 0 & 306 \\
\hline
10 & 0 & 36 & 35 & 0 & 261 \\
11 & 0 & 62 & 36 & 0 & 219\\
12 & 0 & 89 & 37 & 2 & 179 \\
13 & 0 & 114 & 38 & 2 & 140 \\
14 & 0 & 143 & 39 & 2 & 104 \\
\hline
15 & 0 & 186 & 40 & 2 & 74 \\
16 & 0 & 249 & 41 & 5 & 53 \\
17 & 0 & 324 & 42 & 9 & 39 \\
18 & 0 & 410 & 43 & 9 & 30 \\
19 & 0 & 496 & 44 & 9 & 30 \\
\hline
20 & 0 & 567 & 45 & 9 & 30 \\
21 & 0 & 609 & 46 & 9 & 30 \\
22 & 0 & 612 & 47 & 9 & 30 \\
23 & 0 & 603 & 48 & 9 & 30 \\
24 & 0 & 591 & 49 & 9 & 30 \\
   &   &     & 50 & 9 & 30 \\
\hline
\end{tabular}}
\end{center}
\caption{$\text{dim}_{\mathbb{Q}} \left( \text{Hom}_S \left( B_\Sigma^{(e)}, M_4 \right)_{0} \right)$ and $\text{dim}_{\mathbb{Q}} \left( \text{Ext}^1_S \left( B_\Sigma^{(e)}, M_4 \right)_{0} \right)$ for $0 \leq e \leq 50$.}
\label{table1004}
\end{table}

\begin{figure}
\begin{center}

  \subfloat[$\text{dim}_{\mathbb{Q}} \left( \text{Hom}_S \left( B_\Sigma^{(e)}, M_4 \right)_{0} \right)$ for $0 \leq e \leq 50$.]{\label{figure 90001}

    \begin{tikzpicture}[scale = 0.45, x=.8cm,font=\sffamily]

      \draw[-stealth] (0,0) -- coordinate (x axis mid) (18,0) node [below] {$e$};
      \draw[-stealth] (0,0) -- coordinate (y axis mid) (0,12);

      \foreach \x in {0,6,12,...,48}
      \draw (\x/3,1pt) -- (\x/3,-3pt)
      node[anchor=north] {\x};

      \foreach \y in {0,4,8,...,20}
      \draw (1pt,\y/2) -- (-3pt,\y/2)
      node[anchor=east] {\y};

      \draw plot[mark=+,only marks] coordinates {
       (0/3, 0)
       (1/3, 0)
       (2/3, 0)
       (3/3, 0)
       (4/3, 0)
       (5/3, 0)
       (6/3, 0)
       (7/3, 0)
       (8/3, 0)
       (9/3, 0)
       (10/3, 0)
       (11/3, 0)
       (12/3, 0)
       (13/3, 0)
       (14/3, 0)
       (15/3, 0)
       (16/3, 0)
       (17/3, 0)
       (18/3, 0)
       (19/3, 0)
       (20/3, 0)
       (21/3, 0)
       (22/3, 0)
       (23/3, 0)
       (24/3, 0)
       (25/3, 0)
       (26/3, 0)
       (27/3, 0)
       (28/3, 0)
       (29/3, 0)
       (30/3, 0)
       (31/3, 0)
       (32/3, 0)
       (33/3, 0)
       (34/3, 0)
       (35/3, 0)
       (36/3, 0/2)
       (37/3, 2/2)
       (38/3, 2/2)
       (39/3, 2/2)
       (40/3, 2/2)
       (41/3, 5/2)
       (42/3, 9/2)
       (43/3, 9/2)
       (44/3, 9/2)
       (45/3, 9/2)
       (46/3, 9/2)
       (47/3, 9/2)
       (48/3, 9/2)
       (49/3, 9/2)
       (50/3, 9/2)
      };

    \end{tikzpicture}
  }
  \subfloat[$\text{dim}_{\mathbb{Q}} \left( \text{Ext}^1_S \left( B_\Sigma^{(e)}, M_4 \right)_{0} \right)$ for $0 \leq e \leq 50$.]{\label{figure 90002}

    \begin{tikzpicture}[scale = 0.45, x=.8cm,font=\sffamily]

      \draw[-stealth] (0,0) -- coordinate (x axis mid) (18,0) node [below] {$e$};
      \draw[-stealth] (0,0) -- coordinate (y axis mid) (0,12);

      \foreach \x in {0,6,12,...,48}
      \draw (\x/3,1pt) -- (\x/3,-3pt)
      node[anchor=north] {\x};

      \foreach \y in {0,100,...,600}
      \draw (1pt,\y/55) -- (-3pt,\y/55)
      node[anchor=east] {\y};

      \draw plot[mark=+,only marks] coordinates {
       (0/3, 0)
       (1/3, 0)
       (2/3, 0)
       (3/3, 0)
       (4/3, 0)
       (5/3, 0)
       (6/3, 0)
       (7/3, 0)
       (8/3, 3/55)
       (9/3, 15/55)
       (10/3, 36/55)
       (11/3, 62/55)
       (12/3, 89/55)
       (13/3, 114/55)
       (14/3, 143/55)
       (15/3, 186/55)
       (16/3, 249/55)
       (17/3, 324/55)
       (18/3, 410/55)
       (19/3, 496/55)
       (20/3, 567/55)
       (21/3, 609/55)
       (22/3, 612/55)
       (23/3, 603/55)
       (24/3, 591/55)
       (25/3, 576/55)
       (26/3, 558/55)
       (27/3, 537/55)
       (28/3, 513/55)
       (29/3, 486/55)
       (30/3, 456/55)
       (31/3, 423/55)
       (32/3, 387/55)
       (33/3, 348/55)
       (34/3, 306/55)
       (35/3, 261/55)
       (36/3, 219/55)
       (37/3, 179/55)
       (38/3, 140/55)
       (39/3, 104/55)
       (40/3, 74/55)
       (41/3, 53/55)
       (42/3, 39/55)
       (43/3, 30/55)
       (44/3, 30/55)
       (45/3, 30/55)
       (46/3, 30/55)
       (47/3, 30/55)
       (48/3, 30/55)
       (49/3, 30/55)
       (50/3, 30/55)
      };

    \end{tikzpicture}
  }
  \end{center}
  \caption{Computing the sheaf cohomologies of $\tilde{M_4}$ with \texttt{plesken}.}
  \label{fig1005}
\end{figure}

\subsection{The Module \texorpdfstring{$\mathbf{M_5}$}{M5}}

The module $M_5$ admits the following minimal free resolution:
\begin{align}
\begin{split}
0 & \to S \left( -36 \right) \oplus S \left( -47 \right) \oplus S \left( -48 \right) \\
  &\to S \left( -34 \right)^{\oplus 2} \oplus S \left( -35 \right) \oplus S \left( -45 \right) \oplus S \left( -23 \right)  \oplus S \left( -38 \right) \\
  &\to S \left( -6 \right) \oplus S \left( -21 \right) \twoheadrightarrow M_5 \to 0 \, .
\end{split}
\end{align}
The Betti numbers of $M_5$ let \texttt{plesken} to compute $e_0 = e_1 = 46$, which implies $h^i ( C_{\mathbf{5}_{-2}}, \tilde{M_5} ) = ( 7, 28 )$. Explicitly we computed with \texttt{plesken} the values listed in \autoref{table1005}, which are visualised in \autoref{fig10016}. Thereby we conclude that the ideal values are $\check{e}_0 = 38$ and $\check{e}_1 = 44$ for this module.

\begin{table}
\begin{center}
{\footnotesize
\begin{tabular}{|c|c|c||c|c|c|}
\hline
$e$ & $\text{Hom}_S \left( B_\Sigma^{(e)}, M_5 \right)_{0}$ & $\text{Ext}^1_S \left( B_\Sigma^{(e)}, M_5 \right)_{0}$ & $e$ & $\text{Hom}_S \left( B_\Sigma^{(e)}, M_5 \right)_{0}$ & $\text{Ext}^1_S \left( B_\Sigma^{(e)}, M_5 \right)_{0}$ \\
\hline \hline
0 & 0 & 0 & 25 & 0 & 576 \\
1 & 0 & 0 & 26 & 0 & 558 \\
2 & 0 & 0 & 27 & 0 & 537 \\
3 & 0 & 0 & 28 & 1 & 514 \\
4 & 0 & 0 & 29 & 1 & 487 \\
\hline
5 & 0 & 0 & 30 & 1 & 457 \\
6 & 0 & 0 & 31 & 1 & 424 \\
7 & 0 & 0 & 32 & 1 & 388 \\
8 & 0 & 3 & 33 & 1 & 349 \\
9 & 0 & 15 & 34 & 2 & 308 \\
\hline
10 & 0 & 34 & 35 & 4 & 265 \\
11 & 0 & 60 & 36 & 5 & 221 \\
12 & 0 & 87 & 37 & 6 & 177 \\
13 & 0 & 113 & 38 & 7 & 136 \\
14 & 0 & 145 & 39 & 7 & 100 \\
\hline
15 & 0 & 203 & 40 & 7 & 70 \\
16 & 0 & 273 & 41 & 7 & 49 \\
17 & 0 & 350 & 42 & 7 & 37 \\
18 & 0 & 432 & 43 & 7 & 31 \\
19 & 0 & 515 & 44 & 7 & 28 \\
\hline
20 & 0 & 579 & 45 & 7 & 28 \\
21 & 0 & 609 & 46 & 7 & 28 \\
22 & 0 & 612 & 47 & 7 & 28 \\
23 & 0 & 603 & 48 & 7 & 28 \\
24 & 0 & 591 & 49 & 7 & 28 \\
   &   &     & 50 & 7 & 28 \\
\hline
\end{tabular}}
\end{center}
\caption{$\text{dim}_{\mathbb{Q}} \left( \text{Hom}_S \left( B_\Sigma^{(e)}, M_5 \right)_{0} \right)$ and $\text{dim}_{\mathbb{Q}} \left( \text{Ext}^1_S \left( B_\Sigma^{(e)}, M_5 \right)_{0} \right)$ for $0 \leq e \leq 50$.}
\label{table1005}
\end{table}

\begin{figure}
\begin{center}

  \subfloat[$\text{dim}_{\mathbb{Q}} \left( \text{Hom}_S \left( B_\Sigma^{(e)}, M_5 \right)_{0} \right)$ for $0 \leq e \leq 50$.]{\label{figure 910001}

    \begin{tikzpicture}[scale = 0.45, x=.8cm,font=\sffamily]

      \draw[-stealth] (0,0) -- coordinate (x axis mid) (18,0) node [below] {$e$};
      \draw[-stealth] (0,0) -- coordinate (y axis mid) (0,12);

      \foreach \x in {0,6,12,...,48}
      \draw (\x/3,1pt) -- (\x/3,-3pt)
      node[anchor=north] {\x};

      \foreach \y in {0,4,8,...,20}
      \draw (1pt,\y/2) -- (-3pt,\y/2)
      node[anchor=east] {\y};

      \draw plot[mark=+,only marks] coordinates {
       (0/3, 0)
       (1/3, 0)
       (2/3, 0)
       (3/3, 0)
       (4/3, 0)
       (5/3, 0)
       (6/3, 0)
       (7/3, 0)
       (8/3, 0)
       (9/3, 0)
       (10/3, 0)
       (11/3, 0)
       (12/3, 0)
       (13/3, 0)
       (14/3, 0)
       (15/3, 0)
       (16/3, 0)
       (17/3, 0)
       (18/3, 0)
       (19/3, 0)
       (20/3, 0)
       (21/3, 0)
       (22/3, 0)
       (23/3, 0)
       (24/3, 0)
       (25/3, 0)
       (26/3, 0)
       (27/3, 0)
       (28/3, 1/2)
       (29/3, 1/2)
       (30/3, 1/2)
       (31/3, 1/2)
       (32/3, 1/2)
       (33/3, 1/2)
       (34/3, 2/2)
       (35/3, 4/2)
       (36/3, 5/2)
       (37/3, 6/2)
       (38/3, 7/2)
       (39/3, 7/2)
       (40/3, 7/2)
       (41/3, 7/2)
       (42/3, 7/2)
       (43/3, 7/2)
       (44/3, 7/2)
       (45/3, 7/2)
       (46/3, 7/2)
       (47/3, 7/2)
       (48/3, 7/2)
       (49/3, 7/2)
       (50/3, 7/2)
      };

    \end{tikzpicture}
  }
  \subfloat[$\text{dim}_{\mathbb{Q}} \left( \text{Ext}^1_S \left( B_\Sigma^{(e)}, M_5 \right)_{0} \right)$ for $0 \leq e \leq 50$.]{\label{figure 910002}

    \begin{tikzpicture}[scale = 0.45, x=.8cm,font=\sffamily]

      \draw[-stealth] (0,0) -- coordinate (x axis mid) (18,0) node [below] {$e$};
      \draw[-stealth] (0,0) -- coordinate (y axis mid) (0,12);

      \foreach \x in {0,6,12,...,48}
      \draw (\x/3,1pt) -- (\x/3,-3pt)
      node[anchor=north] {\x};

      \foreach \y in {0,100,...,600}
      \draw (1pt,\y/55) -- (-3pt,\y/55)
      node[anchor=east] {\y};

      \draw plot[mark=+,only marks] coordinates{ 
       (0/3, 0)
       (1/3, 0)
       (2/3, 0)
       (3/3, 0)
       (4/3, 0)
       (5/3, 0)
       (6/3, 0)
       (7/3, 0)
       (8/3, 3/55)
       (9/3, 15/55)
       (10/3, 34/55)
       (11/3, 60/55)
       (12/3, 87/55)
       (13/3, 113/55)
       (14/3, 145/55)
       (15/3, 203/55)
       (16/3, 273/55)
       (17/3, 350/55)
       (18/3, 432/55)
       (19/3, 515/55)
       (20/3, 579/55)
       (21/3, 609/55)
       (22/3, 612/55)
       (23/3, 603/55)
       (24/3, 591/55)
       (25/3, 576/55)
       (26/3, 558/55)
       (27/3, 537/55)
       (28/3, 514/55)
       (29/3, 487/55)
       (30/3, 457/55)
       (31/3, 424/55)
       (32/3, 388/55)
       (33/3, 349/55)
       (34/3, 308/55)
       (35/3, 265/55)
       (36/3, 221/55)
       (37/3, 177/55)
       (38/3, 136/55)
       (39/3, 100/55)
       (40/3, 70/55)
       (41/3, 49/55)
       (42/3, 37/55)
       (43/3, 31/55)
       (44/3, 28/55)
       (45/3, 28/55)
       (46/3, 28/55)
       (47/3, 28/55)
       (48/3, 28/55)
       (49/3, 28/55)
       (50/3, 28/55)
      };

    \end{tikzpicture}
  }
  \end{center}
  \caption{Computing the sheaf cohomologies of $\tilde{M_5}$ with \texttt{plesken}.}
  \label{fig10016}
\end{figure}

\subsection{The Module \texorpdfstring{$\mathbf{M_6}$}{M6}}

We compute with the help of \texttt{plesken} the following minimal free resolution of $M_6$:
\begin{align}
\begin{split}
0 & \to S \left( -43 \right) \oplus S \left( -45 \right)^{\oplus 2} \\
  &\to S \left( -37 \right)^{\oplus 2} \oplus S \left( -39 \right) \oplus S \left( -41 \right) \oplus S \left( -23 \right) \\
  &\to S \left( -6 \right) \oplus S \left( -21 \right) \twoheadrightarrow M_6 \to 0 \, .
\end{split}
\end{align}
From its Betti numbers we computed $e_0 = e_1 = 43$ by the help of \texttt{plesken}. Consequently $h^i ( C_{\mathbf{5}_{-2}}, \tilde{M_6} ) = ( 6, 27 )$. Explicitly we computed with \texttt{plesken} the values listed in \autoref{table1006}, which are visualised in \autoref{fig1006}. Thereby we conclude that the ideal values are $\check{e}_0 = 35$ and $\check{e}_1 = 43$.

\begin{table}
\begin{center}
{\footnotesize
\begin{tabular}{|c|c|c||c|c|c|}
\hline
$e$ & $\text{Hom}_S \left( B_\Sigma^{(e)}, M_6 \right)_{0}$ & $\text{Ext}^1_S \left( B_\Sigma^{(e)}, M_6 \right)_{0}$ & $e$ & $\text{Hom}_S \left( B_\Sigma^{(e)}, M_6 \right)_{0}$ & $\text{Ext}^1_S \left( B_\Sigma^{(e)}, M_6 \right)_{0}$ \\
\hline \hline
0 & 0 & 0 & 25 & 1 & 577 \\
1 & 0 & 0 & 26 & 1 & 559 \\
2 & 0 & 0 & 27 & 1 & 538 \\
3 & 0 & 0 & 28 & 1 & 514 \\
4 & 0 & 0 & 29 & 1 & 487 \\
\hline
5 & 0 & 0 & 30 & 1 & 457 \\
6 & 0 & 0 & 31 & 1 & 424 \\
7 & 0 & 0 & 32 & 1 & 388 \\
8 & 0 & 3 & 33 & 2 & 350 \\
9 & 0 & 15 & 34 & 4 & 310 \\
\hline
10 & 0 & 36 & 35 & 6 & 267 \\
11 & 0 & 62 & 36 & 6 & 219 \\
12 & 0 & 89 & 37 & 6 & 174 \\
13 & 0 & 114 & 38 & 6 & 132 \\
14 & 0 & 149 & 39 & 6 & 96 \\
\hline
15 & 0 & 204 & 40 & 6 & 66 \\
16 & 0 & 276 & 41 & 6 & 45 \\
17 & 0 & 356 & 42 & 6 & 33 \\
18 & 0 & 445 & 43 & 6 & 27 \\
19 & 0 & 529 & 44 & 6 & 27 \\
\hline
20 & 0 & 592 & 45 & 6 & 27 \\
21 & 1 & 615 & 46 & 6 & 27 \\
22 & 1 & 613 & 47 & 6 & 27 \\
23 & 1 & 604 & 48 & 6 & 27 \\
24 & 1 & 592 & 49 & 6 & 27 \\
   &   &     & 50 & 6 & 27 \\
\hline
\end{tabular}}
\end{center}
\caption{$\text{dim}_{\mathbb{Q}} \left( \text{Hom}_S \left( B_\Sigma^{(e)}, M_6 \right)_{0} \right)$ and $\text{dim}_{\mathbb{Q}} \left( \text{Ext}^1_S \left( B_\Sigma^{(e)}, M_6 \right)_{0} \right)$ for $0 \leq e \leq 50$.}
\label{table1006}
\end{table}

\begin{figure}
\begin{center}

  \subfloat[$\text{dim}_{\mathbb{Q}} \left( \text{Hom}_S \left( B_\Sigma^{(e)}, M_6 \right)_{0} \right)$ for $0 \leq e \leq 50$.]{\label{figure 100901}

    \begin{tikzpicture}[scale = 0.45, x=.8cm,font=\sffamily]

      \draw[-stealth] (0,0) -- coordinate (x axis mid) (18,0) node [below] {$e$};
      \draw[-stealth] (0,0) -- coordinate (y axis mid) (0,12);

      \foreach \x in {0,6,12,...,48}
      \draw (\x/3,1pt) -- (\x/3,-3pt)
      node[anchor=north] {\x};

      \foreach \y in {0,4,8,...,20}
      \draw (1pt,\y/2) -- (-3pt,\y/2)
      node[anchor=east] {\y};

      \draw plot[mark=+,only marks] coordinates{ 
       (0/3, 0)
       (1/3, 0)
       (2/3, 0)
       (3/3, 0)
       (4/3, 0)
       (5/3, 0)
       (6/3, 0)
       (7/3, 0)
       (8/3, 0)
       (9/3, 0)
       (10/3, 0)
       (11/3, 0)
       (12/3, 0)
       (13/3, 0)
       (14/3, 0)
       (15/3, 0)
       (16/3, 0)
       (17/3, 0)
       (18/3, 0)
       (19/3, 0)
       (20/3, 0)
       (21/3, 1/2)
       (22/3, 1/2)
       (23/3, 1/2)
       (24/3, 1/2)
       (25/3, 1/2)
       (26/3, 1/2)
       (27/3, 1/2)
       (28/3, 1/2)
       (29/3, 1/2)
       (30/3, 1/2)
       (31/3, 1/2)
       (32/3, 1/2)
       (33/3, 2/2)
       (34/3, 4/2)
       (35/3, 6/2)
       (36/3, 6/2)
       (37/3, 6/2)
       (38/3, 6/2)
       (39/3, 6/2)
       (40/3, 6/2)
       (41/3, 6/2)
       (42/3, 6/2)
       (43/3, 6/2)
       (44/3, 6/2)
       (45/3, 6/2)
       (46/3, 6/2)
       (47/3, 6/2)
       (48/3, 6/2)
       (49/3, 6/2)
       (50/3, 6/2)
      };

    \end{tikzpicture}
  }
  \subfloat[$\text{dim}_{\mathbb{Q}} \left( \text{Ext}^1_S \left( B_\Sigma^{(e)}, M_6 \right)_{0} \right)$ for $0 \leq e \leq 50$.]{\label{figure 100902}

    \begin{tikzpicture}[scale = 0.45, x=.8cm,font=\sffamily]

      \draw[-stealth] (0,0) -- coordinate (x axis mid) (18,0) node [below] {$e$};
      \draw[-stealth] (0,0) -- coordinate (y axis mid) (0,12);

      \foreach \x in {0,6,12,...,48}
      \draw (\x/3,1pt) -- (\x/3,-3pt)
      node[anchor=north] {\x};

      \foreach \y in {0,100,...,600}
      \draw (1pt,\y/55) -- (-3pt,\y/55)
      node[anchor=east] {\y};

      \draw plot[mark=+,only marks] coordinates{ 
       (0/3, 0)
       (1/3, 0)
       (2/3, 0)
       (3/3, 0)
       (4/3, 0)
       (5/3, 0)
       (6/3, 0)
       (7/3, 0)
       (8/3, 3/55)
       (9/3, 15/55)
       (10/3, 36/55)
       (11/3, 62/55)
       (12/3, 89/55)
       (13/3, 114/55)
       (14/3, 149/55)
       (15/3, 204/55)
       (16/3, 276/55)
       (17/3, 356/55)
       (18/3, 445/55)
       (19/3, 529/55)
       (20/3, 592/55)
       (21/3, 615/55)
       (22/3, 613/55)
       (23/3, 604/55)
       (24/3, 592/55)
       (25/3, 577/55)
       (26/3, 559/55)
       (27/3, 538/55)
       (28/3, 514/55)
       (29/3, 487/55)
       (30/3, 457/55)
       (31/3, 424/55)
       (32/3, 388/55)
       (33/3, 350/55)
       (34/3, 310/55)
       (35/3, 267/55)
       (36/3, 219/55)
       (37/3, 174/55)
       (38/3, 132/55)
       (39/3, 96/55)
       (40/3, 66/55)
       (41/3, 45/55)
       (42/3, 33/55)
       (43/3, 27/55)
       (44/3, 27/55)
       (45/3, 27/55)
       (46/3, 27/55)
       (47/3, 27/55)
       (48/3, 27/55)
       (49/3, 27/55)
       (50/3, 27/55)
      }; 

    \end{tikzpicture}
  }
  \end{center}
  \caption{Computing the sheaf cohomologies of $\tilde{M_6}$ with \texttt{plesken}.}
  \label{fig1006}
\end{figure}

\subsection{The Module \texorpdfstring{$\mathbf{M_7}$}{M7}}

The module $M_7$ admits the following minimal free resolution:
\begin{align}
\begin{split}
0 & \to S \left( -44 \right) \oplus S \left( -45 \right) \oplus S \left( -44 \right) \\
  &\to S \left( -37 \right) \oplus S \left( -38 \right) \oplus S \left( -41 \right) \oplus S \left( -23 \right) \oplus S \left( -38 \right) \\
  &\to S \left( -6 \right) \oplus S \left( -21 \right) \twoheadrightarrow M_7 \to 0 \, .
\end{split}
\end{align}
Its Betti numbers led \texttt{plesken} to conclude $e_0 = e_1 = 43$. Consequently $h^i ( C_{\mathbf{5}_{-2}}, \tilde{M_7} ) = ( 5, 26 )$. Explicitly we computed with \texttt{plesken} the values listed in \autoref{table1007}, which are visualised in \autoref{fig1007}. Thereby we conclude that the ideal values are $\check{e}_0 = 37$ and $\check{e}_1 = 43$.

\begin{table}
\begin{center}
{\footnotesize
\begin{tabular}{|c|c|c||c|c|c|}
\hline
$e$ & $\text{Hom}_S \left( B_\Sigma^{(e)}, M_7 \right)_{0}$ & $\text{Ext}^1_S \left( B_\Sigma^{(e)}, M_7 \right)_{0}$ & $e$ & $\text{Hom}_S \left( B_\Sigma^{(e)}, M_7 \right)_{0}$ & $\text{Ext}^1_S \left( B_\Sigma^{(e)}, M_7 \right)_{0}$ \\
\hline \hline
0 & 0 & 0 & 25 & 0 & 576 \\
1 & 0 & 0 & 26 & 0 & 558 \\
2 & 0 & 0 & 27 & 0 & 537 \\
3 & 0 & 0 & 28 & 1 & 514 \\
4 & 0 & 0 & 29 & 1 & 487 \\
\hline
5 & 0 & 0 & 30 & 1 & 457 \\
6 & 0 & 0 & 31 & 1 & 424 \\
7 & 0 & 0 & 32 & 1 & 388 \\
8 & 0 & 3 & 33 & 1 & 349 \\
9 & 0 & 15 & 34 & 1 & 307 \\
\hline
10 & 0 & 34 & 35 & 1 & 262 \\
11 & 0 & 60 & 36 & 2 & 215 \\
12 & 0 & 87 & 37 & 5 & 170 \\
13 & 0 & 113 & 38 & 5 & 128 \\
14 & 0 & 150 & 39 & 5 & 92 \\
\hline
15 & 0 & 204 & 40 & 5 & 62 \\
16 & 0 & 275 & 41 & 5 & 41 \\
17 & 0 & 354 & 42 & 5 & 29 \\
18 & 0 & 443 & 43 & 5 & 26 \\
19 & 0 & 526 & 44 & 5 & 26 \\
\hline
20 & 0 & 587 & 45 & 5 & 26 \\
21 & 0 & 615 & 46 & 5 & 26 \\
22 & 0 & 612 & 47 & 5 & 26 \\
23 & 0 & 603 & 48 & 5 & 26 \\
24 & 0 & 591 & 49 & 5 & 26 \\
   &   &     & 50 & 5 & 26 \\
\hline
\end{tabular}}
\end{center}
\caption{$\text{dim}_{\mathbb{Q}} \left( \text{Hom}_S \left( B_\Sigma^{(e)}, M_7 \right)_{0} \right)$ and $\text{dim}_{\mathbb{Q}} \left( \text{Ext}^1_S \left( B_\Sigma^{(e)}, M_7 \right)_{0} \right)$ for $0 \leq e \leq 50$.}
\label{table1007}
\end{table}

\begin{figure}
\begin{center}

  \subfloat[$\text{dim}_{\mathbb{Q}} \left( \text{Hom}_S \left( B_\Sigma^{(e)}, M_7 \right)_{0} \right)$ for $0 \leq e \leq 50$.]{\label{figure 130001}

    \begin{tikzpicture}[scale = 0.45, x=.8cm,font=\sffamily]

      \draw[-stealth] (0,0) -- coordinate (x axis mid) (18,0) node [below] {$e$};
      \draw[-stealth] (0,0) -- coordinate (y axis mid) (0,12);

      \foreach \x in {0,6,12,...,48}
      \draw (\x/3,1pt) -- (\x/3,-3pt)
      node[anchor=north] {\x};

      \foreach \y in {0,4,8,...,20}
      \draw (1pt,\y/2) -- (-3pt,\y/2)
      node[anchor=east] {\y};

      \draw plot[mark=+,only marks] coordinates{ 
       (0/3, 0)
       (1/3, 0)
       (2/3, 0)
       (3/3, 0)
       (4/3, 0)
       (5/3, 0)
       (6/3, 0)
       (7/3, 0)
       (8/3, 0)
       (9/3, 0)
       (10/3, 0)
       (11/3, 0)
       (12/3, 0)
       (13/3, 0)
       (14/3, 0)
       (15/3, 0)
       (16/3, 0)
       (17/3, 0)
       (18/3, 0)
       (19/3, 0)
       (20/3, 0)
       (21/3, 0/2)
       (22/3, 0/2)
       (23/3, 0/2)
       (24/3, 0/2)
       (25/3, 0/2)
       (26/3, 0/2)
       (27/3, 0/2)
       (28/3, 1/2)
       (29/3, 1/2)
       (30/3, 1/2)
       (31/3, 1/2)
       (32/3, 1/2)
       (33/3, 1/2)
       (34/3, 1/2)
       (35/3, 1/2)
       (36/3, 2/2)
       (37/3, 5/2)
       (38/3, 5/2)
       (39/3, 5/2)
       (40/3, 5/2)
       (41/3, 5/2)
       (42/3, 5/2)
       (43/3, 5/2)
       (44/3, 5/2)
       (45/3, 5/2)
       (46/3, 5/2)
       (47/3, 5/2)
       (48/3, 5/2)
       (49/3, 5/2)
       (50/3, 5/2)
      };

    \end{tikzpicture}
  }
  \subfloat[$\text{dim}_{\mathbb{Q}} \left( \text{Ext}^1_S \left( B_\Sigma^{(e)}, M_7 \right)_{0} \right)$ for $0 \leq e \leq 50$.]{\label{figure 130002}

    \begin{tikzpicture}[scale = 0.45, x=.8cm,font=\sffamily]

      \draw[-stealth] (0,0) -- coordinate (x axis mid) (18,0) node [below] {$e$};
      \draw[-stealth] (0,0) -- coordinate (y axis mid) (0,12);

      \foreach \x in {0,6,12,...,48}
      \draw (\x/3,1pt) -- (\x/3,-3pt)
      node[anchor=north] {\x};

      \foreach \y in {0,100,...,600}
      \draw (1pt,\y/55) -- (-3pt,\y/55)
      node[anchor=east] {\y};

      \draw plot[mark=+,only marks] coordinates{ 
       (0/3, 0)
       (1/3, 0)
       (2/3, 0)
       (3/3, 0)
       (4/3, 0)
       (5/3, 0)
       (6/3, 0)
       (7/3, 0)
       (8/3, 3/55)
       (9/3, 15/55)
       (10/3, 34/55)
       (11/3, 60/55)
       (12/3, 87/55)
       (13/3, 113/55)
       (14/3, 150/55)
       (15/3, 204/55)
       (16/3, 275/55)
       (17/3, 354/55)
       (18/3, 443/55)
       (19/3, 526/55)
       (20/3, 587/55)
       (21/3, 615/55)
       (22/3, 612/55)
       (23/3, 603/55)
       (24/3, 591/55)
       (25/3, 576/55)
       (26/3, 558/55)
       (27/3, 537/55)
       (28/3, 514/55)
       (29/3, 487/55)
       (30/3, 457/55)
       (31/3, 424/55)
       (32/3, 388/55)
       (33/3, 349/55)
       (34/3, 307/55)
       (35/3, 262/55)
       (36/3, 215/55)
       (37/3, 170/55)
       (38/3, 128/55)
       (39/3, 92/55)
       (40/3, 62/55)
       (41/3, 41/55)
       (42/3, 29/55)
       (43/3, 26/55)
       (44/3, 26/55)
       (45/3, 26/55)
       (46/3, 26/55)
       (47/3, 26/55)
       (48/3, 26/55)
       (49/3, 26/55)
       (50/3, 26/55)
      };

    \end{tikzpicture}
  }
  \end{center}
  \caption{Computing the sheaf cohomologies of $\tilde{M_7}$ with \texttt{plesken}.}
  \label{fig1007}
\end{figure}

\newpage
\bibliography{papers}
\bibliographystyle{custom1}

\end{document}